\documentclass[12pt]{article}
\pdfoutput=1

\usepackage[a4paper,text={16.8cm,22.4cm}]{geometry}
\usepackage{amsmath,amsfonts,slashed,amssymb,tikz,bm,psfrag,graphicx,color,dsfont}
\usepackage{multicol,multirow}
\usepackage{float}
\usepackage{hyperref}     
\usepackage{color}
\usepackage{xcolor}
\usepackage{framed}

\RequirePackage[sort&compress,square,comma,numbers]{natbib}
\allowdisplaybreaks
\renewcommand{\arraystretch}{1.2}

\hypersetup{
  colorlinks   = true,    
  urlcolor     = blue,    
  linkcolor    = blue,    
  citecolor    = red      
}

\definecolor{mygray}{gray}{0.3}
\definecolor{HumboldtGreen}{rgb}{0,.4,0}

\newcommand{\blue}[1]{\textcolor[rgb]{0.00,0.00,1.00}{#1}}

\begin{document}

\begin{titlepage}

\begin{flushright}
\normalsize
December  22, 2022
\end{flushright}

\vspace{0.1cm}
\begin{center}
\Large\bf
Precision calculations of $B_{d, s}  \to \pi, K$  decay form factors \\
in soft-collinear effective theory
\end{center}

\vspace{0.5cm}
\begin{center}
{\bf \small Bo-Yan Cui$^a$,
Yong-Kang Huang$^a$ \footnote{correspondence author: huangyongkang@mail.nankai.edu.cn},
Yue-Long Shen$^b$,
Chao Wang$^c$,
Yu-Ming Wang$^a$ \footnote{correspondence author: wangyuming@nankai.edu.cn} }\\
\vspace{0.7cm}
{\sl  $^a$ \, School of Physics, Nankai University, Weijin Road 94, 300071 Tianjin, China } \\
{ \sl $^b$ College of Information Science and Engineering, Ocean  University of China,
\\  Qingdao 266100,  Shandong, China } \\
{ \sl $^c$ College of Mathematics and Physics, Huaiyin Institute of Technology,
\\  Huaian 223003, Jiangsu, China }\\
\end{center}

\vspace{0.2cm}
\begin{abstract}

We improve QCD calculations of the semileptonic $B_{d, s}  \to \pi, K$  decay form factors at large hadronic recoil
by implementing the next-to-leading-logarithmic resummation for the obtained leading-power light-cone sum rules
in the soft-collinear effective theory (SCET) framework and by computing for the first time the non-vanishing spectator-quark mass correction
dictating the SU(3)-flavour symmetry breaking effects between  these fundamental quantities at the one-loop accuracy.
Additionally, we endeavour to investigate a variety of the subleading-power contributions to these heavy-to-light form factors
at ${\cal O}(\alpha_s^0)$ with the same methodology, by including the higher-order terms in the heavy-quark expansion of the hard-collinear quark propagator, by evaluating the desired effective matrix element of  the next-to-leading-order term
$(\bar \xi_{\rm hc} \, W_{\rm hc}) \,
\gamma_{\mu} \,   \left [  i \,  {\slashed D}_{\top} /  \left (  2 \, m_b \right ) \right ] \,   h_v$
in the ${\rm SCET_{I}}$ representation of the weak transition current,
by taking into account the off-light-cone contributions of the two-body heavy-quark effective theory (HQET) matrix elements
as well as  the three-particle higher-twist corrections from the subleading bottom-meson light-cone distribution amplitudes (LCDAs),
and by computing the twist-five and twist-six four-body higher-twist effects with the aid of the factorization approximation.
Having at our disposal the SCET sum rules for the exclusive $B$-meson decay form factors under discussion,
we further explore in detail  numerical implications of the newly computed subleading-power corrections
by employing the three-parameter model for both the leading-twist and higher-twist $B$-meson distribution amplitudes.
Taking advantage of the customary  Bourrely-Caprini-Lellouch (BCL) parametrization for the complete set
of the semileptonic $B_{d, s}  \to \pi, K$  form factors, we then determine the correlated numerical results for the interesting
series coefficients, by carrying out the simultaneous fit of the exclusive $B$-meson decay form factors to
both the achieved SCET sum rule predictions at small momentum transfer ($q^2$) and the available lattice QCD results at large momentum transfer.
Subsequently, we perform a comprehensive phenomenological analysis of the full angular observables,
the lepton-flavour university ratios and the lepton polarization asymmetries for the flavour-changing charged-current
$B \to \pi \ell \bar \nu_{\ell}$  and $B_s \to K \ell \bar \nu_{\ell}$ decays  (with  $\ell= \mu, \, \tau$)
together with the differential $q^2$-distribution for the exclusive rare $B \to K \nu_{\ell} \bar \nu_{\ell}$ decays
in the Standard Model.

\end{abstract}

\vfil

\end{titlepage}

\tableofcontents

\newpage

\section{Introduction}

Precision calculations of the semileptonic $B_{d, s}  \to \pi, K$  decay form factors   are of paramount importance
for exploring the celebrated Cabibbo-Kobayashi-Maskawa (CKM) mechanism in the Standard Model (SM)
and for sharpening our understanding towards diverse facets of the strong interaction dynamics
encoded in the exclusive heavy-hadron decay processes.
Particularly, the longstanding discrepancy between the dedicated $|V_{ub}|$ determinations from
the exclusive $B \to \pi \ell \bar \nu_{\ell}$ decays
and the inclusive $B \to X_{u} \ell \bar \nu_{\ell}$ processes \cite{HFLAV:2022pwe}  has been continually triggering
the enormous theoretical efforts of determining such heavy-to-light $B$-meson form factors
with an ever-increasing accuracy. In the small hadronic recoil region,
the first-principles calculations for a rich variety of the non-perturbative matrix elements
appearing in the theory descriptions of the semileptonic heavy-meson decays  $B \to \pi \ell \bar \nu_{\ell}$
\cite{Dalgic:2006dt,FermilabLattice:2015mwy,Flynn:2015mha,Colquhoun:2015mfa,Colquhoun:2022atw,FermilabLattice:2015cdh}
and $B_s \to K \ell \bar \nu_{\ell}$ decays \cite{Flynn:2015mha,Bouchard:2014ypa,Bahr:2016ayy,Monahan:2018lzv,FermilabLattice:2019ikx,Bahr:2019eom}
 and of the exclusive electroweak penguin $B \to K \nu_{\ell} \bar \nu_{\ell}$ transitions \cite{Bouchard:2013eph,Bailey:2015dka,Parrott:2022rgu}
have been pursued with the numerical lattice gauge theory by different groups
(see  also  \cite{Boyle:2022uba,DiCanto:2022icc,FlavourLatticeAveragingGroupFLAG:2021npn} for an overview).
Moreover, numerous analytical QCD frameworks with distinct approximations have been constructed to address
the semileptonic $B$-meson decay form factors at large hadronic recoil systematically based upon the heavy quark expansion techniques.

Employing the perturbative QCD factorization theorem for the vacuum-to-pseudoscalar-meson correlation function
and implementing further the parton-hadron duality ans\"{a}tz enables us to derive the desired light-cone sum rules (LCSR)
for the heavy-to-light $B_{d, s}  \to \pi, K$  form factors at the leading-order (LO) accuracy \cite{Belyaev:1993wp,Rusov:2017chr}
and in the next-to-leading-order (NLO) approximation \cite{Khodjamirian:1997ub,Bagan:1997bp,Ball:2004ye,Duplancic:2008ix,Khodjamirian:2011ub}
(see also \cite{Bharucha:2012wy} for the twist-two ${\cal O}(\alpha_s^2 \, \beta_0)$ correction to the vector form factor
$f_{B \to \pi}^{+}(q^2)$), taking advantage of the light-meson light-cone distribution amplitudes (LCDAs) with  definite collinear twists
\cite{Braun:2003rp,Colangelo:2000dp} as the fundamental  non-perturbative ingredients.
Alternatively, the light-cone QCD sum rules for the exclusive  bottom-meson decay form factors
can be derived from the vacuum-to-$B$-meson correlation function with the pseudoscalar meson state
interpolated by an appropriate partonic current \cite{Khodjamirian:2005ea,Khodjamirian:2006st}
(see \cite{DeFazio:2005dx,DeFazio:2007hw} for an equivalent and independent formulation
in the soft-collinear effective theory (SCET) framework),
following the analogous theory prescriptions as described above.
An attractive advantage of this alternative version of QCD sum rules on the light-cone
consists in the very appearance of the universal $B$-meson distribution amplitudes in heavy quark effective theory (HQET)
for the obtained expressions of all the bottom-meson decay form factors,  independent of the particular light-hadron
in the final states. Along the same vein, both the higher-order perturbative corrections and the subleading-power contributions
to the heavy-to-light $B$-meson decay form factors \cite{Wang:2015vgv,Shen:2016hyv,Lu:2018cfc,Gubernari:2018wyi,Gao:2019lta,Shen:2021yhe}, the heavy-to-heavy  $B$-meson decay form factors \cite{Wang:2017jow,Gubernari:2018wyi,Gao:2021sav},
as well as the semileptonic  heavy-baryon decay form factors \cite{Wang:2009hra,Feldmann:2011xf,Wang:2015ndk} have been computed from the LCSR method with the heavy-hadron distribution amplitudes.

Yet another theory framework to evaluate the exclusive heavy-hadron decay matrix elements has been developed
to  regularize the emerged end-point divergences in the conventional collinear factorization formalism
by introducing the intrinsic transverse momenta of the associated soft and collinear partons participating
the short-distance scattering process \cite{Keum:2000wi,Lu:2000em,Ali:2007ff}
(see, however, \cite{Descotes-Genon:2001rya} for additional discussions),
motivated from the theory of the on-shell Sudakov form factor \cite{Collins:1989bt}
and the asymptotic behaviour of elastic meson-meson scattering at high energy \cite{Botts:1989kf}.
Applying such transverse-momentum-dependent (TMD) factorization approach further  allows for the higher-order QCD computations
of a large number of exclusive hadronic matrix elements \cite{Nandi:2007qx,Li:2010nn,Li:2012nk,Li:2012md,Li:2013xna,Cheng:2014gba,Cheng:2014fwa,Cheng:2017vna,Cheng:2020fcx}
 including both the leading-twist and higher-twist contributions simultaneously.
We mention in passing that constructing the factorization-compatible definitions of the  TMD wavefunctions
free of both the rapidity divergence and the pinch singularity  becomes tremendously delicate,
demanding the introduction  of an  intricate soft substraction function
defined with the non-dipolar off-light-cone Wilson lines \cite{Li:2014xda,Wang:2015qqr,Li:2016amo}.

Inspired by the encouraging experimental progresses on measuring the differential $B \to \pi \ell \nu$ decay rates
from the BaBar \cite{BaBar:2010efp,BaBar:2012thb}, Belle \cite{Belle:2010hep,Belle:2013hlo}
and Belle II \cite{Belle-II:2022imn} Collaborations,
on the first observation of the semileptonic $B_s \to K \ell \bar \nu_{\ell}$ decays
at the LHCb experiment \cite{LHCb:2020ist}, and on the anticipated discovery of the flavour-changing neutral current (FCNC)
$B \to K \nu_{\ell} \bar \nu_{\ell}$ decay process at Belle II \cite{Belle-II:2018jsg},
we aim at improving further theory predictions for the semileptonic $B_{d, s}  \to \pi, K$  decay form factors
from the LCSR technique with the HQET $B$-meson distribution amplitudes as previously achieved in \cite{Wang:2015vgv,Lu:2018cfc},
by incorporating a various types of  newly computed  subleading-power contributions
into the next-to-leading-logarithmic (NLL) resummation improved leading-power effects
in the heavy quark expansion.
More specifically, the major new  ingredients of the present paper can be summarized in the following.

\begin{itemize}

\item{Applying the method of the SCET sum rules we evaluate the non-vanishing spectator-quark mass corrections
to the exclusive $B_{d, s}  \to \pi, K$  transition form factors at large hadronic recoil at ${\cal O}(\alpha_s)$,
which will be demonstrated to generate the  SU(3)-flavour symmetry breaking effect
not suppressed in the heavy quark expansion and to preserve the large-recoil symmetry relations
for the soft contributions to the heavy-to-light bottom-meson  decay  form factors.
We then proceed to perform the complete NLL summation of the enhanced logarithms of $m_b/\Lambda_{\rm QCD}$
appearing in the leading-power factorization formulae for the vacuum-to-$B$-meson correlation functions
defined with an interpolating current for the energetic pseudoscalar meson,
by employing the standard renormalization-group (RG) formalism.}

\item{We compute for the first time the subleading-power terms from the heavy quark expansion of the hard-collinear quark propagator
including further two distinct sources of the light-quark mass corrections at tree level
by taking advantage of the classical equations of motion for both the soft light quark and the effective heavy quark
as well as the two-particle and three-particle light-cone bottom-meson distribution amplitudes in HQET.
Moreover, we will verify explicitly that such particular power-suppressed contributions can bring about
the notable symmetry-breaking corrections to the semileptonic $B_{d, s}  \to \pi, K$ form factors.  }

\item{We construct the SCET sum rules for the subleading matrix element of the effective heavy-to-light current
$\langle M(p) | (\bar \xi_{\rm hc} \, W_{\rm hc}) \,
\gamma_{\mu} \,   \left [  i \,  {\slashed D}_{\top} /  \left (  2 \, m_b \right ) \right ] \,   h_v | \bar B_{q^{\prime}}(v)\rangle$
from the  ${\rm SCET_{I}}$  expansion of the  flavour-changing weak current $\bar q \, \Gamma_i \, b$ at the LO accuracy,
by employing the established relations between the relevant light-ray HQET operators \cite{Kawamura:2001jm,Kawamura:2001bp,Braun:2017liq}.}

\item{We derive the tree-level sum rules for the twist-five and twist-six four-body higher-twist corrections
to the exclusive $B_{d, s}  \to \pi, K$ form factors with the factorization ans\"{a}tz,
which allows for expressing these subleading-twist  distribution amplitudes in terms of the quark condensate
and the appropriate two-particle HQET distribution amplitudes (see for instance \cite{Agaev:2010aq,Beneke:2018wjp} for further discussions).  }

\item{We update the previous theory predictions for the complete set of  the semileptonic $B_{d, s}  \to \pi, K$ form factors
in the entire kinematic regime by interpolating  between the improved SCET sum rule computations at small momentum transfer
and the numerical lattice QCD determinations at large momentum transfer with the conventional  Bourrely-Caprini-Lellouch (BCL) parametrization \cite{Lellouch:1995yv,Bourrely:2005hp,Bourrely:2008za}.
Our numerical explorations will evidently  reveal that
implementing the obtained LCSR constraints of the heavy-to-light bottom-meson
form factors in the theory analysis is indeed beneficial for pinning down
the yielding uncertainties of the extracted  series coefficients. }

\end{itemize}

The remainder of this paper is structured as follows.
We will set up the computational framework in Section \ref{section: LP sum rules}
by defining the exclusive heavy-to-light $B$-meson decay form factors of our interest,
by summarizing the general strategies of constructing the desired SCET sum rules
with the bottom-meson distribution amplitudes,
and by presenting the manifest expressions of the short-distance matching coefficients
entering the perturbative factorization formulae for the considered vacuum-to-$B$-meson correlation functions
at the leading-power accuracy. In particular, we derive the spectator-quark mass corrections
to the perturbative hard-collinear functions  at ${\cal O}(\alpha_s)$
and demonstrate further the factorization-scale independence of the obtained expressions for the correlation functions.
Applying the RG evolution equations for both the hard matching coefficients
and the two-particle twist-two and twist-three $B$-meson distribution amplitudes enables us to carry out
the NLL summation of the parametrically enhanced logarithms in the factorized expressions of the vacuum-to-$B$-meson
correlation functions.
We turn to investigate four different classes of the next-to-leading-power (NLP) corrections to
the exclusive $B_{d, s}  \to \pi, K$  transition form factors
at tree level on the basis of  the SCET sum rules
in Section \ref{section: NLP sum rules}, by invoking the appropriate operator identities
between the two-body and three-body light-cone HQET operators
and by employing the higher-twist $B$-meson distribution amplitudes up to the twist-six accuracy.
Numerical explorations of the resulting SCET sum rules for the $B_{d, s}  \to \pi, K$ form factors
at large hadronic recoil including both the updated leading-power contributions in the  $\Lambda_{\rm QCD}/m_b$ expansion
with the non-vanishing light-quark masses and the newly derived subleading power corrections
will be displayed in Section \ref{section: numerical analysis}.
We then proceed to perform the simultaneous fit of the  customary BCL parametrization for
these transition form factors to  the obtained LCSR predictions and the available lattice QCD results,
yielding the correlated numerical values of the nonperturbative form-factor parameters.
Taking advantage of the improved determinations of the bottom-meson decay form factors in the full kinematic region,
we also provide the SM predictions for a variety  of  phenomenologically interesting observables
for the semileptonic $B \to \pi \ell \bar \nu_{\ell}$  and $B_s \to K \ell \bar \nu_{\ell}$ decay processes,
such as  the differential branching fractions, the lepton-flavour universality ratios
the forward-backward asymmetries as well as the lepton polarization asymmetries,
and for the  theoretically cleanest electroweak penguin $B \to K \nu_{\ell} \bar \nu_{\ell}$ decays
in this section.
We will conclude in Section \ref{section: summary} with a summary of our major observations
and perspectives on the future developments.

\section{The NLL LCSR for the exclusive $B_{d, s}  \to \pi, K$ form factors  at leading power}
\label{section: LP sum rules}

We  employ the standard  definitions of the semileptonic heavy-to-light decay form factors
according to the Lorentz decompositions of the following bilinear quark current matrix elements
\begin{eqnarray}
\langle M(p)|  \bar q \gamma_{\mu} b| \bar B (p_B)\rangle
&=& f_{B M}^{+}(q^2) \, \left [ p_B + p -\frac{m_B^2-m_{M}^2}{q^2} q  \right ]_{\mu}
+  f_{B M}^{0}(q^2) \, \frac{m_B^2-m_{M}^2}{q^2} q_{\mu} \,,
\nonumber \\
\langle M(p)|  \bar q \sigma_{\mu \nu} q^{\nu} b| \bar B (p_B)\rangle
&=& \frac{i \, f_{B M}^{T}(q^2)}{m_B+m_M} \,
\left [ q^2 \, (2 \, p + q)_{\mu} - \left (m_B^2-m_{M}^2 \right ) q_{\mu} \right ] \,,
\label{form-factor definition}
\end{eqnarray}
where $m_M$ and $p$ correspond to the mass and the four-momentum of the light pseudoscalar meson,
and $q$ stands for the transfer momentum of the flavour-changing weak current.
Applying the procedure displayed in \cite{DeFazio:2005dx,DeFazio:2007hw,Wang:2015vgv,Lu:2018cfc},
the SCET sum rules for these form factors can be constructed with the vacuum-to-$B$-meson correlation function
\begin{eqnarray}
	\begin{aligned}
		\Pi_{\mu} (n \cdot p, \bar n \cdot p) & =  \int d^4 x \, e^{i p \cdot x} \langle 0 | {\rm T} \left \{
		\bar q^{\prime}(x)\, \slashed {n} \gamma_5 \,q(x),\, \bar q(0)\, \Gamma_\mu \,b(0) \right  \} | \bar B (p_B) \rangle
		\\
		& =  \left\{
		\begin{array}{l}
			\Pi(n \cdot p, \bar n \cdot p)\, n_\mu + \tilde \Pi (n \cdot p, \bar n \cdot p)\, \bar n_\mu,
\hspace{2.0 cm} \Gamma_\mu = \gamma_\mu \vspace{0.5 cm} \\
			\left ( - {i \over 2}  \right) \, \Pi_{\rm T}(n \cdot p, \bar n \cdot p)\,
\left [ \bar n \cdot q \, n_\mu - n \cdot q \, \bar n_\mu \right ] ,
\hspace{1.16cm} \Gamma_\mu = \sigma_{\mu \nu} q^\nu
\end{array}
 \hspace{0.5 cm} \right.
 \label{correlation_function}
\end{aligned}
\end{eqnarray}
where the local QCD current $\bar q^{\prime}(x)\, \slashed {n} \gamma_5 \,q(x)$ interpolates the  pseudoscalar meson.
We further introduce the two light-cone vectors $n_{\mu}$ and $\bar n_{\mu}$ satisfying the constraints
$n^2=\bar n^2=0$ and $n \cdot \bar n=2$, which allow us to write down the four-velocity vector of the heavy bottom-meson
$v_{\mu} = p_B /m_B = (n_{\mu} + \bar n_{\mu})/2$.
In order to facilitate the derivation of the soft-collinear factorization formulae for
the three emerged invariant functions $\Pi$, $\tilde \Pi$ and $\Pi_{\rm T}$,
we adopt the following power counting  scheme for the  interpolating-current momentum  and the light-quark masses
\begin{eqnarray}
n \cdot p \sim {\cal O}(m_b) \,, \qquad \bar n \cdot p \sim {\cal O}(\Lambda_{\rm QCD}) \,,
\qquad m_q \sim m_{q^{\prime}} \sim {\cal O}(\Lambda_{\rm QCD}) \,.
\label{power-counting scheme}
\end{eqnarray}
Performing the two-step matching program ${\rm QCD} \to {\rm SCET_{I}} \to {\rm SCET_{II}}$
for the correlation function (\ref{correlation_function}) in sequence leads to  the familiar factorization formulae
at leading power in the heavy quark expansion \cite{Wang:2015vgv,Lu:2018cfc}
\begin{eqnarray}
\Pi &=& {\cal F}_B(\mu) \, m_B \sum \limits_{k=\pm} \,
{\cal C}^{(k)}(n \cdot p, \mu) \, \int_0^{\infty} {d \omega \over \omega- \bar n \cdot p - i 0}~
{\cal J}^{(k)}\left({\mu^2 \over n \cdot p \, \omega},{\omega \over \bar n \cdot p}\right) \,
\phi_B^{k}(\omega,\mu)  \,, \nonumber \\
\tilde \Pi &=& {\cal F}_B(\mu) \, m_B \sum \limits_{k=\pm} \,
{\cal \widetilde{C}}^{(k)}(n \cdot p, \mu) \, \int_0^{\infty} {d \omega \over \omega- \bar n \cdot p - i 0}~
{\cal \widetilde{J}}^{(k)}\left({\mu^2 \over n \cdot p \, \omega},{\omega \over \bar n \cdot p}\right) \,
\phi_B^{k}(\omega,\mu)  \,, \nonumber \\
\Pi_{\rm T} &=&  {\cal F}_B(\mu) \, m_B \sum \limits_{k=\pm} \,
{\cal C}^{(k)}_{\rm T}(n \cdot p, \mu, \nu) \, \int_0^{\infty} {d \omega \over \omega- \bar n \cdot p - i 0}~
{\cal J}^{(k)}_{\rm T}\left({\mu^2 \over n \cdot p \, \omega},{\omega \over \bar n \cdot p}\right) \,
\phi_B^{k}(\omega,\mu) \,.
\label{LP factorization formulae for the correlators}
\end{eqnarray}
Including the light spectator-quark mass corrections in the factorized expressions  apparently cannot affect
the hard coefficients from the first-step matching  ${\rm QCD} \to {\rm SCET_{I}}$
in the leading-power approximation,
which can be actually determined from the perturbative matching of the heavy-to-light current
by integrating out the hard fluctuation modes with virtualities of order $m_b^2$ \cite{Bauer:2000yr,Beneke:2004rc}.
However, the non-vanishing spectator-quark mass can indeed result in the leading-power contributions
to the hard-collinear functions from the ${\rm SCET_{I}} \to {\rm SCET_{II}}$ matching
with the adopted power-counting scheme (\ref{power-counting scheme})
\begin{eqnarray}
{\cal \widetilde{J}}^{(+)}  &=& {\alpha_s \, C_F \over 4 \pi} 
\bigg  \{ \left [  r \, \left (1 -  {\bar n \cdot p \over \omega} \right )  + {m_q + 2 \,m_{q^{\prime}} \over \omega} \right ] 
\ln \left (1- {\omega \over \bar n \cdot p} \right )
- {m_{q^{\prime}} \over \omega} \,   \ln^2 \left (1- {\omega \over \bar n \cdot p} \right ) 
{\omega -  \bar n \cdot p \over \omega}
\nonumber \\
&& - {m_{q^{\prime}} \over \omega} \, \left [ 2 \, \ln { \mu^2 \over  n \cdot p (\omega- \bar n \cdot p) } + 5 \right ] \,
\left [  \ln \left (1- {\omega \over \bar n \cdot p} \right )  \,
{\omega -  \bar n \cdot p \over \omega} - 1 \right ]
 \bigg \} \,,
\nonumber \\
{\cal J}^{(+)}_{\rm T}  &=&  {\alpha_s \, C_F \over 4 \pi} 
\bigg  \{ \left [ - \, \left (1 -  {\bar n \cdot p \over \omega} \right ) + {m_q + 2 \,m_{q^{\prime}} \over \omega} \right ] 
\ln \left (1- {\omega \over \bar n \cdot p} \right )
- {m_{q^{\prime}} \over \omega} \,   \ln^2 \left (1- {\omega \over \bar n \cdot p} \right )  
{\omega -  \bar n \cdot p \over \omega}
\nonumber \\
&& - {m_{q^{\prime}} \over \omega} \, \left [ 2 \, \ln { \mu^2 \over  n \cdot p (\omega- \bar n \cdot p) } + 5 \right ] \,
\left [  \ln \left (1- {\omega \over \bar n \cdot p} \right )  \,
{\omega -  \bar n \cdot p \over \omega} - 1 \right ]
 \bigg \} \,,
\label{spectataor-quark mass corrections}
\end{eqnarray}
where we have introduced the dimensionless kinematic variable $r=n \cdot p /m_b$.
It is straightforward to verify the factorization-scale independence of
the derived factorization formulae by employing the RG evolution equation for
the twist-three $B_{q^{\prime}}$-meson  distribution amplitude $\phi_B^{-}(\omega, \mu)$
in the absence of the three-particle LCDA contribution  at one loop \cite{Bell:2008er}
\begin{eqnarray}
{d \over d \ln \mu} \phi_B^{-}(\omega, \mu) &=&
- {\alpha_s \over 4 \pi}  \, \int_0^{\infty} d \omega^{\prime} \,
\left [ \gamma_{+}^{(1)}(\omega, \omega^{\prime}, \mu)
- \Gamma_{\rm cusp}^{(0)} \, \frac{\theta(\omega^{\prime} - \omega)}{\omega^{\prime} } \right ]
\phi_B^{-}(\omega, \mu) \,
\nonumber \\
&& -  {\alpha_s \over 4 \pi}  \, \int_0^{\infty} d \omega^{\prime} \, \Gamma_{\rm cusp}^{(0)} \,
\left [ {m_{q^{\prime}} \, \theta(\omega^{\prime} - \omega) \over \omega^{\prime 2}}  \right ]_{\oplus}  \,
\phi_B^{+}(\omega, \mu) + {\cal O}(\alpha_s^2) \,,
\end{eqnarray}
where the perturbative kernel $\gamma_{+}^{(1)}$ and the cusp anomalous dimension $\Gamma_{\rm cusp}^{(0)}$
can be written as
\begin{eqnarray}
&& \gamma_{+}^{(1)} = \left \{  \left ( \Gamma_{\rm cusp}^{(0)} \, \ln {\mu \over \omega} -2  \right ) \, \delta( \omega - \omega^{\prime} )
- \Gamma_{\rm cusp}^{(0)} \, \left [ \frac{\omega \,  \theta(\omega^{\prime}-\omega)}{\omega^{\prime} \, (\omega^{\prime}-\omega)}
+ \frac{\theta(\omega-\omega^{\prime})}{(\omega-\omega^{\prime})}  \right ]_{\oplus} \right \} \,  \, C_F \,,
\nonumber \\
&& \Gamma_{\rm cusp}^{(0)} = 4 \, C_F  \,,
\end{eqnarray}
with the $\oplus$-function defined by
\begin{eqnarray}
\int_0^{\infty} \, d \omega^{\prime} \, \left [ f(\omega,\omega^{\prime}) \right ]_{\oplus} \, g(\omega^{\prime})
= \int_0^{\infty} \, d \omega^{\prime} \, f(\omega,\omega^{\prime})  \,
\left [ g(\omega^{\prime}) - g(\omega) \right ] \,.
\end{eqnarray}
Importantly, the yielding spectator-quark mass corrections to the hard-collinear matching coefficients
appear to be universal for the two vacuum-to-$B$-meson correlation functions
with distinct weak transition currents.
This interesting pattern can be attributed to the very fact that only the one-loop correction to
the light-pseudoscalar-meson vertex diagram can bring about the non-vanishing spectator-quark mass effect
at leading power, thus validating the earlier speculation on the differences between
the  heavy-to-light form factors for exclusive $B$-meson and $B_s$-meson decays \cite{Leibovich:2003jd}.
The remaining short-distance functions in the factorized correlation functions
(\ref{LP factorization formulae for the correlators}) have been determined in \cite{Lu:2018cfc}
at the one-loop accuracy.

Inspecting the obtained soft-collinear factorization formulae for the considered invariant functions
indicates that there is no common choice of the factorization scale to get rid of the parametrically
large logarithms of $m_b/\Lambda_{\rm QCD}$.
Adopting the factorization scale of order $\sqrt{m_b \, \Lambda_{\rm QCD}}$, we are then required  to
perform an all-order summation of such enhanced logarithms entering in both the hard matching coefficients
and the two-particle bottom-meson distribution amplitudes.
Taking advantage of the momentum-space RG equations for ${\cal \widetilde{C}}^{(-)}$,
${\cal C}^{(-)}_{\rm T}$ and the HQET decay constant ${\cal F}_B$
(expressible in terms of the QCD decay constant $f_B$ and the matching coefficient $K(\mu)$ \cite{Eichten:1989zv})
allows us to  derive the desired scale dependence of these quantities in the following form
\begin{eqnarray}
&& {\cal F}_B (\mu)  =  \hat{U}_2(\mu_{h2}, \, \mu) \,\,  {\cal F}_B (\mu_{h2})  \,,
\qquad
{\cal \widetilde{C}}^{(-)}(n \cdot p, \mu) = \hat{U}_1(n \cdot p, \, \mu_{h1}, \, \mu)
\,\,  {\cal \widetilde{C}}^{(-)}(n \cdot p, \mu_{h1}) \,,
\nonumber \\
&& {\cal C}^{(-)}_{\rm T}(n \cdot p, \mu, \nu)  =  U_1(n \cdot p, \, \mu_{h1}, \, \mu)
\,\,  \hat{U}_3(\nu_h, \nu)  \, {\cal C}_{\rm T}^{(-)}(n \cdot p, \mu_{h1}, \nu_h) \,.
\end{eqnarray}
The manifest expressions of the QCD evolution function $\hat{U}_1(n \cdot p, \, \mu_{h1}, \, \mu)$
can be obtained from the expanded result of $U_1(E_{\gamma}, \, \mu_{h}, \, \mu)$ presented in \cite{Beneke:2011nf}
with the replacement $E_{\gamma} \to n \cdot p /2$.
The two additional  RG functions $\hat{U}_2$ and $\hat{U}_3$ at the NLL accuracy are given by
\begin{eqnarray}
\hat{U}_2(\mu_{h2}, \, \mu) &=& z_2^{- \frac{\gamma_K^{(0)}}{2 \,\beta_0}} \bigg [1+ \frac{\alpha_s(\mu_{h2})}{4 \pi}  \,
\left (  {\gamma_K^{(1)} \over 2 \, \beta_0} - {\gamma_K^{(0)} \, \beta_1 \over 2 \, \beta_0^2 } \right ) (1-z_2)
+{\cal O}(\alpha_s^2) \bigg ]  \,,
\nonumber  \\
 \hat{U}_3(\nu_h, \nu) &=& z_3^{- \frac{\gamma_{\rm T}^{(0)}}{2 \,\beta_0}} \bigg [1+ \frac{\alpha_s(\nu_{h})}{4 \pi}  \,
\left (  {\gamma_{\rm T}^{(1)} \over 2 \, \beta_0} - {\gamma_{\rm T}^{(0)} \, \beta_1 \over 2 \, \beta_0^2 } \right ) (1-z_3)
+{\cal O}(\alpha_s^2) \bigg ]  \,,
\end{eqnarray}
where the necessary anomalous dimensions for the HQET heavy-to-light current
and for the QCD tensor current are  \cite{Ji:1991pr,Broadhurst:1991fz,Bell:2010mg}
\begin{eqnarray}
\gamma_K^{(0)} =  3 \, C_F  \,,  & \qquad &
\gamma_K^{(1)} =  C_F \, \left [ {127 \over 6}  + {14 \, \pi^4 \over 9} - {5 \over 3} \, n_f \right ]  \,,
\nonumber \\
\gamma_{\rm T}^{(0)} =  -2 \, C_F  \,,   & \qquad &
\gamma_{\rm T}^{(1)} =   C_F \, \left [ 19 \, C_F  -  {257 \over 9} \, C_A  + {26 \over 9} \, n_f \,\right ]  \,,
\end{eqnarray}
with the conventions $z_2=\alpha_s(\mu)/\alpha_s(\mu_{h2})$ and $z_3=\alpha_s(\nu)/\alpha_s(\nu_{h})$.
Applying the dual-space representations for the $B$-meson distribution amplitudes constructed in \cite{Bell:2013tfa,Braun:2014owa}
\begin{eqnarray}
\phi_B^{+}(\omega, \mu) &=& \int_0^{+\infty} d s  \, \sqrt{\omega \, s } \, J_1 (2 \,\sqrt{\omega \, s } ) \, \eta_{+}(s, \mu) \,,
\nonumber \\
\phi_B^{-}(\omega, \mu) &=& \int_0^{+\infty} d s  \, \sqrt{\omega \, s } \, J_0 (2 \,\sqrt{\omega \, s } ) \,
\left [  \eta_{+}(s, \mu) + \eta_{3}^{(0)}(s, \mu)\right ] \,,
\end{eqnarray}
both the twist-two and twist-three coefficient functions $\eta_{+}(s, \mu)$ and $\eta_{3}^{(0)}(s, \mu)$ are observed to
possess the autonomous scale dependence \cite{Braun:2015pha}
\begin{eqnarray}
\eta_{+}(s, \mu) &=& U_{\phi}^{\rm tw2}(s, \mu, \mu_0) \, \eta_{+}(s, \mu_0) \,,
\nonumber \\
\eta_{3}^{(0)}(s, \mu) &=& U_{\phi}^{\rm tw3}(s, \mu, \mu_0) \, \eta_{3}^{(0)}(s, \mu_0) \,.
\end{eqnarray}
The analytical results of the two evolution factors $U_{\phi}^{\rm tw2}(s, \mu, \mu_0)$ and $ U_{\phi}^{\rm tw3}(s, \mu, \mu_0)$
can be further written as \cite{Braun:2017liq,Braun:2015pha}
\begin{eqnarray}
U_{\phi}^{\rm tw2}(s, \mu, \mu_0) &=&  {\rm exp} \left \{ -{ \Gamma_{\rm cusp}^{(0)}  \over 4 \, \beta_0^2} \,
\left [ \ln z_0 - 1 +{1 \over z_0}  \right ] - {\beta_1 \over 2 \beta_0^2} \, \ln^2 z_0
+ \left ( {\Gamma_{\rm cusp}^{(1)} \over \Gamma_{\rm cusp}^{(0)}} - {\beta_1 \over 2 \beta_0} \right ) \,
\left [ z_0 - 1 - \ln z_0 \right ]  \right \}
\nonumber \\
&& \times \, \left ( s \, e^{2 \, \gamma_E} \, \mu_0 \right )^{\Gamma_{\rm cusp}^{(0)} \, \ln z_0 / (2 \, \beta_0)} \,\,
z_0^{\gamma_{\rm tw2}^{(0)} / (2 \, \beta_0)}\,,
\nonumber \\
U_{\phi}^{\rm tw3}(s, \mu, \mu_0) &=& z_0^{\gamma_{\rm tw3}^{(0)} / (2 \, \beta_0)} \, U_{\phi}^{\rm tw2}(s, \mu, \mu_0)  \,,
\end{eqnarray}
where  the dimensionless quantity $z_0=\alpha_s(\mu) /  \alpha_s(\mu_0)$
and the newly appeared  anomalous dimensions $\Gamma_{\rm cusp}^{(0)}$, $\Gamma_{\rm cusp}^{(1)}$,
$\gamma_{\rm tw2}^{(0)}$ and  $\gamma_{\rm tw3}^{(0)}$ are explicitly given by
\begin{eqnarray}
\Gamma_{\rm cusp}^{(0)} = 4 \, C_F  \,,
\hspace{0.5 cm}
\Gamma_{\rm cusp}^{(1)} =  C_F \, \left [ {268 \over 3}  - 4 \, \pi^2  - {40 \over 9} \, n_f \right ] \,,
\hspace{0.5 cm}
 \gamma_{\rm tw2}^{(0)} = -2 \, C_F \,,
\hspace{0.5 cm}
\gamma_{\rm tw3}^{(0)} = 2 \, N_c \,.
\end{eqnarray}

In order to construct  the SCET sum rules for the exclusive heavy-to-light form factors,
we  proceed to derive the hadronic dispersion relations for the vacuum-to-$B$-meson correlation functions
with the aid of the conventional parameterizations for the bottom-meson decay matrix elements
collected in  (\ref{form-factor definition})
\begin{eqnarray}
\Pi_{\mu, {\rm V}} (n \cdot p, \bar n \cdot p)  &=& \left ( {1 \over 2} \right ) \,
\frac{f_{P} \, m_B}{m_{M}^2/ n \cdot p - \bar n \cdot p}
\bigg \{  \bar n_{\mu} \, \left [ \frac{n \cdot p}{m_B} \, f_{BM}^{+} (q^2) + f_{BM}^{0} (q^2)  \right ]
\nonumber \\
&& +  \,  n_{\mu} \, \frac{m_B}{n \cdot p-m_B}  \, \,
\left [ \frac{n \cdot p}{m_B} \, f_{BM}^{+} (q^2) -  f_{B M}^{0} (q^2)  \right ] \bigg \} \, \nonumber \\
&& + \, \int_{\omega_s}^{+\infty}   \, \frac{d \omega^{\prime} }{\omega^{\prime} - \bar n \cdot p - i 0} \,
\left [ \rho_{V, 1}^{h}(\omega^{\prime}, n \cdot p)  \, n_{\mu} \,
+\rho_{V, 2}^{h}(\omega^{\prime}, n \cdot p)  \, \bar{n}_{\mu}  \right ] \,,
\nonumber \\
\Pi_{\mu, {\rm T}} (n \cdot p, \bar n \cdot p)  &=&
\left ( - {i \over 2}  \right) \,  \left [ \bar n \cdot q \, n_\mu - n \cdot q \, \bar n_\mu  \right ]  \,
\bigg \{  \frac{f_{M} \, n \cdot p }{m_{M}^2/ n \cdot p - \bar n \cdot p}
\, \left [ {m_B \over m_B+m_M} \,  f_{B M}^{T} (q^2) \right ] \nonumber \\
&& + \, \int_{\omega_s}^{+\infty}   \, \frac{d \omega^{\prime} }{\omega^{\prime} - \bar n \cdot p - i 0} \,\,
\rho_{\rm T}^{h}(\omega^{\prime}, n \cdot p) \bigg \}  \,,
\label{hadronic dispersion relations}
\end{eqnarray}
where we adopt  the standard definition for  the decay constant of the pseudoscalar meson \cite{ParticleDataGroup:2022pth}
\begin{eqnarray}
\langle 0 |\bar q^{\prime}\, \slashed {n} \gamma_5 \,q | M(p)\rangle = i \, n \cdot p \, f_M.
\end{eqnarray}
Evidently,  $\Pi_{\mu, {\rm V}}$ and $\Pi_{\mu, {\rm T}}$ correspond to $\Gamma_\mu = \gamma_\mu$
and $\Gamma_\mu = \sigma_{\mu \nu} q^\nu$ for the particular spin structure of the weak current
$\bar q(0)\, \Gamma_\mu \,b(0)$ in the definition (\ref{correlation_function}), respectively.
Matching the spectral representations of the NLL resummation improved factorization formulae
with the obtained hadronic dispersion relation (\ref{hadronic dispersion relations})
and implementing further the Borel transformation in the variable $\bar n \cdot p \to \omega_M$,
we can readily derive the NLL sum rules for the semileptonic $B_{d, s}  \to \pi, K$  decay form factors
in the leading-power approximation
\begin{eqnarray}
&& f_{M} \,\, {\rm exp} \left [- {m_{M}^2 \over n \cdot p \,\, \omega_M} \right ] \,\,
\left \{ \frac{n \cdot p} {m_B} \, f_{B M, \, \rm LP}^{+}(q^2)
\,, \,\,\,   f_{B M, \, \rm LP}^{0}(q^2)  \right \}  \,  \nonumber \\
&& =   \left [ \hat{U}_2(\mu_{h2}, \mu) \, \mathcal{F}_B(\mu_{h2}) \right ]
\,\, \int_0^{\omega_s} \,\, d \omega^{\prime} \, e^{-\omega^{\prime}/\omega_M} \,
\nonumber \\
&&  \hspace{0.4 cm} \times \, \bigg \{ \widetilde{\bf \Phi}_{B, \, \rm {eff}}^{+} (\omega^{\prime}, \mu)
+  \, \left [  \hat{U}_1(n \cdot p, \mu_{h1}, \mu) \,\, \widetilde{\cal{C}}^{(-)}(n \cdot p, \mu_{h1}) \right ]\,
\widetilde{\bf \Phi}_{B, \, \rm {eff}}^{-} (\omega^{\prime}, \mu) \nonumber \\
&& \hspace{1.0 cm} \pm \, { n \cdot p - m_B \over m_B} \,
\left [{\bf \Phi}_{B, \, \rm {eff}}^{+} (\omega^{\prime}, \mu)
+ {\cal C}^{(-)}(n \cdot p, \mu_{h1}) \, {\bf \Phi}_{B, \, \rm {eff}}^{-} (\omega^{\prime}, \mu)   \right ]  \bigg \} \,,
\label{NLL sum rules at LP I}
\\
&& f_{M} \,\,  {\rm exp} \left [- {m_{M}^2 \over n \cdot p \,\, \omega_M} \right ]  \,\,
\frac{n \cdot p} {m_B+m_M} \, f_{B M, \, \rm LP}^{T}(q^2) \,  \nonumber \\
&& =   \left [ \hat{U}_2(\mu_{h2}, \mu) \, {\cal F}_B(\mu_{h2}) \right ]
\,\, \int_0^{\omega_s} \,\, d \omega^{\prime} \, e^{-\omega^{\prime}/\omega_M} \,  \nonumber \\
&&  \hspace{0.4 cm} \times \bigg \{ \widehat{\bf \Phi}_{B, \, \rm {eff}}^{+} (\omega^{\prime}, \mu)
+  \,\left [ \hat{U}_1(n \cdot p, \mu_{h1}, \mu) \,\, \hat{U}_3(\nu_{h}, \nu) \,\,
{\cal C}^{(-)}_{\rm T}(n \cdot p, \mu_{h1}, \nu_{h})  \right ] \,
\widetilde{\bf \Phi}_{B, \, \rm {eff}}^{-} (\omega^{\prime}, \mu)  \bigg \} \,.
\hspace{0.8 cm}
\label{NLL sum rules at LP II}
\end{eqnarray}
For brevity we have introduced the effective bottom-meson ``distribution amplitudes"
absorbing the hard-collinear strong interaction dynamics into the standard HQET soft functions
\begin{framed}
\begin{eqnarray}
\widetilde{\bf \Phi}_{B, \, \rm {eff}}^{+} &=&
{\alpha_s \, C_F \over 4 \, \pi} \,
\bigg \{   r \, \int_{\omega^{\prime}}^{\infty} \, d \omega \,
{\phi_B^{+}(\omega, \mu) \over \omega}
- \left ( m_q + 2 \, m_{q^{\prime}} \right ) \, \int_{\omega^{\prime}}^{\infty} \, d \omega \,
\ln \left ( {\omega -  \omega^{\prime} \over \omega^{\prime}} \right )   \,\,
{d \over d \omega} \, {\phi_B^{+}(\omega, \mu) \over \omega}
\nonumber \\
&&  \hspace{-0.8 cm}
- \, 2 \, m_{q^{\prime}}  \, \int_{0}^{\infty} \, {d \omega \over \omega } \,
\left [ \theta(\omega - \omega^{\prime}) \, \left ( \ln{\mu^2  \over n \cdot p \, \omega^{\prime} }
+ {5 \over 2} \right ) + \theta(\omega^{\prime} - \omega) \,
\ln \left ( {\omega^{\prime}  - \omega \over \omega^{\prime} } \right ) \right ]  \,
 \, {d \phi_B^{+}(\omega, \mu)  \over d \omega}   \bigg \},
\nonumber \\
\widetilde{\bf \Phi}_{B, \, \rm {eff}}^{-} &=&
\phi_{B}^{-}(\omega^{\prime}, \mu)
+  \frac{\alpha_s \, C_F}{4 \, \pi} \,\, \bigg \{ \int_0^{\omega^{\prime}} \,\, d \omega \,\,\,
\left [ {2 \over \omega - \omega^{\prime}}  \,\,\, \left (\ln {\mu^2 \over n \cdot p \, \omega^{\prime}}
- 2 \, \ln {\omega^{\prime} - \omega \over \omega^{\prime}} \right )\right ]_{\oplus} \,
\phi_{B}^{-}(\omega, \mu)
 \nonumber \\
&&  \hspace{-1.5 cm}
- \int_{\omega^{\prime}}^{\infty} \,\, d \omega \,\,\,
\bigg [ \ln^2 {\mu^2 \over n \cdot p \, \omega^{\prime}}
- \left ( 2 \, \ln {\mu^2 \over n \cdot p \, \omega^{\prime}}  + 3 \right ) \,\,
\ln {\omega - \omega^{\prime} \over \omega^{\prime}}
+ \, 2 \,\, \ln {\omega \over \omega^{\prime}}   + {\pi^2 \over 6} - 1 \bigg ]
\, {d \phi_{B}^{-}(\omega, \mu) \over d \omega}  \bigg \},
\nonumber \\
{\bf \Phi}_{B, \, \rm {eff}}^{+} &=&  {\alpha_s \, C_F \over 4 \, \pi} \,
\int_{\omega^{\prime}}^{\infty} \, d \omega \, {\phi_B^{+}(\omega, \mu) \over \omega}  \,,
\hspace{3.0 cm}
{\bf \Phi}_{B, \, \rm {eff}}^{-}  = \phi_B^{-}(\omega, \mu),
 \\
\widehat{\bf \Phi}_{B, \, \rm {eff}}^{+}  &=&  {\alpha_s \, C_F \over 4 \, \pi} \,
\bigg \{  - \int_{\omega^{\prime}}^{\infty} \, d \omega \,
{\phi_B^{+}(\omega, \mu) \over \omega}
- \left ( m_q + 2 \, m_{q^{\prime}} \right ) \, \int_{\omega^{\prime}}^{\infty} \, d \omega \,
\ln \left ( {\omega -  \omega^{\prime} \over \omega^{\prime}} \right )   \,\,
{d \over d \omega} \, {\phi_B^{+}(\omega, \mu) \over \omega}
\nonumber \\
&&  \hspace{-0.8 cm}
- \, 2 \, m_{q^{\prime}}  \, \int_{0}^{\infty} \, {d \omega \over \omega } \,
\left [ \theta(\omega - \omega^{\prime}) \, \left ( \ln{\mu^2  \over n \cdot p \, \omega^{\prime} }
+ {5 \over 2} \right ) + \theta(\omega^{\prime} - \omega) \,
\ln \left ( {\omega^{\prime}  - \omega \over \omega^{\prime} } \right ) \right ]  \,
 \, {d \phi_B^{+}(\omega, \mu)  \over d \omega}   \bigg \}.
 \nonumber  
\end{eqnarray}
\end{framed}
It remains interesting to point out that the newly computed spectator-quark mass corrections
preserve the so-called large-recoil symmetry relations for the soft contributions to the exclusive
$B_{d, s}  \to \pi, K$ form factors at leading power in the heavy quark expansion
(see \cite{Beneke:2000wa} for further discussions).
Bearing in  mind the scaling behaviour of the light-cone variable
$\omega^{\prime} \sim \omega_s \sim {\cal O}(\Lambda_{\rm QCD}^2/m_b)$
in the established sum rules (\ref{NLL sum rules at LP I}) and (\ref{NLL sum rules at LP II}),
we can immediately observe that $\ln \left [(\omega-\omega^{\prime}) / \omega^{\prime} \right ]$
entering in the nonperturbative functions $\widetilde{\bf \Phi}_{B, \, \rm {eff}}^{+}$
and $\widehat{\bf \Phi}_{B, \, \rm {eff}}^{+}$
must be counted as the large logarithm $\ln \left ( m_b / \Lambda_{\rm QCD} \right )$ in the heavy quark limit.
The very appearance of such logarithmic term  in the yielding SCET sum rules further implies that
evaluating the spectator-quark mass contribution with the perturbative factorization technique straightforwardly
will  give rise to  the soft-collinear convolution integrals with unwanted end-point singularities
(see \cite{Liu:2019oav,Liu:2020wbn,Beneke:2022obx,Bell:2022ott,Feldmann:2022ixt,Lu:2022fgz}
for the interesting progress on exploring the end-point dynamics
and tackling the rapidity logarithms in the different contexts).

\section{The  LCSR for the exclusive $B_{d, s}  \to \pi, K$ form factors beyond leading power}
\label{section: NLP sum rules}

We are now in a position to investigate the power-suppressed corrections to the exclusive bottom-meson decay form factors
from a variety of distinct sources  by applying the LCSR method with
the  higher-twist HQET distribution amplitudes.
To this end, we will need to construct the subleading-power factorization formulae for the vacuum-to-$B$-meson
correlation functions and take advantage of the non-trivial  identities for the two-body and three-body light-ray HQET operators
due to the classical equations of motion \cite{Kawamura:2001jm,Kawamura:2001bp,Braun:2017liq}.

\subsection{The NLP contribution form the hard-collinear propagator}

\begin{figure}
\begin{center}
\includegraphics[width=0.65 \columnwidth]{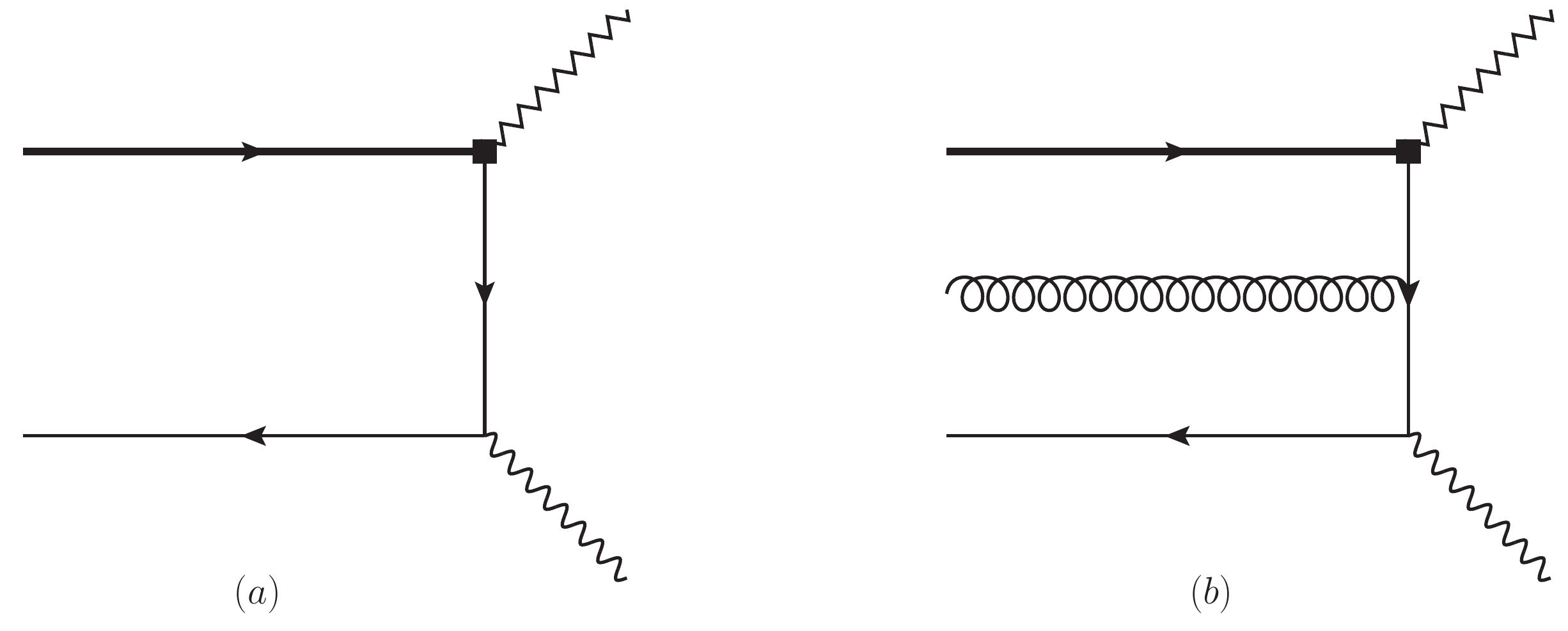}
\vspace*{0.1cm}
\caption{The subleading power two-particle and three-particle corrections
to the vacuum-to-bottom-meson correlation function (\ref{correlation_function})
at the tree-level accuracy,
where the square box indicates an insertion of the weak vertex $\bar{q} \, \Gamma_\mu  \, b$
and the waveline stands for the interpolating current
$\bar q^{\prime} \, \slashed {n} \, \gamma_5 \, q$ for the light pseudoscalar meson.  }
\label{fig: LO-2P-3P-Correlator}
\end{center}
\end{figure}

The first class of the subleading power contribution arises from retaining the higher-order terms
in the heavy quark expansion of the hard-collinear quark propagator as depicted in Figure \ref{fig: LO-2P-3P-Correlator}.
Applying the computational strategy discussed in \cite{Shen:2020hfq,Wang:2021yrr},
we proceed  to write down the two-particle contribution to the QCD correlation function (\ref{correlation_function})
at the tree-level accuracy
\begin{eqnarray}
\Pi_{\mu, {\rm V}, \rm NLP}^{\rm hc}  (n \cdot p, \bar n \cdot p) & = &  i \int d^4x \int \frac{d^4 k}{(2 \pi)^4}
{\rm exp} \left (i \, k \cdot x \right )  \, \frac{1} {(p-k)^2 - m_q^2 +i0} \,
\nonumber \\
&& \times \, \left \langle 0 \left | \overline{q}^{\prime}(x) \, \slashed{n} \, \gamma_5 \, ( \slashed{p} - \slashed{k} + m_q ) \,
\gamma_\mu \, h_v(0) \right | \bar B(p_B) \right \rangle \,.
\label{2P NLP correction at LO}
\end{eqnarray}
Implementing an expansion in powers of $\Lambda_{\rm QCD}/m_b$ for  the  hard-collinear quark propagator
in (\ref{2P NLP correction at LO}) immediately leads to
\begin{eqnarray}
\frac{(\slashed{p} - \slashed{k}) + m_q} {(p-k)^2 - m_q^2 + i0}
&=&\underbrace{\frac{1}{\bar n \cdot (p-k)} \, {\slashed{\bar n}  \over 2}}
+ \underbrace{{1 \over (p-k)^2 } \, \left [ \bar n \cdot p \,  {\slashed{n}  \over 2}
- \slashed{k} + \frac{n \cdot k \,\,  \bar n \cdot p}{\bar n \cdot (p-k)} \,  {\slashed{\bar n}  \over 2}  \right ]}
\nonumber \\
&& \hspace{1.0 cm} {\rm LP}  \hspace{4.2 cm} {\rm NLP}
\nonumber \\
&& + \underbrace{{1 \over (p-k)^2 } \,
\left [ m_q + {m_q^2 - m_{q^{\prime}}^2 \over \bar n \cdot (p-k)} \,  {\slashed{\bar n}  \over 2}   \right ]}
+ \, ...,
\nonumber \\
&& \hspace{1.5 cm} {m_{q^{(\prime)}} \,\, \rm  NLP}
\label{NLP terms of hard-collinear propagator}
\end{eqnarray}
where the abbreviation ``LP" represents the  leading-power effect discussed in Section \ref{section: LP sum rules}.
We can further cast the subleading-power terms displayed in the first line of (\ref{NLP terms of hard-collinear propagator})
in the form
\begin{eqnarray}
\Pi_{\mu, {\rm V}, \rm NLP}^{\rm hc, \, I} (n \cdot p, \bar n \cdot p)
& = &  \int d^4x \int \frac{d^4 k}{(2 \pi)^4} \,
{\rm exp} \left (i \, k \cdot x \right )  \, \frac{1} {(p-k)^2 + i0} \,
\nonumber \\
&& \times \, \bigg \{ {\partial \over \partial x_{\rho}} \,
\left \langle 0 \left | \overline{q}^{\prime}(x) \, \slashed{n} \, \gamma_5 \, \gamma_{\rho} \,
\gamma_\mu \, h_v(0) \right | \bar B(p_B) \right \rangle
\nonumber \\
&& \hspace{0.2 cm} - { \bar n \cdot p \over \bar n \cdot (p-k)} \,
(2 \, v_{\rho} - \bar n_{\rho}) \,
{\partial \over \partial x_{\rho}} \,
\left \langle 0 \left | \overline{q}^{\prime}(x) \, \slashed{n} \, \gamma_5 \, {\slashed{\bar n} \over 2} \,
\gamma_\mu \, h_v(0) \right | \bar B(p_B) \right \rangle  \bigg \},
\hspace{1.0 cm}
\end{eqnarray}
by employing the standard technique of the integration by parts (IBP) as well as the precise relation between
the four-vectors  $n_{\alpha} = 2 \, v_{\alpha} - \bar n_{\alpha}$.
Taking advantage of the well-known operator identities due to the HQET equations of motion
\begin{eqnarray}
v_{\rho} \,  {\partial \over \partial x_{\rho}} \,
\left [  \bar q^{\prime}(x) \,\Gamma \,  h_v(0) \right ]
&=& i \, \int_0^1 d u \, \bar u \, \bar q^{\prime}(x) \,g_s \, G_{\alpha \beta}(u x) \,
x^{\alpha} \, v^{\beta} \, \Gamma \,  h_v(0)
+ (v \cdot \partial) \,  \left [  \bar q^{\prime}(x) \,\Gamma \,  h_v(0) \right ]  \,,
\hspace{0.8 cm}  
\label{the first HQET EOMs}
\\
{\partial \over \partial x_{\rho}} \bar q^{\prime}(x) \, \gamma_{\rho} \, \Gamma \,   h_v(0)
&=& - i \, \int_0^1 d u \, u \,\, \bar q^{\prime}(x) \,g_s \, G^{\lambda \rho}(u x) \,
x_{\lambda} \, \gamma_{\rho} \, \Gamma \,  h_v(0)
+ i \, m_{q^{\prime}} \, \bar q^{\prime}(x) \Gamma \,  h_v(0)  \,,
\label{the second HQET EOMs}
\end{eqnarray}
where the total translation operator $\partial_{\rho}$ acting on an arbitrary composite operator
${\cal O}(x_1, ..., x_n)$ with $n$ space-time arguments is defined by
\begin{eqnarray}
\partial_{\rho} \, {\cal O}(x_1, ..., x_n)
= {\partial \over \partial y^{\rho}} \,
{\cal O}(x_1+y, ..., x_n+y) \bigg|_{y=0}\,,
\end{eqnarray}
we can then readily derive the factorized expressions for the  effective NLP  matrix elements
of our interest at tree level
\begin{eqnarray}
\Pi_{\rm NLP}^{\rm hc, \, I}
& = &  \left [ { 2 \, {\cal F}_B(\mu) \, m_B \over n \cdot p } \right ] \,
\bigg \{ \int_0^{\infty} d \omega_1 \, \int_0^{\infty} d \omega_2 \, \int_0^{1}  d u \,
\frac{u \, \Phi_4(\omega_1, \omega_2, \mu) + \Psi_4(\omega_1, \omega_2, \mu)}
{\left [ \bar n \cdot p - \omega_1 - u \, \omega_2 \right ]^2} \,
\nonumber \\
&& + \int_0^{\infty} d \omega \, \left ( \bar{\Lambda}  - {\omega + m_{q^{\prime}} \over 2}  \right ) \,
\frac{\phi_B^{+}(\omega, \mu)}{\left [ \bar n \cdot p - \omega \right ]} \bigg \}
+  {\cal O}(\alpha_s)\,,
\hspace{1.0 cm} \\
\tilde{\Pi}_{\rm NLP}^{\rm hc, \, I}
& = &  \left [ - { 2 \, {\cal F}_B(\mu) \, m_B \over n \cdot p } \right ] \,
\bigg \{ \int_0^{\infty} d \omega_1 \, \int_0^{\infty} d \omega_2 \, \int_0^{1}  d u \,
\frac{\bar u \, (\bar n \cdot p + \omega_1 + u \, \omega_2 ) }
{\left [ \bar n \cdot p - \omega_1 - u \, \omega_2 \right ]^3}
\,  \Psi_5(\omega_1, \omega_2, \mu)
\nonumber \\
&& + \int_0^{\infty} d \omega \, \left ( \bar{\Lambda}  - {\omega \over 2}  \right ) \,
\frac{\omega \, \phi_B^{+}(\omega, \mu)}{\left [ \bar n \cdot p - \omega \right ]^2} \bigg \}
+  {\cal O}(\alpha_s) \,.
\end{eqnarray}
The hadronic parameter $\bar{\Lambda}$ characterizing the ``effective mass" of the bottom-meson state in HQET
can be defined in an explicitly covariant and gauge invariant manner \cite{Falk:1992fm}
\begin{eqnarray}
\bar \Lambda \equiv \frac{\langle 0 | \bar q \,\,  i \, v \cdot \overleftarrow{D} \, \Gamma \,   h_v | \bar B_q (v) \rangle}
{\langle 0 | \bar q \,  \Gamma \,   h_v | \bar B_q (v) \rangle} \,.
\label{def: effective bottom-meson mass}
\end{eqnarray}
Additionally, we have adopted the systematic parametrization of the three-body light-cone HQET matrix element
at the twist-six accuracy \cite{Braun:2017liq}
\begin{eqnarray}
&& \langle 0 | \bar q_{\alpha}(\tau_1 \, \bar n) \, g_s \, G_{\mu \nu}(\tau_2 \, \bar n) \,
h_{v \, \beta}(0) | \bar B_v \rangle \nonumber \\
&& = {{\cal F}_B(\mu) \, m_B \over 4} \,
\bigg [ (1 + \slashed{v}) \, \bigg \{ (v_{\mu} \gamma_{\nu} - v_{\nu} \gamma_{\mu})  \,
\left [\hat{\Psi}_A(\tau_1, \tau_2, \mu) - \hat{\Psi}_V(\tau_1, \tau_2, \mu) \right ]
- i \, \sigma_{\mu \nu} \, \hat{\Psi}_V(\tau_1, \tau_2, \mu) \nonumber  \\
&& \hspace{0.4 cm}
- (\bar n_{\mu} \, v_{\nu} - \bar n_{\nu} \, v_{\mu} ) \, \hat{X}_A(\tau_1, \tau_2, \mu)
+ (\bar n_{\mu} \, \gamma_{\nu} - \bar n_{\nu} \, \gamma_{\mu} ) \,
\left [ \hat{W}(\tau_1, \tau_2, \mu)  + \hat{Y}_A(\tau_1, \tau_2, \mu)   \right ] \nonumber \\
&& \hspace{0.4 cm} + \, i \, \epsilon_{\mu \nu \alpha \beta} \,
\bar n^{\alpha} \, v^{\beta}  \, \gamma_5 \, \hat{\tilde{X}}_A(\tau_1, \tau_2, \mu)
- \, i \, \epsilon_{\mu \nu \alpha \beta} \,
\bar n^{\alpha} \, \gamma^{\beta}  \, \gamma_5 \, \hat{\tilde{Y}}_A(\tau_1, \tau_2, \mu)  \nonumber \\
&&  \hspace{0.4 cm}  - \, ( \bar n_{\mu} \, v_{\nu} -  \bar n_{\nu} \, v_{\mu} ) \,
\slashed{\bar  n } \, \hat{W}(\tau_1, \tau_2, \mu)
+ \, ( \bar n_{\mu} \, \gamma_{\nu} - \bar n_{\nu} \, \gamma_{\mu} ) \,
\slashed{\bar  n } \, \hat{Z}(\tau_1, \tau_2, \mu)   \bigg \}  \, \gamma_5 \bigg ]_{\beta \, \alpha}  \,.
\label{def: 3-particle HQET  DAs}
\end{eqnarray}
The momentum-space distribution amplitudes can be obtained by carrying out the Fourier transformation
in the two light-cone variables $\tau_{1, 2}$  \cite{Braun:2017liq,Braun:2003wx}
\begin{eqnarray}
&& \Psi_{X}(\omega_1, \omega_2, \mu) =  \int_{-\infty}^{+\infty} {d \tau_1 \over 2 \pi}
\, \int_{-\infty}^{+\infty} {d \tau_2 \over 2 \pi} \,\,
{\rm exp} \left [ i (\omega_1 \, \tau_1 + \omega_2 \, \tau_2) \right ]  \,
\hat{\Psi}_{X}(\tau_1 - i 0, \tau_2 - i 0, \mu)  \,,
\nonumber \\
&& \Psi_{X}  \in  \left \{\Psi_V, \, \Psi_A, \, X_A, \, Y_A, \, \tilde{X}_A, \, \tilde{Y}_A, \, W, \, Z\right \},
\,\,
\hat{\Psi}_{X}  \in  \left \{\hat{\Psi}_V, \, \hat{\Psi}_A, \, \hat{X}_A, \, \hat{Y}_A, \, \hat{\tilde{X}}_A, \,
\hat{\tilde{Y}}_A, \, \hat{W}, \, \hat{Z}\right \} \,.
\hspace{1.0 cm}
\end{eqnarray}
To facilitate the construction of the perturbative factorization formulae,
it turns out to be more advantageous to introduce  the three-particle HQET distribution amplitudes  with the definite collinear twist
by virtue of the appearing invariant functions  (see  \cite{Geyer:2005fb,Geyer:2007yh} for further discussions
on the comparison between dynamical twist and  geometric twist )
\begin{eqnarray}
\Phi_3 &=& \Phi_A - \Phi_V \,,
\qquad  \hspace{3.8 cm}
\Phi_4 = \Phi_A + \Phi_V \,,
\nonumber \\
\Psi_4  &=& \Psi_A + X_A \,,
\qquad  \hspace{3.8 cm}
\tilde{\Psi}_4  = \Psi_V - \tilde{X}_A \,,
\nonumber \\
\tilde{\Phi}_5 &=&  \Psi_A + \Psi_V + 2 \, Y_A - 2 \, \tilde{Y}_A + 2 \, W \,,
\qquad
\Psi_5 =  - \Psi_A + X_A - 2 \, Y_A  \,,
\nonumber \\
\tilde{\Psi}_5 &=&  - \Psi_V -  \tilde{X}_A + 2 \, \tilde{Y}_A  \,,
\qquad  \hspace{2.2 cm}
\Phi_6 =  \Phi_A - \Phi_V + 2 \, Y_A  + 2 \, W
+ 2 \, \tilde{Y}_A - 4 \, Z \,.
\hspace{1.0 cm}
\end{eqnarray}

Applying the standard factorization method, we can compute the power-suppressed contributions
presented in the second line of (\ref{NLP terms of hard-collinear propagator})
due to the non-vanishing quark masses
\begin{eqnarray}
\Pi_{\rm NLP}^{\rm hc, \, II}
& = &  \left [ - { {\cal F}_B(\mu) \, m_B \over n \cdot p } \right ] \,
 \int_0^{\infty} d \omega \,
\frac{m_q} {\bar n \cdot p - \omega }  \, \phi_B^{-}(\omega, \mu) +  {\cal O}(\alpha_s)  \,,
 \\
\tilde{\Pi}_{\rm NLP}^{\rm hc, \, II}
& = &  \left [ - { {\cal F}_B(\mu) \, m_B \over n \cdot p } \right ] \,
 \int_0^{\infty} d \omega \,
\frac{\left (m_q^2  - m_{q^{\prime}}^2 \right  )} {\left [ \bar n \cdot p - \omega \right ]^2}  \,
\, \phi_B^{-}(\omega, \mu) +  {\cal O}(\alpha_s) \,.
\end{eqnarray}
In contrast to the NLP active-quark mass corrections,  the newly identified spectator-quark mass contributions
cannot generate the large-recoil symmetry breaking effects at ${\cal O}(\alpha_s^0)$.

Along the same vein, we can derive the soft-collinear factorization formulae
for the particular subleading-power corrections to the correlation function
$\Pi_{\mu, {\rm T}}$ from the higher-order terms in the heavy quark expansion of the hard-collinear quark propagator
\begin{eqnarray}
\Pi_{\rm T, \, NLP}^{\rm hc, \, I}
=   \left ( \tilde{\Pi}_{\rm NLP}^{\rm hc, \, I} - \Pi_{\rm NLP}^{\rm hc, \, I} \right )  + {\cal O}(\alpha_s)\,,
\qquad
\Pi_{\rm T, \, NLP}^{\rm hc, \, II}
=  \left (  \tilde{\Pi}_{\rm NLP}^{\rm hc, \, II} - \Pi_{\rm NLP}^{\rm hc, \, II} \right ) + {\cal O}(\alpha_s) \,.
\label{relation of the hc propagator corrections}
\end{eqnarray}
These interesting constraints can be attributed to the classical HQET equation of motion
\begin{eqnarray}
\left ( \sigma_{\mu \nu} \, q^{\nu} \right ) \, \mathds{1}  \, h_v
= \left ( \sigma_{\mu \nu} \, q^{\nu} \right ) \, \slashed{v}  \, h_v
= \left ( - {i \over 2}  \right) \, \left [ \bar n \cdot q \, n_\mu - n \cdot q \, \bar n_\mu \right ] \,
\left ( {\slashed {n} \over 2 } - {\slashed {\bar n} \over 2 }  \right )  \, h_v \,.
\end{eqnarray}
Including the non-Eikonal gluonic interaction with the bottom-quark field
in the higher-order corrections to the vacuum-to-bottom-meson correlation functions will generally
invalidate such symmetry relations.

Expressing the established NLP factorization formulae in the dispersion forms
and equating the achieved spectral representations for the correlation functions
with the corresponding  hadronic dispersion relations (\ref{hadronic dispersion relations})
allows us to construct the desired  LCSR for the power-suppressed contributions
from the expanded hard-collinear quark propagator
\begin{eqnarray}
&& f_{M} \,\, {\rm exp} \left [- {m_{M}^2 \over n \cdot p \,\, \omega_M} \right ] \,\,
\left \{ \frac{n \cdot p} {m_B} \, f_{B M, \, \rm NLP}^{+, \, \rm hc}(q^2)
\,, \,\,\,   f_{B M, \, \rm NLP}^{0, \, \rm hc}(q^2) \,,
\,\,\, \frac{n \cdot p} {m_B + m_M} \, f_{B M, \, \rm NLP}^{T, \, \rm hc}(q^2) \right \}  \,
\nonumber \\
&& =   { \mathcal{F}_B(\mu) \over n \cdot p} \,
\bigg [ \int_0^{\omega_s}  \, d \omega_1 \,\, \int_0^{\infty}  \, {d \omega_2 \over \omega_2}
\,\, \bigg \{ e^ {-{\omega_1 + \omega_2 \over \omega_M} }  \,
\varrho_{\rm NLP}^{\rm hc, \, I}(\omega_1, \omega_2,  \mu) \, \theta(\omega_s - \omega_1 - \omega_2)
\nonumber \\
&& \hspace{0.5 cm}
+  \, \left [ \left (  e^ {-{\omega_1 \over \omega_M} } -   e^ {-{\omega_s  \over \omega_M} } \right ) \,
\theta(\omega_1 + \omega_2 -\omega_s)
+ \left ( e^ {-{\omega_1 \over \omega_M}} -   e^ {-{\omega_1 + \omega_2 \over \omega_M} } \right ) \,
\theta(\omega_s - \omega_1 - \omega_2 )  \right ]
\nonumber \\
&& \hspace{0.5 cm}
\times \, \varrho_{\rm NLP}^{\rm hc, \, II}(\omega_1, \omega_2,  \mu)
+  e^ {-{\omega_1  \over \omega_M} }  \,  \varrho_{\rm NLP}^{\rm hc, \, III}(\omega_1, \omega_2,  \mu)  \bigg \}
+ \int_0^{\omega_s}  \, d \omega  \,\, e^ {-{\omega \over \omega_M} }  \,\,
 \tilde{\varrho}_{\rm NLP}^{\rm hc}(\omega, \mu) \bigg ]
 +  {\cal O}(\alpha_s) \,.
 \hspace{1.0 cm}
 \label{NLP LCSR from the hc propagator}
\end{eqnarray}
The yielding expressions for the emerged coefficient functions $\varrho_{\rm NLP}^{\rm hc, \, (I, II, III)}$
and $\tilde{\varrho}_{\rm NLP}^{\rm hc}$ can be explicitly written as
\begin{eqnarray}
\varrho_{\rm NLP}^{\rm hc, \, I}
&=& 2 \, \left \{ {\omega_1 + \omega_2 \over \omega_2} \, \Psi_5(\omega_1, \omega_2, \mu) \,
- \, \kappa_{i} \, \left (\Phi_4 + \Psi_4)(\omega_1, \omega_2, \mu \right )  \right \} \,,
 \\
\varrho_{\rm NLP}^{\rm hc, \, II}
&=&  -2 \, \left ( {\omega_M \over \omega_2} \right ) \,
\left [ \Psi_5(\omega_1, \omega_2, \mu) + \kappa_i \, \Phi_4(\omega_1, \omega_2, \mu)  \right ] \,,
\\
\varrho_{\rm NLP}^{\rm hc, \, III}
&=& 2 \, \left \{ \left ({d \over d \omega_1} - {1 \over \omega_2 } \right ) \,
\left [ \omega_1 \, \Psi_5(\omega_1, \omega_2, \mu) \right ] \,
- \, \kappa_{i}\, \Phi_4(\omega_1, \omega_2, \mu)  \right \} \,,
 \\
\tilde{\varrho}_{\rm NLP}^{\rm hc} &=&  \left [ 1 - \left ( {\omega_s - \omega \over \omega} \right ) \,
{m_q^2 - m_{q^{\prime}}^2 \over \omega_s \, \omega_M} \,
+ \left(\omega - 2 \, \bar \Lambda - {m_q^2 - m_{q^{\prime}}^2 \over \omega_s}  \right ) \,
 {d \over d \omega}   \right ] \,
\left [ \omega \, \phi_B^{-}(\omega, \mu) \right ]
\nonumber  \\
&& + \, \kappa_i \, \left [ \left (\omega - 2 \, \bar \Lambda \right ) \, \phi_B^{+}(\omega, \mu)  +
\left (m_q + m_{q^{\prime}} \right ) \, \phi_B^{-}(\omega, \mu) \right ] \,,
\end{eqnarray}
where the non-universal $\kappa_i$-factors are responsible for  the symmetry-breaking effects
\begin{eqnarray}
\kappa_{+} = - \kappa_{0} = {(n \cdot p - m_B) / m_B}  \,,
\qquad \kappa_{T} = -1 \,.
\end{eqnarray}
Importantly, we have further verified that the newly derived  sum rules (\ref{NLP LCSR from the hc propagator})
for the light-quark-mass insensitive NLP corrections from the heavy quark expansion of
the hard-collinear propagator are consistent with the previous computations
for the semileptonic $B \to D \ell \bar \nu_{\ell}$ form factors accomplished  in \cite{Gao:2021sav}.

\subsection{The NLP contribution from the subleading effective current}

We now proceed to determine the second class of the  subleading-power correction
to the exclusive $B_{d, s}  \to \pi, K$   form factors
arising from the peculiar higher-order term
in the ${\rm SCET}_{\rm I}$ representation of the heavy-to-light transition current
$\bar q \, \Gamma_\mu \,b$ \cite{Beneke:2002ni}
\begin{eqnarray}
J^{\rm (A2)} \supset
(\bar \xi_{\rm hc} \, W_{\rm hc}) \,\,
\Gamma \,\, \left ( { i \, \overrightarrow{\slashed D}_{\top} \over 2 \, m_b} \right )
\, \,   h_v   + ... \,,
\qquad
D_{\top}^{\mu} &\equiv&  D^{\mu} - (v\cdot D)  \, v^{\mu} \,,
\label{NLP SCET current}
\end{eqnarray}
which  corresponds to the standard ${\rm QCD} \to {\rm HQET}$ matching for the bottom-quark field
\begin{eqnarray}
b(x) =  {\rm exp}\left ( - i \, m_b \, v \cdot x  \right ) \,
\bigg [ 1 + { i \, \overrightarrow{{\slashed D}}_{\top}  \over 2 \, m_b}
+  { (v\cdot \overrightarrow{D}) \, \overrightarrow{{\slashed D}}_{\top}  \over 4 \, m_b^2}
- { \overrightarrow{{\slashed D}}_{\top} \, \overrightarrow{{\slashed D}}_{\top} \over 8 \, m_b^2}
+ {\cal O} \left ({1 \over m_b^3} \right ) \bigg ] \, h_v(x) \,.
\end{eqnarray}
It is then straightforward to express the resulting NLP contribution to the vacuum-to-$B$-meson
correlation function (\ref{correlation_function}) in the following form
\begin{eqnarray}
\Pi_{\mu, {\rm V}, \rm NLP}^{\rm (A2)}   (n \cdot p, \bar n \cdot p) & = &
- \left ( {n \cdot p \over 4 \, m_b} \right ) \, \int d^4x \int \frac{d^4 k}{(2 \pi)^4}
{\rm exp} \left (i \, k \cdot x \right )  \, \frac{1} {(p-k)^2 - m_q^2 +i0} \,
\nonumber \\
&& \times \, \left \langle 0 \left | \overline{q}^{\prime}(x) \, \slashed{n} \, \gamma_5 \, \slashed{\bar n}  \,
\gamma_\mu \, \overrightarrow{{\slashed D}}_{\top} \, h_v(0) \right | \bar B(p_B) \right \rangle
\nonumber \\
&=&  - \left ( {n \cdot p \over 2 \, m_b} \right ) \, \int d^4x \int \frac{d^4 k}{(2 \pi)^4}
{\rm exp} \left (i \, k \cdot x \right )  \, \frac{\bar n_{\mu} } {(p-k)^2 - m_q^2 +i0} \,
\nonumber \\
&& \times \, \left \langle 0 \left | \overline{q}^{\prime}(x) \,
\left ( \overrightarrow{{\slashed D}}_{\top} \,\,  \slashed{n} \,
 +  \,2 \, \bar n \cdot \overrightarrow{D}_{\top} \right ) \,
\gamma_5 \,  h_v(0) \right | \bar B(p_B) \right \rangle \,,
\label{NLP SCET current correction at LO}
\end{eqnarray}
where we have employed the lowest-order equation of motion of the effective heavy-quark field
\begin{eqnarray}
i \, v \cdot \overrightarrow{D} \, h_v = 0 \,.
\end{eqnarray}
This evidently permits the  replacement
$i\, \overrightarrow{\slashed D}_{\top} \, h_v \to i \, \overrightarrow{\slashed D} \, h_v$
in the effective weak current at the NLP accuracy.
Taking advantage of an additional HQET operator identity \cite{Kawamura:2001jm,Kawamura:2001bp,Braun:2017liq}
 \begin{eqnarray}
\bar q^{\prime}(x) \, \Gamma \, {\overrightarrow{D}}_{\rho} \,  h_v(0)
&=& \partial_{\rho} \left [ \bar q^{\prime}(x) \, \Gamma \, h_v(0) \right ] \,
+ \, i \, \int_0^1 d u \, \bar u \,\, \bar q^{\prime}(x) \,g_s \, G_{\lambda \rho}(u x) \,
x^{\lambda} \, \Gamma \,  h_v(0) \nonumber \\
&& - \, {\partial \over \partial x^{\rho}} \bar q^{\prime}(x) \, \Gamma \, h_v(0)  \,,
\end{eqnarray}
in combination with the two operator relations displayed in (\ref{the first HQET EOMs})
and (\ref{the second HQET EOMs}),
we can construct the perturbative factorization formula for the non-local hadronic matrix element
in (\ref{NLP SCET current correction at LO})
\begin{eqnarray}
\Pi_{\rm NLP}^{\rm (A2)}    & = &   {\cal O}(\alpha_s) \,,
 \\
\tilde{\Pi}_{\rm NLP}^{\rm (A2)}   & = &
\left [ {\mathcal{F}_B(\mu)  \, m_B  \over 2 \, m_b} \right ] \,
\bigg \{  \int_0^{\infty} \, d \omega_1 \, \int_0^{\infty} \, d \omega_2 \,
{2 \, (\Psi_4 + \Phi_4)(\omega_1, \omega_2, \mu) \over (\omega_1 - \bar n \cdot p) \, (\omega_1 + \omega_2 - \bar n \cdot p) } \,
\nonumber \\
&& \hspace{-0.5 cm}
+ \, \int_0^{\infty} \, { d \omega \over \omega - \bar n \cdot p} \,
\left [ \left(\omega - 2 \, \bar \Lambda \right ) \, \phi_B^{+}(\omega, \mu)
+ \left (\omega - \bar \Lambda + m_{q^{\prime}} \right )  \, \phi_B^{-}(\omega, \mu)  \right ] \bigg \}
+ {\cal O}(\alpha_s)  \,.
\label{factorization formulae for the NLP SCET current correction}
\end{eqnarray}
Applying the analogous computational strategy,  we can further  compute the yielding NLP correction to
 the correlation function $\Pi_{\mu, {\rm T}}$ at tree level
\begin{eqnarray}
\Pi_{\rm T, \, NLP}^{\rm (A2)}
=   - \left ( \tilde{\Pi}_{\rm NLP}^{\rm (A2)} - \Pi_{\rm NLP}^{\rm (A2)} \right )  + {\cal O}(\alpha_s) \,.
\label{relation of the NLP SCET corrections}
\end{eqnarray}
We remark in passing that  such an interesting constraint (\ref{relation of the NLP SCET corrections})
differs from the previously established relations (\ref{relation of the hc propagator corrections})
for the NLP corrections from the HQET expansion of the hard-collinear quark propagator
by an overall factor of ``$-1$".
This observation can be traced back to the anti-commutation relation
$\left \{ \slashed{v},   \overrightarrow{\slashed D}_{\top} \right \} \, h_v = 0$,
thus ensuring an exact algebraic identity
\begin{eqnarray}
\left ( \sigma_{\mu \nu} \, q^{\nu} \right ) \, \left ( { i \, \overrightarrow{\slashed D}_{\top} \over 2 \, m_b} \right )  \, h_v
&=& \left (  {i \over 2}  \right) \, \left [ \bar n \cdot q \, n_\mu - n \cdot q \, \bar n_\mu \right ] \,
\left ( {\slashed {n} \over 2 } - {\slashed {\bar n} \over 2 }  \right )  \,
\left ( { i \, \overrightarrow{\slashed D}_{\top} \over 2 \, m_b} \right )  \, h_v \,.
\end{eqnarray}

Matching the spectral representations for the established soft-collinear factorization formulae
(\ref{factorization formulae for the NLP SCET current correction})
and (\ref{relation of the NLP SCET corrections})
with the corresponding hadronic dispersion relations (\ref{hadronic dispersion relations})
enables us to derive the final expressions for the  NLP sum rules of the $B_{d, s}  \to \pi, K$ decay form factors
\begin{eqnarray}
&& f_{M} \,\, {\rm exp} \left [- {m_{M}^2 \over n \cdot p \,\, \omega_M} \right ] \,\,
\left \{ \frac{n \cdot p} {m_B} \, f_{B M, \, \rm NLP}^{+, \, \rm (A2)}(q^2)
\,, \,\,\,   f_{B M, \, \rm NLP}^{0, \, \rm (A2)}(q^2) \,,
\,\,\, \blue{\left [ - \frac{n \cdot p} {m_B + m_M} \right ]} \, f_{B M, \, \rm NLP}^{T, \, \rm (A2)}(q^2) \right \}  \,
\nonumber \\
&& =   { \mathcal{F}_B(\mu) \over n \cdot p} \,
\bigg [ \int_0^{\omega_s}  \, d \omega_1 \,\, \int_0^{\infty}  \, {d \omega_2 \over \omega_2}
\,\, \bigg \{ e^ {-{\omega_1 + \omega_2 \over \omega_M} }  \, \varrho_{\rm NLP}^{\rm (A2), \, I}(\omega_1, \omega_2,  \mu) \,
\theta(\omega_s - \omega_1 - \omega_2 )
\nonumber \\
&& \hspace{1.8 cm}
+  \, e^ {-{\omega_1  \over \omega_M} }  \,  \varrho_{\rm NLP}^{\rm (A2), \, II}(\omega_1, \omega_2,  \mu) \bigg \} \,\,
+ \int_0^{\omega_s}  \, d \omega  \,\, e^ {-{\omega \over \omega_M} }  \,\,
 \tilde{\varrho}_{\rm NLP}^{\rm (A2)}(\omega, \mu) \bigg ]
 +  {\cal O}(\alpha_s) \,.
 \label{NLP LCSR from the subleading SCET current}
\end{eqnarray}
The non-perturbative coefficient functions $\varrho_{\rm NLP}^{\rm (A2), \, (I, II)}$
and $\tilde{\varrho}_{\rm NLP}^{\rm (A2)}$ can be expressed in terms of the two-particle and three-particle
HQET distribution amplitudes
\begin{eqnarray}
&& \varrho_{\rm NLP}^{\rm (A2), \, I}
= - {n \cdot p \over m_b} \, \left [  \left (\Phi_4 + \Psi_4)(\omega_1, \omega_2, \mu \right ) \right ]  \,,
 \qquad
\varrho_{\rm NLP}^{\rm (A2), \, II}
=  {n \cdot p \over m_b} \,  \left (\Phi_4 + \Psi_4)(\omega_1, \omega_2, \mu \right ) \,,
\\
&&  \tilde{\varrho}_{\rm NLP}^{\rm (A2)} =  {n \cdot p \over 2 \, m_b} \,
\left [ \left (\omega - 2 \, \bar \Lambda \right ) \, \phi_B^{+}(\omega, \mu)  +
\left (\omega -  \bar \Lambda + m_{q^{\prime}} \right )  \, \phi_B^{-}(\omega, \mu) \right ] \,.
\hspace{1.0 cm}
\end{eqnarray}
Remarkably, the considered NLP corrections to the heavy-to-light form factors
from the effective matrix elements of the subleading ${\rm SCET}_{\rm I}$ current $J^{\rm (A2)}$
preserve the well-known symmetry relation between the vector and scalar form factors,
but violate the large-recoil symmetry of the vector and tensor form factors already at ${\cal O}(\alpha_s^0)$.

\subsection{The NLP contribution form the higher-twist two-particle and three-particle LCDAs}

As emphasized repeatedly in \cite{Ball:1998sk,Ball:1998ff}, the systematic and consistent  description of
 the higher-twist corrections to exclusive hard reactions in QCD
will require us to  simultaneously take into account the transverse-momentum dependence of the valence (anti)-quarks in the leading
Fock-state wavefunction and the subleading distribution amplitudes of the non-minimal partonic configuration
 with additional gluons and/or quark-antiquark pairs.
Including the off-light-cone corrections to the renormalized two-body  non-local HQET matrix element
at the ${\cal O}(x^2)$ accuracy discussed in \cite{Braun:2017liq}
\begin{eqnarray}
&& \langle  0 | \left (\bar q^{\prime} \, Y_s \right)_{\beta} (x) \,
\left (Y_s^{\dag} \, h_v \right )_{\alpha}(0)| \bar B(v)\rangle \nonumber \\
&& = - \frac{i \, {\cal F}_B(\mu) \, m_B}{4}  \,
\int_0^{\infty} \, d \omega \,\, {\rm exp} \left [ {- i \, \omega \, v \cdot x} \right ] \,
\bigg \{  \frac{1+ \slashed{v}}{2} \, \,
\bigg [ 2 \, \left ( \phi_{B}^{+}(\omega, \mu) + \blue{x^2 \, g_B^{+}(\omega, \mu)}  \right ) \nonumber \\
&& \hspace{0.5 cm} - {1 \over v \cdot x}  \,
\left  [ \left ( \phi_{B}^{+}(\omega, \mu) - \phi_{B}^{-}(\omega, \mu)  \right )
+ \blue{ x^2 \, \left ( g_{B}^{+}(\omega, \mu) - g_{B}^{-}(\omega, \mu)  \right ) }  \right ]  \,
\slashed{x} \bigg ] \, \gamma_5 \bigg \}_{\alpha \beta}\,,
\label{def: off-light-cone two-particle B-meson DAs}
\end{eqnarray}
we can proceed to construct the tree-level factorization formulae for the two-particle higher-twist corrections
to the three invariant functions $\Pi$, $\tilde \Pi$ and $\Pi_{\rm T}$ \cite{Lu:2018cfc}
\begin{eqnarray}
\Pi_{\rm NLP}^{\rm 2PHT}  & = &   {\cal O}(\alpha_s) \,,
 \\
\tilde{\Pi}_{\rm NLP}^{\rm 2PHT}   & = &
\left [ {2 \, \mathcal{F}_B(\mu)  \, m_B  \over n \cdot p} \right ] \,
\bigg \{  \int_0^{\infty} \, d \omega_1 \, \int_0^{\infty} \, d \omega_2 \, \int_0^{1} \, d u  \,\,
{\bar u \, \Psi_5(\omega_1, \omega_2, \mu) \over  \left [ \bar n \cdot p - \omega_1 - u \, \omega_2 \right ]^2 } \,
\nonumber \\
&& 
- \, 2  \, \int_0^{\infty} \, d \omega \,\,
{\hat{g}_B^{-}(\omega, \mu)  \over  \left [ \bar n \cdot p - \omega \right ]^2  } \, \bigg \}
+ {\cal O}(\alpha_s)  \,,
 \\
\Pi_{\rm T, \, NLP}^{\rm 2PHT}
&=&    \left ( \tilde{\Pi}_{\rm NLP}^{\rm 2PHT} - \Pi_{\rm NLP}^{\rm 2PHT} \right )  + {\cal O}(\alpha_s) \,.
\label{factorization formulae for the 2P higher-twist corrections}
\end{eqnarray}
In the above derivation we have employed the nontrivial constraint on the momentum-space distribution amplitudes
 due to the HQET equations of motion
\begin{eqnarray}
-2 \, {d^2 g_B^{-}(\omega, \mu)  \over d \omega^2} &=&
\left [ {3 \over 2} + (\omega - \bar \Lambda) \, {d \over d \omega}   \right ] \, \phi_B^{-}(\omega, \mu)
- \left ( {1 \over 2}  \right ) \, \phi_B^{+}(\omega, \mu)
\nonumber \\
&& + \int_0^{\infty} \, {d \omega_2 \over \omega_2 } \,
\left ( {d \over d \omega} - {1 \over \omega_2} \right ) \, \Psi_5(\omega, \omega_2, \mu)
+ \int_0^{\omega} \, {d \omega_2 \over \omega_2^2 } \, \Psi_5(\omega-\omega_2, \omega_2, \mu) \,.
\hspace{1.0 cm}
\label{the LO EOM for gBm}
\end{eqnarray}
This allows us to decompose the higher-twist LCDA $g_B^{-}(\omega, \mu)$ into the ``genuine" twist-five
three-particle distribution amplitude $\Psi_5(\omega_1, \omega_2, \mu)$
and the lower-twist ``Wandzura-Wilczek" contribution labelled as $\hat{g}_B^{-}(\omega, \mu)$,
which can be explicitly expressed in terms of the customary two-particle $B$-meson distribution amplitudes
\begin{eqnarray}
\hat{g}_B^{-}(\omega, \mu) =  \left ( {1 \over 4} \right ) \, \int_{\omega}^{\infty}  \, d \rho \,
\left  \{ (\rho - \omega) \, \left [ \phi_B^{+}(\rho) -  \phi_B^{-}(\rho) \right ]
- 2 \, (\bar \Lambda - \rho)  \, \phi_B^{-}(\rho)  \right \}  \,.
\label{def: WW part of gBminus}
\end{eqnarray}

Applying further the light-cone expansion for the massive-quark propagator in the background gluon field
up to the gluon field strength terms without the covariant derivatives  \cite{Balitsky:1987bk,Rusov:2017chr}
(see also \cite{Khodjamirian:2010vf,Khodjamirian:2012rm} for an alternative representation)
\begin{eqnarray}
S(x, 0, m_q) &\equiv& \langle 0 | {\rm T} \, \{q(x), \, \bar q (0)  \} | 0\rangle
\nonumber \\
& \supset  &  i \, g_s \, \int_{-\infty}^{+\infty} \,\, {d^4 \ell \over (2 \pi)^4} \,
{\rm exp} \left [ {- i \, \ell \cdot x} \right ] \,
\int_0^1 \, d u \, \left  [ {u \, x_{\mu} \, \gamma_{\nu} \over \ell^2 - m_q^2}
 - \frac{(\slashed{\ell} + m_q) \, \sigma_{\mu \nu}}{2 \, (\ell^2 - m_q^2)^2}  \right ]
\, G^{\mu \nu}(u \, x) \,,
\hspace{1.0 cm}
\end{eqnarray}
we can continue to  write down the tree-level factorized expressions for the yielding three-particle higher-twist corrections
to the vacuum-to-$B$-meson correlation functions (\ref{correlation_function})
displayed in Figure \ref{fig: LO-2P-3P-Correlator}(b) \cite{Lu:2018cfc}
\begin{eqnarray}
\Pi_{\rm NLP}^{\rm 3PHT}    & = &
\left [  { {\cal F}_B(\mu) \, m_B \over n \cdot p } \right ] \,
 \int_0^{\infty} d \omega_1 \, \int_0^{\infty} d \omega_2 \, \int_0^{1}  d u \,
\frac{1} {\left [ \bar n \cdot p - \omega_1 - u \, \omega_2 \right ]^2} \,
\nonumber \\
&& \times \, \left \{  2 \,  \bar u \, \Phi_4(\omega_1, \omega_2, \mu)
+ {m_q \over n \cdot p} \left (\Psi_5 - \tilde{\Psi}_5 \right )(\omega_1, \omega_2, \mu)  \right \}
 + {\cal O}(\alpha_s)  \,,
 \\
\tilde{\Pi}_{\rm NLP}^{\rm 3PHT}   & = &
\left [  { {\cal F}_B(\mu) \, m_B \over n \cdot p } \right ] \,
 \int_0^{\infty} d \omega_1 \, \int_0^{\infty} d \omega_2 \, \int_0^{1}  d u \,
\frac{1} {\left [ \bar n \cdot p - \omega_1 - u \, \omega_2 \right ]^2} \,
\nonumber \\
&& \times \, \left \{ \left [ (2 \, u -1) \, \Psi_5 - \tilde{\Psi}_5 \right ] (\omega_1, \omega_2, \mu)
- {2 \, m_q \over n \cdot p} \, \Phi_6(\omega_1, \omega_2, \mu)  \right \}
 + {\cal O}(\alpha_s)   \,,
\hspace{1.0 cm}  \\
\Pi_{\rm T, \, NLP}^{\rm 3PHT}
&=&    \left ( \tilde{\Pi}_{\rm NLP}^{\rm 3PHT} - \Pi_{\rm NLP}^{\rm 3PHT} \right )  + {\cal O}(\alpha_s) \,.
\label{factorization formulae for the3P higher-twist corrections}
\end{eqnarray}

Adding together the obtained two-particle and three-particle subleading twist corrections
in the dispersion form and implementing the standard continuum subtraction procedure with the parton-hadron duality ans\"{a}tz
leads to the following sum rules for the NLP contributions to the semileptonic $B_{d, s}  \to \pi, K$  form factors
\begin{eqnarray}
&& f_{M} \,\, {\rm exp} \left [- {m_{M}^2 \over n \cdot p \,\, \omega_M} \right ] \,\,
\left \{ \frac{n \cdot p} {m_B} \, f_{B M, \, \rm NLP}^{+, \, \rm HT}(q^2)
\,, \,\,\,   f_{B M, \, \rm NLP}^{0, \, \rm HT}(q^2) \,,
\,\,\, \frac{n \cdot p} {m_B + m_M} \, f_{B M, \, \rm NLP}^{T, \, \rm HT}(q^2) \right \}  \,
\nonumber \\
&& =   { \mathcal{F}_B(\mu) \over n \cdot p} \,
\bigg [ \int_0^{\omega_s}  \, d \omega_1 \,\, \int_0^{\infty}  \, {d \omega_2 \over \omega_2}
\,\, \bigg \{ e^ {-{\omega_1 + \omega_2 \over \omega_M} }  \,
\varrho_{\rm NLP}^{\rm HT, \, I}(\omega_1, \omega_2,  \mu) \,
\theta(\omega_s - \omega_1 - \omega_2 )
\nonumber \\
&& \hspace{0.5 cm}
+  \, \left [ \left (  e^ {-{\omega_1 \over \omega_M} } -   e^ {-{\omega_s  \over \omega_M} } \right ) \,
\theta(\omega_1 + \omega_2 -\omega_s)
+ \left ( e^ {-{\omega_1 \over \omega_M}} -   e^ {-{\omega_1 + \omega_2 \over \omega_M} } \right ) \,
\theta(\omega_s - \omega_1 - \omega_2 )  \right ]
\nonumber \\
&& \hspace{0.5 cm}
\times \, \varrho_{\rm NLP}^{\rm HT, \, II}(\omega_1, \omega_2,  \mu)
+  e^ {-{\omega_1  \over \omega_M} }  \,  \varrho_{\rm NLP}^{\rm HT, \, III}(\omega_1, \omega_2,  \mu) \bigg \} \,\,
+ \int_0^{\omega_s}  \, d \omega  \,\, e^ {-{\omega \over \omega_M} }  \,\,
 \tilde{\varrho}_{\rm NLP}^{\rm HT}(\omega, \mu) \bigg ]
 +  {\cal O}(\alpha_s) \,.
 \hspace{1.0 cm}
 \label{NLP LCSR from the 2P and 3P higher-twist effects}
\end{eqnarray}
The explicit expressions for the  coefficient functions $\varrho_{\rm NLP}^{\rm HT, \, (I, II, III)}$
and $\tilde{\varrho}_{\rm NLP}^{\rm HT}$ are given by
\begin{eqnarray}
\varrho_{\rm NLP}^{\rm HT, \, I} &=&  \left [ - \left (1 +  {m_q \over n \cdot p} \, \kappa_i \right ) \,
\left (\Psi_5 - \tilde{\Psi}_5 \right )(\omega_1, \omega_2, \mu) \right ]
+ {2 \, m_q \over n \cdot p}  \, \Phi_6(\omega_1, \omega_2, \mu),
 \\
\varrho_{\rm NLP}^{\rm HT, \, III} &=&  \left [ 2 \,  \kappa_i \, \Phi_4(\omega_1, \omega_2, \mu)
+ \left (1 +  {m_q \over n \cdot p} \, \kappa_i \right ) \,
\left (\Psi_5 - \tilde{\Psi}_5 \right )(\omega_1, \omega_2, \mu) \right ]
- {2 \, m_q \over n \cdot p}  \, \Phi_6(\omega_1, \omega_2, \mu),
\hspace{0.70 cm} \\
\varrho_{\rm NLP}^{\rm HT, \, II} &=& -2 \,  \left ( {\omega_M \over \omega_2} \right ) \,
\kappa_i \,  \Phi_4(\omega_1, \omega_2, \mu)  \,,
\qquad  \hspace{2.0 cm}
\tilde{\varrho}_{\rm NLP}^{\rm HT} =  - 4 \, {d \hat{g}_B^{-}(\omega, \mu) \over d \omega}.
\end{eqnarray}
Interestingly, the obtained tree-level sum rules for the two-particle higher-twist contributions are independent of  the non-universal $\kappa_i$-factors, thus maintaining the large-recoil symmetry relations of the considered bottom-meson decay form factors.
According to our power-counting  scheme for the intrinsic LCSR parameters
$\omega_s \sim \omega_M \sim {\cal O}(\Lambda_{\rm QCD}^2/m_b)$,
we can immediately determine the desired scaling behaviours of the two-particle and three-particle higher-twist corrections
at ${\cal O}(\alpha_s^0)$ in the heavy quark limit \cite{Lu:2018cfc}
\begin{eqnarray}
f_{B M, \, \rm NLP}^{+, \, \rm HT} \sim f_{B M, \, \rm NLP}^{0, \, \rm HT} \sim f_{B M, \, \rm NLP}^{T, \, \rm HT}
\sim {\cal O} ((\Lambda_{\rm QCD}/m_b)^{5/2}) \,,
\end{eqnarray}
which  turn out to be  suppressed by one power of $\Lambda_{\rm QCD}/m_b$ in comparison with
the leading-power SCET sum rules (\ref{NLL sum rules at LP I}) and (\ref{NLL sum rules at LP II}).
It is however important to  emphasize that evaluating  the (currently unknown) higher-order  radiative corrections to
the three-particle twist-three $B$-meson LCDA contributions can actually bring about
the unsuppressed symmetry-preserving effects for the exclusive $B_{d, s}  \to \pi, K$ form factors
in the heavy quark expansion.
This observation can be understood from the very fact that  the two-particle twist-three distribution amplitude
$\phi_B^{-}(\omega, \mu)$ appearing in the tree-level sum rules  can be generated by the one-loop renormalization
of the three-particle $B$-meson LCDA  $\Phi_3(\omega_1, \omega_2, \mu)$ \cite{Descotes-Genon:2009jif,Braun:2015pha}
(see  \cite{Beneke:2003pa} for further discussions in the SCET framework).

\subsection{The NLP contribution form the higher-twist four-particle effects }

\begin{figure}
\begin{center}
\includegraphics[width=0.75 \columnwidth]{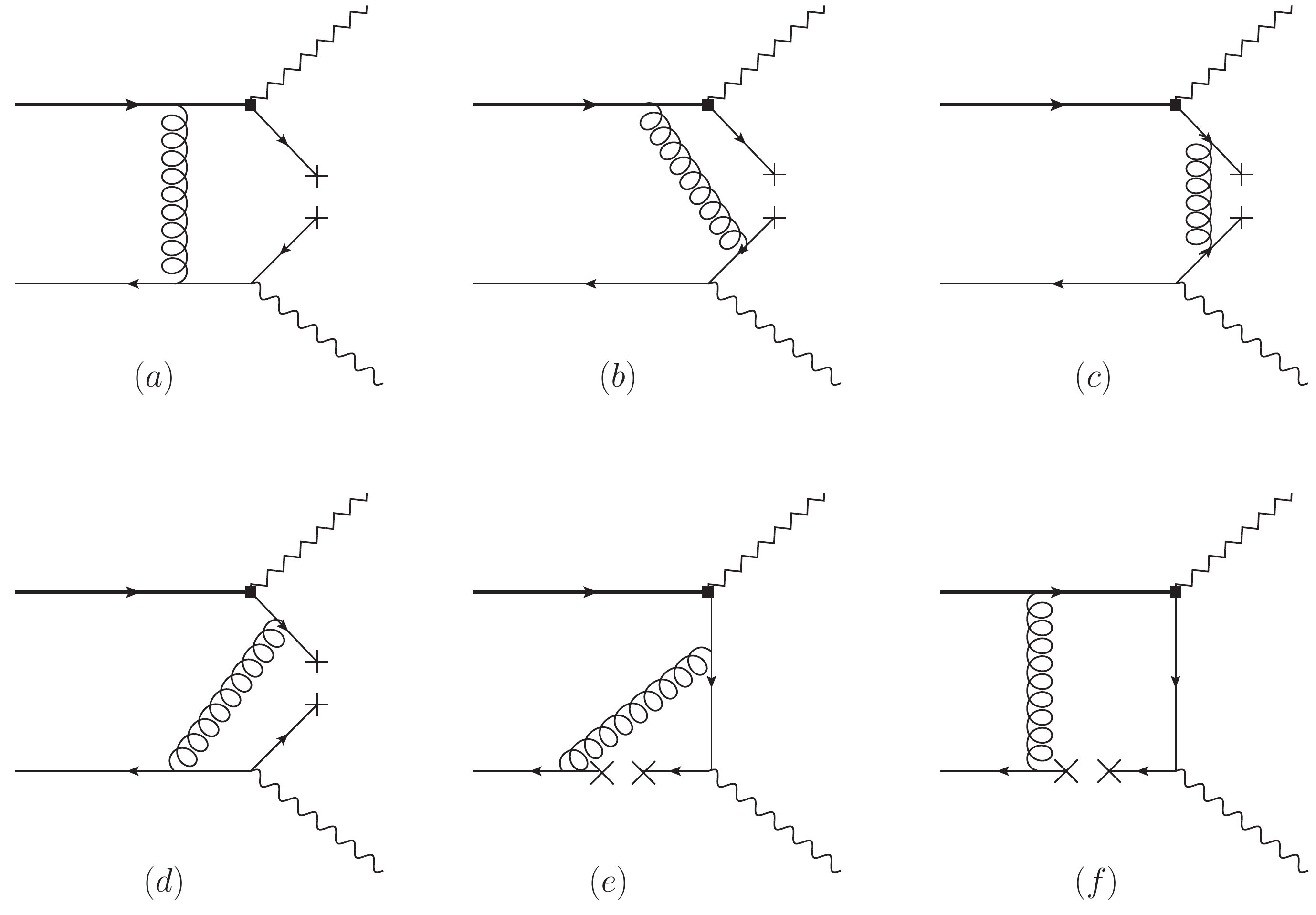}
\vspace*{0.1cm}
\caption{Diagrammatic representations of the twist-five and twist-six four-particle corrections
to the considered vacuum-to-bottom-meson correlation functions (\ref{correlation_function})
at the ${\cal O}(\alpha_s)$ accuracy.  }
\label{fig: 4P higher-twist correlator at LO}
\end{center}
\end{figure}

We are now in a position to compute the NLP corrections to the heavy-to-light bottom-meson decay form factors
from the twist-five and twist-six four-particle LCDA contributions in the factorization approximation,
following the computational prescriptions for the electromagnetic pion form factor at intermediate momentum transfer \cite{Braun:1999uj},
the exclusive photon-pion transition form factor  $\gamma^{\ast} \gamma \to \pi$  \cite{Agaev:2010aq}
and  the radiative leptonic $B \to \gamma \ell \bar \nu_{\ell}$ decay amplitude \cite{Beneke:2018wjp}.
Evaluating the lowest-order Feynman diagrams in Figure \ref{fig: 4P higher-twist correlator at LO}
straightforwardly leads to the factorized expressions for such non-leading Fock-state corrections
\begin{eqnarray}
\Pi_{\rm NLP}^{\rm 4P}   & = & {\cal O}(\alpha_s^2)
 \\
\tilde{\Pi}_{\rm NLP}^{\rm 4P}  & = &
{2 \, \pi \over 3} \, \frac{\alpha_s(\mu)  \, C_F  \, \langle \bar q^{\prime} q^{\prime} \rangle}{n \cdot p \, \bar n \cdot p} \,
{\cal F}_B(\mu) \, m_B \, \int_0^{\infty} \frac{d \omega}{\omega - \bar n \cdot p} \,
 {\phi_B^{+}(\omega, \mu)   \over \omega} \,
\nonumber \\
&& \times \,  \left \{  
 {2 \, \bar n \cdot p \over \omega} \,
\left [ 1 + {\bar n \cdot p - \omega \over \omega}  \, \ln {\bar n \cdot p - \omega  \over \bar n \cdot p} \right ]
- 1 +  {\langle \bar q q \rangle \over \langle \bar q^{\prime} q^{\prime} \rangle } \,
{\omega \over \omega - \bar n \cdot p }    \right \}  \,
+ {\cal O}(\alpha_s^2)\,,
\hspace{0.8 cm}
\\
\Pi_{\rm T, \, NLP}^{\rm 4P}   & = &
\left (\tilde{\Pi}_{\rm NLP}^{\rm 4P} - \Pi_{\rm NLP}^{\rm 4P} \right )
+ {\cal O}(\alpha_s^2) \,.
\end{eqnarray}
It is perhaps worth mentioning that the  determined  four-particle contributions
to the vacuum-to-bottom-meson correlation functions (\ref{correlation_function}) from
the particular diagram (e) in Figure \ref{fig: 4P higher-twist correlator at LO}
can be most conveniently  computed  with the familiar background-field expansion of the quark propagator
on the light-cone \cite{Balitsky:1987bk}
\begin{eqnarray}
\langle 0 | {\rm T} \, \{q(x), \, \bar q (0)  \} | 0\rangle
& \supset & \frac{\Gamma({d / 2}-1)}{8 \, \pi^{d/2} \, (-x^2)^{d/2-1}}  \,
\int_0^1 d u \, u \, \bar u \, \slashed{x} \, x_{\nu}  \, D^{\mu} \, g_s \, G^{\mu \nu}(u x)
\nonumber \\
&& + \, \frac{\Gamma({d / 2}-2)}{16 \, \pi^{d/2} \, (-x^2)^{d/2-2}}  \,
\int_0^1 d u \, \left ( u \, \bar u  - {1 \over 2} \right ) \,
 D^{\mu} \, g_s \, G^{\mu \nu}(u x) \, \gamma^{\nu} \,,
\end{eqnarray}
together with  the classical equation of motion in QCD
\begin{eqnarray}
D^{\mu} \, G_{\mu \nu}^{a} = - i\, g_s \, \sum_{q} \, \bar q \, \gamma_{\nu} \, T^a \, q.
\end{eqnarray}
Additionally, our explicit calculations of the three diagrams (a), (b) and (c) in Figure \ref{fig: 4P higher-twist correlator at LO}
indicate that they can only  generate the yet higher-power corrections in the heavy quark expansion
when compared with the dominating  contributions from the  diagrams (d) and (e).
These enlightening pattern differs drastically from the counterpart NLP contributions to
the two helicity form factors  of $B \to \gamma \ell \bar \nu_{\ell}$,
due to the longitudinally polarized pseudoscalar-meson current in the former
and the transversely polarized on-shell photon state in the latter.
Moreover, the remaining  diagram (f) in Figure \ref{fig: 4P higher-twist correlator at LO}
turns out to be insensitive to both the hard and hard-collinear QCD dynamics.

We can proceed to work out the dispersion representations for the four-particle corrections
to the invariant functions $\Pi$, $\tilde \Pi$ and $\Pi_{\rm T}$ at ${\cal O}(\alpha_s)$
and further derive the NLP sum rules for the yielding twist-five and twist-six contributions with the standard strategy
\begin{eqnarray}
&& f_{M} \,\, {\rm exp} \left [- {m_{M}^2 \over n \cdot p \,\, \omega_M} \right ] \,\,
\left \{ \frac{n \cdot p} {m_B} \, f_{B M, \, \rm NLP}^{+, \, \rm 4P}(q^2)
\,, \,\,\,   f_{B M, \, \rm NLP}^{0, \, \rm 4P}(q^2) \,,
\,\,\, \frac{n \cdot p} {m_B + m_M} \, f_{B M, \, \rm NLP}^{T, \, \rm 4P}(q^2) \right \}  \,
\nonumber \\
&& =  {2 \, \pi \over 3} \, \frac{\alpha_s(\mu)  \, C_F  \, \langle \bar q^{\prime} q^{\prime}  \rangle}{n \cdot p}  \,
{\cal F}_B(\mu) \,
\bigg \{  \int_{\omega_s}^{+\infty} \,  {d \omega  \over \omega^2}  \,
\left [  {2 \, \omega_M \over \omega} \, \left (  {\rm exp} \left (- {\omega_s \over  \omega_M} \right )  -1 \right )
+ 1 - {\langle \bar q  q  \rangle \over \langle \bar q^{\prime} q^{\prime}  \rangle}   \right ]
 \phi_B^{+}(\omega, \mu)
\nonumber \\
&& \hspace{0.5 cm}
+ \, \int_0^{\omega_s} \,  {d \omega  \over \omega^2}  \,
\left [ \left ( 1 - { 2 \, \omega_M \over \omega}
- {\langle \bar q  q  \rangle \over \langle \bar q^{\prime} q^{\prime}  \rangle} \right )
+ \left (  1 + {2 \, \omega_M \over \omega}
+ {\langle \bar q  q  \rangle \over \langle \bar q^{\prime} q^{\prime}  \rangle} \,
\left (1 + {\omega \over \omega_M} \right ) \right ) \,
{\rm exp} \left (- {\omega \over  \omega_M} \right )   \right ] \,
\nonumber \\
&& \hspace{0.5 cm}
\,\, \times \, \phi_B^{+}(\omega, \mu)  \,
+  \, {\langle \bar q  q  \rangle \over \langle \bar q^{\prime} q^{\prime}  \rangle} \,
\left [ {\rm exp} \left (- {\omega_s \over  \omega_M} \right ) \,
{\phi_B^{+}(\omega_s, \mu)  \over \omega_s} \right ]
\bigg \}   + {\cal O}(\alpha_s^2) \,.
\label{LCSR for the 4P corrections}
\end{eqnarray}
In contrast with the previously determined subleading-twist corrections to
the exclusive  $B \to \gamma \ell \bar \nu_{\ell}$  decay form factors \cite{Beneke:2018wjp},
these non-valence  Fock-state contributions  appear to preserve  the (classical) large-recoil symmetry relations
between the semileptonic $B_{d, s}  \to \pi, K$  form factors,
according to the newly established LCSR (\ref{LCSR for the 4P corrections}).

Collecting the  different pieces together, we can now summarize the eventual NLP sum rules
for the exclusive bottom-meson decay form factors at small momentum transfer
\begin{eqnarray}
f_{B M, \, \rm NLP}^{i}  =  f_{B M, \, \rm NLP}^{i, \, \rm hc}
+  f_{B M, \, \rm NLP}^{i, \, \rm (A2)}
+ f_{B M, \, \rm NLP}^{i, \, \rm HT}
+  f_{B M, \, \rm NLP}^{i, \, \rm 4P} \,,
\hspace{0.5 cm}
(i=+, \, -, \, T)
\label{NLP sum rules: summary}
\end{eqnarray}
where the analytical expressions for the individual terms on the right-handed side
have been displayed in (\ref{NLP LCSR from the hc propagator}),
(\ref{NLP LCSR from the subleading SCET current}),
(\ref{NLP LCSR from the 2P and 3P higher-twist effects})
and (\ref{LCSR for the 4P corrections}), respectively.

\section{Numerical analysis}
\label{section: numerical analysis}

Having at our disposal the improved sum rules for the exclusive $B_{d, s}  \to \pi, K$ form factors
at large hadronic recoil including both the leading-power spectator-quark mass corrections at ${\cal O}(\alpha_s)$
and the newly derived NLP contributions from four distinct dynamical sources,
we are now prepared to explore their numerical implications on a variety of the phenomenological observables
for the semileptonic $B \to \pi \ell \bar \nu_{\ell}$  and $B_s \to K \ell \bar \nu_{\ell}$ decays  (with  $\ell= \mu, \, \tau$)
as well as the theoretically cleanest electroweak penguin $B \to K \nu_{\ell} \bar \nu_{\ell}$ decay processes.
To achieve this goal, we will first specify the essential theory inputs (for instance, the electroweak SM parameters,
the bottom-quark mass,  the pseudoscalar-meson decay constants,
both the leading-twist and higher-twist HQET distribution amplitudes,
the intrinsic sum rule parameters) appearing in the obtained expressions for the heavy-to-light bottom-meson form factors.
In particular, we will extrapolate the updated LCSR predictions of the considered form factors to the entire kinematic region
by performing the numerical fits of the series coefficients in the conventional BCL expansions \cite{Lellouch:1995yv,Bourrely:2005hp,Bourrely:2008za},
taking into account further the available lattice QCD results at large momentum transfer.
An emphasis will be placed on the very impacts of the newly achieved LCSR predictions
on pinning down the theory uncertainties  of the exclusive $B_{d, s}  \to \pi, K$ form factors
by  carrying out the analogous BCL  fits merely to the numerical lattice QCD determinations
in the lower-recoil region.

\subsection{Theory inputs}

\begin{table}
\centering
\renewcommand{\arraystretch}{2.0}
\resizebox{\columnwidth}{!}{
\begin{tabular}{|l|ll||l|ll|}
\hline
\hline
  Parameter
& Value
& Ref.
&  Parameter
& Value
& Ref.
\\
\hline
\hline
  $G_F$                                         & $1.166379 \times 10^{-5} \,\, {\rm GeV}^{-2} $
                                                                                                  & \cite{ParticleDataGroup:2022pth} 
& $\alpha_s^{(5)}(m_Z)$                          & $0.1179 \pm 0.0009$                            & \cite{ParticleDataGroup:2022pth} 
\\
   $m_{\mu}$                                     &  $105.658$ MeV                                 &  \cite{ParticleDataGroup:2022pth}
&  $m_{\tau}$                                    &  $1776.86 \pm 0.12$ MeV                        &  \cite{ParticleDataGroup:2022pth}
\\
\hline
\hline
  $\overline{m}_b (\overline{m}_b)$              & $4.203 \pm 0.011$  GeV                         & \cite{ParticleDataGroup:2022pth} 
& $m_b^{\rm PS} (2 \, {\rm GeV})$                & $4.532^{+0.013}_{-0.039}$  GeV                 & \cite{Beneke:2014pta} 
\\
  $m_{B_d}$                                     &  $5279.66 \pm 0.12$ MeV                         &  \cite{ParticleDataGroup:2022pth}
& $\tau_{B_d}$                                  &  $(1.519 \pm 0.004)$ ps                         &  \cite{ParticleDataGroup:2022pth}
\\
  $m_{B_s}$                                     &  $5366.92 \pm 0.10$ MeV                         &  \cite{ParticleDataGroup:2022pth}
& $\tau_{B_s}$                                  &  $(1.527 \pm 0.011)$ ps                         &  \cite{ParticleDataGroup:2022pth}
\\
  $f_{B_d}|_{N_f = 2+1+1}$                      &  $190.0 \pm 1.3$ MeV                            & \cite{FlavourLatticeAveragingGroupFLAG:2021npn}
& $f_{B_s}|_{N_f = 2+1+1}$                      &  $230.3 \pm 1.3$ MeV                            &  \cite{FlavourLatticeAveragingGroupFLAG:2021npn}
\\
  $m_{B^{\ast} (1-)}$                          &  $5324.70 \pm 0.21$ MeV                          & \cite{ParticleDataGroup:2022pth}
& $m_{B_s^{\ast} (0+)}$                        &  $5415.4^{+1.8}_{-1.5}$ MeV                      &  \cite{ParticleDataGroup:2022pth}
\\
  $m_{B^{\ast} (0+)}$                          &  $5627 \pm 35$ MeV                            & \cite{Bardeen:2003kt}
& $m_{B_s^{\ast} (0+)}$                        &  $5718 \pm 35$ MeV                            &  \cite{Bardeen:2003kt}
\\
\hline
\hline
  $\overline{m}_u (2 \, {\rm GeV})$              & $2.20 \pm 0.08$  MeV                         & \cite{ParticleDataGroup:2022pth} 
& $\overline{m}_d (2 \, {\rm GeV})$              & $4.69 \pm 0.05$  MeV                         & \cite{ParticleDataGroup:2022pth} 
\\
  $\overline{m}_s (2 \, {\rm GeV})$              & $93.1 \pm 0.6$  MeV                         & \cite{ParticleDataGroup:2022pth} 
&              &                                 & 
\\
$m_{\pi}$              & $139.57$  MeV                         & \cite{ParticleDataGroup:2022pth} 
&  $f_{\pi}$           & $130.2 \pm 1.2$  MeV                  & \cite{ParticleDataGroup:2022pth}
\\
$m_{K}$              & $493.677$  MeV                         & \cite{ParticleDataGroup:2022pth} 
&  $f_{K}$           & $155.7 \pm 0.3$  MeV                  & \cite{ParticleDataGroup:2022pth}
\\
\hline
\hline
  $\lambda_{B_d}(\mu_0)$                        & $(350 \pm 150)$ MeV                             & \cite{Beneke:2020fot,Wang:2021yrr}  
 &                                              & $\{0.7, \, 6.0\}$                                &
\\
  $\lambda_E^2(\mu_0)/\lambda_H^2(\mu_0)$       & $0.50 \pm 0.10$                                & \cite{Beneke:2018wjp}
& $ \{\widehat{\sigma}_{B_{d, s}}^{(1)}(\mu_0), \,
\widehat{\sigma}_{B_{d, s}}^{(2)}(\mu_0)\}$          & $\{0.0, \, \pi^2/6\}$                            & \cite{Beneke:2020fot}
\\
  $2 \, \lambda_E^2(\mu_0) + \lambda_H^2(\mu_0)$       & $(0.25 \pm 0.15) \, {\rm GeV^2}$         & \cite{Beneke:2018wjp}
&                                                &  $\{-0.7, \, -6.0\}$                             &
\\
  $\lambda_{B_s}(\mu_0)$                        & $(400 \pm 150)$ MeV                             & \cite{Beneke:2020fot}  
 &                                              &                                &
 \\
\hline
\hline
$\lambda$  & $0.2250 \pm   0.0006$   &   \cite{ParticleDataGroup:2022pth}  &   $A$
&  $ 0.826^{+0.018}_{-0.015}$    & \cite{ParticleDataGroup:2022pth}
\\
$\bar \rho$ & $ 0.159 \pm 0.010$    &    \cite{ParticleDataGroup:2022pth}   & $\bar \eta$
& $0.348 \pm 0.01$  &\cite{ParticleDataGroup:2022pth}
\\
\hline
\hline
$s_{0}^{\pi}$          & $\left \{ (0.70 \pm 0.05)  \, {\rm GeV^2} \right \} $                   & \cite{Lu:2018cfc,Khodjamirian:2006st} 
&  $M^2$               &  $(1.25 \pm 0.25) \, {\rm GeV^2} $                   & \cite{Lu:2018cfc,Khodjamirian:2006st}
\\
$s_{0}^{K}$            & $\left \{ (1.05 \pm 0.05)  \, {\rm GeV^2} \right \} $                   & \cite{Lu:2018cfc,Khodjamirian:2006st} 
&                      &                  & 
\\
$ \langle \bar q  q \rangle  (2 \, {\rm GeV})$          & $ -(286 \pm 23) \, {\rm MeV^3}$                   & \cite{FlavourLatticeAveragingGroupFLAG:2021npn} 
&  $ \langle \bar s s \rangle : \langle \bar q  q \rangle $              &  $0.8 \pm 0.1$                   & \cite{Ioffe:2002ee,Gelhausen:2013wia}
\\
\hline
\hline
\end{tabular}
}
\renewcommand{\arraystretch}{1.0}
\caption{Numerical values of the theory input parameters employed in the LCSR determinations
of the exclusive $B_{d, s}  \to \pi, K$ form factors as well as  the subsequent phenomenological analysis
for the semileptonic bottom-meson decay observables. }
\label{table: theory inputs}
\end{table}

We summarize explicitly the numerical values of the necessary SM inputs and  the  hadronic parameters in
Table \ref{table: theory inputs}.
We will adopt the three-loop evolution of the strong coupling constant $\alpha_s(\mu)$ in the ${\rm \overline{MS}}$ scheme
by taking the determined  interval $\alpha_s^{(5)}(m_Z)$ from \cite{ParticleDataGroup:2022pth}
and employing the perturbative matching scales $\mu_{4} =4.8 \, {\rm GeV}$ and $\mu_{4} =1.3 \, {\rm GeV}$
for crossing  $n_f=4$ and $n_f=3$, respectively \cite{Shen:2020hfq,Beneke:2020fot}.
In addition, the bottom-quark mass entering the short-distance coefficient functions of the obtained SCET sum rules
is generally understood to be the pole mass on account of the on-shell kinematics.
Converting the precisely known ${\rm \overline{MS}}$ mass to the counterpart  pole scheme will, however, bring about
the numerical results sensitive to the truncation order of  the perturbative matching relation
due to the existence of an infrared renormalon \cite{Bigi:1994em,Beneke:1994sw}.
Consequently, we will take advantage of  the potential-subtracted (PS) renormalization scheme \cite{Beneke:1998rk}
for the bottom-quark mass (see \cite{El-Khadra:2002zzm} for an overview of several popular definitions for the heavy-quark mass)
and then perform the scheme conversion  of the hard functions from the pole mass
to the PS mass scheme.
In addition, we will employ the four-flavour lattice-computation results \cite{FlavourLatticeAveragingGroupFLAG:2021npn}
for the leptonic decay constants of bottom mesons
in the ${\rm SU}(2)$ isospin-symmetric limit
(see \cite{Bazavov:2017lyh}  for additional discussions on the strong-isospin breaking corrections).
Following the theory prescription displayed in \cite{ParticleDataGroup:2022pth},
the adopted pion decay constant $f_{\pi}$ corresponds to  the three-flavour FLAG 2021 average
\cite{FlavourLatticeAveragingGroupFLAG:2021npn}
with the uncertainty increased by including the $0.7 \, \%$ charm sea-quark contribution.

We now turn to discuss the acceptable phenomenological models for the two-particle and three-particle
bottom-meson distribution amplitudes in HQET, fulfilling the nontrivial constraints from the classical equations of motion
and the expected asymptotic behaviours at small quark and gluon momenta from the conformal symmetry analysis.
For definiteness, we will employ the newly proposed three-parameter ans\"{a}tz for the two-particle distribution amplitudes
in  coordinate space \cite{Beneke:2018wjp}
(see \cite{Feldmann:2022uok} for an alternative parametrization in terms of an expansion in associated Laguerre polynomials)
\begin{eqnarray}
\eta_{+}(s, \mu_0)= {}_{1}F_{1}(\alpha; \beta; -s \, \omega_0) \,,
\qquad
\eta_{3}^{(0)}(s, \mu_0)= - {\lambda_E^2 - \lambda_H^2  \over 18} \, s^2 \,
\left [  {}_{1}F_{1}(\alpha+2; \beta+2; -s \, \omega_0) \right ]\,,
\hspace{1.0 cm}
\label{three-parameter model for the 2P DAs}
\end{eqnarray}
which allows us to construct the analytical solutions to the Lange-Neubert evolution equations in the one-loop approximation
\cite{Gao:2021sav}
\begin{eqnarray}
\phi_B^{+}(\omega, \mu) &=& \hat{U}_{\phi}^{\rm tw2}(\mu, \mu_0) \,\, {1 \over \omega^{\kappa_s+1}} \,\,
{\Gamma(\beta) \over \Gamma(\alpha) } \, {\cal G}(\omega, \alpha, \beta;  0, 2, 1)\,,
 \\
\phi_B^{-}(\omega, \mu) &=& \hat{U}_{\phi}^{\rm tw2}(\mu, \mu_0) \,\, {1 \over \omega^{\kappa_s+1}} \,\,
{\Gamma(\beta) \over \Gamma(\alpha) } \, {\cal G}(\omega, \alpha, \beta;  0, 1, 1)
\nonumber \\
&& + \, \hat{U}_{\phi}^{\rm tw3}(\mu, \mu_0) \,\,
\left [ - {\lambda_E^2(\mu) - \lambda_H^2(\mu)  \over 18 }  \right ]
\,\, {1 \over \omega^{\kappa_s+3}} \,\, 
{\Gamma(\beta+2) \over \Gamma(\alpha+2) } \,\,
\bigg \{ {\cal G}(\omega, \alpha, \beta;  0, 3, 3)
 \\
&& + \, \left (\beta - \alpha \right ) \,
\left [ {\omega \over \omega_0} \, {\cal G}(\omega, \alpha, \beta;  0, 2, 2)
- \beta \, {\omega  \over \omega_0} \,  {\cal G}(\omega, \alpha, \beta;  1, 2, 2)
- {\cal G}(\omega, \alpha, \beta;  1, 3, 3) \right ] \bigg \}  \,.
\nonumber
\end{eqnarray}
The manifest expressions for the two evolution functions $\hat{U}_{\phi}^{\rm tw2}$
and $\hat{U}_{\phi}^{\rm tw3}$ in momentum space can be further written as
\begin{eqnarray}
\hat{U}_{\phi}^{\rm tw2}(\mu, \mu_0) &=&  {\rm exp} \bigg  \{ -{ \Gamma_{\rm cusp}^{(0)}  \over 4 \, \beta_0^2} \,
\bigg  [ {4 \, \pi \over \alpha_s(\mu_0)} \,
\left ( \ln z_0 - 1 +{1 \over z_0}  \right ) - {\beta_1 \over 2 \beta_0^2} \, \ln^2 z_0
\nonumber \\
&& + \left ( {\Gamma_{\rm cusp}^{(1)} \over \Gamma_{\rm cusp}^{(0)}} - {\beta_1 \over 2 \beta_0} \right ) \,
\left [ z_0 - 1 - \ln z_0 \right ] \bigg  ]  \bigg  \}
\, \left ( e^{2 \, \gamma_E} \, \mu_0 \right )^{\Gamma_{\rm cusp}^{(0)} \, \ln z_0 / (2 \, \beta_0)} \,\,
z_0^{\gamma_{\rm tw2}^{(0)} / (2 \, \beta_0)} \,,
\hspace{1.0 cm}
 \\
\hat{U}_{\phi}^{\rm tw3}(\mu, \mu_0) &=& z_0^{\gamma_{\rm tw3}^{(0)} / (2 \, \beta_0)} \,
\hat{U}_{\phi}^{\rm tw2}(\mu, \mu_0)  \,.
\end{eqnarray}
Moreover, we have introduced the following conventions for the expansion coefficient $\kappa_s$
as well as the Meijer ${\cal G}$-function \cite{Luke:1969}
\begin{eqnarray}
\kappa_s = {\Gamma_{\rm cusp}^{(0)}  \over 2 \, \beta_0}  \,
\ln {\alpha_s(\mu) \over \alpha_s(\mu_0)} \,,
\qquad
{\cal G}(\omega, \alpha, \beta;  l, m, n)
= G^{21}_{23}\bigg({}^{1, \, \beta+l}_{\kappa_s + m, \, \alpha, \, \kappa_s+n} \, \bigg| \frac{\omega}{\omega_0}\, \bigg )  \,.
\end{eqnarray}
The appearing HQET parameters $\lambda_E^2$ and  $\lambda_H^2$ can be defined by the effective matrix elements
of the chromo-electric and chromo-magnetic operators \cite{Grozin:1996pq}
\begin{eqnarray}
&& \langle 0 | \bar q^{\prime} (0) \, g_s \, G_{\mu \nu} \, \Gamma \, h_v(0)| \bar B(v) \rangle
\nonumber \\
&& = - {{\cal F}_{B}(\mu) \, m_{B} \over 6} \,
{\rm Tr} \left \{ \gamma_5 \,\, \Gamma \,\,
\left (  {1 + \slashed v \over 2} \right ) \,
\left [ \lambda_H^2  \, \left (  i \, \sigma_{\mu \nu} \right )
+ (\lambda_H^2 - \lambda_E^2) \,
\left ( v_{\mu} \, \gamma_{\nu} - v_{\nu} \, \gamma_{\mu} \right ) \right ]  \right \}   \,.
\end{eqnarray}
Solving the  RG evolution equations for these two hadronic quantities  $\lambda_E^2$ and  $\lambda_H^2$ at one loop
\begin{eqnarray}
{d \over d \ln \mu} \,
\left(
\begin{array}{c}
\lambda_E^2(\mu) \\
\lambda_H^2(\mu) \\
\end{array}
\right)
+  {\alpha_s(\mu) \over 4 \pi} \, \gamma_{\rm EH}^{(0)}  \,
\left(
\begin{array}{c}
\lambda_E^2(\mu) \\
\lambda_H^2(\mu) \\
\end{array}
\right) = 0
\,,
\label{1-loop evolution equations of lambdaE and lamdbaH}
\end{eqnarray}
with the available anomalous dimension matrix from \cite{Grozin:1996hk,Nishikawa:2011qk}
\begin{eqnarray}
\gamma_{\rm EH}^{(0)} =
\left(
                 \begin{array}{cc}
                   {8 \over 3} \, C_F + {3 \over 2}  \, N_c \,\,  &  \,\,  {4 \over 3} \, C_F - {3 \over 2}  \, N_c \\
                   {4 \over 3} \, C_F - {3 \over 2}  \, N_c \,\,  & \,\,  {8 \over 3} \, C_F + {5 \over 2}  \, N_c \\
                 \end{array}
               \right)
\,,
\end{eqnarray}
we can readily determine their renormalization-scale dependencies by diagonalizing the renormalization kernel
at the leading-logarithmic accuracy.
The method of two-point QCD sum rules has been applied to estimate these HQET parameters repeatedly,
yielding the numerical predictions significantly different each other even with the sizeable theory uncertainties
\begin{eqnarray}
\left \{  \lambda_E^2({1\, \rm GeV}),  \,\, \lambda_H^2({1\, \rm GeV}) \right \}
= \left\{
\begin{array}{l}
\left \{  (0.11 \pm 0.06) \, {\rm GeV^2}, \,\, (0.18 \pm 0.07) \, {\rm GeV^2}  \right \}  \,,
\qquad \hspace{0.5 cm} \text{\cite{Grozin:1996pq}}
\vspace{0.5 cm} \\
\left \{  (0.03 \pm 0.02) \, {\rm GeV^2}, \,\, (0.06 \pm 0.03) \, {\rm GeV^2}  \right \}    \,,
 \qquad  \hspace{0.5 cm}
\text{\cite{Nishikawa:2011qk}}
\vspace{0.5 cm} \\
\left \{  (0.01 \pm 0.01) \, {\rm GeV^2}, \,\, (0.15 \pm 0.05) \, {\rm GeV^2}  \right \}    \,.
 \qquad  \hspace{0.5 cm}
\text{\cite{Rahimi:2020zzo}}
\end{array}
\hspace{0.8 cm} \right.
\label{sum rule results  for lambdaE and lambdaH}
\end{eqnarray}
The substantial discrepancies of the obtained  numerical results between \cite{Grozin:1996pq}
and \cite{Nishikawa:2011qk} can be traced back to the remarkable perturbative corrections
to the short-distance Wilson coefficients for the dimension-five quark-gluon mixed condensate
$\langle \bar q \, \sigma_{\mu \nu} \, G^{\mu \nu} q \rangle$
and to the further inclusion of the particular higher-order power corrections at tree level
from  the dimension-six vacuum condensates in the factorization approximation in the latter.
On the other hand, the authors of \cite{Rahimi:2020zzo} proposed to
employ an alternative diagonal correlation function of the two three-body HQET currents
(instead of the non-diagonal current-current correlator implemented in \cite{Grozin:1996pq,Nishikawa:2011qk})
for constructing the desired sum rules of the essential ingredients  $\lambda_E^2$ and  $\lambda_H^2$,
in an attempt to pin down  the systematic uncertainty from the parton-hadron duality.
However, such attractive benefits are unfortunately  achieved at the price of worsening
the operator-product-expansion (OPE) convergence in the partonic computation of the new correlation function
and enhancing the intricate contributions from the continuum and higher excited states
(see \cite{Rahimi:2020zzo} for more detailed discussions).
In the absence of the satisfactory determinations of these two HQET quantities,
we will therefore take the numerical intervals for the two combinations $2 \, \lambda_E^2 + \lambda_H^2$
and $\lambda_E^2/\lambda_H^2$ displayed in Table \ref{table: theory inputs},
covering the allowed ranges of the previously obtained results from \cite{Grozin:1996pq,Nishikawa:2011qk}
and simultaneously satisfying the  derived (conservative) upper bounds from \cite{Rahimi:2020zzo}
due to the positive definite spectral densities.

Applying the customary  definitions of the inverse logarithmic moments for
the leading-twist bottom-meson distribution amplitude \cite{Beneke:2018wjp,Beneke:2020fot,Shen:2020hfq}
\begin{eqnarray}
\frac{1}{\lambda_{B}(\mu)} &=& \int_0^{\infty} \, {d \omega \over \omega} \, \phi_B^{+}(\omega, \mu)  \,,
\qquad
\frac{\widehat{\sigma}_{B}^{(n)} (\mu)}{\lambda_{B}(\mu)} =
\int_{0}^{\infty} \, {d \omega \over \omega} \,
\left [ \ln \left ( {\lambda_{B}(\mu_0) \over \omega} \right )
- \gamma_E \right ]^{n} \, \phi_{B}^{+}(\omega, \mu) \,,
\hspace{1.0 cm}
\end{eqnarray}
we can immediately determine these fundamental nonperturbative quantities in terms of
the three shape parameters in our model (\ref{three-parameter model for the 2P DAs})
\begin{eqnarray}
\lambda_{B}(\mu_0) &=& \left ( {\alpha-1 \over \beta-1} \right ) \, \omega_0 \,,
\qquad
\widehat{\sigma}_{B}^{(1)} (\mu_0) =  \psi (\beta-1) - \psi (\alpha-1)
+ \ln  \left ( {\alpha-1 \over \beta-1} \right ) \,,
\nonumber \\
\widehat{\sigma}_{B}^{(2)} (\mu_0) &=& \left [  \widehat{\sigma}_{B}^{(1)} (\mu_0)  \right ]^2 +
\psi^{(1)}(\alpha-1) - \psi^{(1)}(\beta-1)  + {\pi^2 \over 6} \,,
\end{eqnarray}
where $\psi(z)$ stands for the digamma function defined by the logarithmic derivative of the $\Gamma$-function.
Adopting the one-loop RG equation for the twist-two bottom-meson distribution amplitude
enables us to derive the leading-logarithmic evolutions for the first few momentums
\begin{eqnarray}
\lambda_{B}(\mu) &=& \lambda_{B}(\mu_0) \,
\left \{ 1 + {\alpha_s(\mu_0) \, C_F \over  \pi} \, \ln {\mu \over \mu_0}  \,
\left [  \widehat{\sigma}_{B}^{(1)} (\mu_0) +  \ln {\sqrt{\mu \mu_0} \, e^{\gamma_E} \over \lambda_{B}(\mu_0)}
 - \left ( {1 \over 2} \right )   \right ] \right \}
+ {\cal O}(\alpha_s^2) \,, \hspace{1.0 cm}
 \\
\widehat{\sigma}_{B}^{(1)} (\mu) &=&  \widehat{\sigma}_{B}^{(1)} (\mu_0)
\, \left \{ 1 +   {\alpha_s(\mu_0) \, C_F \over  \pi} \, \ln {\mu \over \mu_0}  \,
\left [ \widehat{\sigma}_{B}^{(1)} (\mu_0)  -
\frac{\widehat{\sigma}_{B}^{(2)} (\mu_0) }{\widehat{\sigma}_{B}^{(1)} (\mu_0) }  \right ]  \right \}
+ {\cal O}(\alpha_s^2) \,,
 \\
\widehat{\sigma}_{B}^{(2)} (\mu) &=&  \widehat{\sigma}_{B}^{(2)} (\mu_0)
\, \left \{ 1 +   {\alpha_s(\mu_0) \, C_F \over  \pi} \, \ln {\mu \over \mu_0}  \,
\left [  \widehat{\sigma}_{B}^{(1)} (\mu_0)  -
\frac{\widehat{\sigma}_{B}^{(3)} (\mu_0) - 4 \, \zeta_3 }{\widehat{\sigma}_{B}^{(2)} (\mu_0) }   \right ]  \right \}
+ {\cal O}(\alpha_s^2) \,.
\end{eqnarray}
In spite of  the enormous efforts undertaken to determine the first inverse moment $\lambda_B^{-1}(\mu)$
with  different theory prescriptions, we are still unable to accomplish the robust computations of
this fundamental parameter  from the first field-theoretical principles at present
(see however \cite{Wang:2019msf,Kane:2019jtj} for interesting discussions from the lattice QCD perspectives).
Consequently,  we prefer to employ the conservative interval $\lambda_{B_d}(\mu_0) = (350 \pm 150) \, {\rm MeV}$
accommodating the indirect extractions from the radiative leptonic $B \to \gamma \ell \bar \nu$ decay rates
\cite{Wang:2016qii,Wang:2018wfj,Janowski:2021yvz} in the subsequent numerical analysis
and assign further ${\cal O}(15  \, \%)$ SU(3)-flavour symmetry breaking effect for the ratio $\lambda_{B_s}/\lambda_{B_d}$
(characterizing the typical splitting between the constituent down-quark and strange-quark masses).
We mention in passing that these two HQET  parameters have been recently computed from the traditional QCD sum rule approach
by investigating the appropriate correlation function of an effective light-cone heavy-to-light current
and an interpolating current for the pseudoscalar heavy-meson state \cite{Khodjamirian:2020hob},
following closely the theory strategy suggested in \cite{Braun:2003wx,Grozin:1996pq}.
The yielding numerical predictions $\lambda_{B_d}(\mu_0) = (383 \pm 153) \, {\rm MeV}$
and $\lambda_{B_s}/\lambda_{B_d} = 1.19 \pm 0.14$ \cite{Khodjamirian:2020hob}
are apparently in the same ballpark as the corresponding intervals displayed in Table \ref{table: theory inputs}.

The general  ans\"{a}tz for the higher twist distribution amplitudes
incorporating both the anticipated low-momentum behaviour
and the tree-level equations-of-motion constraints
can be constructed by introducing a unique   profile function  \cite{Beneke:2018wjp,Shen:2020hfq,Wang:2021yrr}
\begin{eqnarray}
\Phi_3(\omega_1, \omega_2, \mu_0) &=&
- \left ( {1 \over 2} \right ) \, \varkappa(\mu_0) \,
\left [ \lambda_E^2 (\mu_0) - \lambda_H^2 (\mu_0) \right ] \,
\omega_1 \, \omega_2^2 \, f^{\prime}(\omega_1 + \omega_2) \,,
 \\
\Phi_4(\omega_1, \omega_2, \mu_0) &=&
 \left ( {1 \over 2} \right )  \, \varkappa(\mu_0) \, \left [ \lambda_E^2 (\mu_0) + \lambda_H^2 (\mu_0) \right ] \,
\omega_2^2 \, f(\omega_1 + \omega_2) \,,
 \\
\Psi_4(\omega_1, \omega_2, \mu_0) &=&
\varkappa(\mu_0) \,  \lambda_E^2 (\mu_0) \, \omega_1 \,  \omega_2 \, f(\omega_1 + \omega_2) \,,
 \\
\tilde{\Psi}_4(\omega_1, \omega_2, \mu_0) &=&
\varkappa(\mu_0) \,  \lambda_H^2(\mu_0) \, \omega_1 \,  \omega_2 \, f(\omega_1 + \omega_2) \,,
 \\
\Psi_5(\omega_1,\omega_2, \mu_0) &=&
- \varkappa(\mu_0) \,  \lambda_E^2(\mu_0) \, \omega_2  \, \int_{\omega_1 + \omega_2}^\infty d \eta \, f(\eta),
 \\
\Phi_6(\omega_1, \omega_2, \mu_0) &=&
\varkappa(\mu_0) \, \left [ \lambda_E^2 (\mu_0) - \lambda_H^2 (\mu_0) \right ] \,
\int_{\omega_1 + \omega_2}^\infty d \eta_1 \,  \int_{\eta_1}^{\infty} \, d \eta_2 \, f(\eta_2)\,.
\end{eqnarray}
We further collect the explicit expressions of the two-particle subleading-twist HQET distribution amplitudes
from  the off-light-cone corrections for completeness
\begin{eqnarray}
g_B^{+}(\omega, \mu) &=& \hat{g}_B^{+}(\omega, \mu)
- {1 \over 2} \, \int_0^{\omega} \, d \omega_1 \, \int_0^1 d u \,
{\bar u \over u} \,\, \psi_4 \left (\omega, {\omega - \omega_1 \over u}, \mu \right ) \,,
\nonumber \\
g_B^{-}(\omega, \mu) &=& \hat{g}_B^{-}(\omega, \mu)
- {1 \over 2} \, \int_0^{\omega} \, d \omega_1 \, \int_0^1 d u \,
{\bar u \over u} \,\, \psi_5 \left (\omega, {\omega - \omega_1 \over u}, \mu \right ) \,,
\label{full expressions of gBpm}
\end{eqnarray}
where the twist-five ``Wandzura-Wilczek" term $ \hat{g}_B^{-}(\omega, \mu)$
has been presented in (\ref{def: WW part of gBminus}) and
\begin{eqnarray}
\hat{g}_B^{-}(\omega, \mu) =  \left ( {1 \over 4} \right ) \, \int_{\omega}^{\infty}  \, d \rho \,
\left  \{ (\rho - \omega) \, \left [ \phi_B^{-}(\rho) -  \phi_B^{+}(\rho) \right ]
- 2 \, (\bar \Lambda - \rho)  \, \phi_B^{+}(\rho)  \right \}  \,.
\label{def: WW part of gBplus}
\end{eqnarray}
The particular three-parameter model for the twist-two coordinate-space  LCDA $\eta_{+}(s, \mu_0)$
in (\ref{three-parameter model for the 2P DAs}) implies the following nonperturbative function $f(\omega)$
and the normalization constant $\varkappa(\mu_0)$
\begin{eqnarray}
f(\omega) &=& {\Gamma(\beta) \over \Gamma(\alpha)}  \,
U \left (\beta-\alpha, 3-\alpha, {\omega \over \omega_0} \right ) \,
{1 \over \omega_0^2} \,\, {\rm exp} \left ( - {\omega \over \omega_0} \right ) \,,
\nonumber \\
\varkappa^{-1}(\mu_0) &=& \left ( {1 \over 2} \right ) \, \int_{0}^{\infty} d \omega \, \omega^3 \, f(\omega)
= \left [ {3 \, \alpha \, (\alpha-1) \over  \beta \, (\beta-1) } \right ] \, \omega_0^2  \,.
\end{eqnarray}
Furthermore, we will take the ``effective mass" of the bottom-meson state entering the NLP sum rules
(\ref{NLP LCSR from the hc propagator}) and  (\ref{NLP LCSR from the subleading SCET current})
as $\bar \Lambda = m_{B_q} - m_{b} +{\cal O}(\Lambda_{\rm QCD}^2/m_b)$ with $m_{b} = (4.8 \pm 0.1) \, {\rm GeV}$
numerically (see \cite{Manohar:2000dt} for the yet higher-order correction to this essential mass relation and
\cite{Bosch:2004bt,Feldmann:2014ika} for further discussions on the scheme dependence of this hadronic quantity).

Additionally, we will vary the hard-matching scales $\mu_{h1}$ and  $\mu_{h2}$
appearing in the NLL SCET sum rules for the heavy-to-light bottom-meson transition  form factors (\ref{NLL sum rules at LP I})
and (\ref{NLL sum rules at LP II}) in the interval $[m_b/2, \, 2\, m_b]$ with the central value $m_b$,
as widely implemented in the exclusive heavy-hadron decay phenomenologies \cite{Beneke:2018wjp,Beneke:2020fot,Shen:2020hfq}.
The renormalization  scale $\nu$ of the QCD tensor current will be taken as the hard scale of order of the $b$-quark mass.
By contrast, the factorization scale $\mu$ will be treated as the hard-collinear scale
in the  range of $(1.5 \pm 0.5) \, {\rm GeV}$.

Following the standard procedure outlined in \cite{Wang:2015vgv},
the determinations of two intrinsic  LCSR parameters $\omega_M$ and $\omega_s$ can be routinely achieved by
imposing the necessary constraints on the smallness of the continuum contributions in the dispersion integrals
of the three invariant functions $\Pi$, $\tilde{\Pi}$ and $\Pi_{\rm T}$
and on the stability of the obtained sum rules against the variation of the Borel mass $\omega_M$.
Proceeding with this practical prescription leads to the following intervals
\begin{eqnarray}
M^2 &=& n \cdot p \, \omega_M = (1.25 \pm 0.25) \, {\rm GeV^2}  \,,
\qquad
s_0^{\pi} =  n \cdot p \, \omega_s = (0.70 \pm 0.05) \, {\rm GeV^2} \,,
\nonumber \\
s_0^{K} &=&  n \cdot p \, \omega_s = (1.05 \pm 0.05) \, {\rm GeV^2} \,,
\end{eqnarray}
which are in excellent agreement with the numerical results employed in the LCSR computations
of the pion-photon form factor \cite{Wang:2017ijn} as well as the pion electromagnetic form factor \cite{Braun:1999uj},
and in the two-point QCD sum rules for the $K$-meson decay constant \cite{Khodjamirian:2003xk}.

\subsection{Numerical  predictions for the $B_{d, s}  \to \pi, K$ form factors}
\label{subsection: numerics for the bottom-meson form factors}

\begin{figure}
\begin{center}
\includegraphics[width=0.45 \columnwidth]{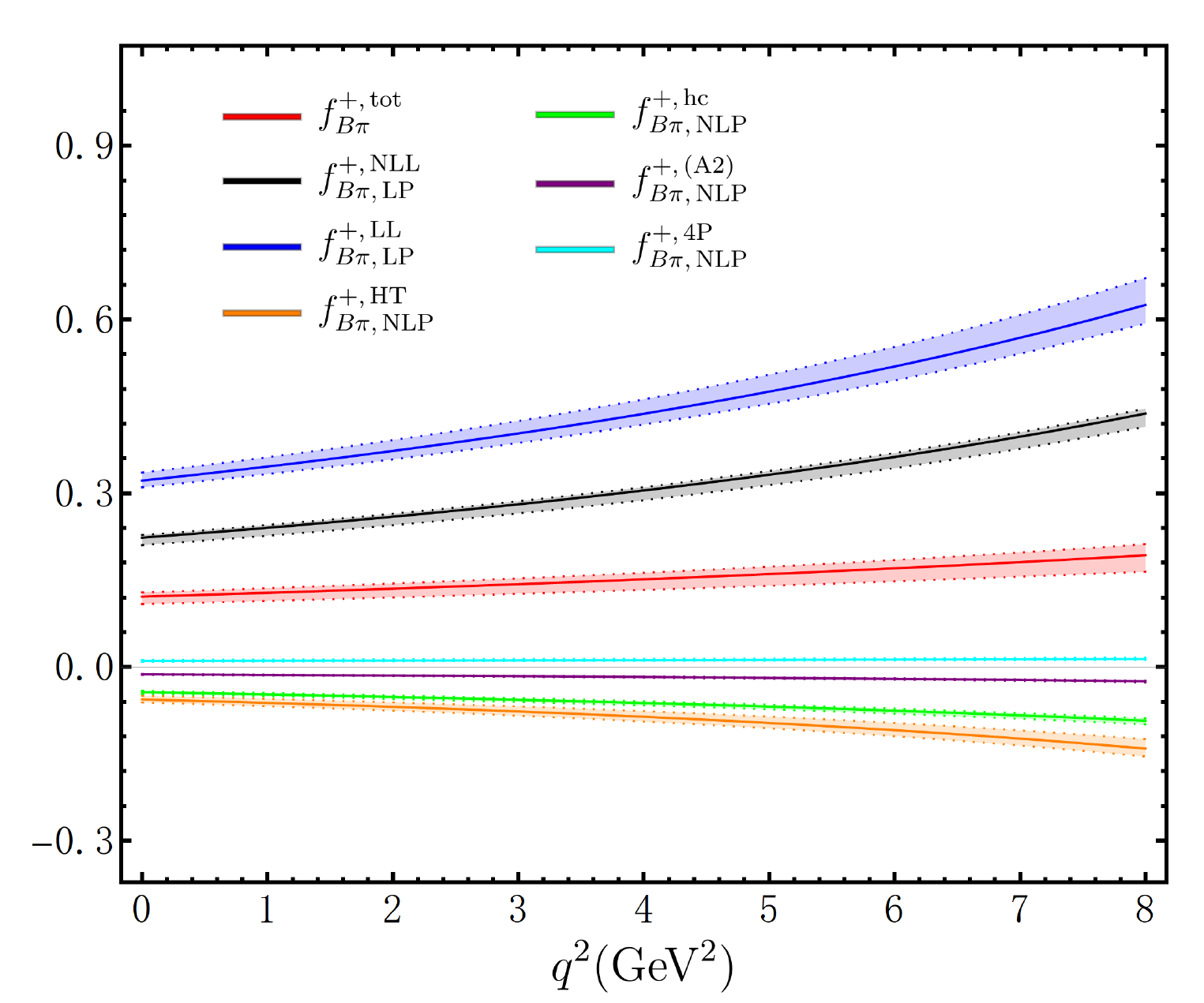}
\hspace{1.0 cm}
\includegraphics[width=0.45 \columnwidth]{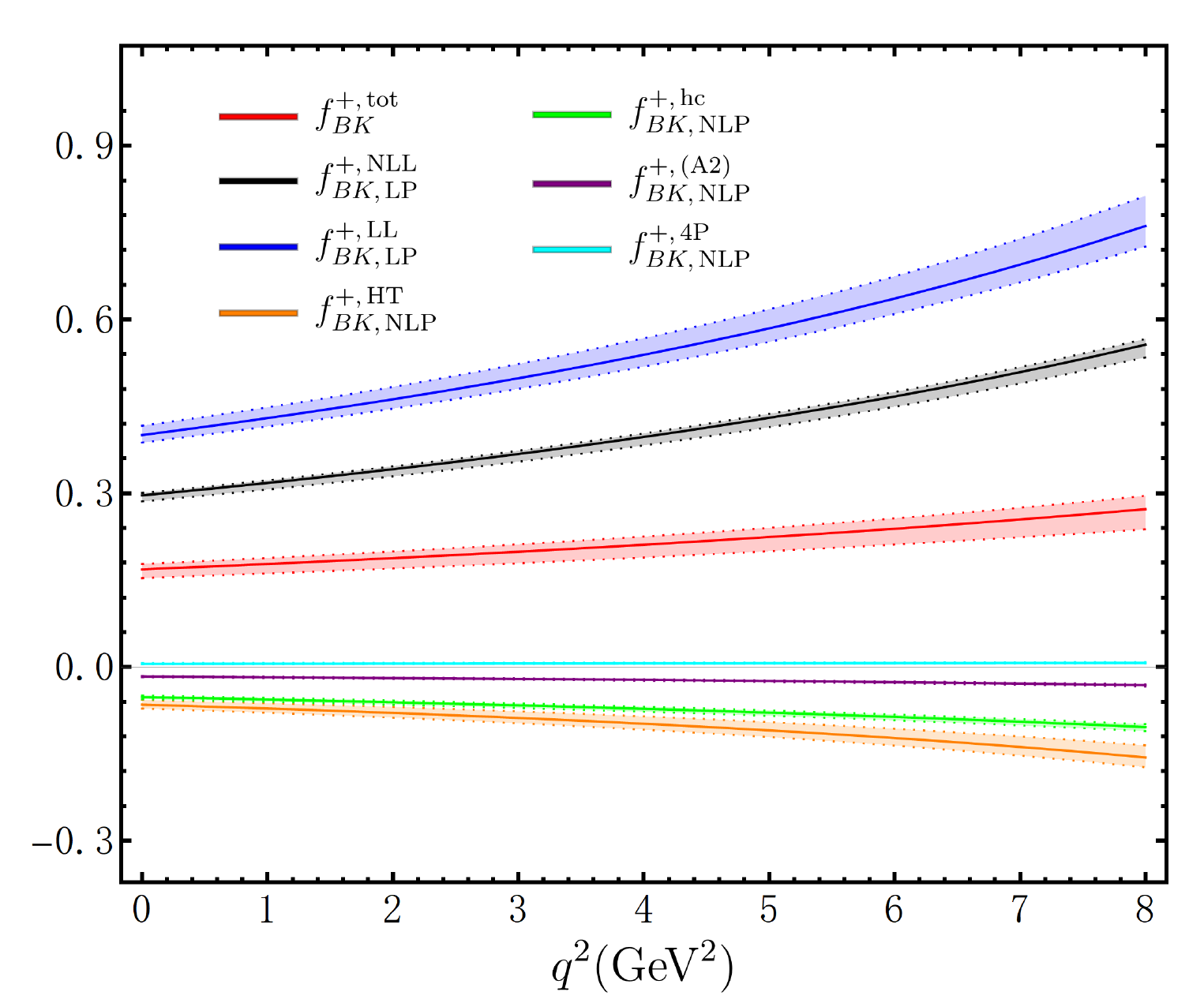}
\\
\includegraphics[width=0.45 \columnwidth]{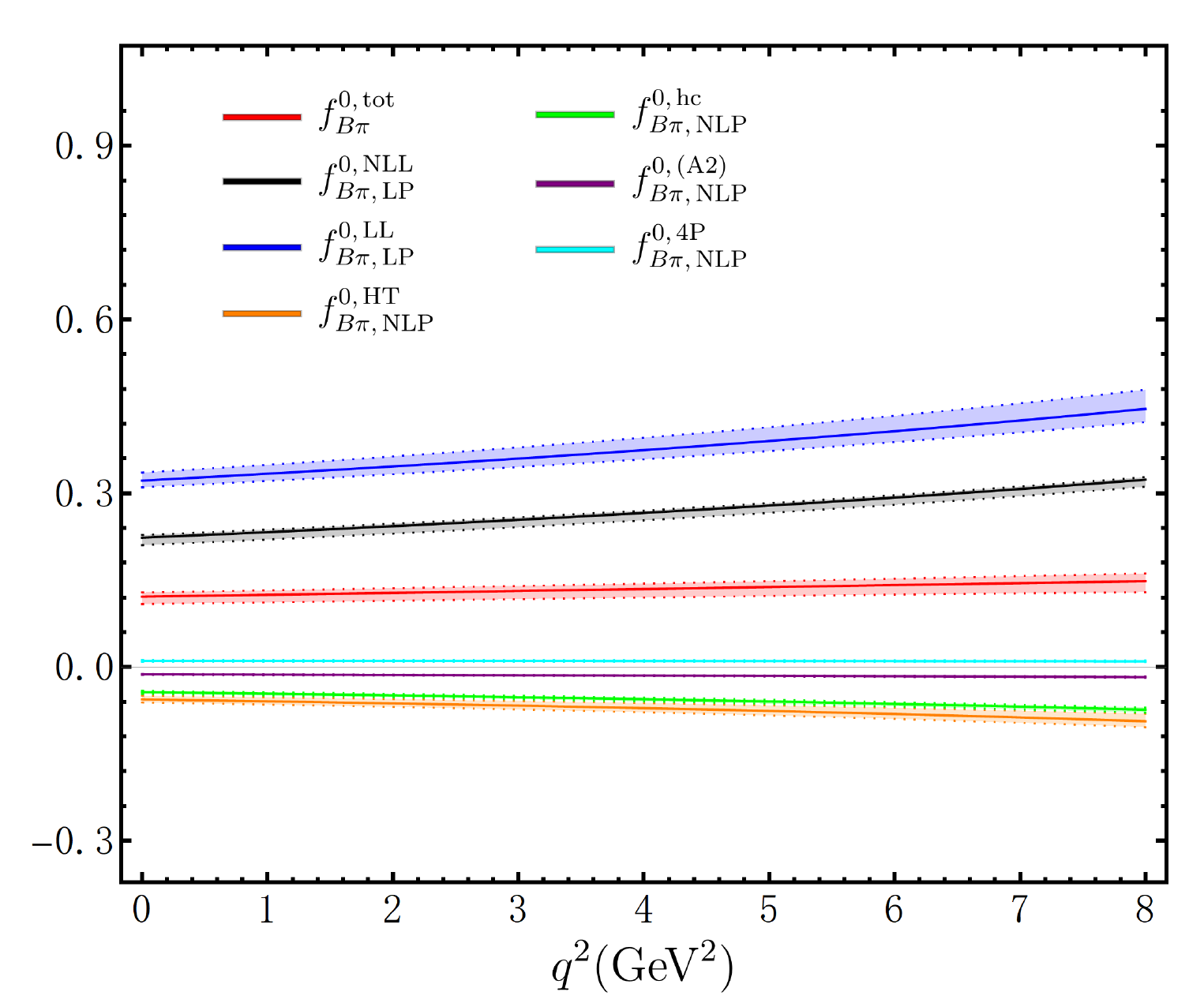}
\hspace{1.0 cm}
\includegraphics[width=0.45 \columnwidth]{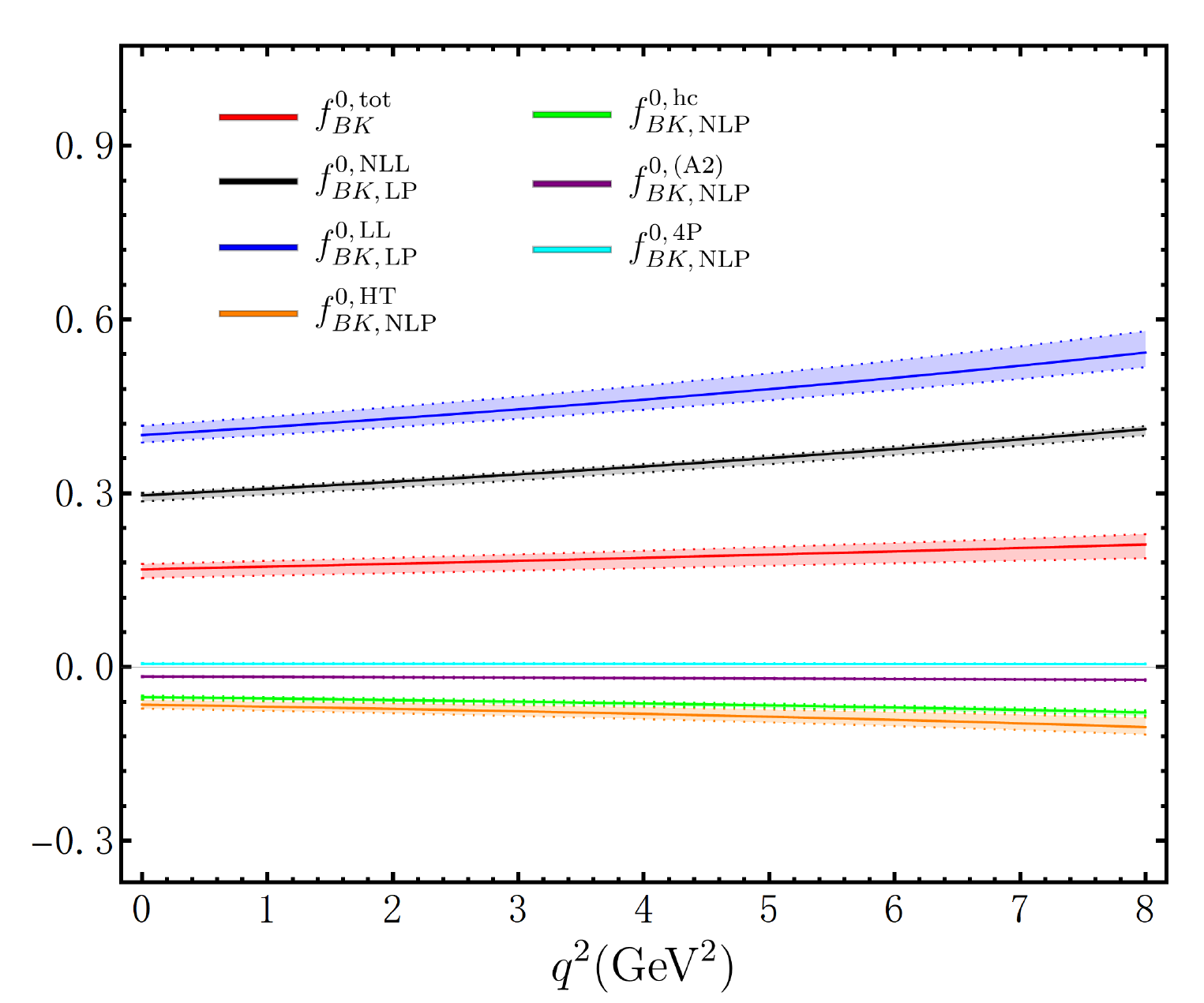}
\\
\includegraphics[width=0.45 \columnwidth]{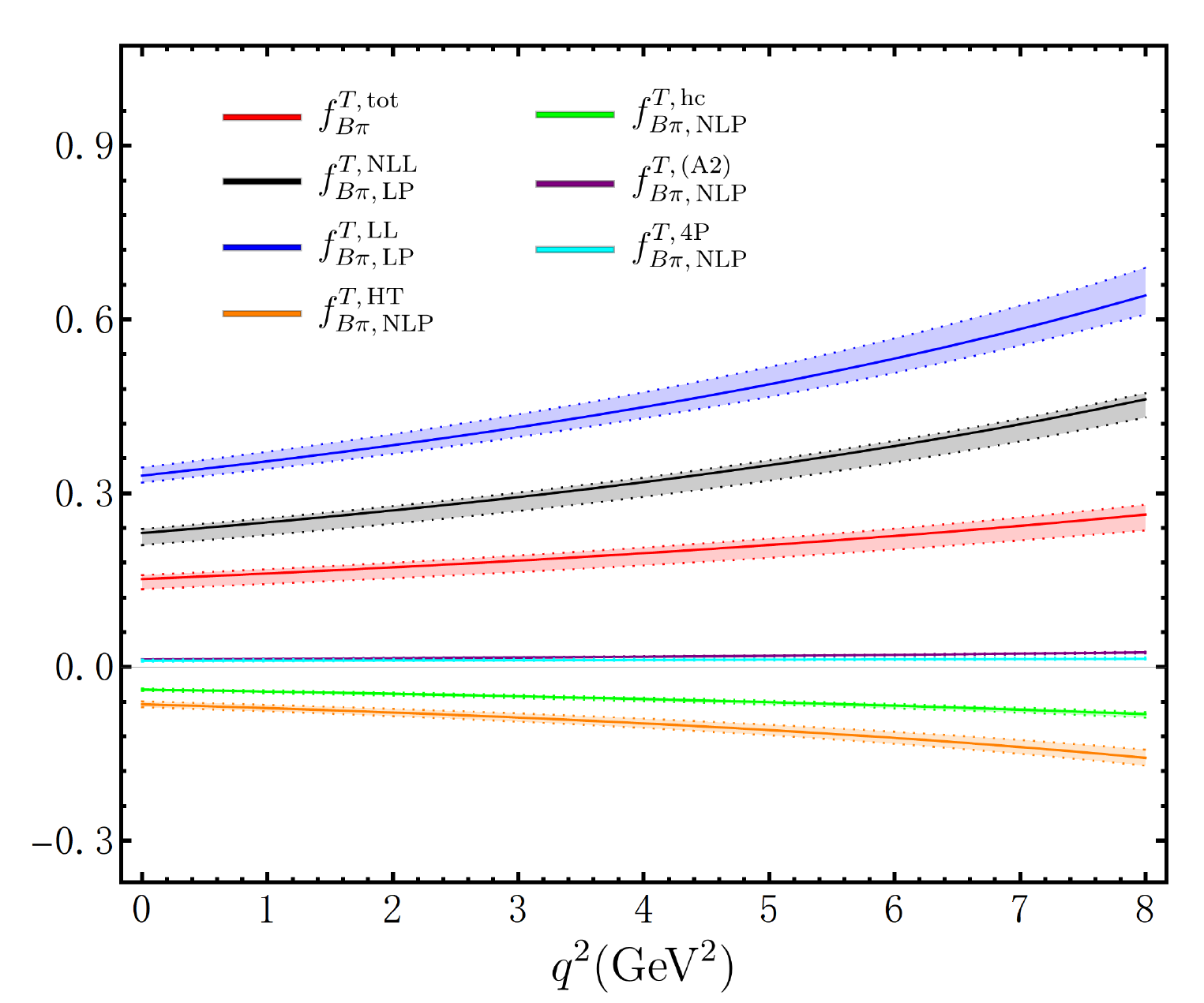}
\hspace{1.0 cm}
\includegraphics[width=0.45 \columnwidth]{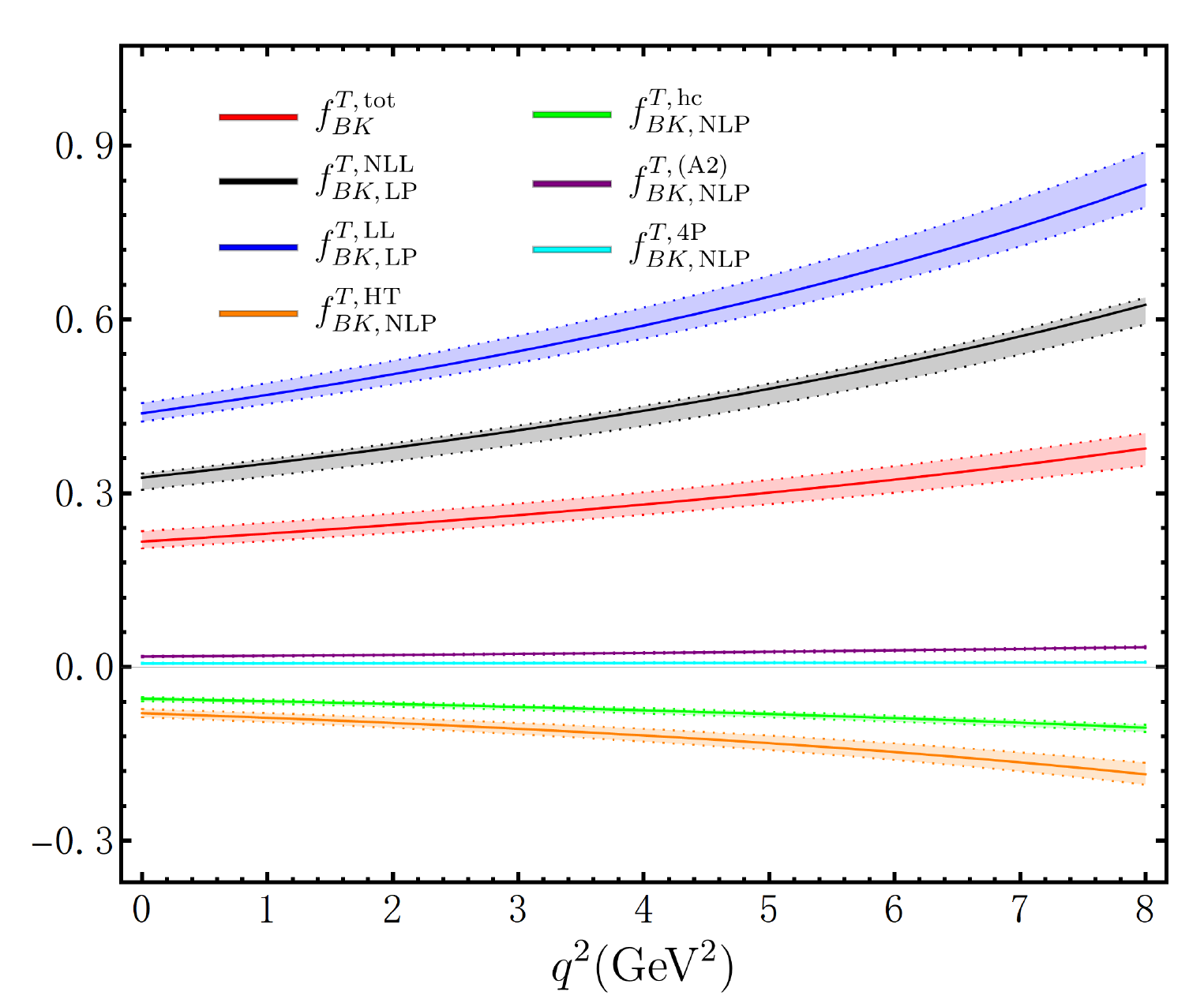}
\vspace*{0.1cm}
\caption{Breakdown of the distinct dynamical mechanisms contributing to
the semileptonic $B \to \pi \ell \bar \nu_{\ell}$ form factors (left panel)
and to the electroweak penguin  $B \to K \nu_{\ell} \bar \nu_{\ell}$ decay form factors (right panel)
in the kinematic region $0 \leq q^2 \leq 8 \, {\rm GeV^2}$
from the updated LCSR computations with the HQET bottom-meson distribution amplitudes,
where the perturbative uncertainties due to the variations of both the hard and hard-collinear matching scales
are indicated by the individual error bands.}
\label{fig: distinct pieces of B to pi and B to K form factors in LCSR}
\end{center}
\end{figure}

\begin{figure}
\begin{center}
\includegraphics[width=0.45 \columnwidth]{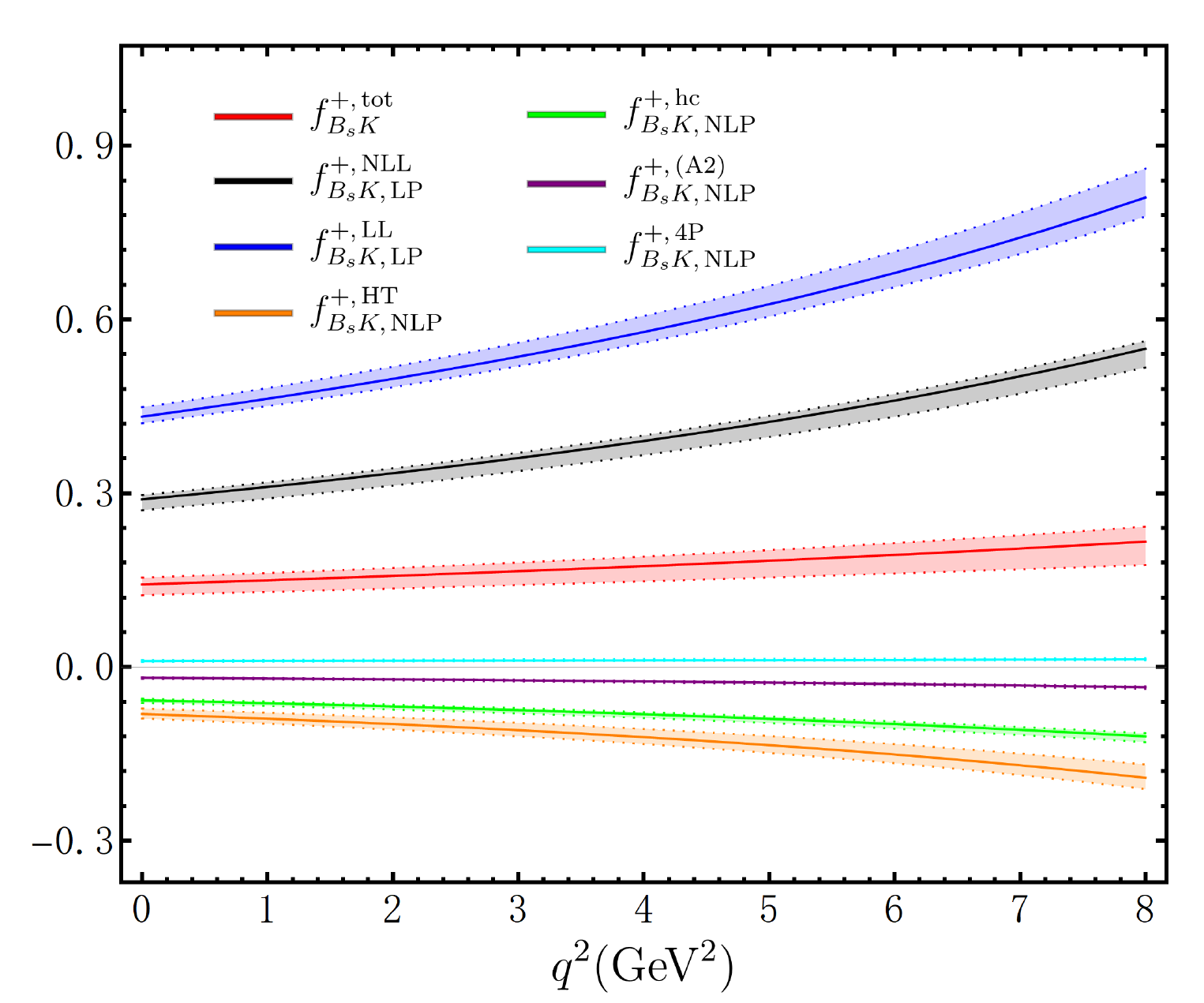}
\hspace{1.0 cm}
\includegraphics[width=0.45 \columnwidth]{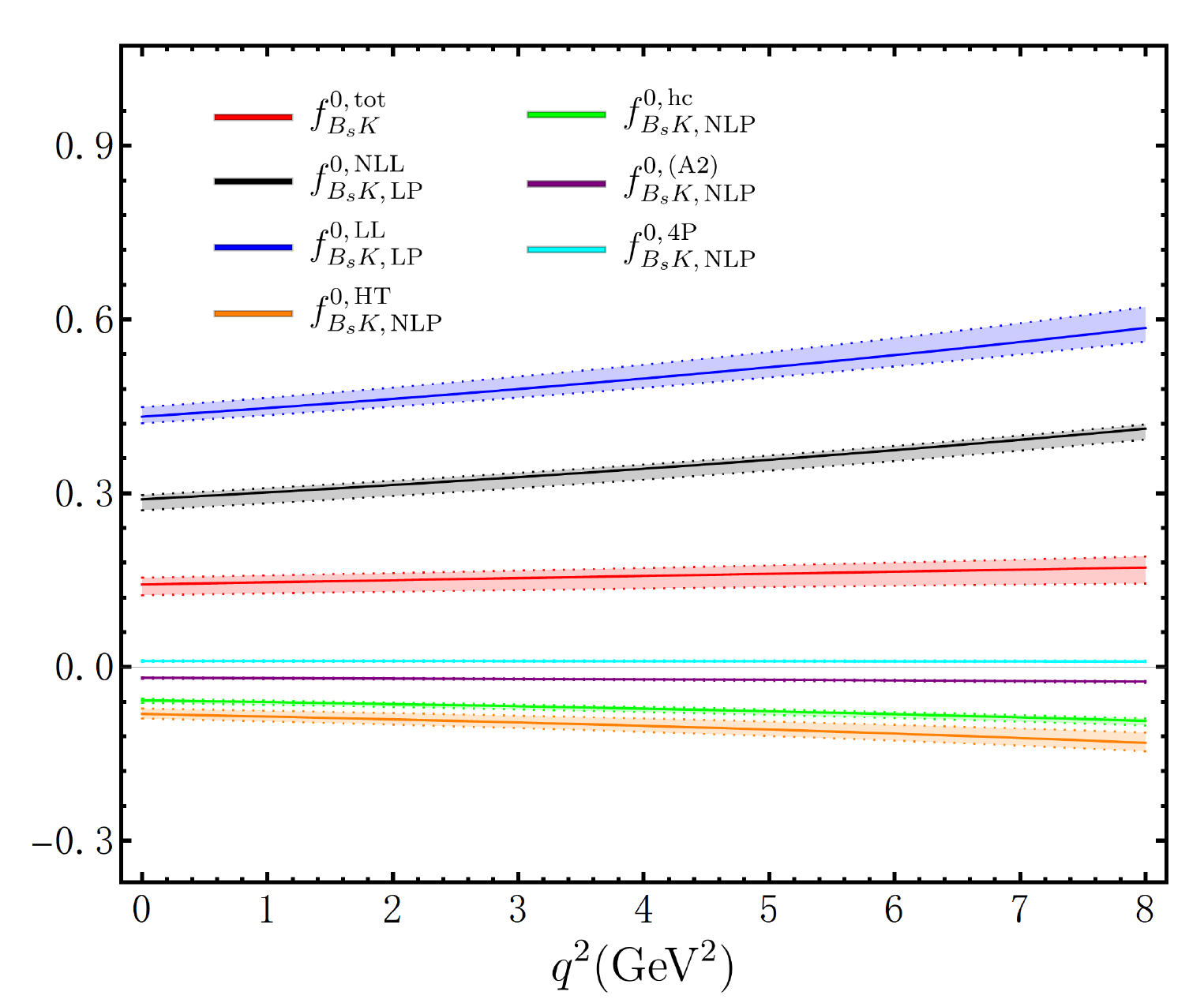}
\\
\includegraphics[width=0.45 \columnwidth]{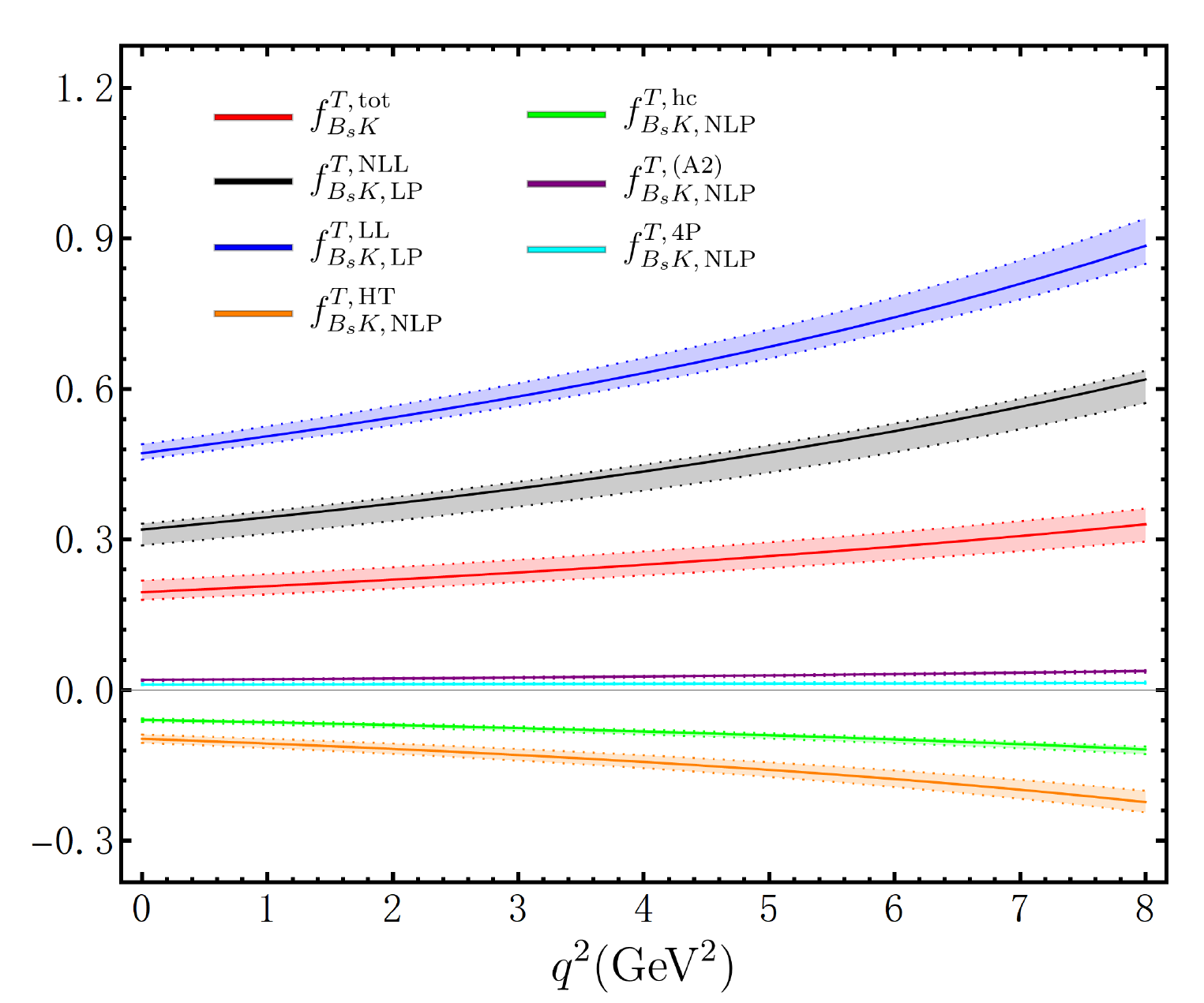}
\vspace*{0.1cm}
\caption{Breakdown of the distinct dynamical mechanisms contributing to the semileptonic
 $B_s \to K \ell \bar \nu_{\ell}$ form factors  in the kinematic region $0 \leq q^2 \leq 8 \, {\rm GeV^2}$
from the updated LCSR computations with the HQET bottom-meson distribution amplitudes,
where the perturbative uncertainties due to the variations of both the hard and hard-collinear matching scales
are indicated by the individual error bands.}
\label{fig: distinct pieces of Bs to K form factors in LCSR}
\end{center}
\end{figure}

We are now ready  to explore the phenomenological impacts of the NLL resummation improved leading-power contributions
(including further the light spectator-quark mass effects) and  the newly derived NLP corrections at the tree-level accuracy
on the semileptonic  $B_{d, s}  \to \pi, K$ decay form factors at large hadronic recoil.
In order to develop a transparent understanding of the dynamical patterns dictating these intricate form factors,
we display explicitly the yielding  leading-power contributions to the complete set of
the exclusive bottom-meson decay  form factors at the NLL accuracy,
the NLP contribution from expanding the hard-collinear quark propagator in the small parameter $\Lambda_{\rm QCD}/m_b$,
the NLP contribution from the power suppressed term in the ${\rm SCET}_{\rm I}$ representation of the weak transition current,
the subleading-twist contribution from the two-particle and three-particle HQET distribution amplitudes,
together with  the twist-five and twist-six four-particle bottom-meson distribution amplitudes in Figures
\ref{fig: distinct pieces of B to pi and B to K form factors in LCSR}
and \ref{fig: distinct pieces of Bs to K form factors in LCSR}
in the kinematic range $0 \leq q^2 \leq 8 \, {\rm GeV^2}$.
In particular, we have included the perturbative uncertainties for the individual pieces under discussion
from varying both  the hard and hard-collinear matching scales as indicated by the separate error bands.
The NLL resummation improved sum rules on the light-cone are indeed  beneficial for pinning down
the obtained theory uncertainties when compared with  the counterpart leading-logarithmic computations.
Generally, the perturbative QCD corrections to the short-distance matching coefficients in the SCET sum rules
can  shift the corresponding leading-power contributions  by an amount of ${\cal O} (30 \, \%)$ numerically.
It is evident from Figures \ref{fig: distinct pieces of B to pi and B to K form factors in LCSR}
and \ref{fig: distinct pieces of Bs to K form factors in LCSR}
that the most prominent subleading-power corrections to the heavy-to-light
bottom-meson decay form factors  arise from  the peculiar  higher-twist contributions of the two-particle and three-particle
HQET  distribution amplitudes  at ${\cal O}(\alpha_s^0)$ (more precisely from the two-particle twist-five
off-light-cone correction as already noticed in \cite{Gao:2019lta,Gao:2021sav}) 
yielding consistently $(25-30) \, \%$ reduction of the corresponding leading-power LCSR predictions at NLL
for $q^2 \in \left [0, \,  8 \right] \, {\rm GeV^2}$.
By contrast, the  estimated  four-particle twist-five and twist-six corrections in the factorization approximation
can only bring about insignificant impacts on the leading-power contributions to the exclusive $B_{d, s}  \to \pi, K$ form factors:
$(2-4) \, \%$ numerically, which can be attributed to the smallness of the  normalization constant
$|\langle \bar q q \rangle : (\lambda_B \, s_0)| \simeq 10 \, \%$ in the tree-level sum rules (\ref{LCSR for the 4P corrections}).
This interesting observation indicates that the higher-twist four-particle contributions will actually be   suppressed
by an extra power of $\Lambda_{\rm QCD}^2/s_0$ (rather than the additional powers of $\Lambda_{\rm QCD}/m_b$
in the heavy quark expansion) in analogy to the previous discussions \cite{Agaev:2010aq,Wang:2018wfj} in  different contexts.
Moreover, the newly determined subleading-power contributions from the hard-collinear quark propagator
shown in (\ref{NLP LCSR from the hc propagator}) can further generate the sizeable
destructive interferences (as large as ${\cal O} (20 \, \%)$  numerically) with the counterpart leading-power contributions.
It remains important to emphasize that the obtained hierarchy relations for the exclusive bottom-meson decay form factors
at maximal recoil
\begin{eqnarray}
f_{B K}^{+}(0) > f_{B_s K}^{+}(0) >  f_{B \pi}^{+}(0) \,,
\qquad
f_{B K}^{T}(0) > f_{B_s K}^{T}(0) >  f_{B \pi}^{T}(0)\,,
\end{eqnarray}
coincide precisely with the emerged patterns predicted by the two independent LCSR computations
with the final-state pseudoscalar-meson distribution amplitudes \cite{Duplancic:2008tk,Khodjamirian:2017fxg}
(see  \cite{Zeng:2008xr} for the alternative estimates with the TMD factorization approach and
\cite{Yao:2021pdy} for the quantitative analysis in the framework of Dyson-Schwinger equations).

\begin{figure}
\begin{center}
\includegraphics[width=0.45 \columnwidth]{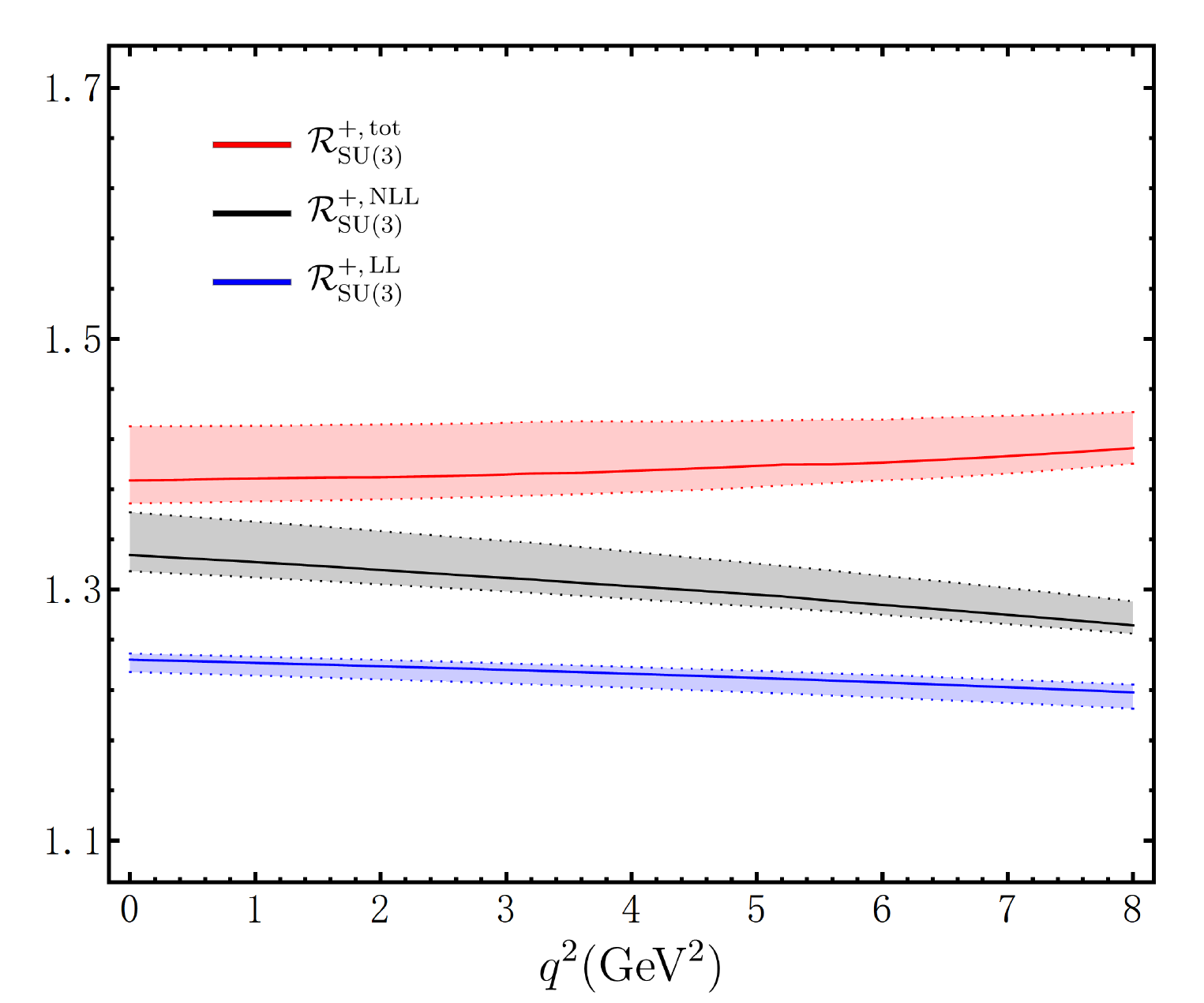}
\hspace{1.0 cm}
\includegraphics[width=0.45 \columnwidth]{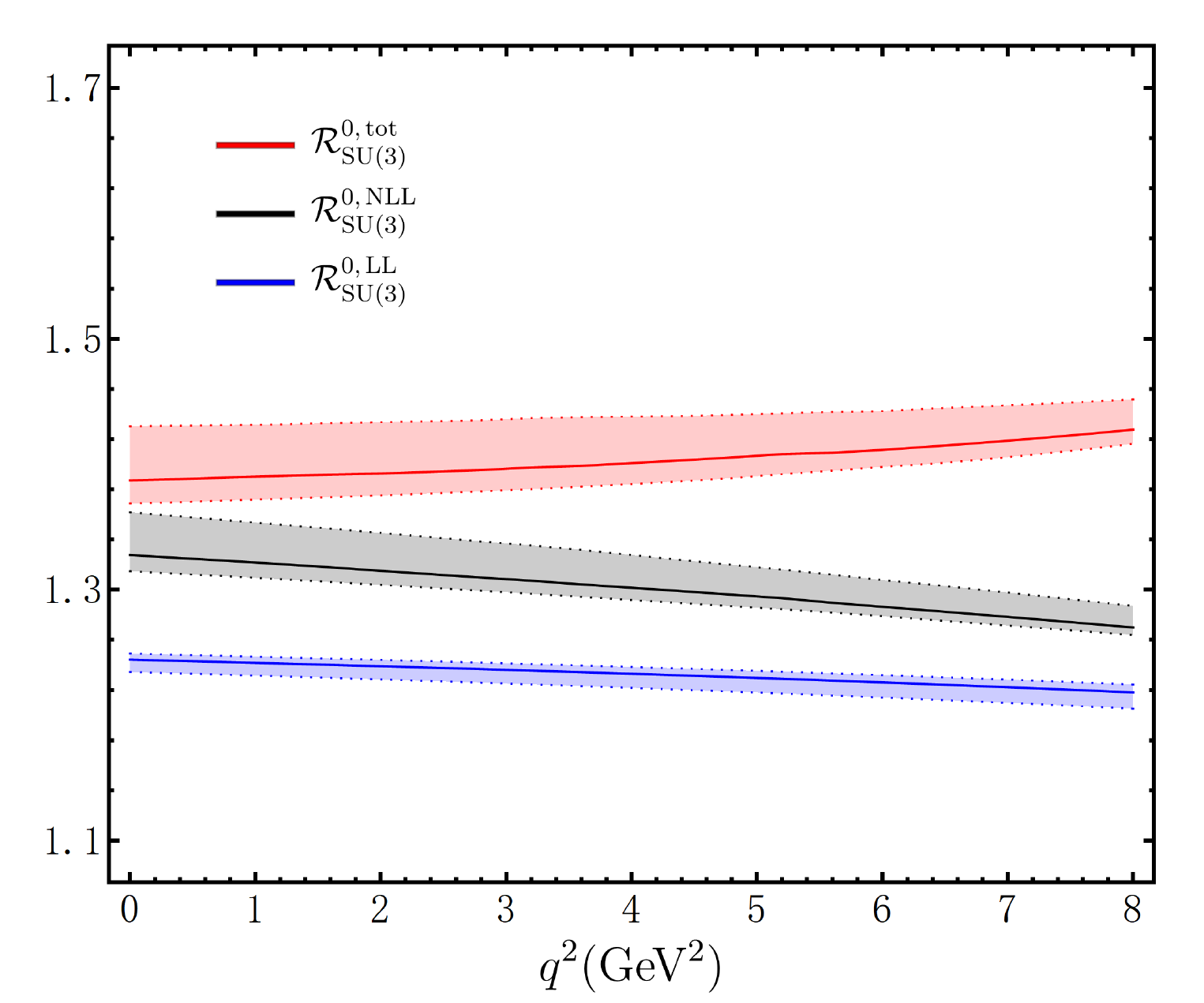}
\\
\includegraphics[width=0.45 \columnwidth]{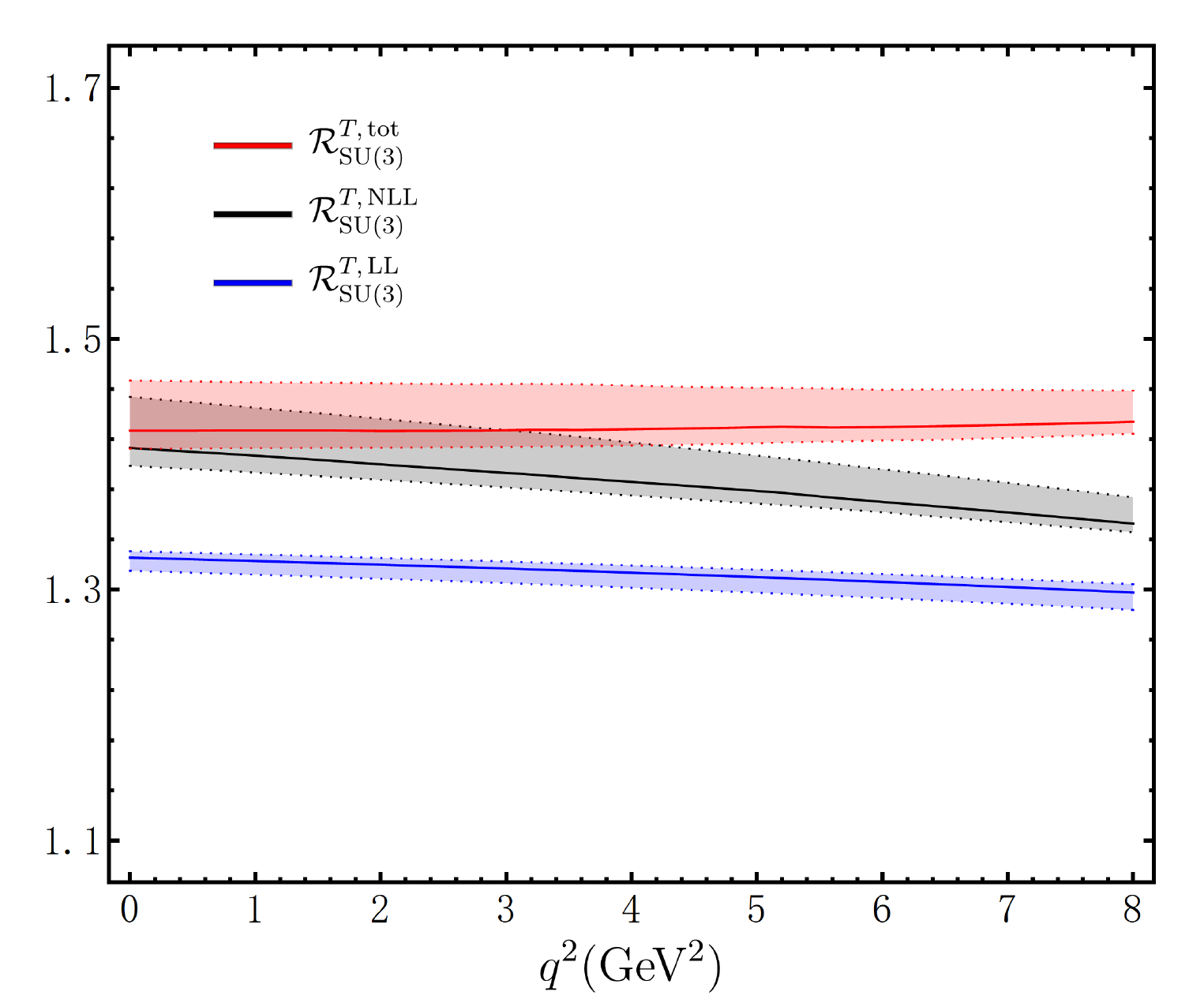}
\vspace*{0.1cm}
\caption{Theory predictions for the SU(3)-flavour symmetry breaking effects between
the semileptonic $B \to \pi$ and $B \to K$ form factors from the updated SCET sum rules
with the bottom-meson distribution amplitudes at the twist-six accuracy,
where the perturbative uncertainties due to the variations of both the hard and hard-collinear matching scales
are indicated by the individual error bands.}
\label{fig: SU(3) symmetry breaking effects from the LCSR method}
\end{center}
\end{figure}

We are now in a position to investigate the SU(3)-flavour symmetry breaking effects between
the exclusive $B \to \pi$ and $B \to K$ form factors, on the basis of the established SCET sum rules
with the bottom-meson distribution amplitudes up to the twist-six accuracy,
by introducing further the following  quantities \cite{Lu:2018cfc,Gao:2019lta}
\begin{eqnarray}
{\cal R}_{\rm SU(3)}^{i}(q^2) = \left[  f_{B \to K}^{i}(q^2) \right ] : \left [ f_{B \to \pi}^{i}(q^2) \right ] \,,
\qquad
({\rm with } \,\, i = +, \, 0, \, T )\,.
\end{eqnarray}
In our theoretical framework such flavour-symmetry violations arise from the apparent discrepancies
in the light-quark masses, in  the light-flavour hadron masses,
in the leptonic decay constants $f_{\pi}$ and $f_{K}$,
in the threshold parameters for the pion and kaon channels,
and finally in the nonperturbative quark-condensate densities
$\langle \bar u u \rangle$ and $\langle \bar s s \rangle$.
It can be observed from Figure \ref{fig: SU(3) symmetry breaking effects from the LCSR method}
 that the leading-power LCSR predictions
for the SU(3)-flavour symmetry breaking effects give rise to  numerically
${\cal O} (30 \, \%)$ for the two ratios ${\cal R}_{\rm SU(3)}^{+, \, 0}$
and ${\cal O} (40 \, \%)$ for the tensor-form-factor ratio ${\cal R}_{\rm SU(3)}^{T}$
in the large recoil region.
In particular,  the newly determined subleading-power contributions from the same LCSR method
can only bring about the insignificant numerical impacts on the SU(3)-flavour symmetry violating effects
(see \cite{Gao:2019lta} for the analogous observation
for the exclusive semileptonic $B \to V \ell \bar \nu_{\ell}$ form factors).
It is worthwhile mentioning that  we have not taken into account the remaining sources
of the SU(3)-flavour symmetry breaking effects due to the electromagnetic corrections
and the systematic uncertainties owing to the parton-hadron duality approximation.

\begin{figure}
\begin{center}
\includegraphics[width=0.45 \columnwidth]{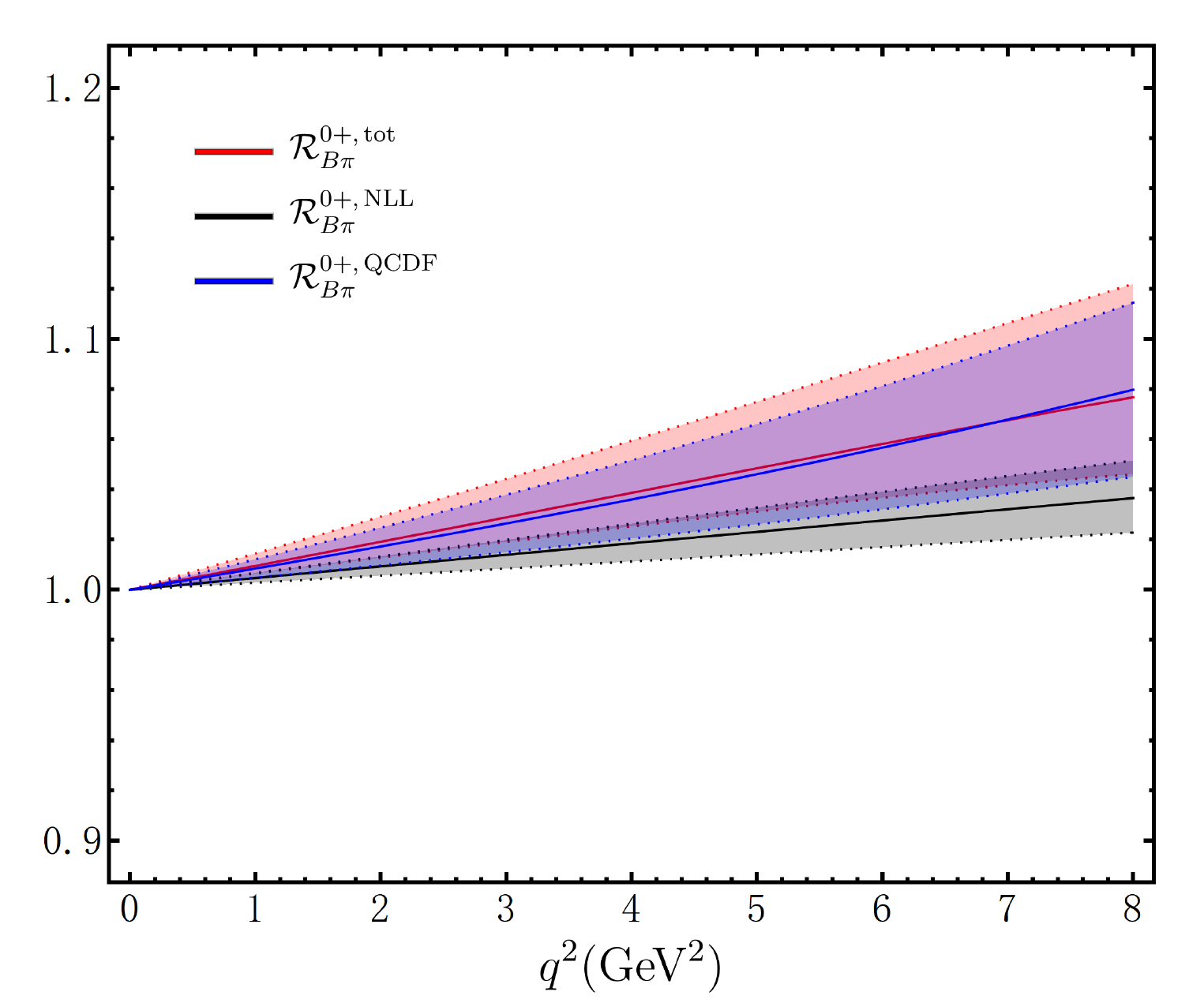}
\hspace{1.0 cm}
\includegraphics[width=0.45 \columnwidth]{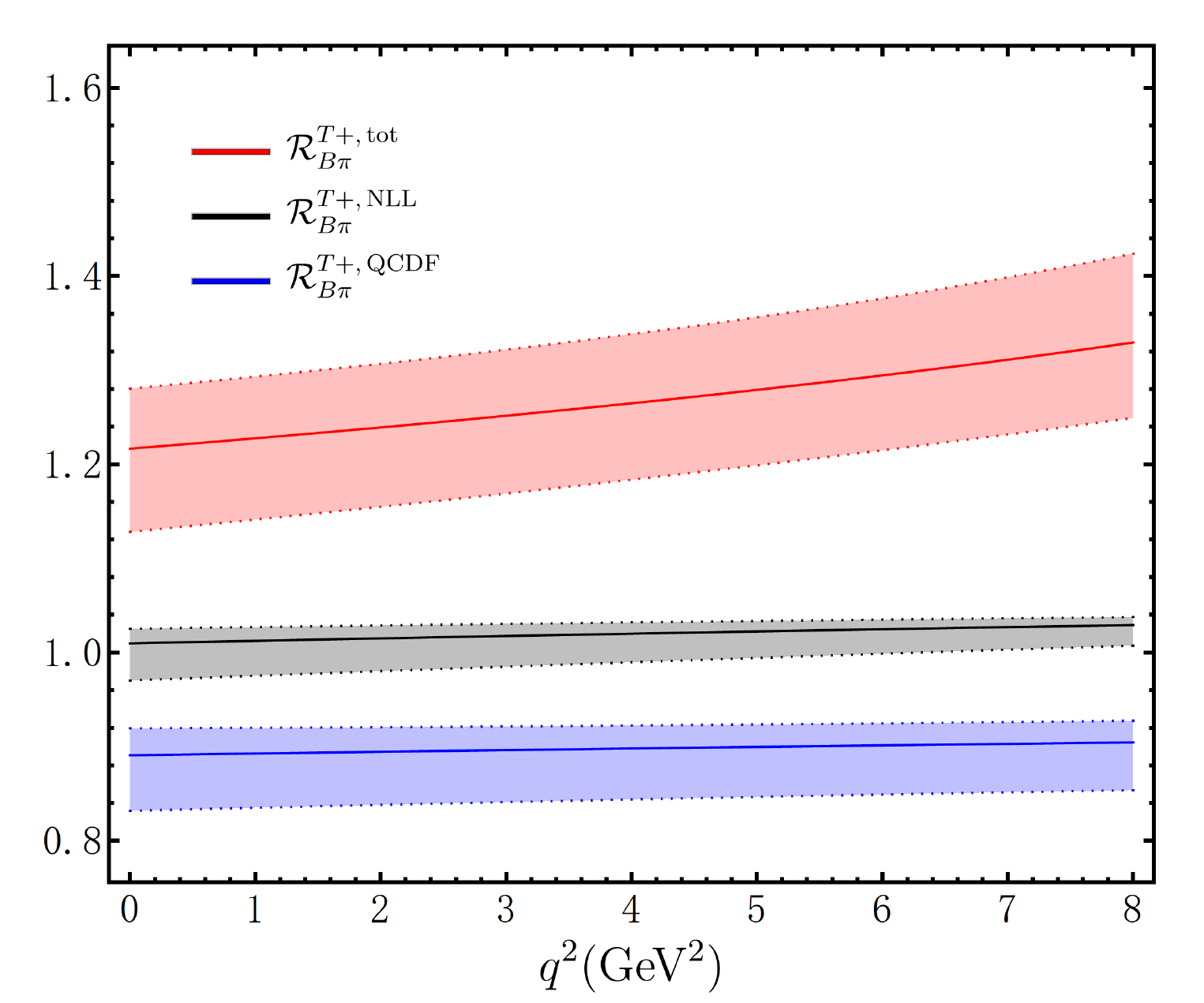}
\vspace*{0.1cm}
\caption{Theory predictions for the large-recoil symmetry breaking corrections  between
the exclusive  $B \to \pi$ form factors from the updated SCET sum rules
with the bottom-meson distribution amplitudes at the twist-six accuracy.
For a comparison, we further display the numerical results from the QCD factorization approach
with the so-called physical form factor scheme \cite{Beneke:2000wa}
by taking the light-meson LCSR computation of the vector form factor $f_{B \pi}^{+}(q^2)$ \cite{Khodjamirian:2017fxg}
as the fundamental theory input.}
\label{fig: large-recoil symmetry breaking effects}
\end{center}
\end{figure}

We proceed to explore the celebrated large-recoil symmetry breaking effects between
the exclusive heavy-to-light bottom-meson decay form factors
due to the higher-order perturbative corrections
and the intricate subleading-power corrections in the $\Lambda_{\rm QCD}/m_b$ expansion.
In order to facilitate the straightforward comparison with the theory predictions
from the QCD factorization approach, it proves convenient to investigate the two particular
form-factor ratios for the semileptonic $B \to \pi \ell \bar \nu_{\ell}$
and $B \to \pi \ell \bar {\ell}$  decay processes \cite{Beneke:2000wa,Lu:2018cfc}
\begin{eqnarray}
{\cal R}_{B \pi}^{0 +}(q^2) = \left(  {m_B  \over n \cdot p} \right ) \,  \frac{f_{B \pi}^{0}(q^2)}{f_{B \pi}^{+}(q^2)} \,,
\qquad
{\cal R}_{B \pi}^{T +}(q^2) =  \left ( {m_B  \over m_B + m_{\pi}} \right ) \, \frac{f_{B \pi}^{T}(q^2)}{f_{B \pi}^{+}(q^2)} \,.
\end{eqnarray}
It is interesting to notice that the NLL resummation improved LCSR predictions for the two quantities  ${\cal R}_{B \pi}^{0 +}$
and ${\cal R}_{B \pi}^{T +}$ are in reasonable agreement with the QCD factorization computations at the leading-power accuracy.
On the contrary, confronting  our numerical  results for the form-factor ratio ${\cal R}_{B \pi}^{T +}$
from the bottom-meson LCSR method including four different classes of the  NLP corrections
with the counterpart predictions from QCD factorization reveal an intriguing  tension
on both the magnitude and sign of the large-recoil symmetry breaking corrections
as displayed in Figure \ref{fig: large-recoil symmetry breaking effects}.
Inspecting the individual terms in the obtained subleading-power sum rules (\ref{NLP sum rules: summary})
indicates that the emerged discrepancies between the two different QCD calculations
stem from the newly determined NLP contribution of the power-suppressed ${\rm SCET_{I}}$ current $J^{\rm (A2)}$
with the  LCSR  method as collected in (\ref{NLP LCSR from the subleading SCET current}) explicitly,
which has not been included in the numerical exploration of the current QCD factorization framework \cite{Beneke:2000wa}.
As a consequence, it becomes more and more demanding to construct the appropriate perturbative factorization formula
for this NLP  matrix element directly with the modern effective-field-theory technique.

Apparently, we can only establish the soft-collinear factorization formulae for the desired vacuum-to-$B$-meson
correlation functions (\ref{correlation_function}) at small and intermediate momentum transfers,
$0 \leq q^2 \leq q_{\rm cut}^2$, where the practical choice of $q_{\rm cut}^2$ varies between $8 \, {\rm GeV^2}$
and $10 \, {\rm GeV^2}$ numerically.
We are therefore required to extrapolate the bottom-meson LCSR computations of
the semileptonic $B_{d, s}  \to \pi, K$ form factors towards the large momentum transfer $q^2$
by applying the $z$-series parametrization based upon the positivity and analyticity properties
of these transition form factors \cite{Okubo:1971jf,Okubo:1971my,Bourrely:1980gp}.
Adopting the conformal transformation \cite{Lellouch:1995yv,Bourrely:2008za,Bourrely:2005hp}
\begin{eqnarray}
z(q^2, t_0)  = \frac{\sqrt{t_{+} - q^2} - \sqrt{t_{+} - t_0}}
{\sqrt{t_{+} - q^2} +\sqrt{t_{+} - t_0}} \,,
\label{def: z parameter}
\end{eqnarray}
with the threshold parameter $t_{+} \equiv \left [ m_B + m_{\pi (K)} \right ]^2$  for the exclusive semileptonic
$b \to u (s)$ transitions,
enables us to map the complex cut $q^2$-plane onto the unit disk $|z(q^2, t_0)|  \leq 1$.
On the other hand, the free parameter $t_0 < t_{+}$ corresponds to the value of $q^2$
mapping onto the origin in the $z$-plane, namely $z(t_0, t_0)=0$.
In order to maximally reduce the interval of $z$ obtained after mapping (\ref{def: z parameter})
of the semileptonic domain $q^2 \in \left [0, \, (m_{B_{q^{\prime}}}- m_M)^2 \right ]$,
the auxiliary parameter $t_0$ can be customarily taken as
\begin{eqnarray}
t_0 = t_{+} - \sqrt{t_{+} \, (t_{+} - t_{-})} \,,
\qquad t_{-} = (m_{B_{q^{\prime}}}- m_M)^2 \,.
\end{eqnarray}
following closely the comprehensive discussions presented in  \cite{FlavourLatticeAveragingGroupFLAG:2021npn}.
Taking into account further the asymptotic behaviours of the vector form factors
near threshold due to the angular momentum conservation  implies the simplified series expansion
originally proposed in \cite{Bourrely:2008za} (see \cite{Boyd:1994tt,Boyd:1997kz} for an alternative parametrization)
\begin{eqnarray}
f_{B_{q^{\prime}} M}^{+}(q^2) &=&   \frac{1}{1-q^2/m_{B_{q}^{\ast}}^2 }  \,
 \sum_{k=0}^{N-1} \, b_{k}^{+} \,
\left [  z(q^2, t_0)^k  - (-1)^{k - N} \, {k \over N } \, z(q^2, t_0)^N  \right ]    \,,
\label{z-parametrization for the vector form factor}
\end{eqnarray}
where the lowest  bottom-meson $B_{d}^{\ast}$ ($B_{s}^{\ast}$) is expected to be the only resonance
of the $J^P=1^{-}$ channel below the $B \pi$ ($B K$) production region.
It is evident from the  BCL parametrization (\ref{z-parametrization for the vector form factor}) that
the well-known scaling behaviour $f_{B_{q^{\prime}} M}^{+}(q^2) \sim 1/q^2$ at $|q^2| \to \infty$
from the perturbative QCD analysis \cite{Akhoury:1994tnu} is indeed fulfilled.
For the practical purpose,  we will truncate the $z$-series expansion at $N=3$
in the subsequent fitting program.

Along the same vein, we can proceed to parameterize the remaining  form factors
by adjusting the overall pole functions appropriately and by dropping out the near-threshold constraints
for the scalar transition  form factors
\begin{eqnarray}
f_{B_{(s)} \pi (K)}^{0}(q^2) &=&  \, \sum_{k=0}^{N-1} \, b_{k}^{0} \,\, z(q^2, t_0)^k    \,,
\qquad
f_{B K}^{0}(q^2) =  \frac{1}{1-q^2/m_{B_{s}^{\ast}  (0+)}^2} \, \sum_{k=0}^{N-1} \, b_{k}^{0} \,\, z(q^2, t_0)^k \,,
\nonumber  \\
f_{B_{q^{\prime}} M}^{T}(q^2) &=&  \frac{1}{1-q^2/m_{B_{q}^{\ast}}^2} \,
 \sum_{k=0}^{N-1} \, b_{k}^{T} \,
\left [  z(q^2, t_0)^k  - (-1)^{k - N} \, {k \over N } \, z(q^2, t_0)^N  \right ]    \,.
\label{z-parametrization for the scalar and tensor form factors}
\end{eqnarray}
The very disappearance of the pole factors in the above-mentioned parameterizations
of the two particular form factors $f_{B \pi}^{0}(q^2)$ and $f_{B_{s} K}^{0}(q^2)$
can be attributed to the fact that
the low-lying bottom-resonance in the $J^P=0^{+}$ channel
with the predicted mass  $m_{B^{\ast} (0+)} = (5627 \pm 35) \, {\rm MeV}$ \cite{Bardeen:2003kt}
turns out to be located above the $B \pi$ production threshold
$m_B + m_{\pi} = (5419 \pm 0.12) \, {\rm MeV}$ \cite{ParticleDataGroup:2022pth}.
By contrast, the low-lying bottom-meson resonance
in the BCL parametrization (\ref{z-parametrization for the scalar and tensor form factors})
for the scalar form factor $f_{B K}^{0}(q^2)$,
with the estimated mass $m_{B_s^{\ast} (0+)} = (5718  \pm 35) \, {\rm MeV}$
from the heavy-hadron chiral perturbation theory \cite{Bardeen:2003kt},
appears to be somewhat  below the particle-pair production threshold
$m_B + m_{K} = (5773 \pm 0.12) \, {\rm MeV}$ \cite{ParticleDataGroup:2022pth}.
For convenience, we have also summarized the relevant resonance masses employed
in our $z$-parametrization fits in Table \ref{table: theory inputs}.

\begin{table}
\centering
\renewcommand{\arraystretch}{2.0}
\resizebox{\columnwidth}{!}{
\begin{tabular}{|c|c||cccccccc|}
\hline
\hline
 \multicolumn{2}{|c||}{$B \to \pi$ \, Form Factors}  & \multicolumn{8}{|c|}{Correlation Matrix}
\\
\hline
Parameters & Values & $b_0^+$ & $b_1^+$ & $b_2^+$ & $b_0^0$ & $b_1^0$ & $b_0^T$ & $b_1^T$ & $b_2^T$
 \\
 \hline
 \hline
 $b_0^+$ & $0.404(13)$ & $1$ & $0.276$ & $-0.446$ & $0.291$ & $0.164$ & $0.583$ & $0.266$ & $-0.24$
 \\
 $b_1^+$ & $-0.618(63)$ &  &  $1$ & $-0.374$ & $0.067$ & $0.485$ & $0.234$ & $0.693$ & $-0.193$
 \\
 $b_2^+$ & $-0.473(215)$ &  &  & $1$ & $0.108$ & $0.157$ & $-0.185$ & $-0.185$ & $0.592$
 \\
 $b_0^0$ & $0.496(19)$ &  &  &  & $1$ & $-0.226$ & $0.318$ & $0.06$ & $0.018$
 \\
 $b_1^0$ & $-1.537(56)$ &  &  &  &  & $1$ & $0.04$ & $0.431$ & $0.165$
 \\
 $b_0^T$ & $0.396(15)$ &  &  &  &  &  & $1$ & $0.178$ & $-0.423$
 \\
 $b_1^T$ & $-0.553(73)$ &  &  &  &  &  &  &  $1$ & $-0.307$
 \\
 $b_2^T$ & $-0.248(235)$ &  &  &  &  &  &  &  &  $1$
 \\
 \hline
 \hline
\end{tabular}
}
\renewcommand{\arraystretch}{1.0}
\caption{Theory predictions for the correlated $z$-series coefficients
in the vector, scalar and tensor $B \to \pi$ form factors
determined by fitting the BGL parametrization simultaneously
against our LCSR results including a variety of the subleading-power corrections
and the available  lattice QCD data points from \cite{Flynn:2015mha,FermilabLattice:2015mwy,FermilabLattice:2015cdh}
with the preferred truncation $N=3$. }
\label{BCL fit foe the B to pi form factors}
\end{table}

\begin{figure}
\begin{center}
\includegraphics[width=0.45 \columnwidth]{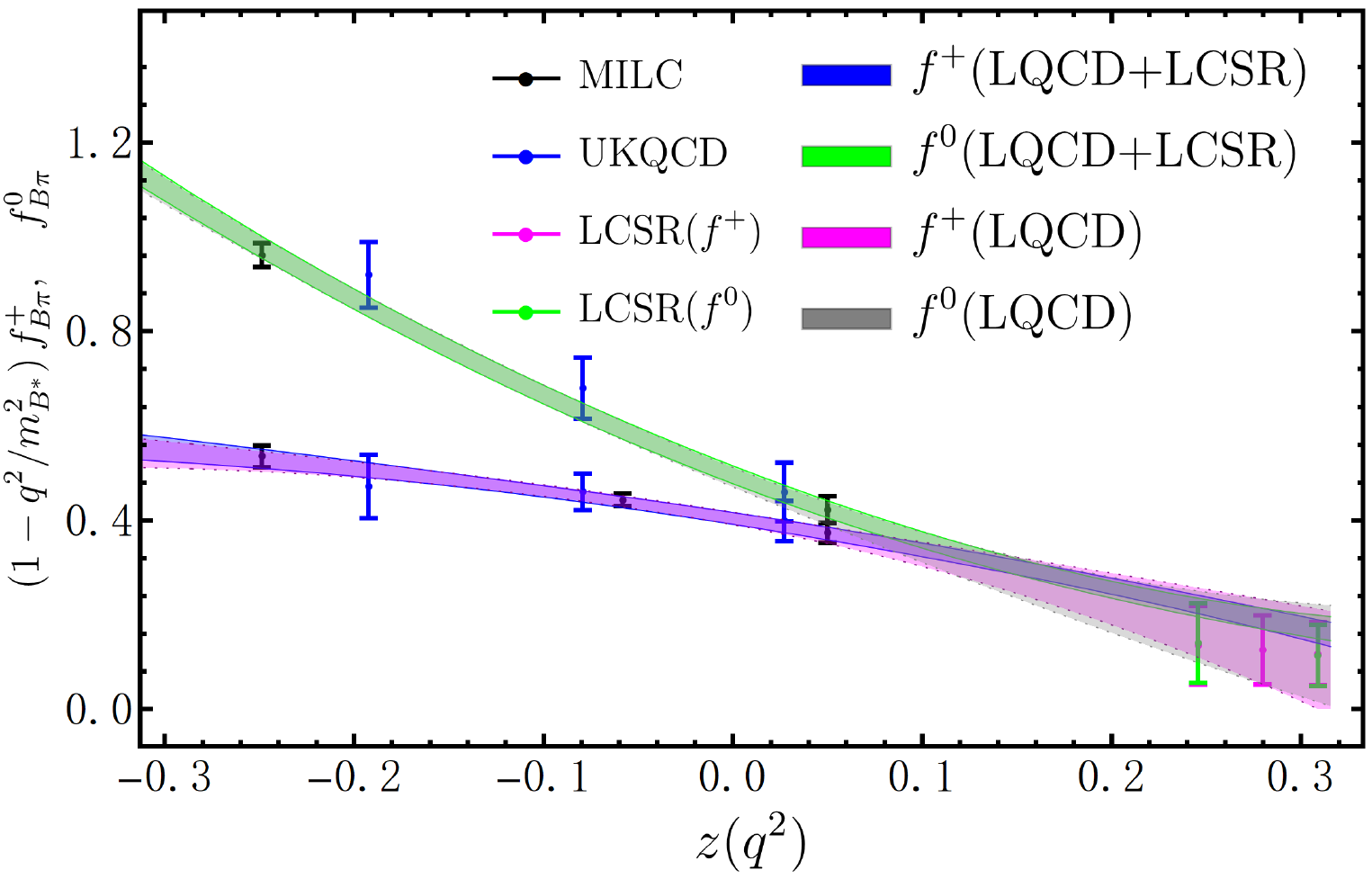}
\hspace{1.0 cm}
\includegraphics[width=0.45 \columnwidth]{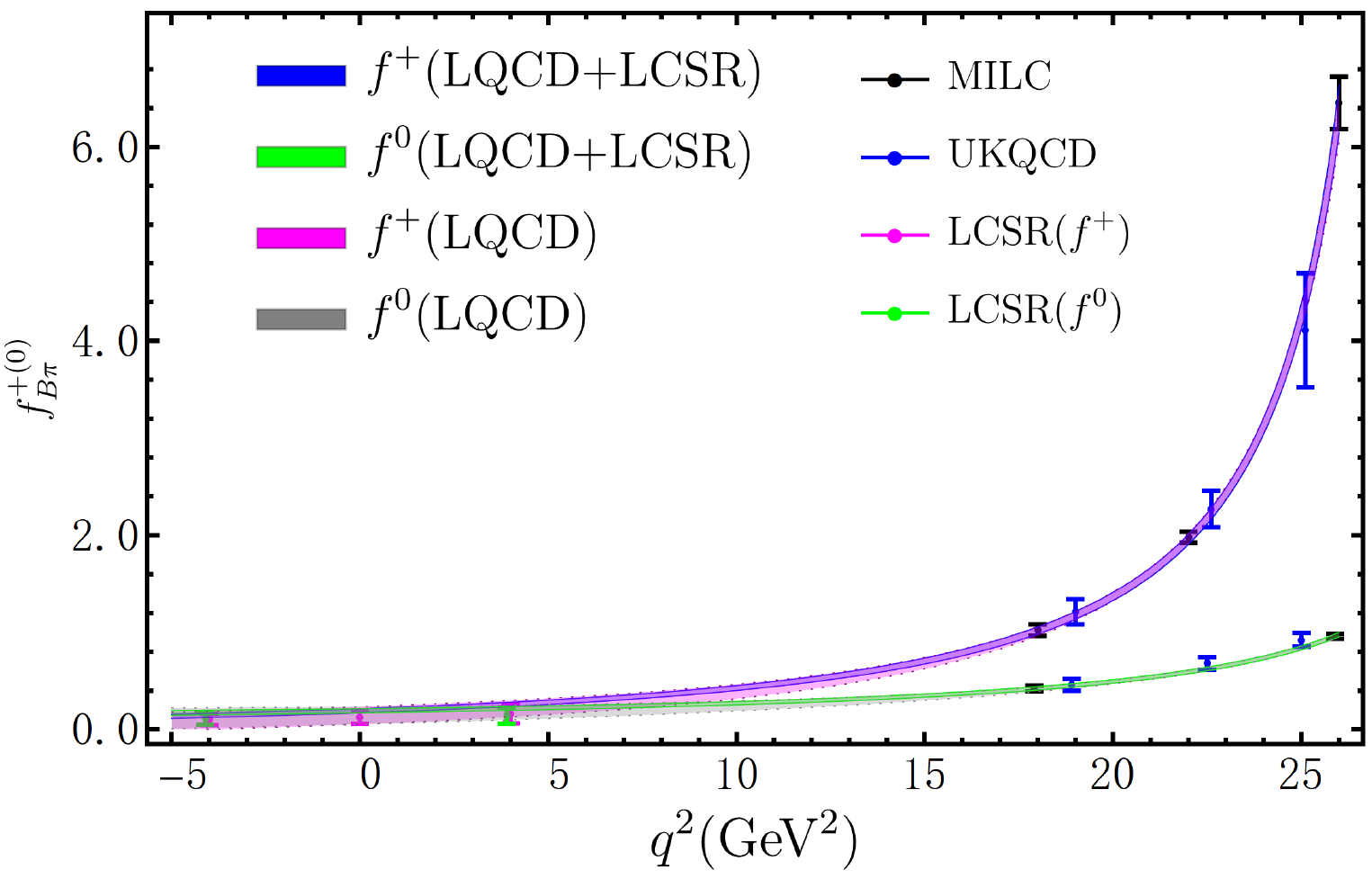}
\\
\vspace*{0.2 cm}
\includegraphics[width=0.45 \columnwidth]{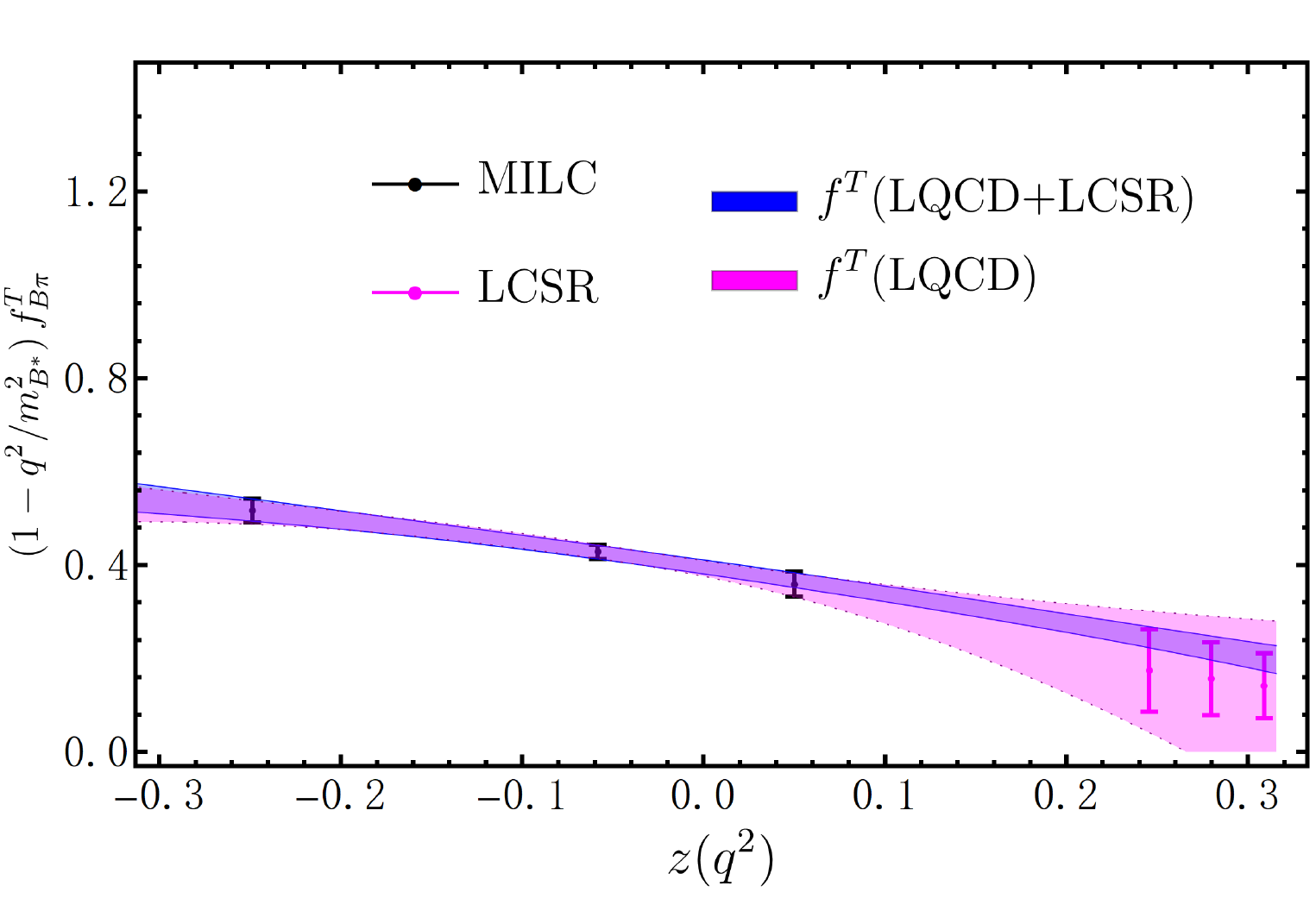}
\hspace{1.0 cm}
\includegraphics[width=0.45 \columnwidth]{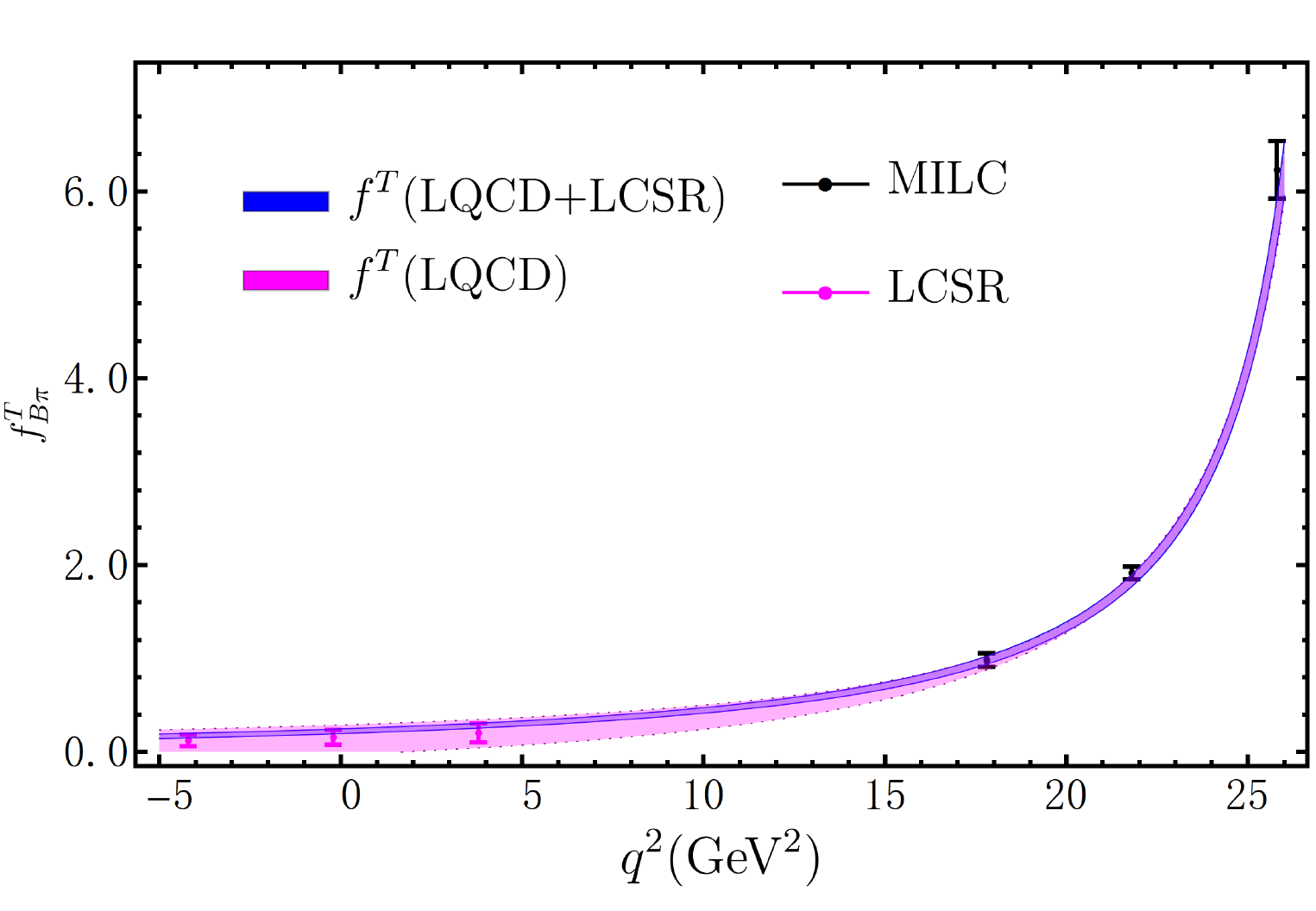}
\vspace*{0.1 cm}
\caption{Theory predictions for the complete set of the semileptonic $B \to \pi$ form factors
versus  $z$ (left panel) and versus $q^2$ (right panel) in the entire kinematic region
obtained by carrying out the combined BCL $z$-fit of the updated LCSR (from this work)
and lattice QCD (from \cite{Flynn:2015mha,FermilabLattice:2015mwy,FermilabLattice:2015cdh}) data points.
We further display the yielding numerical results of these form factors by performing an alternative $z$-series fit
against the ``{\it only lattice QCD}" data points \cite{Flynn:2015mha,FermilabLattice:2015mwy,FermilabLattice:2015cdh}
exclusively for a comparison.   }
\label{fig: BCL fit of B to pi form factors}
\end{center}
\end{figure}

We are now prepared to determine the BCL series coefficients $b_k^{+, \, 0, \, T}$
by performing the binned $\chi^2$ fit of the updated LCSR predictions for the bottom-meson decay  form factors
at three distinct kinematic points, namely $q^2= \left \{-4.0, \, 0, \, 4.0  \right \} \, {\rm GeV^2}$,
in combination with the available lattice data points in the higher-$q^2$ region.
Enforcing the  kinematic constraint between the vector and scalar form factors
$f_{B M}^{+}(0) = f_{BM}^{0}(0) $  allows  us further  to derive the following exact relations
between the expansion coefficients of our interest
\begin{eqnarray}
b_2^{0}  &=& 12.78 \, \left (  b_0^{+} -   b_0^{0}  \right ) + 3.482 \, b_1^{+} + 1.186 \, b_2^{+}
- 3.575 \, b_1^{0} \,, \qquad ({\rm for} \,\, B \to \pi)
\nonumber \\
b_2^{0} &=&  24.06  \, \left ( b_0^{+} -   b_0^{0}  \right ) + 4.837 \, b_1^{+} + 1.136 \, b_2^{+}
- 4.905 \, b_1^{0} \,, \qquad ({\rm for} \,\, B_s \to K)
\nonumber \\
b_2^{0} &=&  48.59  \, \left ( b_0^{+} -   b_0^{0}  \right ) + 6.923 \, b_1^{+} + 1.096 \, b_2^{+}
- 6.971 \, b_1^{0} \,. \qquad ({\rm for} \,\, B \to K)
\label{exact relations of z-series parameters}
\end{eqnarray}
With regard to the lattice QCD results for the semileptonic $B \to \pi$ form factors,
we can  straightforwardly employ the synthetic data points of $f_{B \pi}^{+}(q^2)$ and $f_{B \pi}^{0}(q^2)$
at three representative values of $q^2 = \left \{19.0, \, 22.6, \, 25.1  \right \} \, {\rm GeV^2}$
with the full correlation matrices from the RBC/UKQCD Collaboration \cite{Flynn:2015mha},
by adopting $N_f=2+1$-flavour gauge-field ensembles
with the domain-wall fermion action and Iwasaki gluon action.
However, the FNAL/MILC Collaboration do not provide the yielding data points for the  exclusive $B \to \pi$ form factors
explicitly  in their publications \cite{FermilabLattice:2015mwy,FermilabLattice:2015cdh},
which present the outcome of the combined BCL fit to their data points with the truncation $N=3$ instead.
Consequently, we will take advantage of the BCL fit results to generate the correlated synthetic data points
of the three $B \to \pi$ form factors in the kinematic region $19.0 \, {\rm GeV}^2 \leq q^2 \leq  26.4 \, {\rm GeV}^2$.
Carrying out the simultaneous fit of the conventional  BCL parameterizations
(\ref{z-parametrization for the vector form factor}) and
(\ref{z-parametrization for the scalar and tensor form factors})
to our LCSR pseudo data points as well as the lattice QCD data points gives rise  to
the desired intervals of the $z$-series coefficients and their correlation matrix
for the semileptonic $B \to \pi$ form factors displayed in Table \ref{BCL fit foe the B to pi form factors}.
Furthermore, we observe that this numerical fit yields a minimal $\chi^2=7.93$
for $14$ degrees of freedom in the fitting program,
which corresponds to an excellent $p$-value of $89 \, \%$ in turn.
It has been  verified manifestly  that the fitted BCL coefficients  fulfill
the very dispersive bounds \cite{Bourrely:2008za,Bharucha:2010im} derived from the correlation functions
of two flavour-changing currents with the aid of unitarity and crossing symmetry
(see \cite{Becher:2005bg} for further discussions on the scaling behaviour of the sum of coefficients
$\sum_{k=0}^{N} \left (b_k^{+} \right )^2$  in the heavy quark limit).
In order to develop a transparent understanding towards the eventually predicted momentum-transfer dependence
from interpolating the LCSR and lattice QCD results,
we display the obtained  numerical predictions for  the  three exclusive $B \to \pi$ form factors
versus  $z$ (left panel) and versus $q^2$ (right panel) in the entire kinematic region
in Figure \ref{fig: BCL fit of B to pi form factors},
where the counterpart predictions of these form factors  from implementing an alternative $z$-expansion fit
of  the ``{\it only lattice QCD}" data points \cite{Flynn:2015mha,FermilabLattice:2015mwy,FermilabLattice:2015cdh}
exclusively are further shown for the convenience.
It is evident from this comparative exploration that including the newly derived LCSR data points
at small momentum transfer in our fitting procedure will indeed be highly beneficial for
improving the theory precision for all three $B \to \pi$ form factors
in the kinematic regime  $0.10 \leq z \leq 0.31$
(namely, $q^2 \in \left [ -4.0 ,   15.5 \right ]  \, {\rm GeV^2}$) significantly.
This interesting  observation can be actually understood from the very fact that
extrapolating the current lattice QCD results
towards the lower $q^2$ region solely will bring about the more pronounced uncertainties
for the form-factor shapes  in comparison with the direct LCSR computations
as already discussed in \cite{Flynn:2015mha,FermilabLattice:2015mwy,FermilabLattice:2015cdh}.

\begin{figure}
\begin{center}
\includegraphics[width=0.45 \columnwidth]{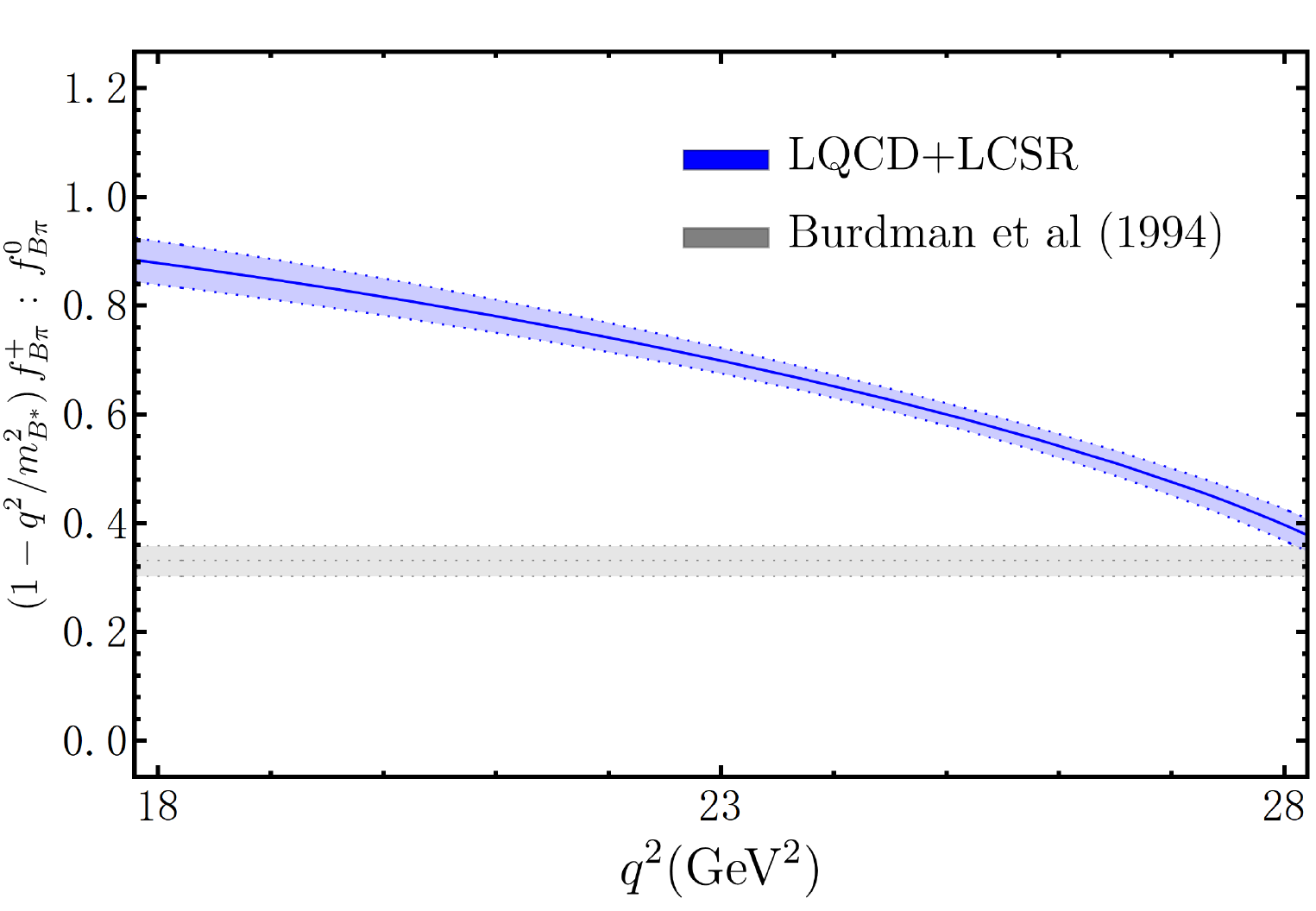}
\hspace{1.0 cm}
\includegraphics[width=0.45 \columnwidth]{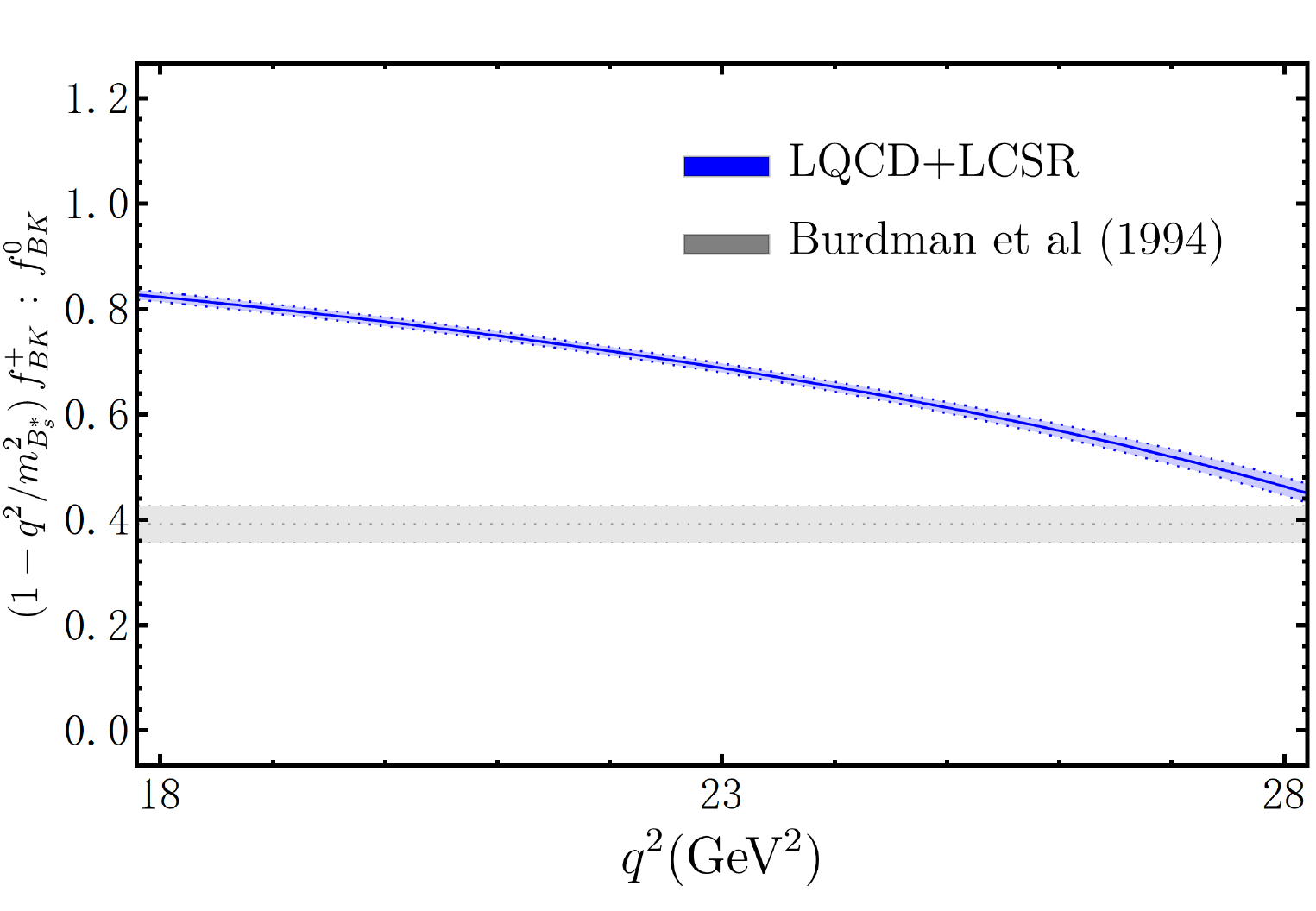}
\\
\vspace*{0.2 cm}
\includegraphics[width=0.45 \columnwidth]{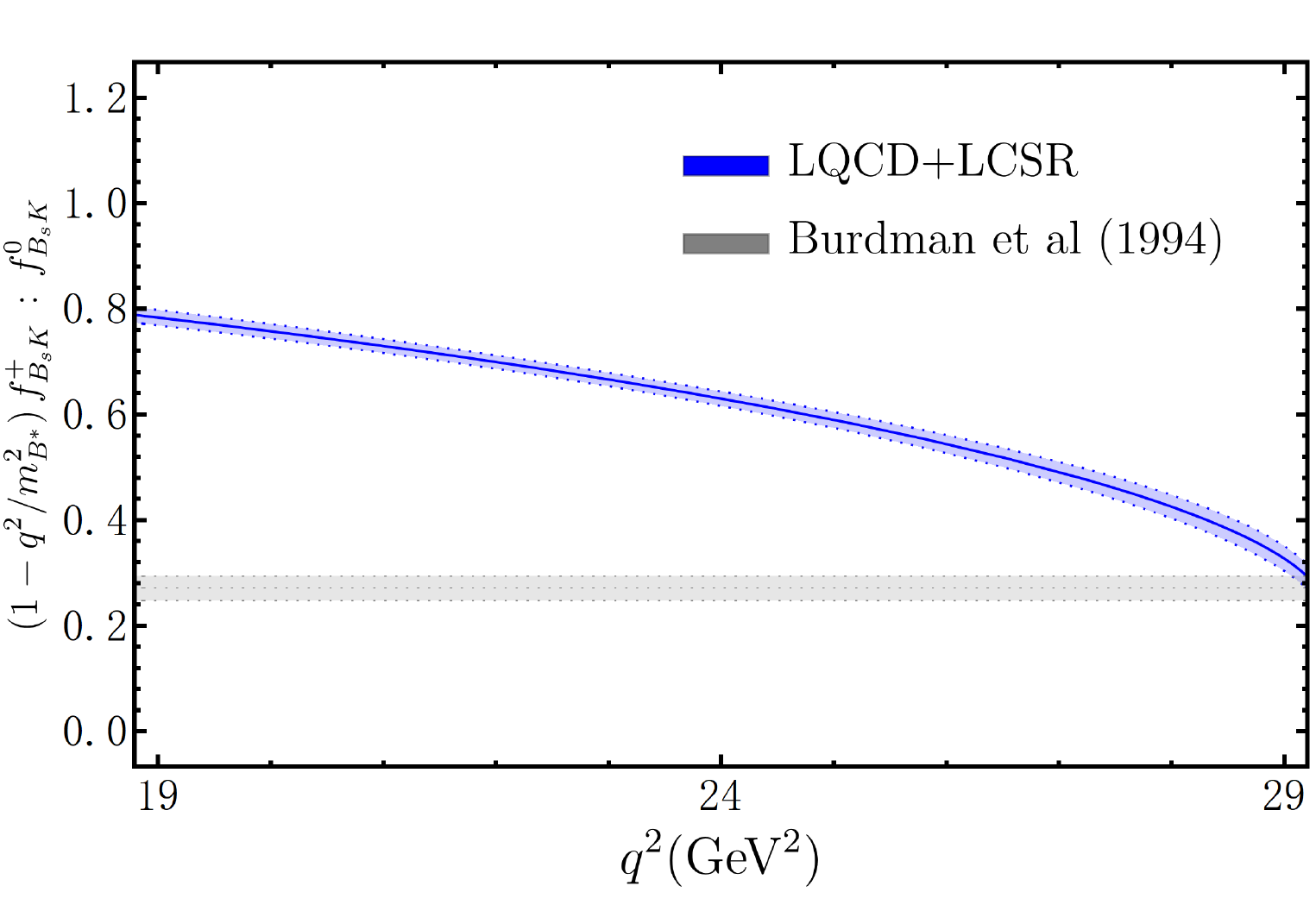}
\vspace*{0.1 cm}
\caption{Theory predictions of the particular form-factor ratio
$\left [ \left (1 - q^2 /  m_{B_q^{\ast}}^2 \right ) \, f_{B_{q^{\prime}} M}^{+}(q^2) \right ]  :
\left [ f_{B_{q^{\prime}} M}^{0}(q^2) \right ]$ for the exclusive semileptonic $B_{d, s}  \to \pi, K$  decays
determined from the combined $z$-expansion fit
against the LCSR and lattice QCD data points (blue bands) and from the theoretical expectations
of the heavy quark spin symmetry and the current algebra technique
at NLO in the $\Lambda_{\rm}/m_b$ expansion \cite{Burdman:1993es} (grey bands).}
\label{fig: soft-pion relation of fplus and fzero}
\end{center}
\end{figure}

Additionally, it is instructive to compare our form-factor predictions from the combined BCL fit
with the theoretical expectations from the heavy quark spin symmetry and the current algebra method.
In the zero-recoil limit we can derive an interesting relation between the vector and scalar form factors
up to the accuracy of ${\cal O}(\Lambda_{\rm QCD}^2/m_b^2)$ \cite{Burdman:1993es}
\begin{eqnarray}
\lim_{q^2 \to m_{B_{q^{\prime}}}^2 } \, \frac{f_{B_{q^{\prime}} M}^{+}(q^2)}{f_{B_{q^{\prime}} M}^{0}(q^2)}
= \left ( \frac{f_{B_q^{\ast}}} {f_{B_{q^{\prime}}}} \right ) \, \left (1- {m_M^2 \over m_{B_{q^{\prime}}}^2}  \right )\,
\left ( \frac{\hat{g}_{\rm eff}}{ 1 - q^2 /  m_{B_q^{\ast}}^2} \right )   \,,
\label{soft-pion relation of fplus and fzero}
\end{eqnarray}
where the static coupling $\hat{g}_{\rm eff}$ entering the Lagrangian density of
the heavy-hadron chiral perturbation theory is independent of the heavy quark mass \cite{Manohar:2000dt}.
We will adopt the nonperturbative determination of this  low-energy constant from the NLO LCSR computations
with the pion distribution amplitudes $\hat{g}_{\rm eff}=0.30 \pm 0.02$ \cite{Khodjamirian:2020mlb}
(see also \cite{Belyaev:1994zk,Khodjamirian:1999hb} for the earlier analyses in the same framework).
Moreover, we will employ the updated numerical results for the leptonic decay constants
of the heavy-light vector mesons \cite{Gelhausen:2013wia}
\begin{eqnarray}
f_{B^{\ast}} = 210^{+10}_{-12} \, {\rm MeV} \,,
\qquad
f_{B_{s}^{\ast}} = 251^{+14}_{-16} \, {\rm MeV}  \,,
\end{eqnarray}
based upon the standard method of the two-point QCD sum rules.
We present the obtained numerical predictions for the form-factor ratio
$\left [ \left (1 - q^2 /  m_{B_q^{\ast}}^2 \right ) \, f_{B_{q^{\prime}} M}^{+}(q^2) \right ]  :
\left [ f_{B_{q^{\prime}} M}^{0}(q^2) \right ]$ from the combined $z$-expansion fitting program
in the lower-recoil region  in Figure \ref{fig: soft-pion relation of fplus and fzero},
where the complementary predictions with uncertainties from the combination of heavy quark and chiral
symmetries are further displayed explicitly for a comparison.

\begin{figure}
\begin{center}
\includegraphics[width=0.45 \columnwidth]{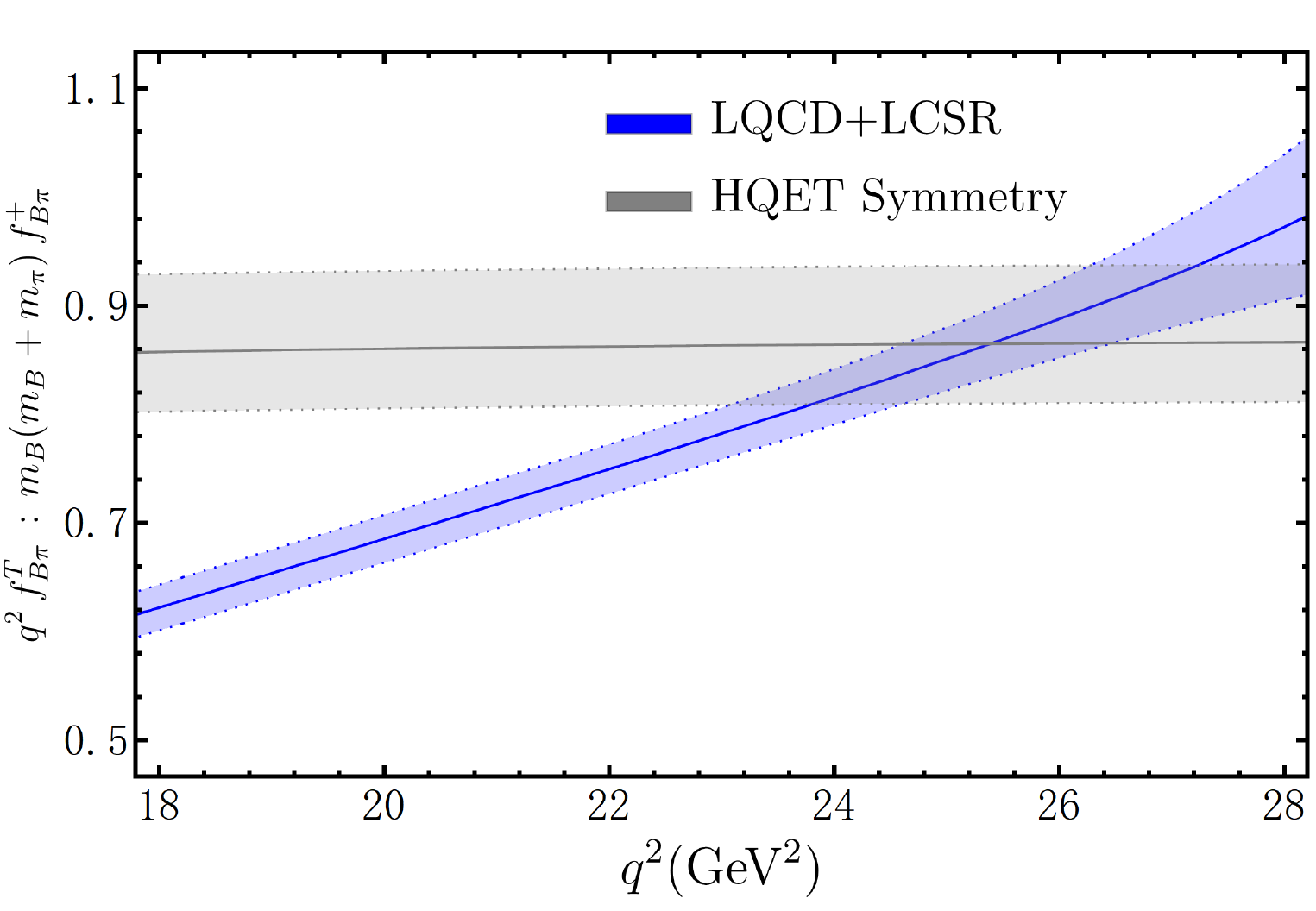}
\hspace{1.0 cm}
\includegraphics[width=0.45 \columnwidth]{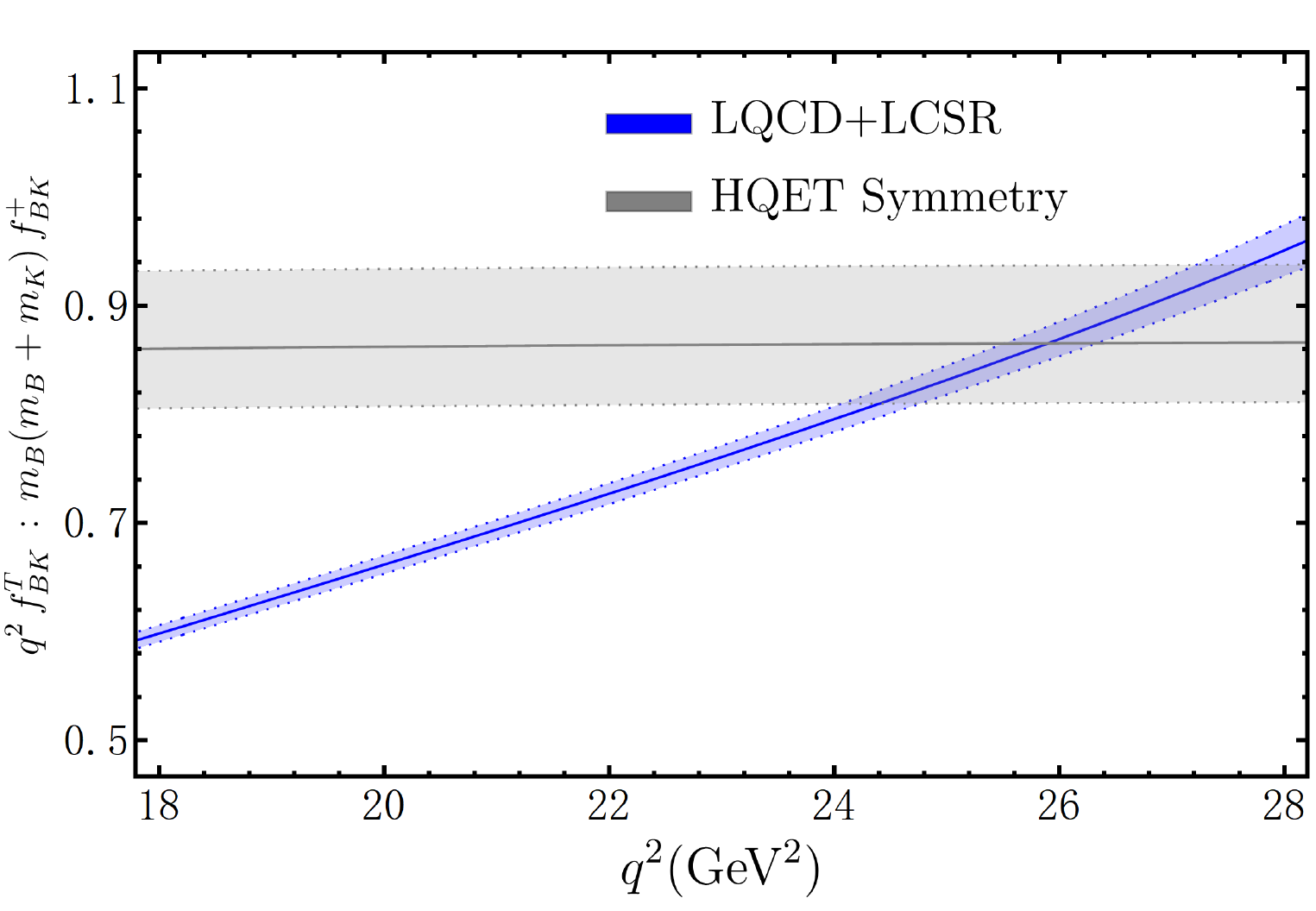}
\\
\vspace*{0.2 cm}
\includegraphics[width=0.45 \columnwidth]{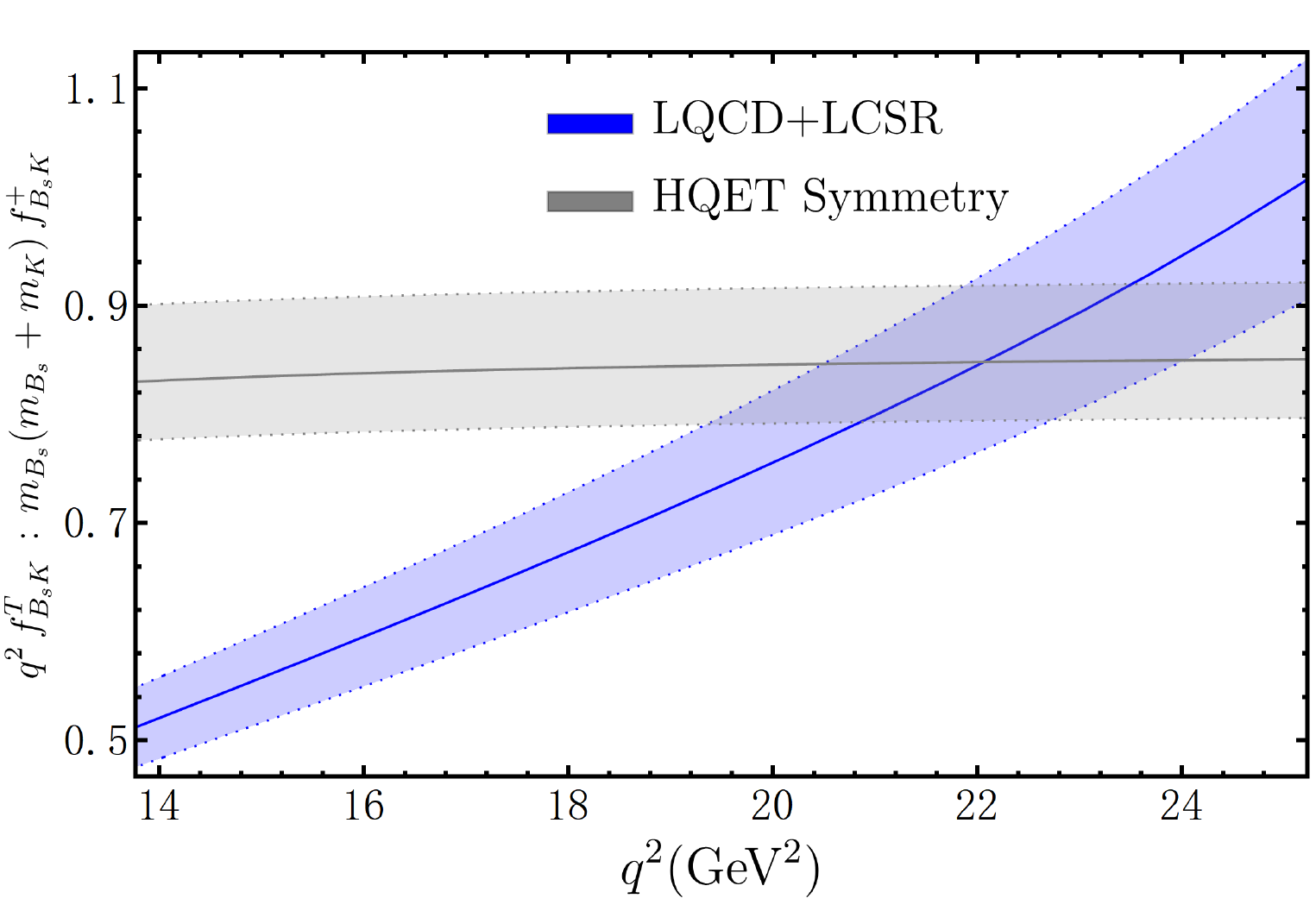}
\vspace*{0.1 cm}
\caption{The low-recoil theory predictions for the intriguing form-factor ratio $\left [  q^2  \, f_{B_{q^{\prime}} M}^{T}(q^2) \right ]
: \left [ m_{B_{q^{\prime}}} \,  (m_{B_{q^{\prime}}} + m_M) \,  f_{B_{q^{\prime}} M}^{+}(q^2) \right ]$
for the exclusive semileptonic $B_{d, s}  \to \pi, K$  decay processes  determined
from the combined BCL $z$-series expansion fitting against both the LCSR and lattice QCD data points
(blue bands) and from the theoretical expectations of the improved Isgur-Wise relation (\ref{soft-pion relation of fplus and fT})
at the NLP accuracy due to the combination of heavy quark and chiral symmetries (grey bands) \cite{Grinstein:2002cz,Grinstein:2004vb}. }
\label{fig: soft-pion relation of fT and fplus}
\end{center}
\end{figure}

The heavy quark spin symmetry relation between the vector and tensor form factors in the low recoil region
can be  derived by exploring  an exact operator identity on account of
the QCD equations of motion for the quark fields
\begin{eqnarray}
 i \, \partial^{\nu} \left (\bar q \, i \, \sigma_{\mu \nu} \, b \right )
 = i \,  \, \partial_{\mu} \left (\bar q  \, b \right )  \,
 - (m_b + m_q) \, \bar q \, \gamma_{\mu} \, b  \,
 - 2 \, \left ( \bar q \, i \, \overleftarrow{D}_{\mu} \, b \right ) \,,
 \label{QCD EOM for the tensor current}
\end{eqnarray}
and by employing the Lorentz decomposition for the subleading heavy-to-light HQET matrix element
of the dimension-four operator \cite{Grinstein:2002cz,Grinstein:2004vb}
\begin{eqnarray}
\langle M(p) | \bar q \, i \, \overleftarrow{D}_{\mu} \, h_v | \bar B(p_B)\rangle
= \delta_{+}(q^2) \, (2 \, p  + q)_{\mu} + \delta_{-}(q^2) \, q_{\mu}  \,.
\label{effective dimension-four HQET matrix element}
\end{eqnarray}
Performing the conventional matching procedure ${\rm QCD} \to {\rm  HQET}$
for the emerged flavour-changing weak currents in the previous identity (\ref{QCD EOM for the tensor current})
\begin{eqnarray}
\bar q \, \gamma_{\mu} \, b  & \to &  C_0^{(v)}(\mu) \,\,  \bar q \, \gamma_{\mu} \, h_v
\, + \, C_1^{(v)}(\mu) \,\,  \bar q \, v_{\mu} \, h_v
\, + \, \left ( {1 \over 2 \, m_b} \right ) \, \bar q \, \gamma_{\mu} \,  i \, \slashed{D}  \, h_v
+ ...\,,
\nonumber \\
\bar q \, i \, \overleftarrow{D}_{\mu} \, b  & \to & D_0^{(v)}(\mu) \, \overline{m}_b(\mu) \,\,  \bar q \, \gamma_{\mu} \, h_v
+ D_1^{(v)}(\mu) \, \overline{m}_b(\mu) \,\,  \bar q \, v_{\mu} \, h_v
+  \bar q \, i \, \overleftarrow{D}_{\mu} \, h_v +  ... \,,
\end{eqnarray}
we can readily derive an improved Isgur-Wise relation between the semileptonic bottom-meson decay form factors
in the small recoil region \cite{Grinstein:2002cz,Bobeth:2011nj}
\begin{eqnarray}
\frac{f_{B_{q^{\prime}} M}^{T}(q^2)} {f_{B_{q^{\prime}} M}^{+}(q^2)}
= \frac{m_{B_{q^{\prime}}}  (m_{B_{q^{\prime}}} + m_M) }{q^2}
\left [{\cal C}_{\rm T+}(\mu) + \left ( {2 \over m_{B_q^{\prime}}} \right )
\frac{\delta_{+}(q^2)}{f_{B_{q^{\prime}} M}^{+}(q^2) } \right ]
+ {\cal O} \left ( \left ({\Lambda_{\rm QCD} \over m_b} \right )^2 \right ),
\hspace{0.8 cm}
\label{soft-pion relation of fplus and fT}
\end{eqnarray}
The short-distance matching function ${\cal C}_{\rm T+}$ can be evidently expressed in terms of the Wilson coefficients
of the HQET currents \cite{Grinstein:2004vb}
\begin{eqnarray}
{\cal C}_{\rm T+}(\mu) = \left [ 1 + \frac{2 \, D_0^{(v)}(\mu)}{C_0^{(v)}(\mu)}  \right ] \,
\frac{\overline{m}_b(\mu)} {m_{B_{q^{\prime}}}} \,,
\end{eqnarray}
where the analytical expressions of $C_0^{(v)}$ and $D_0^{(v)}$ at the  one-loop accuracy
can be written as
\begin{eqnarray}
C_0^{(v)}(\mu) &=&  1 - {\alpha_s(\mu) \, C_F \over 4 \, \pi} \, \left ( 3 \, \ln {\mu \over m_b}  + 4 \right )
+ {\cal O} (\alpha_s^2) \,,
\nonumber \\
D_0^{(v)}(\mu) &=& 0 +   {\alpha_s(\mu) \, C_F \over 4 \, \pi} \, \left ( 2 \, \ln {\mu \over m_b}  + 2 \right )
+ {\cal O} (\alpha_s^2) \,.
\end{eqnarray}
Applying further the HQET equation of motion for the effective field $h_v$ enables us to
derive an important constraint between the two subleading form factors \cite{Grinstein:2002cz}
\begin{eqnarray}
(m_{B_{q^{\prime}}} + v \cdot p) \, \delta_{+}(q^2)
+ (m_{B_{q^{\prime}}} - v \cdot p) \, \delta_{-}(q^2)  = 0   \,.
\end{eqnarray}
Evaluating the effective matrix element (\ref{effective dimension-four HQET matrix element})
with the aid of the heavy-hadron chiral perturbation theory at the lowest order
in the $v \cdot p / \Lambda_{\rm CSB}$ expansion
(with the notation $\Lambda_{\rm CSB}$ characterizing the chiral-symmetry-breaking scale)
leads to the model-independent prediction
\begin{eqnarray}
\delta_{+}(q^2)  - \delta_{-}(q^2) =
\frac{2 \,  \bar \Lambda \, f_{B_{q^{\prime}}} } {3 \, f_M}  \,
\left ( \frac{\hat{g}_{\rm eff}}{ 1 - q^2 /  m_{B_q^{\ast}}^2} \right )  \,,
\end{eqnarray}
where the hadronic parameter $\bar \Lambda$ stands for the ``effective mass" of the bottom-meson state
as previously defined in (\ref{def: effective bottom-meson mass}).
It is then straightforward to determine the desired soft function $\delta_{+}(q^2)$
dictating the considered form-factor ratio (\ref{soft-pion relation of fplus and fT})
\begin{eqnarray}
\delta_{+}(q^2)  =
\frac{\bar \Lambda \, f_{B_{q^{\prime}}} } {3 \, f_M}  \,
\left ( { m_{B_q^{\prime}}^2  + q^2 \over 2 \, m_{B_q^{\prime}}^2} \right ) \,
\left ( \frac{\hat{g}_{\rm eff}}{ 1 - q^2 /  m_{B_q^{\ast}}^2} \right )  \,.
\end{eqnarray}
We present the yielding theory predictions for the very form-factor ratio
$\left [  q^2  \, f_{B_{q^{\prime}} M}^{T}(q^2) \right ]
: \left [ m_{B_{q^{\prime}}}  (m_{B_{q^{\prime}}} + m_M) \,  f_{B_{q^{\prime}} M}^{+}(q^2) \right ]$
from fitting the BCL $z$-series expansion against the LCSR and lattice QCD data points
in the lower-recoil region in Figure \ref{fig: soft-pion relation of fT and fplus},
where the theoretical expectations from  the improved Isgur-Wise relation (\ref{soft-pion relation of fplus and fT})
in the soft final-state meson approximation  in virtue of the heavy quark spin symmetry
are further  shown for a numerical comparison.

\begin{figure}
\begin{center}
\includegraphics[width=0.45 \columnwidth]{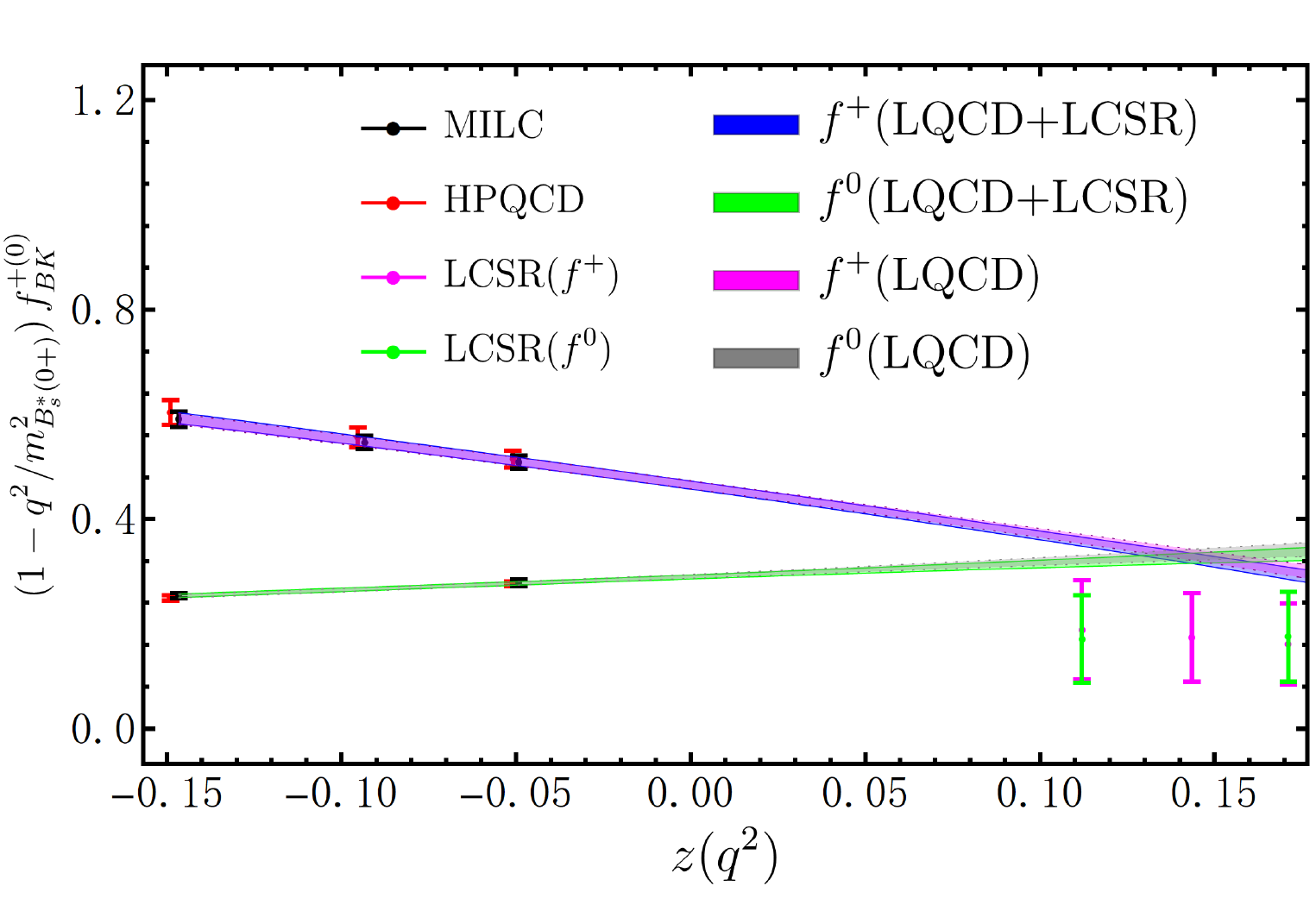}
\hspace{1.0 cm}
\includegraphics[width=0.45 \columnwidth]{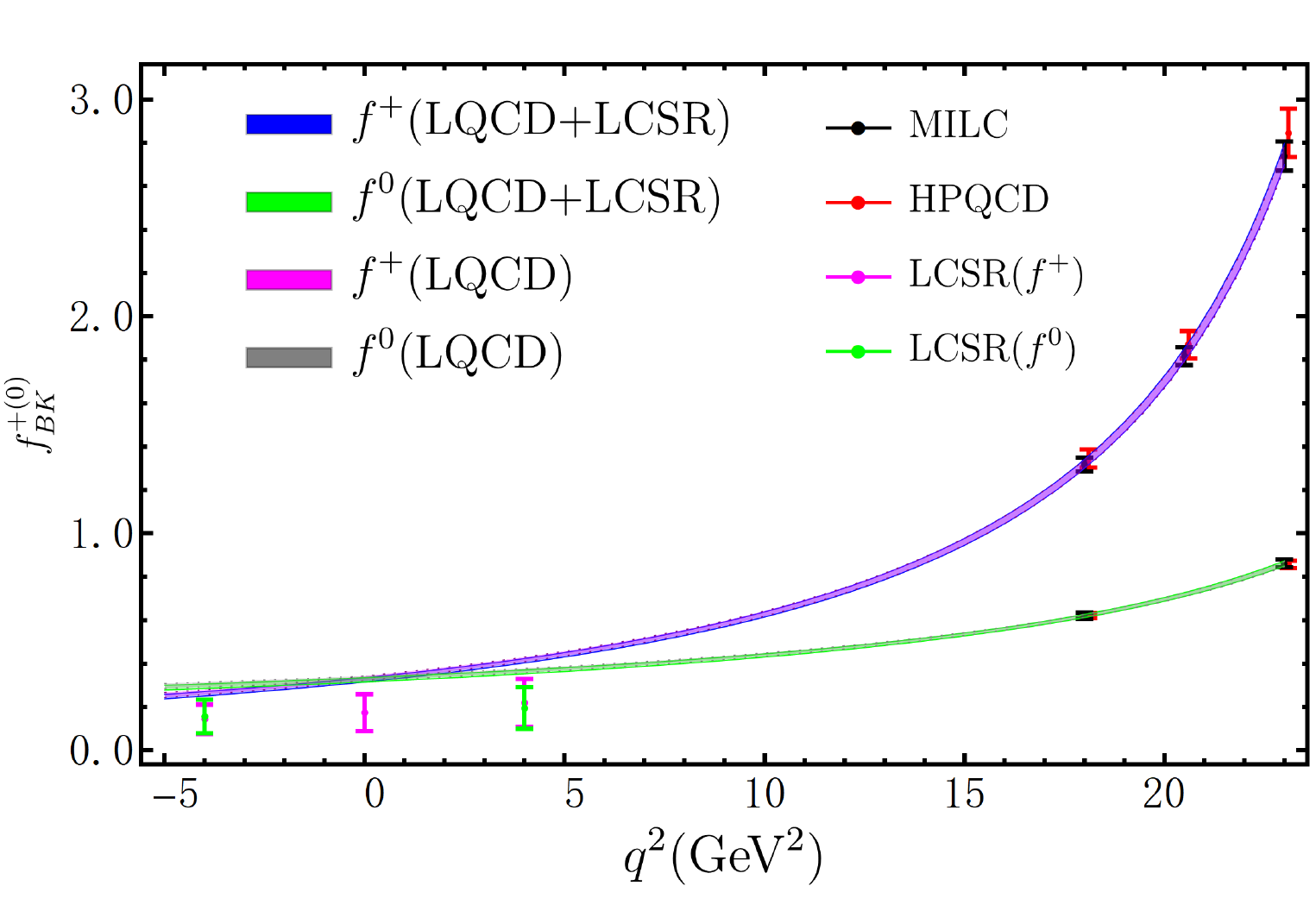}
\\
\vspace*{0.2 cm}
\includegraphics[width=0.45 \columnwidth]{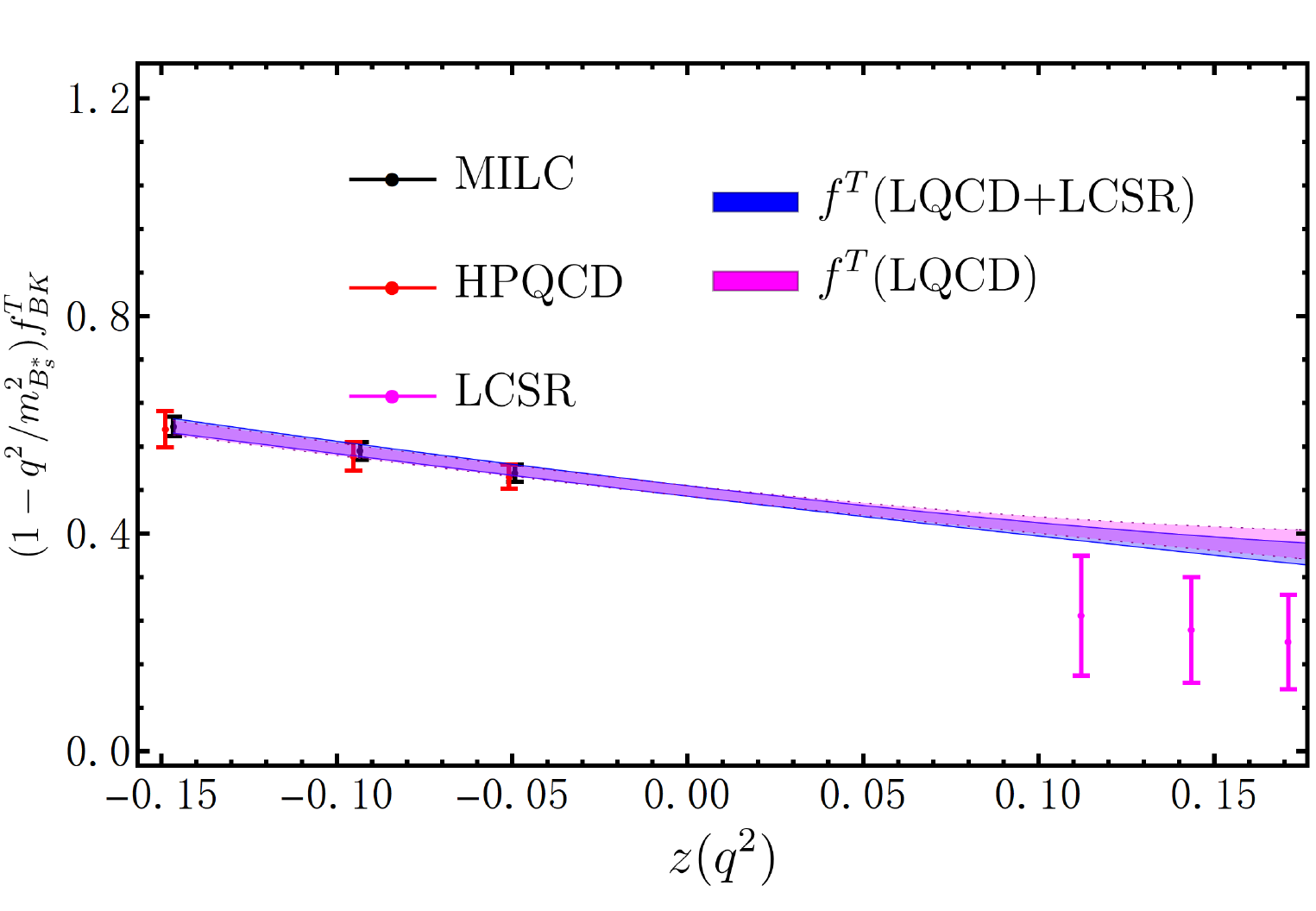}
\hspace{1.0 cm}
\includegraphics[width=0.45 \columnwidth]{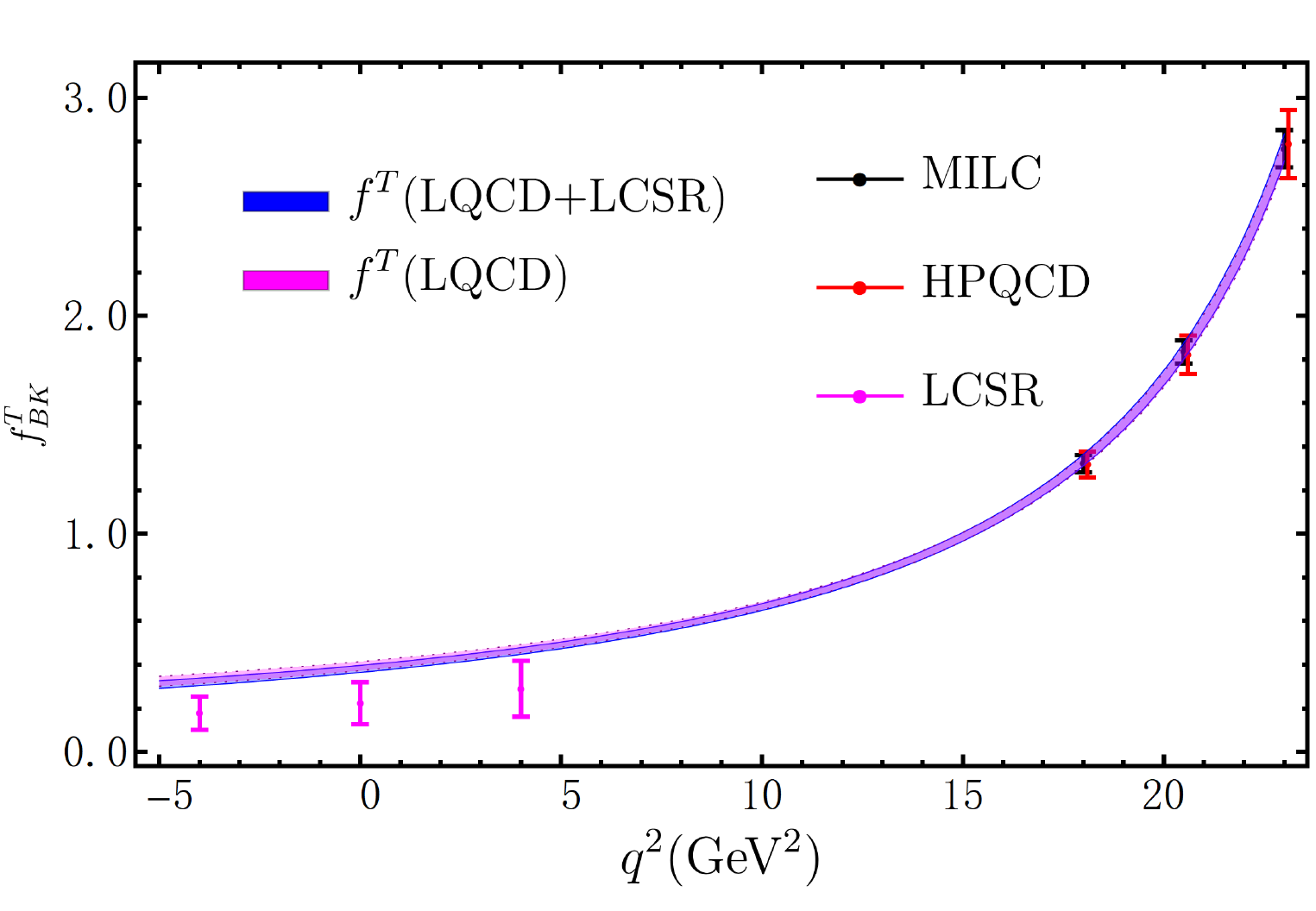}
\vspace*{0.1 cm}
\caption{Theory predictions for the complete set of the semileptonic $B \to K$ form factors
versus  $z$ (left panel) and versus $q^2$ (right panel) in the entire kinematic region
obtained by carrying out the combined BCL $z$-fit of the updated LCSR (from this work)
and lattice simulation (from \cite{Bailey:2015dka,Parrott:2022rgu}) data points.
We further display the yielding numerical results of these essential form factors by performing an independent $z$-series fit
against the ``{\it only lattice QCD}" data points \cite{Bailey:2015dka,Parrott:2022rgu}
exclusively for an instructive  comparison.  }
\label{fig: BCL fit of B to K form factors}
\end{center}
\end{figure}

Subsequently, we  proceed to carry out the combined BCL $z$-expansion fitting of the semileptonic $B \to K$ form factors
against the newly determined LCSR data points at three representative values of
$q^2= \left \{-4.0, \, 0, \, 4.0  \right \} \, {\rm GeV^2}$,  in combination with  the small-recoil lattice QCD results
achieved with the three-flavor gauge-field ensembles  generated by the MILC Collaboration  \cite{Bailey:2015dka}
(employing further the Sheikholeslami-Wohlert (SW) action with the Fermilab interpretation for the bottom quark)
as well as obtained for the first time with $N_f=2+1+1$ gluon field ensembles \cite{Parrott:2022rgu}
(in the meanwhile adopting the highly improved staggered quark (HISQ) formalism for all valence and sea quarks
developed by the HPQCD Collaboration).
Since neither of these lattice collaborations provide explicitly the resulting  physical data points
for the three exclusive $B \to K$ form factors in their publications,
we are then required to generate the  correlated synthetic data points  in the kinematic region
$16.8  \, {\rm GeV}^2 \leq q^2 \leq  22.9 \, {\rm GeV}^2$ from their BCL expansion  fit results
with the truncation $N=3$ as documented in \cite{Bailey:2015dka,Parrott:2022rgu}.
We remark in passing that the HPQCD Collaboration actually adopted the so-called ``modified $z$-expansion" strategy
by simultaneously extrapolate the obtained lattice simulation data to the physical light-quark masses
and zero lattice spacing, and to interpolate their lattice results in the momentum transfer
(see \cite{FlavourLatticeAveragingGroupFLAG:2021npn} for more technical discussions on this alternative approach).
Implementing now the binned $\chi^2$ fit for both the LCSR and lattice simulation results
with the standard BCL parametrization (\ref{z-parametrization for the vector form factor}) and
(\ref{z-parametrization for the scalar and tensor form factors})  leads to our final numerical predictions
of the eight form-factor parameters (with their correlation matrix)
indispensable for the theory description of the electroweak penguin $B \to K \ell \bar \ell$ decays \cite{Khodjamirian:2010vf,Khodjamirian:2012rm} as summarized in Table \ref{BCL fit for the B to K form factors}.
We further observe that our BCL expansion fit brings about a minimal $\chi^2=14.12$
for $16$ degrees of freedom in the fitting procedure, thus corresponding to
an encouraging $p$-value of $59 \, \%$ numerically.
Adopting the tabulated $z$-series coefficients $b_k^{+, 0, T}$ allows us to predict
the desired momentum-transfer dependence for the three semileptonic  $B \to K$ form factors
versus  $z$ (left panel) and versus $q^2$ (right panel) in the entire kinematic region
in Figure \ref{fig: BCL fit of B to K form factors},
where we also display the corresponding theory predictions from performing an independent
$z$-series fitting to the ``{\it only lattice QCD}" data points \cite{Bailey:2015dka,Parrott:2022rgu}
exclusively for the convenience.
It turns out that  the rather remarkable precision for the whole lattice data points
at high momentum transfer from both the FNAL/MILC Collaboration  \cite{Bailey:2015dka}
(with the total uncertainties, including both statistical and systematic errors,
less than $4.0 \, \%$ for all the three $B \to K$ form factors)
and the HPQCD Collaboration \cite{Parrott:2022rgu}
(with the uncertainties below $\{ 2.0 \, \%, 4.0 \, \%,  5.5 \, \% \}$ for
the three form factors $\{f_{B K}^{+}, \,  f_{B K}^{0},  \, f_{B K}^{T} \}$ in consequence)
makes it  challenging to carry out the combined BCL $z$-series fit with high quantity,
by simultaneously  accommodating  the achieved LCSR predictions at low momentum transfer
within the individual $1.0 \, \sigma$ intervals
(albeit with the very sizeable theory uncertainties of the order of $50 \, \%$).
Actually, this intriguing observation  can be attributed to the very fact that
our updated LCSR computations with the bottom-meson distribution amplitudes will bring about
the strongly correlated numerical results for the exclusive $B \to K$ form factors
at the different kinematic points  (despite of the quite uncertain central values as explained above)
in the large hadronic recoil region, which  exhibit the delicacy tension with
the extrapolated lattice QCD predictions with the extraordinary high accuracy in the low recoil region.
In this respect, it would be in high demand to deepen further our understanding,
on the one hand,  towards the momentum dependence of the HQET bottom-meson distribution amplitudes
$\phi_B^{\pm}(\omega, \mu_0)$ at the renormalization scale $\mu_0=1.0 \, {\rm GeV}$
from the first field-theoretical principles on the bottom-meson LCSR aspect,
and on the other hand, towards the unquantified systematic uncertainties of the lattice simulation method
(for instance, several potential concerns with the ``modified $z$-expansion" proposal
as previously discussed in \cite{FlavourLatticeAveragingGroupFLAG:2021npn}).

Furthermore, we confront the combined BCL expansion fit results for the two particular form-factor ratios
(\ref{soft-pion relation of fplus and fzero}) and   (\ref{soft-pion relation of fplus and fT})
at high momentum transfer with the counterpart model-independent  predictions from
the combination of the heavy quark spin symmetry and the current algebra technique
in Figures \ref{fig: soft-pion relation of fplus and fzero}
and \ref{fig: soft-pion relation of fT and fplus}.
Generally, the resulting BCL $z$-fit predictions for the considered low-recoil symmetry breaking corrections
appear to be in reasonable agreement with the theoretical expectations from the HQET symmetry relations
at the unphysical kinematic point $q^2= 27.9 \, {\rm GeV^2}$ within the obtained uncertainties.
By contrast, our BCL fit result for the scalar form-factor ratio at the zero-recoil limit
differs from the counterpart prediction with the heavy quark symmetry and the soft-kaon approximation
by an enormous amount of ${\cal O}(50 \, \%)$,
thus confirming the previous lattice simulation results  from the  FNAL/MILC Collaboration \cite{Bailey:2015dka}.
We are then led to conclude immediately that employing  the derived low-recoil symmetry relations
for the exclusive $B \to K \ell \bar \ell$ phenomenological applications
could result in the substantial derivations from the direct QCD predictions
due to the numerically pronounced NLP corrections in the heavy quark expansion.

\begin{table}
\centering
\renewcommand{\arraystretch}{2.0}
\resizebox{\columnwidth}{!}{
\begin{tabular}{|c|c||cccccccc|}
\hline
\hline
 \multicolumn{2}{|c||}{$B \to K$ \, Form Factors}  & \multicolumn{8}{|c|}{Correlation Matrix}
\\
\hline
Parameters & Values & $b_0^+$ & $b_1^+$ & $b_2^+$ & $b_0^0$ & $b_1^0$ & $b_0^T$ & $b_1^T$ & $b_2^T$
 \\
 \hline
 \hline
 $b_0^+$ & $0.465(8)$ & $1$ & $-0.035$ & $-0.494$ & $0.711$ & $0.283$ & $0.744$ & $0.017$ & $-0.281$
 \\
 $b_1^+$ & $-0.925(53)$ &  &  $1$ & $-0.034$ & $0.160$ & $0.831$ & $-0.111$ & $0.691$ & $-0.056$
 \\
 $b_2^+$ & $-0.355(257)$ &  &  & $1$ & $-0.038$ & $0.166$ & $-0.299$ & $-0.099$ & $0.533$
 \\
 $b_0^0$ & $0.290(4)$ &  &  &  & $1$ & $0.422$ & $0.594$ & $0.061$ & $-0.015$
 \\
 $b_1^0$ & $0.246(38)$ &  &  &  &  & $1$ & $0.141$ & $0.595$ & $0.14$
 \\
 $b_0^T$ & $0.479(10)$ &  &  &  &  &  & $1$ & $-0.023$ & $-0.283$
 \\
 $b_1^T$ & $-0.759(74)$ &  &  &  &  &  &  &  $1$ & $0.305$
 \\
 $b_2^T$ & $-0.479(324)$ &  &  &  &  &  &  &  &  $1$
 \\
 \hline
 \hline
\end{tabular}
}
\renewcommand{\arraystretch}{1.0}
\caption{Theory predictions for the correlated $z$-series coefficients
in the vector, scalar and tensor $B \to K$ form factors
determined by fitting the BGL parametrization simultaneously
against our LCSR results including a variety of the subleading-power corrections
and the available  lattice QCD data points from \cite{Bailey:2015dka,Parrott:2022rgu}
with the preferred truncation $N=3$. }
\label{BCL fit for the B to K form factors}
\end{table}

\begin{figure}
\begin{center}
\includegraphics[width=0.45 \columnwidth]{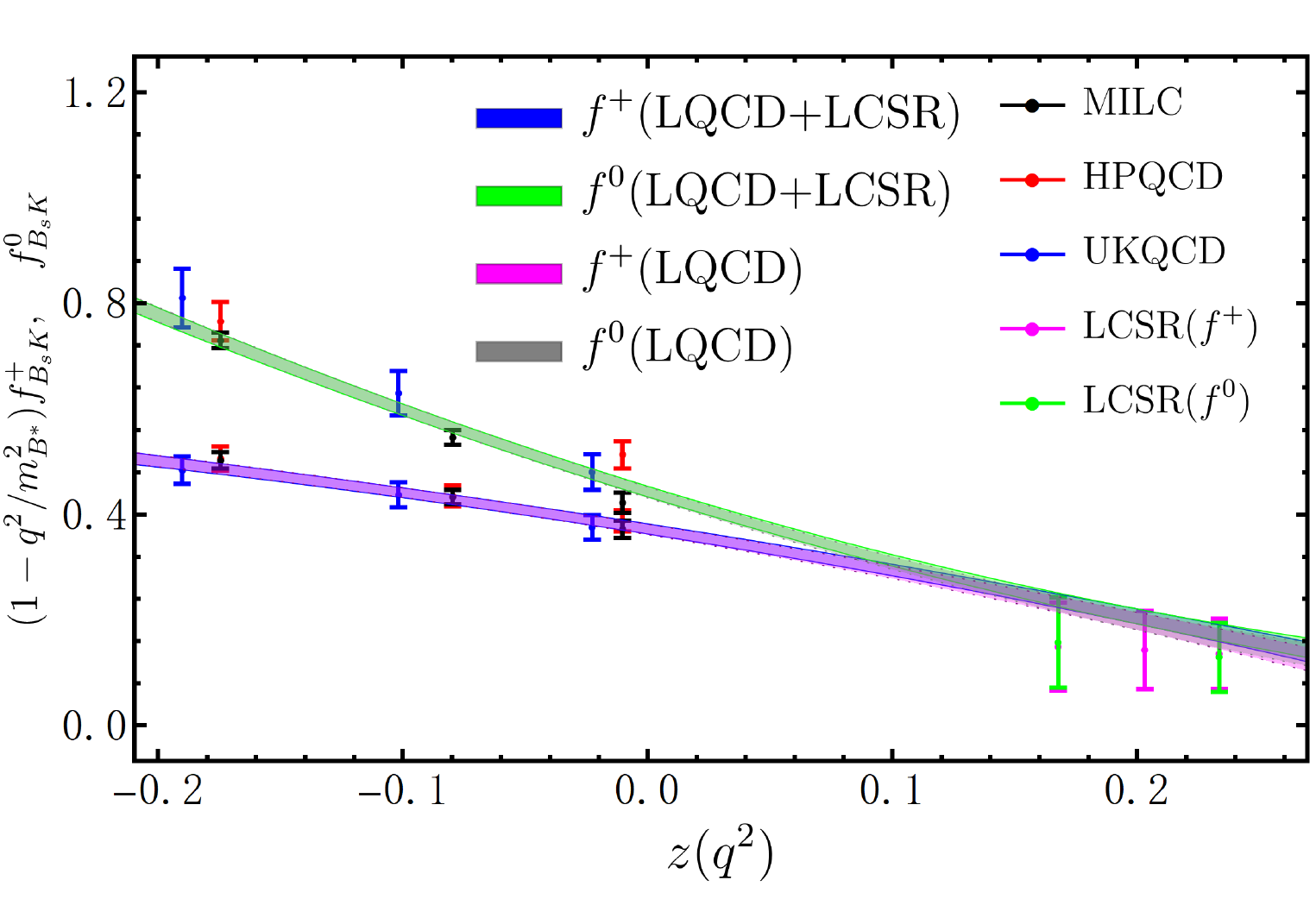}
\hspace{1.0 cm}
\includegraphics[width=0.45 \columnwidth]{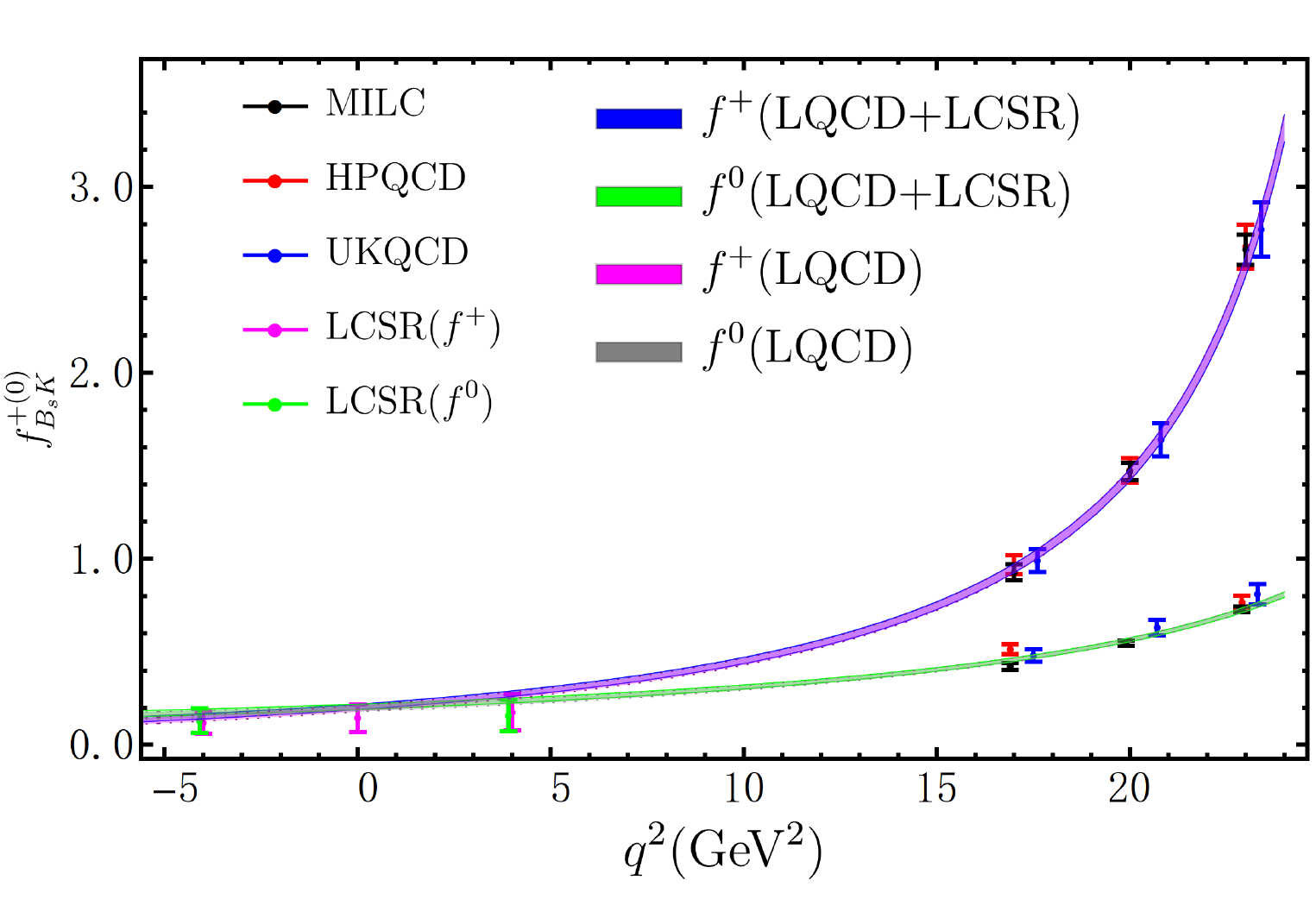}
\\
\vspace*{0.2 cm}
\includegraphics[width=0.45 \columnwidth]{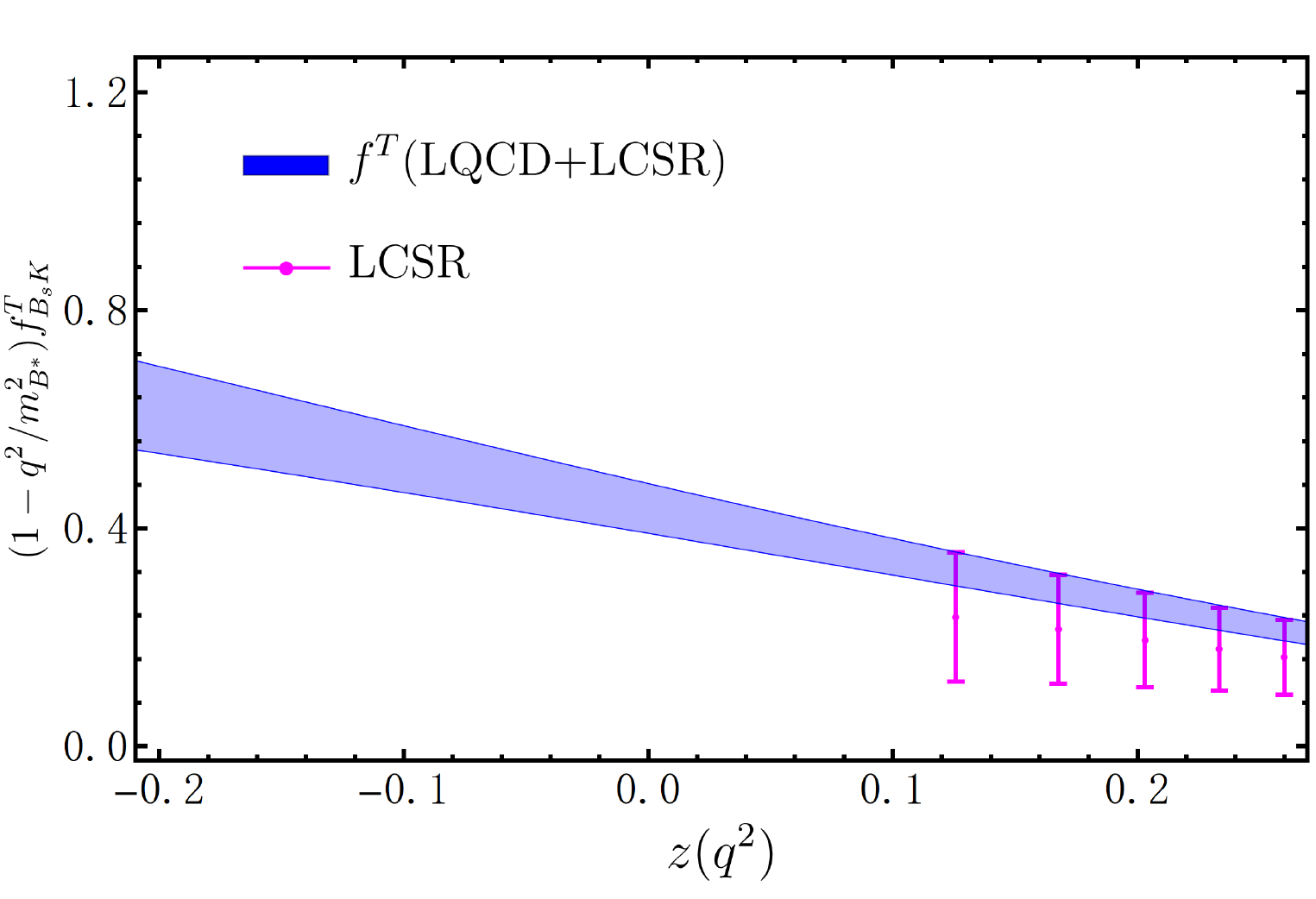}
\hspace{1.0 cm}
\includegraphics[width=0.45 \columnwidth]{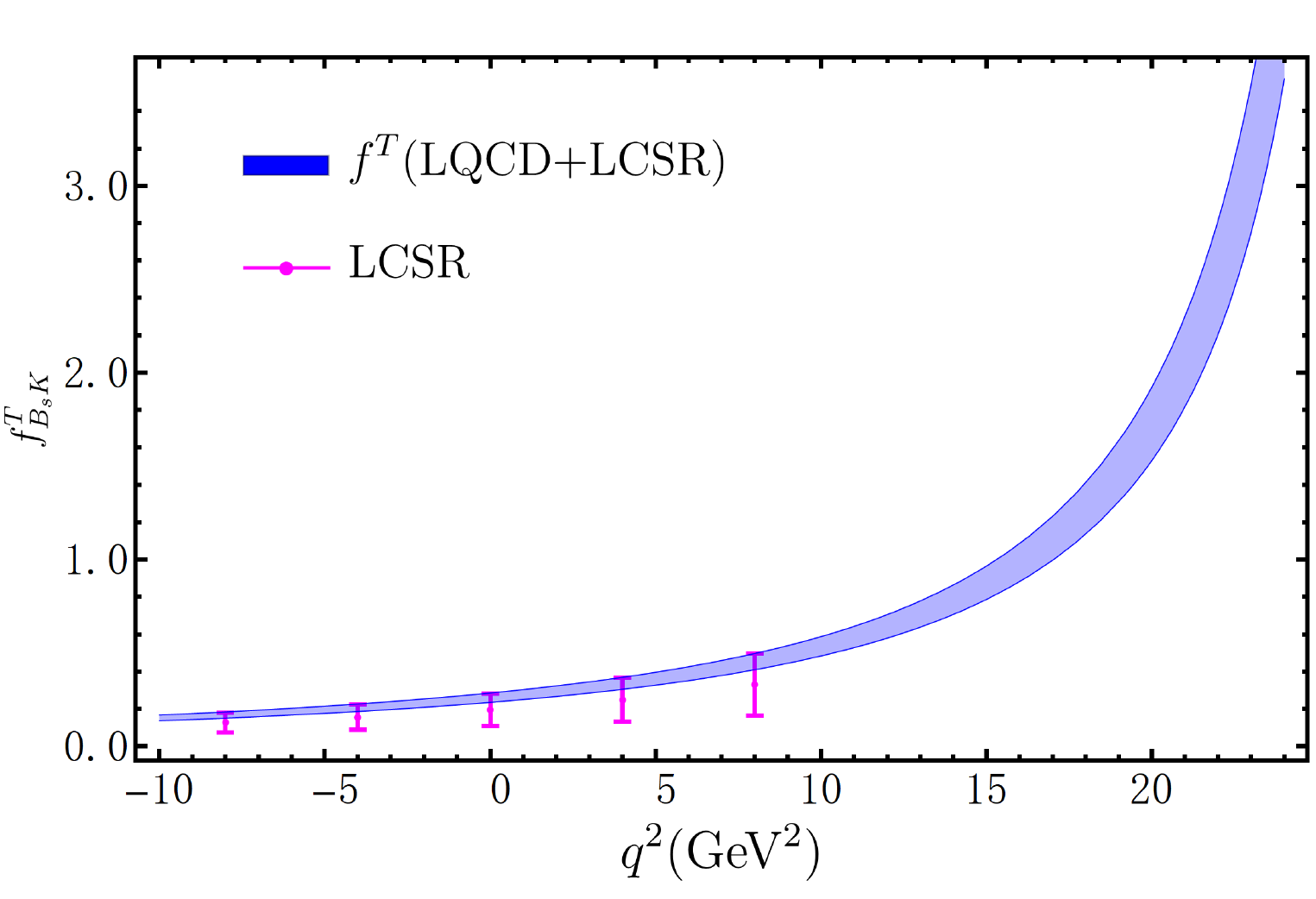}
\vspace*{0.1 cm}
\caption{Theory predictions for the complete set of the semileptonic $B_s \to K$ form factors
versus  $z$ (left panel) and versus $q^2$ (right panel) in the entire kinematic region
obtained by carrying out the combined BCL $z$-fit of the updated LCSR (from this work)
and lattice simulation (merely for the two form factors $f_{B_s K}^{+, \, 0}(q^2)$ 
from \cite{Flynn:2015mha,Bouchard:2014ypa,FermilabLattice:2019ikx}) data points.
We further display the yielding numerical results for the vector and scalar  form factors
by performing an independent $z$-series fit against the
 ``{\it only lattice QCD}" data points \cite{Flynn:2015mha,Bouchard:2014ypa,FermilabLattice:2019ikx}
exclusively for the illustration purpose.  }
\label{fig: BCL fit of Bs to K form factors}
\end{center}
\end{figure}

Along the same vein, we will continue to perform the combined BCL expansion fitting of
the semileptonic $B_s \to K$ form factors against  our improved LCSR predictions
and the available lattice results for the vector and scalar form factors
\cite{Flynn:2015mha,Bouchard:2014ypa,FermilabLattice:2019ikx} simultaneously.
On account of the very absence of the lattice simulation results for the tensor form factor,
we prefer to take advantage of the determined  LCSR data points for $f_{B_s K}^{T}(q^2)$
at five representative values of
$q^2 = \left \{-8.0, \, -4.0, \, 0, \, 4.0, \, 8.0   \right \} \, {\rm GeV^2}$,
while adopting  the large-recoil LCSR  predictions for  the vector and scalar form factors $f_{B_s K}^{+, \, 0}(q^2)$
at three distinct kinematic points of $q^2 = \left \{-4.0, \, 0, \, 4.0  \right \} \, {\rm GeV^2}$
as the same as before.
While the RBC/UKQCD Collaboration \cite{Flynn:2015mha} provides explicitly the synthetic data points of
$f_{B_s K}^{+}(q^2)$ and $f_{B_s K}^{0}(q^2)$
at three representative values of $q^2 = \left \{17.6, \, 20.8, \, 23.4  \right \} \, {\rm GeV^2}$
with the normalized statistical and systematic correlation matrices,
both the HPQCD Collaboration \cite{Bouchard:2014ypa} and the FNAL/MILC Collaboration \cite{FermilabLattice:2019ikx}
only provide their BCL $z$-fit results for the form-factor shape parameters with the truncations
$N=3$ and $N=4$,  respectively.
Consequently, we are then required to produce the necessary lattice data points for $f_{B_s K}^{+, \, 0}(q^2)$ in the kinematic region
$17.0  \, {\rm GeV}^2 \leq q^2 \leq  23.7 \, {\rm GeV}^2$
in order to utilize the complete information of the lattice QCD fits from  \cite{Bouchard:2014ypa,FermilabLattice:2019ikx}.
Performing now the combined BCL fit against both our LCSR predictions in the large recoil region
and the yielding lattice simulation data points in the low recoil region
gives rise to the final predictions for the $z$-expansion coefficients (with their correlation matrix)
for the three semileptonic $B_s \to K$ form factors as collected in Table \ref{BCL fit for the Bs to K form factors}.
In addition,  our numerical fit leads to a slightly larger number  of  $\chi^2=28.58$
for $19$ degrees of freedom in the fitting program.
Unsurprisingly, the obtained BCL fit results for the three coefficients $b_{0, 1, 2}^T$
in Table \ref{BCL fit for the Bs to K form factors}
turn out to be more uncertain compared with the corresponding predictions
for the $z$-expansion parameters of the tensor $B \to \pi$ form factor
collected in Table \ref{BCL fit foe the B to pi form factors},
due to the unavailable lattice data points for the form factor $f_{B_s K}^{T}(q^2)$
at large momentum transfer.
Under such circumstance, achieving the lattice simulation determination
for the very tensor form factor $f_{B_s K}^{T}(q^2)$ at high $q^2$ will be evidently crucial
to pin down the current theory uncertainties from the BCL extrapolation of the LCSR results,
thus providing the fundamental ingredient for the model-independent description
of the exclusive electroweak penguin $\bar B_s \to \bar K^0 \ell \bar \ell$  decays.
Employing the tabulated $z$-series coefficients $b_k^{+, 0, T}$ further enables us
to predict the desired momentum-transfer dependence for the exclusive  $B_s \to K$ form factors
versus  $z$ (left panel) and versus $q^2$ (right panel) in the entire kinematic region
in Figure \ref{fig: BCL fit of Bs to K form factors},
where we also display the alternative BCL fit results for the vector and scalar form factors
with the ``{\it only lattice QCD}" data points \cite{Flynn:2015mha,Bouchard:2014ypa,FermilabLattice:2019ikx}
exclusively.
In the light of the high-precision lattice data points for $f_{B_s K}^{+, \, 0}(q^2)$ at small hadronic recoil
from the RBC/UKQCD Collaboration \cite{Flynn:2015mha} (with the total uncertainties below
$\left \{6.2 \, \%, \, 7.1 \, \% \right \}$ for the vector and scalar form factors, respectively),
from HPQCD Collaboration \cite{Bouchard:2014ypa}
(with the combined uncertainties below $\left \{5.0 \, \%, \, 6.0 \, \% \right \}$
for the two form factors $\{f_{B_s K}^{+}, \,  f_{B_s K}^{0} \}$ in consequence)
and from the FNAL/MILC Collaboration \cite{FermilabLattice:2019ikx}
(with the uncertainties less than $3.0 \, \%$ for both the two form factors),
the resulting theory benefits  from  the combined BCL $z$-expansion fit
to both the LCSR and lattice simulation results  consist in
the rather moderate improvements (at the level of ${\cal O}(20 \, \%)$ numerically)
of the large-recoil form factor predictions  on the counterpart BCL fitting procedure
with the ``{\it only lattice QCD}" data points.

We proceed to compare the combined BCL expansion fitting predictions for the two form-factor ratios
(\ref{soft-pion relation of fplus and fzero}) and   (\ref{soft-pion relation of fplus and fT})
in the low recoil region  with the corresponding model-independent computations based upon
the combination of heavy quark and chiral symmetries
in Figures \ref{fig: soft-pion relation of fplus and fzero}
and \ref{fig: soft-pion relation of fT and fplus}.
We can draw an analogous conclusion (as previously observed in the context of the $B \to K$  form factors)
that the derived  HQET symmetry relations  for the exclusive $B_s \to K$ form factors appear to be well respected
at the unphysical  kinematic point $q^2=28.8 \, {\rm GeV^2}$ within the theory uncertainties.
Apparently, our numerical result for the particular form-factor ratio
$\left [  q^2  \, f_{B_s K}^{T}(q^2) \right ]
: \left [ m_{B_s}  (m_{B_{s}} + m_K) \,  f_{B_s K}^{+}(q^2) \right ]$
suffers from the more pronounced theory uncertainty as displayed
in Figure \ref{fig: soft-pion relation of fT and fplus},
due to the relatively less precise BCL-fitting  prediction for the tensor form factor $f_{B_s K}^{+}(q^2)$.
Moreover, the very low-recoil symmetry breaking correction to the scalar form-factor ratio
$\left [ \left (1 - q^2 /  m_{B^{\ast}}^2 \right ) \, f_{B_{s} K}^{+}(q^2) \right ]  :
\left [ f_{B_{s} K}^{0}(q^2) \right ]$  at the maximal momentum transfer
turns out to be even greater than the counterpart theory predictions
for both the semileptonic $B \to \pi, K$ decay form-factor ratios
as shown in  Figure \ref{fig: soft-pion relation of fplus and fzero}.

\begin{table}
\centering
\renewcommand{\arraystretch}{2.0}
\resizebox{\columnwidth}{!}{
\begin{tabular}{|c|c||cccccccc|}
\hline
\hline
 \multicolumn{2}{|c||}{$B_s \to K$ \, Form Factors}  & \multicolumn{8}{|c|}{Correlation Matrix}
\\
\hline
Parameters & Values & $b_0^+$ & $b_1^+$ & $b_2^+$ & $b_0^0$ & $b_1^0$ & $b_0^T$ & $b_1^T$ & $b_2^T$
 \\
 \hline
 \hline
 $b_0^+$ & $0.373(10)$ & $1$ & $0.202$ & $-0.294$ & $0.653$ & $0.135$ & $0.422$ & $-0.277$ & $0.103$
 \\
 $b_1^+$ & $-0.731(41)$ &  &  $1$ & $0.075$ & $0.255$ & $0.683$ & $0.488$ & $-0.301$ & $0.092$
 \\
 $b_2^+$ & $-0.473(146)$ &  &  & $1$ & $0.099$ & $0.366$ & $0.215$ & $-0.121$ & $0.025$
 \\
 $b_0^0$ & $0.443(10)$ &  &  &  & $1$ & $-0.024$ & $0.407$ & $-0.268$ & $0.100$
 \\
 $b_1^0$ & $-1.427(45)$ &  &  &  &  & $1$ & $0.433$ & $-0.262$ & $0.075$
 \\
 $b_0^T$ & $0.437(46)$ &  &  &  &  &  & $1$ & $-0.835$ & $0.550$
 \\
 $b_1^T$ & $-0.900(167)$ &  &  &  &  &  &  &  $1$ & $-0.914$
 \\
 $b_2^T$ & $0.091(172)$ &  &  &  &  &  &  &  &  $1$
 \\
 \hline
 \hline
\end{tabular}
}
\renewcommand{\arraystretch}{1.0}
\caption{Theory predictions for the correlated $z$-series coefficients
in the vector, scalar and tensor $B_s \to K$ form factors
determined by fitting the BGL parametrization simultaneously
against our LCSR results including a variety of the subleading-power corrections
and the available  lattice QCD data points from
\cite{Flynn:2015mha,Bouchard:2014ypa,FermilabLattice:2019ikx}
with the preferred truncation $N=3$. }
\label{BCL fit for the Bs to K form factors}
\end{table}

\subsection{Phenomenological analysis of the   $B_{(s)}  \to \pi (K) \ell \bar \nu_{\ell}$ observables }

Having at our disposal the combined BCL $z$-fit results  for the exclusive $B \to \pi, K$ form factors,
we are now prepared to explore their phenomenological implications on the semileptonic
$B_{(s)}  \to \pi (K) \ell \bar \nu_{\ell}$ decay observables
constructed from the corresponding full angular distributions,
such as  the differential branching fractions, the normalized forward-backward asymmetries,
the new-physics (NP) sensitive ``flat terms" (vanishing in the massless lepton limit in the SM),
and the lepton-flavour universality ratios, the lepton polarization asymmetries.
In particular,  the ever-increasing precision measurements
on the binned $q^2$ distributions  for  the golden exclusive  $B \to \pi \ell \bar \nu_{\ell}$
(with $\ell=e, \, \mu$) decay processes from the BaBar Collaboration \cite{BaBar:2010efp,BaBar:2012thb},
the Belle Collaboration \cite{Belle:2010hep,Belle:2013hlo}
as well as the  Belle II \cite{Belle-II:2022imn} Collaboration enable us to
further extract the desired CKM matrix element $|V_{ub}|$ straightforwardly
in combination with  our improved determination of the vector form factor $f_{B \pi}^{+}(q^2)$
in the entire kinematic regime.
In order to achieve this goal, we first present the explicit expression
for  the full differential decay distribution of $B_{q^{\prime}} \to M \ell \bar \nu_{\ell}$
with respect to the two kinematic variables $q^2$ and $\cos \theta_{\ell}$
(dropping out the very intricate but numerically subdominant electromagnetic correction)
\begin{eqnarray}
\frac{d^2 \Gamma(B_{q^{\prime}} \to M \ell \bar \nu_{\ell})}
{d q^2 \, d \cos \theta_{\ell} } = a_{\theta_{\ell}}(q^2)
+ b_{\theta_{\ell}}(q^2) \, \cos \theta_{\ell}
+  c_{\theta_{\ell}}(q^2) \, \cos^2 \theta_{\ell}  \,,
\label{angular distribution of B to M l nu}
\end{eqnarray}
where the three $q^2$-dependent angular coefficient functions are given by \cite{Becirevic:2016hea}
\begin{eqnarray}
a_{\theta_{\ell}}(q^2)  &=& {\cal N}_{\rm ew} \,\,  \lambda^{3/2} \,
\left ( 1 - {m_{\ell}^2 \over q^2} \right  )^2 \,
\left [ \left | f_{B_{q^{\prime}} M}^{+}(q^2) \right |^2
+ {1 \over \lambda} \,\, {m_{\ell}^2 \over q^2} \,  \left ( 1 - { m_M^2  \over m_{B_{q^{\prime}} }^2 } \right )^2  \,
\left | f_{B_{q^{\prime}} M}^{0}(q^2) \right |^2 \right ] \,,
\hspace{1.0 cm}
 \\
b_{\theta_{\ell}}(q^2)  &=&  2 \, {\cal N}_{\rm ew} \,  \lambda \, \left ( 1 - {m_{\ell}^2 \over q^2} \right  )^2 \,
{m_{\ell}^2 \over q^2} \,  \left ( 1 - { m_M^2  \over m_{B_{q^{\prime}} }^2 } \right )  \,
{\rm Re} \left [ f_{B_{q^{\prime}} M}^{+}(q^2) \, f_{B_{q^{\prime}} M}^{0 \, \ast}(q^2) \right ] \,,
 \\
c_{\theta_{\ell}}(q^2)  &=& -  {\cal N}_{\rm ew} \,\,  \lambda^{3/2} \,
\left ( 1 - {m_{\ell}^2 \over q^2} \right  )^3 \, \left | f_{B_{q^{\prime}} M}^{+}(q^2) \right |^2 \,,
\end{eqnarray}
and we have introduced the following shorthand notations for convenience
\begin{eqnarray}
&& {\cal N}_{\rm ew} = \frac{G_F^2 \, |V_{ub}|^2 \,  m_{B_{q^{\prime}}}^3 }{256 \, \pi^3} \,,
\qquad
\lambda \equiv \lambda \left (1, \, {m_M^2 \over m_{B_{q^{\prime}}}^2}, \, {q^2 \over m_{B_{q^{\prime}}}^2} \right )\,,
\nonumber \\
&& \lambda(a, b, c) \equiv  a^2 + b^2 + c^2 - 2 \, (ab + ac +bc) \,.
\end{eqnarray}
In addition,  the helicity angle $\theta_{\ell}$ is defined as the angle between the $\ell^{-}$ direction of flight
and the final-state meson  momentum in the dilepton rest frame.
We can immediately observe two interesting algebra relations for the angular functions
$b_{\theta_{\ell}}(q^2) =0$ and  $a_{\theta_{\ell}}(q^2) + c_{\theta_{\ell}}(q^2) =0$
in the massless lepton limit.

\begin{table}
\centering
\renewcommand{\arraystretch}{2.0}
\resizebox{0.80 \columnwidth}{!}{
\begin{tabular}{|c|c||cccccc|}
\hline
\hline
 Parameters  & Values & \multicolumn{6}{|c|}{Correlation Matrix}
 \\ \hline
 $b_0^+$ & $0.409(12)$ & $1$ & $0.111$ & $-0.452$ & $0.298$ & $0.119$ & $-0.87$
 \\
 $b_1^+$ & $-0.507(42)$ &  & $1$ & $-0.793$ & $-0.088$ & $0.203$ & $-0.298$
 \\
 $b_2^+$ & $-0.267(152)$ &  &  & $1$ & $0.008$ & $-0.096$ & $0.405$
 \\
 $b_0^0$ & $0.507(18)$ &  &  &  & $1$ & $-0.445$ & $-0.266$
 \\
 $b_1^0$ & $-1.446(45)$ &  & &  & &  $1$ & $-0.201$
 \\
 $|V_{ub}|$ & $3.76(13) \times 10^{-3}$ &  &  & &  &  & $1$
 \\
 \hline
 \hline
\end{tabular}
}
\renewcommand{\arraystretch}{1.0}
\vspace{0.3 cm}
\caption{Theory predictions for the  $B \to \pi \ell \bar \nu_{\ell}$
form-factor shape parameters and the CKM matrix element $|V_{ub}|$
(with their correlation matrix) from carrying out the simultaneous fit
against the SCET sum rules, lattice QCD and experimental data points
with the aid of the truncated BCL $z$-parameterizations at $N=3$.}
\label{BCL fit results of Vub with N=3}
\end{table}

Integrating over the helicity angle $\theta_{\ell}$ allows for spelling out the expression
for the differential decay rate of $B_{q^{\prime}} \to M \ell \bar \nu_{\ell}$
in the bottom-meson rest frame
\begin{eqnarray}
\frac{d \Gamma(B_{q^{\prime}} \to M \ell \bar \nu_{\ell})}
{d q^2 } &=& \int_{-1}^{1} d \cos \theta_{\ell}  \,\,
\frac{d^2 \Gamma(B_{q^{\prime}} \to M \ell \bar \nu_{\ell})}
{d q^2 \, d \cos \theta_{\ell} }
= 2 \, \left [  a_{\theta_{\ell}}(q^2) +  {1 \over 3} \, c_{\theta_{\ell}}(q^2) \right ]
\nonumber \\
&=& \frac{G_F^2 \, |V_{ub}|^2 \,  m_{B_{q^{\prime}}}^3 }{192 \, \pi^3} \, \lambda^{3/2} \,
\left ( 1 - {m_{\ell}^2 \over q^2} \right  )^2 \,
\bigg \{  \left ( 1 + {m_{\ell}^2 \over 2 \, q^2} \right ) \,
\left | f_{B_{q^{\prime}} M}^{+}(q^2) \right |^2
\nonumber \\
&&  + \, {1 \over \lambda} \,    {3 \, m_{\ell}^2 \over 2 \, q^2} \,
\left ( 1 - { m_M^2  \over m_{B_{q^{\prime}} }^2 } \right )^2  \,
\left | f_{B_{q^{\prime}} M}^{0}(q^2) \right |^2  \bigg \}  \,,
\end{eqnarray}
which can be further employed to determine the  measurable $q^2$-binned branching fractions.
Following the strategy presented in \cite{FlavourLatticeAveragingGroupFLAG:2021npn,Leljak:2021vte,Biswas:2021qyq,Biswas:2021cyd},
we will turn  to  extract the magnitude of the CKM matrix element $|V_{ub}|$
by carrying out a simultaneous fit to the SCET sum rules, lattice QCD and experimental data points
with the aid of the constrained BCL $z$-series parameterizations,
thus leaving their relative normalization $|V_{ub}|$ as a free parameter.
As emphasized previously in \cite{FlavourLatticeAveragingGroupFLAG:2021npn},
this attractive  fitting  strategy combines the theoretical and experimental inputs in a more efficient manner,
yielding a somewhat smaller uncertainty on $|V_{ub}|$ numerically.
Taking advantage of the available state-of-the-art experimental data sets
from the three untagged measurements by the BaBar Collaboration \cite{BaBar:2012thb,BaBar:2012thb}
and the Belle Collaboration \cite{Belle:2010hep} assuming isospin symmetry,
from the two tagged measurements of $\bar B^0 \to \pi^{+} \ell \bar \nu_{\ell}$
and  $B^{-} \to \pi^{0} \ell \bar \nu_{\ell}$ by the Belle Collaboration \cite{Belle:2013hlo},
and from the untagged  $B^0 \to \pi^{-} \ell \bar \nu_{\ell}$ measurements
by the Belle II Collaboration \cite{Belle-II:2022imn},
we display the numerical fit results for both the form-factor shape parameters
with the truncation $N=3$ for the BCL $z$-expansion and $|V_{ub}|$
in Table \ref{BCL fit results of Vub with N=3}, including their correlation matrix.
The quality of the binned maximum-likelihood fit can be understood from
the resulting chi-square per degree of freedom
$\chi^2/{\rm dof} = 86.79 / (73-6) \approx 1.30$.
In particular, the newly achieved predictions for the five BCL parameters $b_k^{+, \, 0}$
entering in the vector and scalar $B \to \pi$ form factors
are compatible with the corresponding numerical results presented in
Table \ref{BCL fit foe the B to pi form factors}, at the $1.0 \, \sigma$ level,
from fitting against only the LCSR and lattice simulation data points.
Additionally,  the thus-far determined interval for $|V_{ub}|$ from our nominal fit model
\begin{eqnarray}
\left |V_{ub} \right |_{B \to \pi \ell \bar \nu_{\ell}} = (3.76 \pm 0.13) \times \, 10^{-3} \,,
\qquad
({\rm BCL \,\, fit \,\, with}  \,\, N=3)
\label{Vub value: N=3}
\end{eqnarray}
appears to be in excellent agreement with the counterpart numerical result
from the analogous  fitting  strategy  but with the LCSR input data points
generated by the traditional dispersive  technique
with the $\pi$-meson distribution amplitudes \cite{Leljak:2021vte}
and from the combined BCL fit against the lattice and experimental results \cite{FlavourLatticeAveragingGroupFLAG:2021npn}.

For the sake of  understanding quantitatively  the systematic uncertainties
from the truncations of the BCL series expansions,
we repeat our numerical fit procedure to the simultaneous  determinations of the vector and scalar form-factor shape parameters
as well as the CKM matrix element $|V_{ub}|$ with the different truncation $N=4$,
yielding the correlated numerical predictions shown in Table  \ref{BCL fit results of Vub with N=4}.
Moreover, this particular BCL expansion fit turns out to generate a minimal $\chi^2=85.31$
for $65$ degrees of freedom, thus corresponding to the equally good  fit quantity
when compared with the former case with the truncation $N=3$.
Unsurprisingly, both the yielding  central value and theory uncertainty
for the numerical result  of $|V_{ub}|$
\begin{eqnarray}
\left |V_{ub} \right |_{B \to \pi \ell \bar \nu_{\ell}} = (3.72 \pm 0.14) \times \, 10^{-3} \,,
\qquad
({\rm BCL \,\, fit \,\, with}  \,\, N=4)
\end{eqnarray}
coincide with the previous BCL fitting results with $N=3$ perfectly.
Apparently, the combined BCL fit results for the $z$-series coefficients
of the semileptonic $B \to \pi$ form factors also stabilize at $N=3$
and do not change notably by increasing the expansion order to $N=4$.
We are therefore led to conclude that truncating the $z$-series expansions at the order $N=3$
in the numerical fit procedure  will be indeed  sufficient to provide us
the reliable and satisfactory theory predictions.

\begin{table}
\centering
\renewcommand{\arraystretch}{2.0}
\resizebox{\columnwidth}{!}{
\begin{tabular}{|c|c||cccccccc|}
\hline
\hline
Parameters  & Values & \multicolumn{8}{|c|}{Correlation Matrix}
 \\
\hline
 $b_0^+$ & $0.409(12)$ & $1$ & $0.095$ & $-0.367$ & $-0.028$ & $0.214$ & $0.075$ & $0.031$ & $-0.840$
 \\
 $b_1^+$ & $-0.477(52)$ & & $1$ & $-0.109$ & $-0.649$ & $0.07$ & $0.089$ & $-0.075$ & $-0.390$
 \\
 $b_2^+$ & $-0.127(211)$ &  &  & $1$ & $-0.565$ & $0.219$ & $-0.089$ & $-0.202$ & $0.153$
 \\
 $b_3^+$ & $-0.732(745)$ &  &  &  &  $1$ & $-0.295$ & $0.062$ & $0.295$ & $0.275$
 \\
 $b_0^0$ & $0.506(22)$ &  & &  & &  $1$ & $-0.579$ & $-0.777$ & $-0.234$
 \\
 $b_1^0$ & $-1.341(172)$ &  & &  & &  & $1$ & $0.715$ & $-0.126$
 \\
 $b_2^0$ & $1.913(324)$ &  & &  & & && $1$ & $-0.027$
 \\
 $|V_{ub}|$ & $3.72(14) \times 10^{-3}$ &  &  & &  &  & & & $1$
 \\
\hline
\hline
\end{tabular}
}
\renewcommand{\arraystretch}{1.0}
\vspace{0.3 cm}
\caption{Theory predictions for the  $B \to \pi \ell \bar \nu_{\ell}$
form-factor shape parameters and the CKM matrix element $|V_{ub}|$
(with their correlation matrix) from carrying out the simultaneous fit
against the SCET sum rules, lattice QCD and experimental data points
with the aid of the truncated BCL $z$-parameterizations at $N=4$.}
\label{BCL fit results of Vub with N=4}
\end{table}

\begin{figure}
\begin{center}
\includegraphics[width=0.45 \columnwidth]{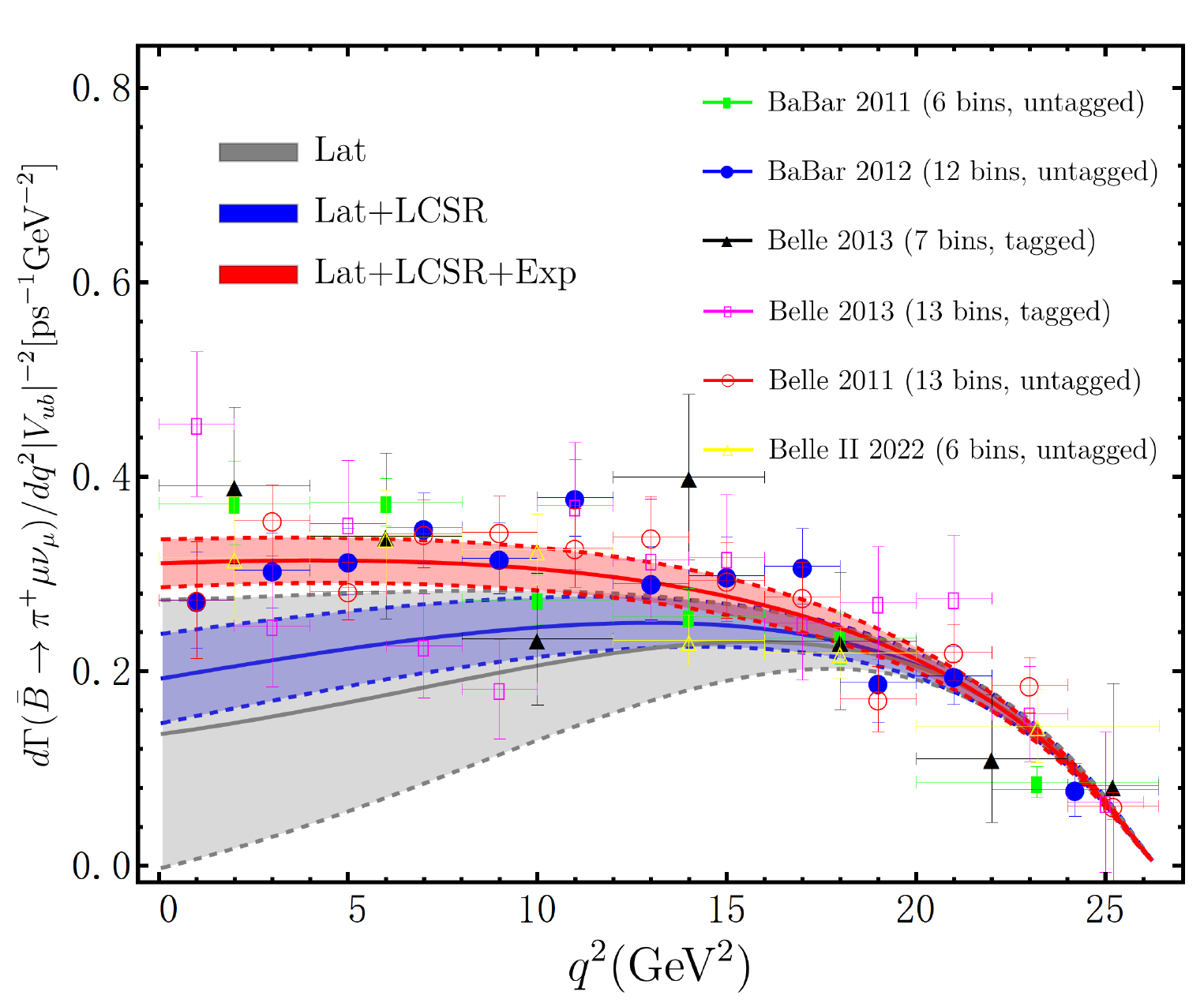}
\hspace{1.0 cm}
\includegraphics[width=0.45 \columnwidth]{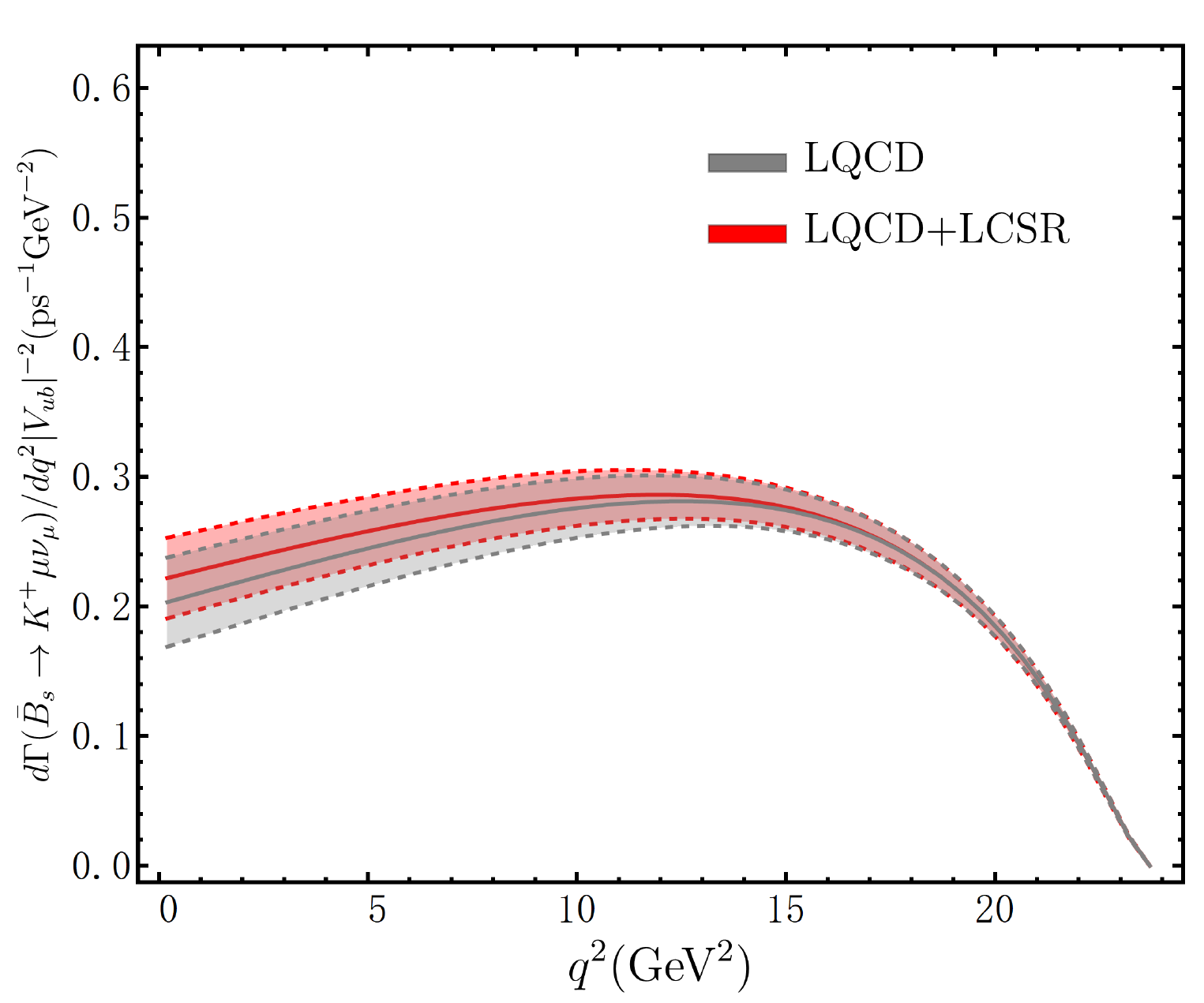}
\\
\vspace*{0.2 cm}
\includegraphics[width=0.45 \columnwidth]{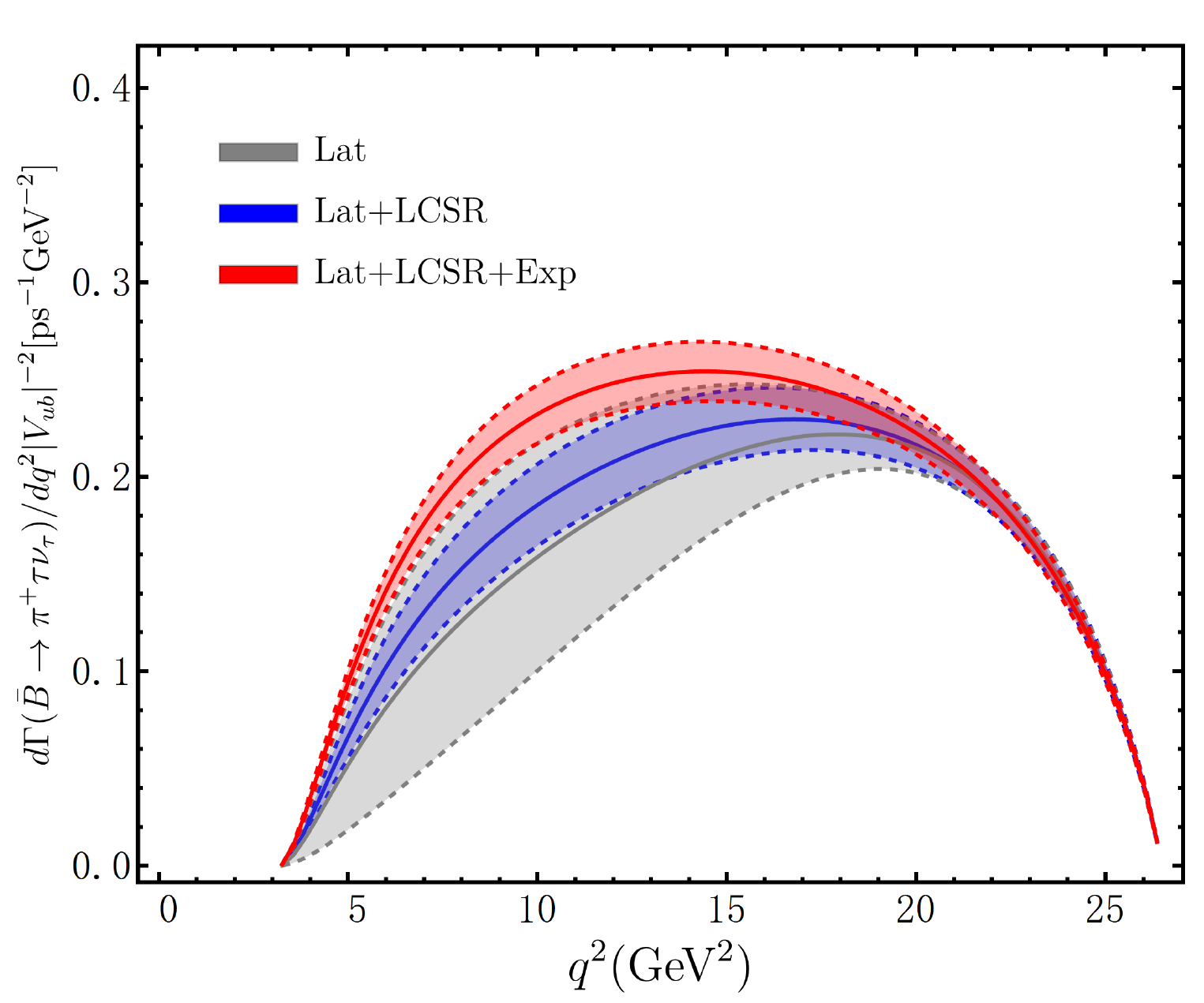}
\hspace{1.0 cm}
\includegraphics[width=0.45 \columnwidth]{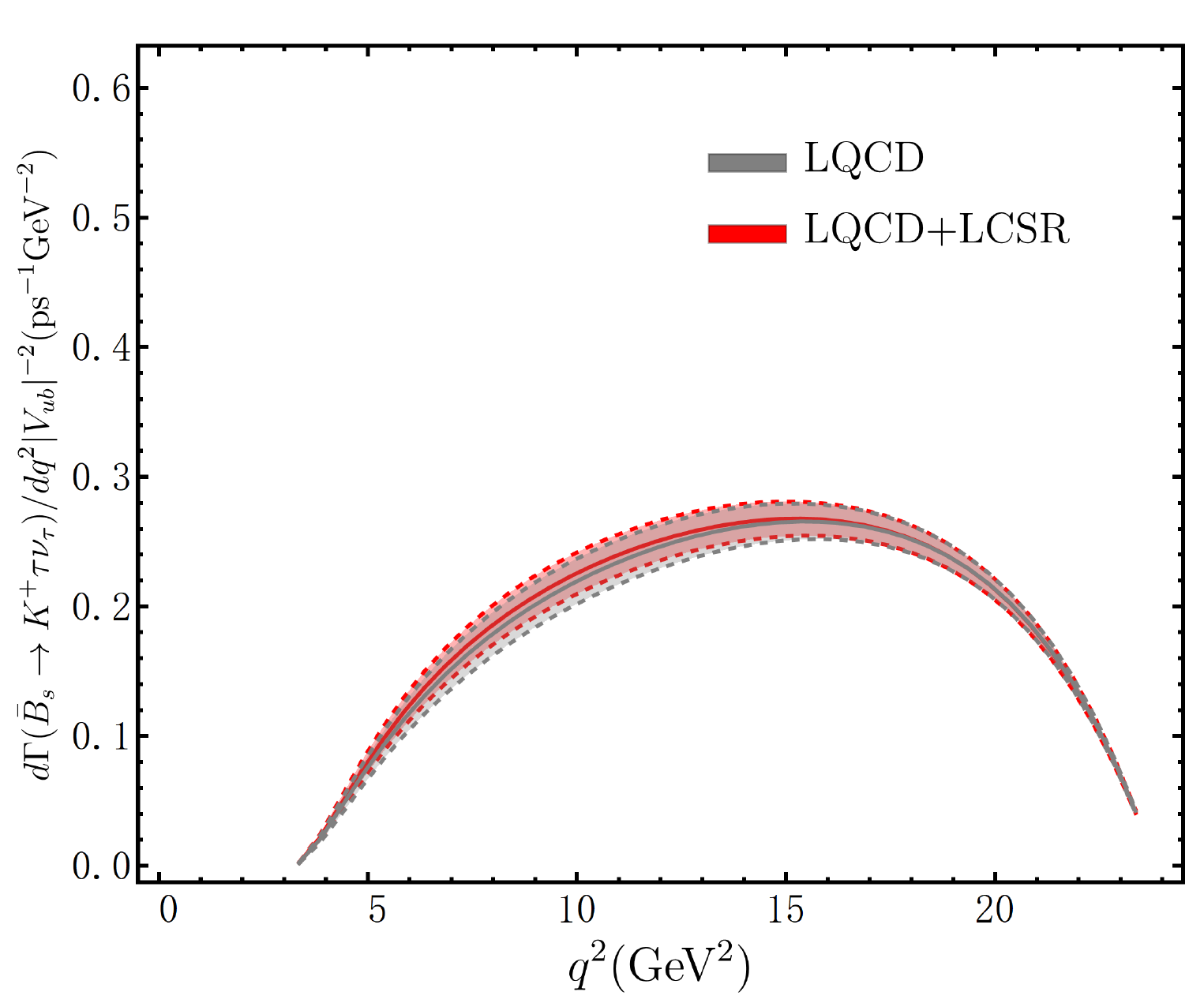}
\vspace*{0.1 cm}
\caption{Theory predictions for the differential $q^2$ distributions
of the exclusive semileptonic $B \to \pi  \ell \bar \nu_{\ell}$  (left panel)
and $B_s \to K \ell \bar \nu_{\ell}$ (right panel) decay  processes
in the entire kinematic region with the distinct BCL $z$-series fits of the form-factor shape parameters.
The available experimental measurements on the binned $q^2$ distributions
of the ``golden"  decay process $B \to \pi  \mu \bar \nu_{\mu}$
from the BaBar \cite{BaBar:2010efp,BaBar:2012thb}, Belle \cite{Belle:2010hep,Belle:2013hlo}
and Belle II \cite{Belle-II:2022imn} Collaborations are further displayed for an exploratory comparison.  }
\label{fig: differential decay rates for B to pi l nu and Bs to K l nu}
\end{center}
\end{figure}

We further display our final theory predictions for the differential $q^2$ distributions
of the semileptonic $B_{(s)} \to \pi (K) \ell \bar \nu_{\ell}$ (with $\ell = \mu, \, \tau$)
decay processes  in the entire kinematic region in Figure \ref{fig: differential decay rates for B to pi l nu and Bs to K l nu},
where the experimental measurements of the  $B \to \pi  \ell \bar \nu_{\ell}$ decay rates
from  the BaBar \cite{BaBar:2010efp,BaBar:2012thb}, Belle \cite{Belle:2010hep,Belle:2013hlo}
and Belle II \cite{Belle-II:2022imn} Collaborations are further shown for a numerical comparison.
In addition, we collect simultaneously the obtained numerical results from  fitting the BCL $z$-series parameterizations
with  three distinct scenarios  of the input data points:
I) only synthetic lattice data points,
II) synthetic lattice data points $\oplus$ LCSR results,
III)  synthetic lattice data points $\oplus$ LCSR results $\oplus$ experimental data.
We can readily observe from Figure \ref{fig: differential decay rates for B to pi l nu and Bs to K l nu}
 that employing our improved  LCSR predictions at large hadronic recoil
in the BCL expansion  fit program  will be highly beneficial for pinning  down  the theory uncertainties
from the particular  fitting strategy with only the synthetic lattice data points.
Moreover, we discover the slight tension of the predicted large-recoil $B \to \pi  \mu \bar \nu_{\mu}$ decay distributions
between the scenarios II) and  III) fitting strategies with the BCL $z$-series parameterizations.
With regard to the counterpart exclusive $B_{s} \to K \ell \bar \nu_{\ell}$ decay channels,
taking into account the newly obtained LCSR data points in the numerical fit
will  bring about the moderate improvements on the resulting partial decay rates
determined from fitting against only the synthetic lattice data points,
as previously discussed in Section \ref{subsection: numerics for the bottom-meson form factors}.
For convenience, we also collect here our theory predictions for the total branching fractions
of $B_{s} \to K \ell \bar \nu_{\ell}$ with the extracted interval of the CKM matrix element $|V_{ub}|$
shown in (\ref{Vub value: N=3})
\begin{eqnarray}
\mathcal{BR}(B_s \to K \mu  \bar \nu_{\mu})  &=& (1.200 \pm 0.128) \times 10^{-4} \,,
\nonumber \\
\mathcal{BR}(B_s \to K \tau \bar \nu_{\tau}) &=& (0.847 \pm 0.078)  \times 10^{-4} \,,
\end{eqnarray}
the former of which coincides well with the first experimental measurement from the LHCb Collaboration
$\mathcal{BR}(B_s \to K \mu  \bar \nu_{\mu}) =
\left [ 1.06 \pm 0.05 (\rm stat) \pm 0.08 (\rm syst)\right ]  \times 10^{-4}$  \cite{LHCb:2020ist}
 by employing the Cabibbo favored semileptonic $B_{s} \to D_s \ell \bar \nu_{\ell}$ decay process
as the normalization  channel.
Unfortunately, both the two  semitauonic bottom-meson decays
$B \to \pi \tau \bar \nu_{\tau}$ and $B_s \to K \tau \bar \nu_{\tau}$
have not been observed in the high luminosity  Belle II and LHCb experiments to date
(see however  the upper limit of  ${\cal BR} (B \to \pi \tau \bar \nu_{\tau}) < 2.5 \times 10^{-4}$
at the $90 \, \%$ confidence level from the Belle Collaboration \cite{Belle:2015qal}).

\begin{table}
\centering
\renewcommand{\arraystretch}{2.0}
\resizebox{\columnwidth}{!}{
\begin{tabular}{|c||c|c|c|c|}
\hline
\hline
Observables  & Lattice QCD & Lattice QCD $\oplus$ LCSR & LCSR  & This work \\
\hline
\multirow{3}{*}{${\cal R}_{\pi}$} & $0.69 \pm 0.19$ \cite{Flynn:2015mha}
& $0.78 \pm 0.10$ \cite{Becirevic:2020rzi}  & $0.69^{+0.03}_{-0.05}$ \cite{Khodjamirian:2011ub}  & $0.720 \pm 0.027 \big |_{N=3}$
\\
& $0.767 \pm 0.145$ \cite{Martinelli:2022tte}  & $0.699 \pm 0.022$ \cite{Leljak:2021vte} &  $0.68^{+0.10}_{-0.09}$ \cite{Zhou:2019jny}  &
\\
&  $0.838 \pm 0.075$  \cite{Martinelli:2022tte} & $0.677 \pm 0.010$ \cite{Biswas:2021cyd}  & $0.65^{+0.13}_{-0.11}$ \cite{Zhou:2019jny}   & $0.746 \pm 0.039 \big |_{N=4}$
\\
\hline
\multirow{3}{*}{${\cal R}_{K}$} & $0.77 \pm 0.12$  \cite{Flynn:2015mha}  & $-$ & $-$
& $0.700 \pm 0.016\big |_{N=3}$
\\
& $ 0.695 \pm 0.050 $ \cite{Bouchard:2014ypa}  &   &   &
\\
& $0.836 \pm 0.034$ \cite{FermilabLattice:2019ikx} & $-$ & $-$  & $0.680 \pm 0.019\big |_{N=4}$
\\
\hline
\hline
\end{tabular}
}
\vspace{0.3 cm}
\caption{Theory predictions for the LFU ratios of the exclusive semileptonic
$B_{(s)}  \to \pi (K) \ell \bar \nu_{\ell}$ decay processes from the combined BCL expansion fitting
against the synthetic lattice data points and the newly obtained  LCSR results. }
\label{table: physical observables for B to pi l nu and Bs to Kl nu}
\end{table}

\begin{figure}
\begin{center}
\includegraphics[width=0.60 \columnwidth]{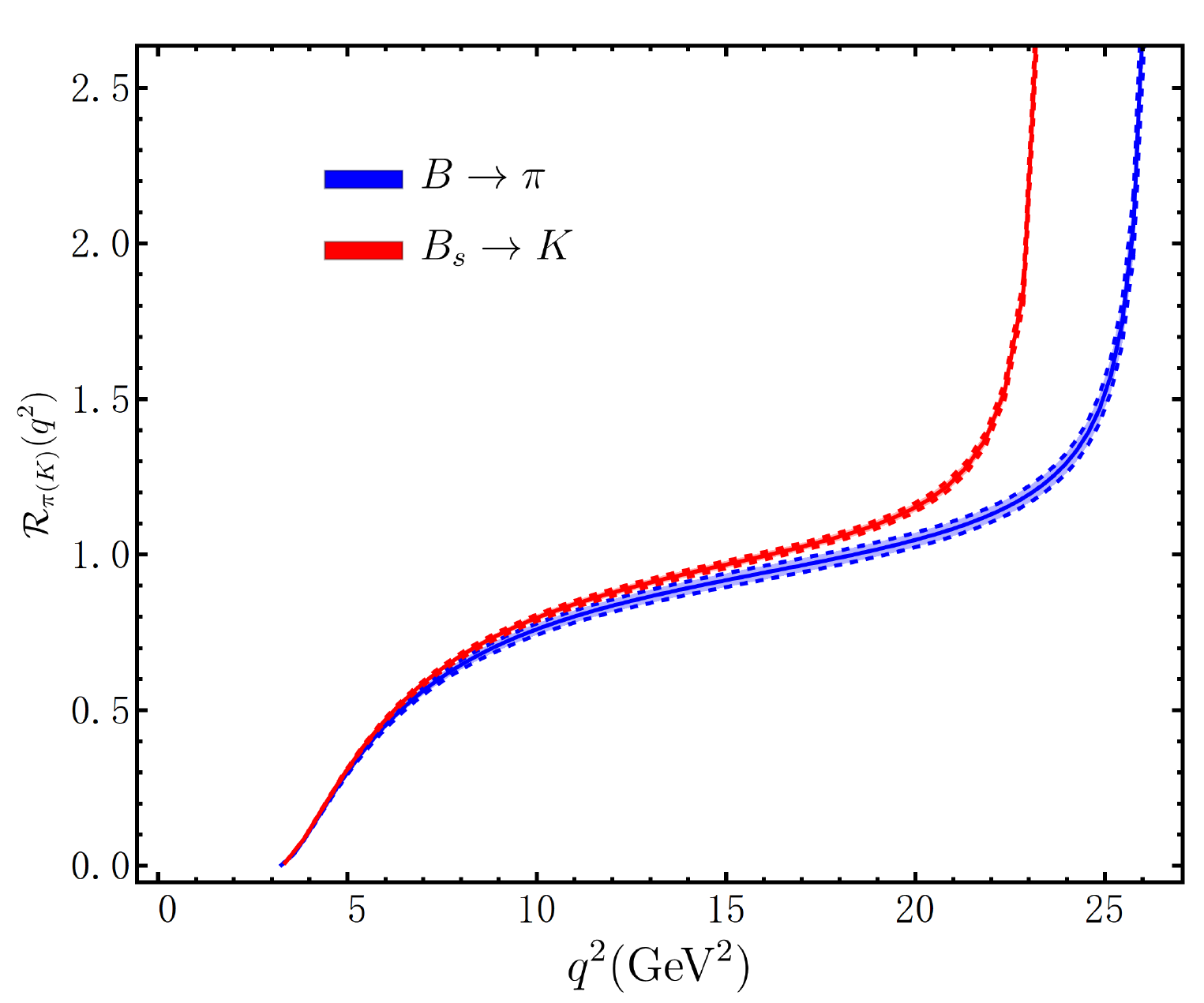}
\vspace*{0.1 cm}
\caption{Theory predictions for the two interesting  LFU ratios ${\cal R}_{\pi(K)}$
of the differential $B_{(s)}  \to \pi (K) \ell \bar \nu_{\ell}$  decay distributions
obtained from the combined BCL expansion fitting against
the synthetic lattice data points and the newly obtained bottom-meson LCSR results. }
\label{fig: differential Rpi and RK rarios}
\end{center}
\end{figure}

In light of the increasing sensitivity of the semitauonic bottom-hadron decays
to the mysterious NP signature  due to the very large $\tau$-lepton  mass,
we proceed to investigate two particular lepton-flavour-universality (LFU) probing observables
for the exclusive $B_{(s)}  \to \pi (K) \ell \bar \nu_{\ell}$ decays
independent of the CKM matrix element $|V_{ub}|$
\begin{eqnarray}
{\cal R}_{\pi(K)}  =  \frac{\Gamma \left (B_{(s)} \to \pi(K) \tau \bar \nu_{\tau} \right )}
{\Gamma \left (B_{(s)} \to \pi(K) \mu  \bar \nu_{\nu} \right )}
=  \frac{\int_{m_{\tau}^2}^{q_{\rm max}^2} \, d q^2 \,
{d \Gamma(B_{(s)} \to \pi(K) \tau \bar \nu_{\tau} ) / d q^2 }}
{\int_{m_{\mu}^2}^{q_{\rm max}^2} \, d q^2 \,
{d \Gamma(B_{(s)} \to \pi(K) \mu \bar \nu_{\mu} ) / d q^2 }} \,.
\end{eqnarray}
Apparently, precision predictions of such interesting LFU quantities would require a good knowledge
of both the vector and scalar form factors in the whole semileptonic regions.
Adopting the combined BCL $z$-series fit results with two distinct truncations $N \in \left \{ 3, \, 4 \right \}$
yields the desired numerical predictions for the two LFU observables
as summarized in Table \ref{table: physical observables for B to pi l nu and Bs to Kl nu}.
Generally, our numerical results for ${\cal R}_{\pi(K)}$ are compatible with the previous theory determinations
based upon the lattice simulation and LCSR methods.
In addition, our BCL expansion  fit result of the LFU ratio ${\cal R}_{\pi}$
can evidently  accommodate the rather loose Belle measurement of
${\cal R}_{\pi} | _{\rm Belle \,\, 2016}= 1.05 \pm 0.51$ \cite{Belle:2015qal}.
We further present the resulting predictions for the two LFU ratios
of the differential $B_{(s)}  \to \pi (K) \ell \bar \nu_{\ell}$  decay distributions
in Figure \ref{fig: differential Rpi and RK rarios},
which can be straightforwardly confronted with the counterpart numerical results
from the  RBC/UKQCD Collaborations \cite{Flynn:2015mha},
 HPQCD \cite{Bouchard:2014ypa} and FNAL/MILC \cite{FermilabLattice:2019ikx} Collaborations.

\begin{figure}
\begin{center}
\includegraphics[width=0.45 \columnwidth]{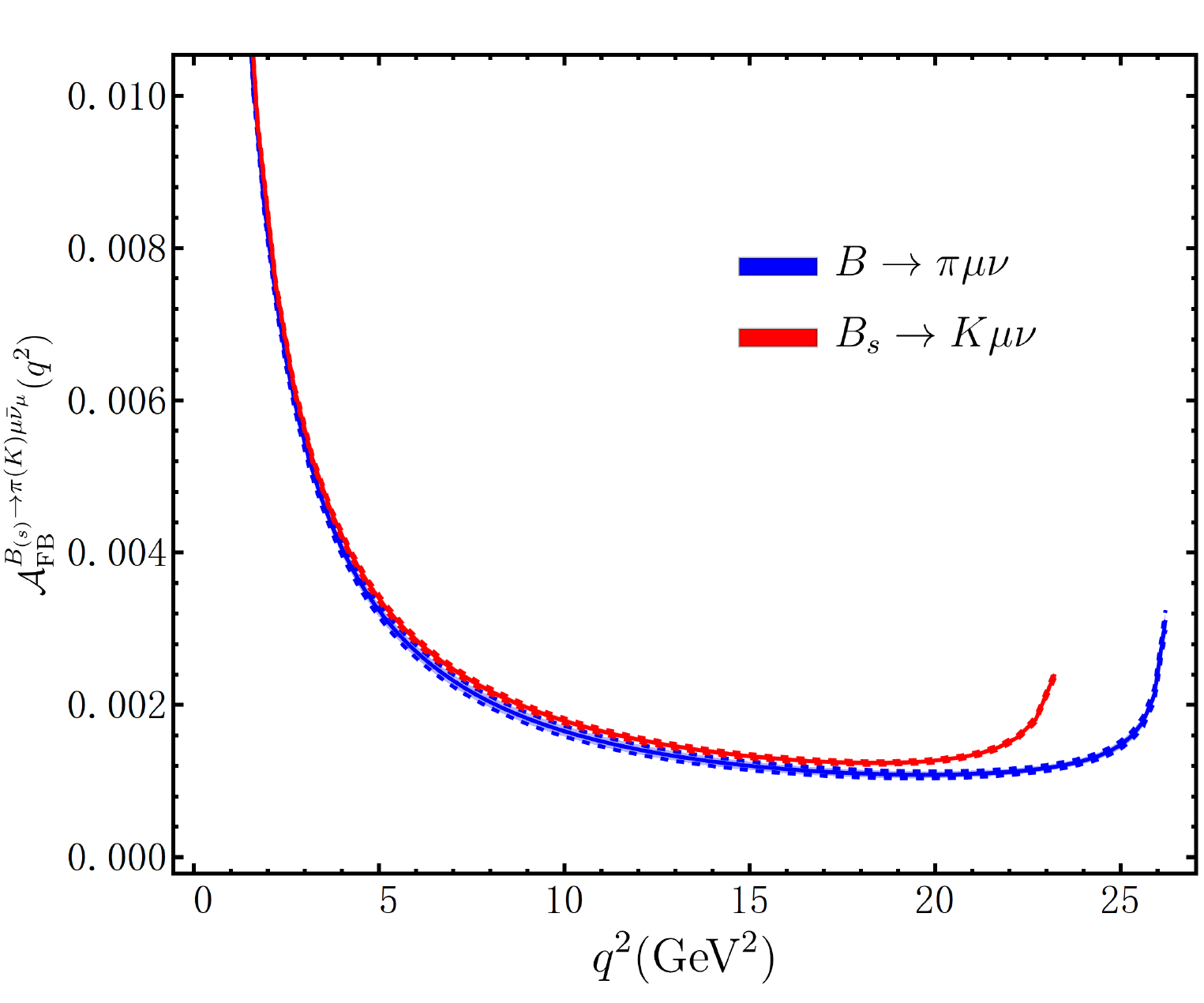}
\hspace{1.0 cm}
\includegraphics[width=0.45 \columnwidth]{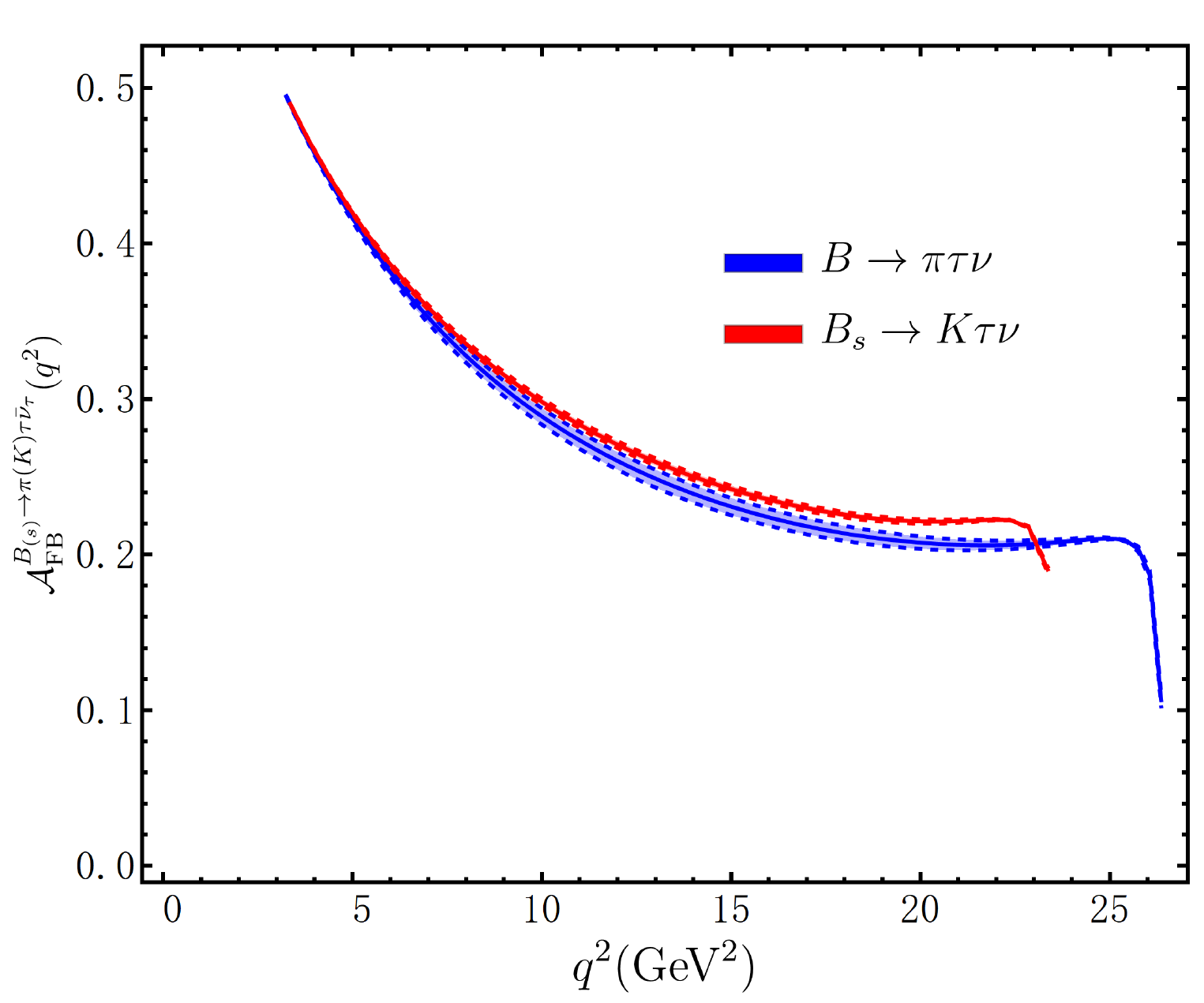}
\\
\vspace*{0.2 cm}
\includegraphics[width=0.45 \columnwidth]{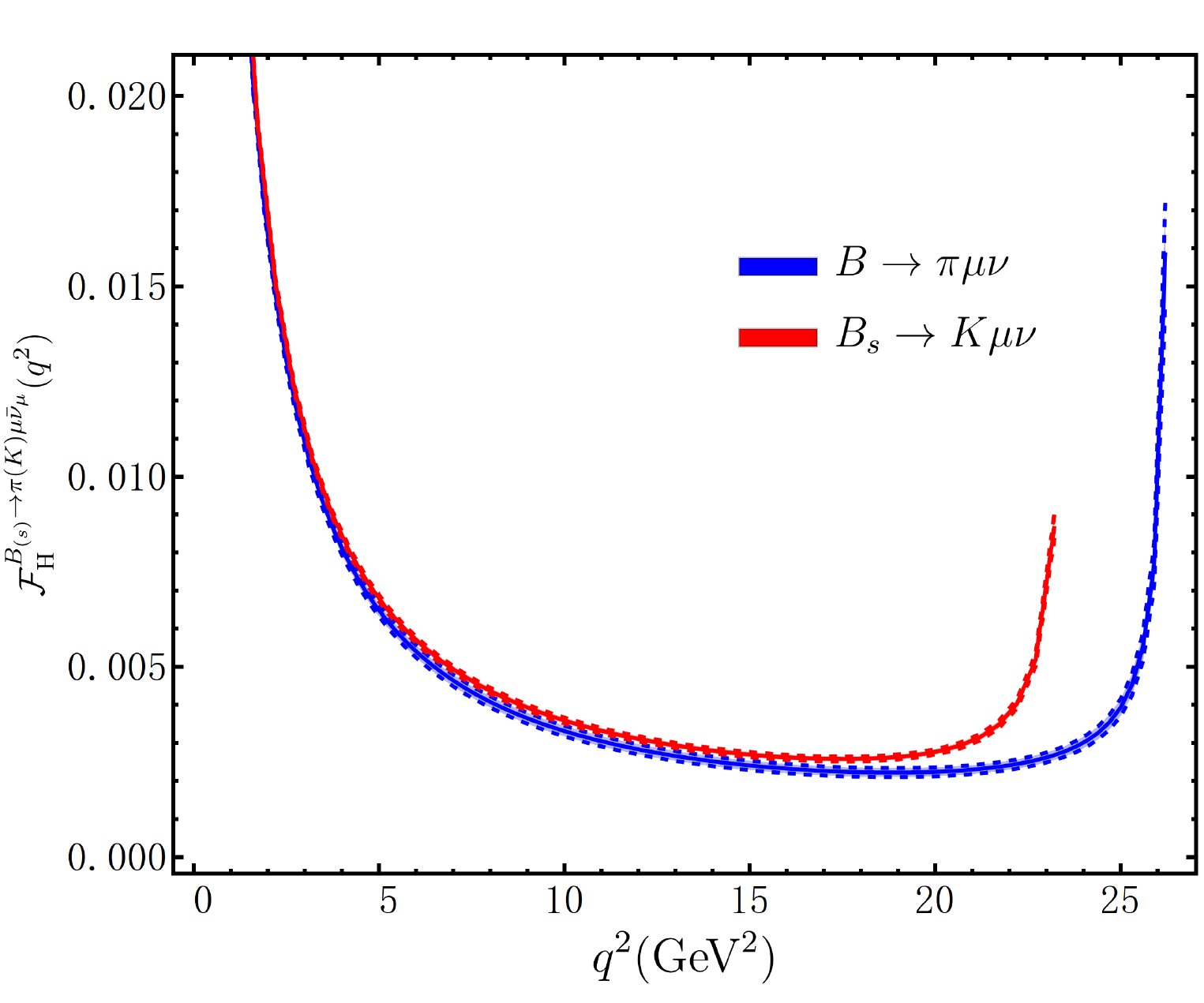}
\hspace{1.0 cm}
\includegraphics[width=0.45 \columnwidth]{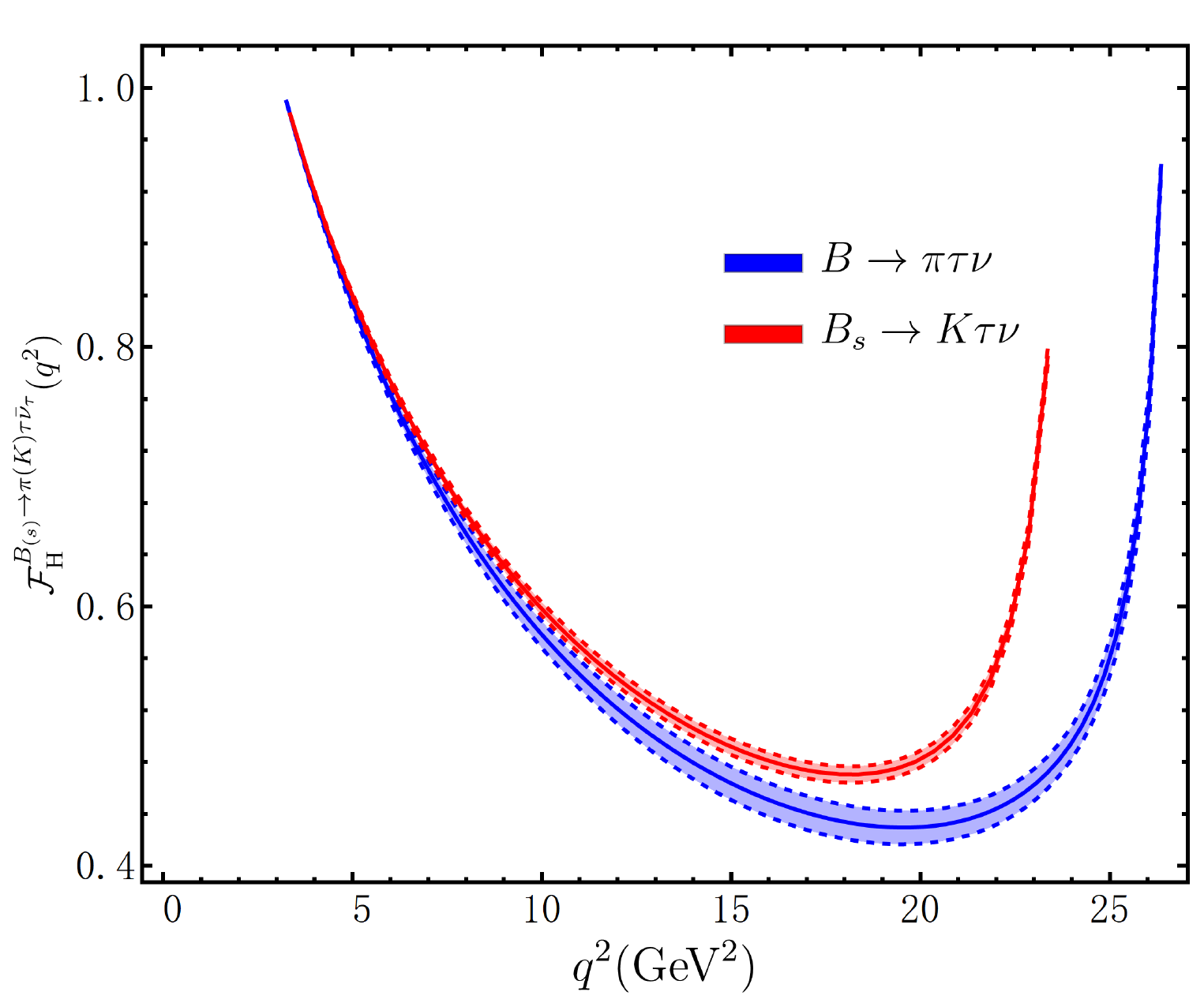}
\\
\vspace*{0.2 cm}
\includegraphics[width=0.45 \columnwidth]{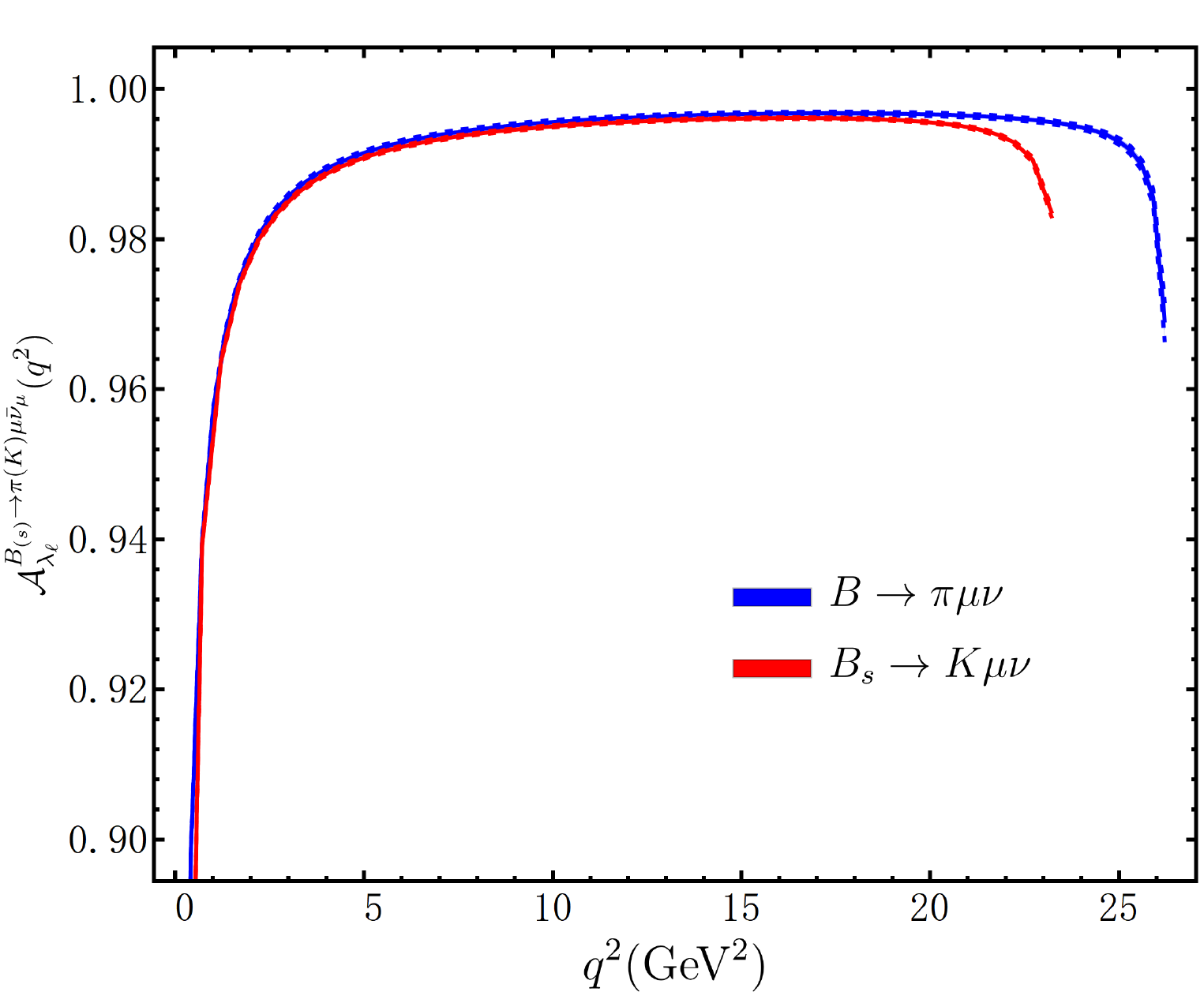}
\hspace{1.0 cm}
\includegraphics[width=0.45 \columnwidth]{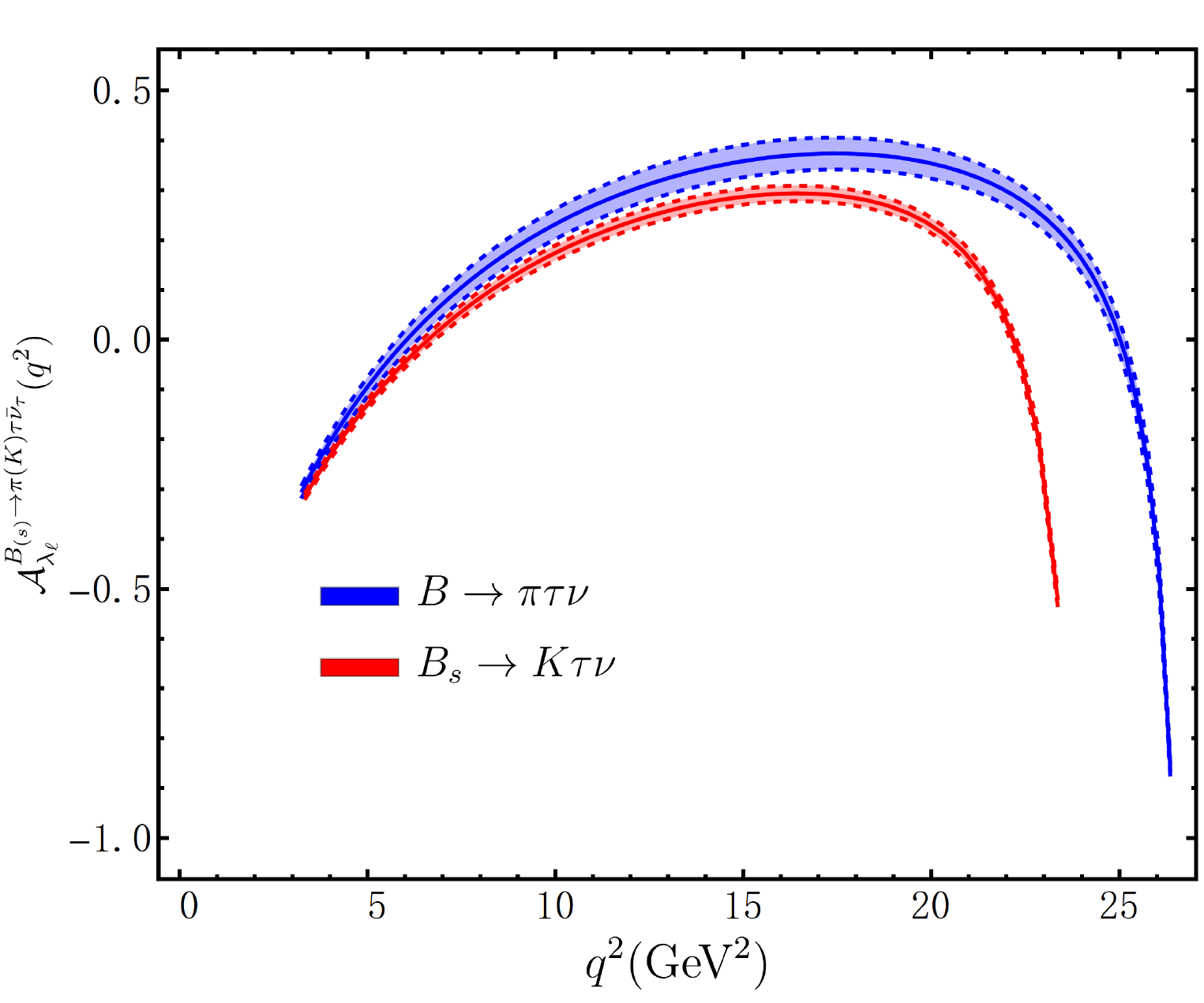}
\vspace*{0.1 cm}
\caption{Theory predictions for the three distinct classes of the angular observables
${\cal A}_{\rm FB}^{B_{(s)}  \to \pi (K) \ell \bar \nu_{\ell}}(q^2)$,
${\cal F}_{\rm H}^{B_{(s)}  \to \pi (K) \ell \bar \nu_{\ell}}(q^2)$ and
${\cal A}_{\lambda_{\ell}}^{B_{(s)}  \to \pi (K) \ell \bar \nu_{\ell}}(q^2)$
obtained from the combined BCL expansion fitting against
the synthetic lattice data points and  the newly obtained bottom-meson LCSR results.}
\label{fig: three angular observables for B to pi l nu and Bs to K l nu}
\end{center}
\end{figure}

Applying the two-fold differential spectrum of the flavour-changing charged-current
$B_{q^{\prime}} \to M \ell \bar \nu_{\ell}$ decay process
displayed in  (\ref{angular distribution of B to M l nu}),
we can  construct two angular observables sensitive to Beyond the Standard Model (BSM) physics
associated with  electroweak  symmetry breaking
\begin{eqnarray}
{\cal A}_{\rm FB}^{B_{(s)}  \to \pi (K) \ell \bar \nu_{\ell}} (q^2)
&=&  \left [ \frac{d \Gamma(B_{(s)}  \to \pi (K) \ell \bar \nu_{\ell})} {d q^2} \right ]^{-1} \,
\int_{-1}^{1} d \cos \theta_{\ell} \,\,  {\rm sgn}(\cos \theta_{\ell}) \,
\frac{d^2 \Gamma(B_{(s)}  \to \pi (K) \ell \bar \nu_{\ell})}
{d q^2 \, d \cos \theta_{\ell} } \,,
\nonumber \\
&=& \left [{1 \over 2} \,  b_{\theta_{\ell}}(q^2) \right ] :
\left [ a_{\theta_{\ell}}(q^2) + {1 \over 3} \, c_{\theta_{\ell}}(q^2) \right ]
 \\
{\cal F}_{\rm H}^{B_{(s)}  \to \pi (K) \ell \bar \nu_{\ell}} (q^2)
 &=&  1 +{2 \over 3} \, \left [ \frac{d \Gamma(B_{(s)}  \to \pi (K) \ell \bar \nu_{\ell})} {d q^2} \right ]^{-1} \,
 {d^2 \over d (\cos \theta_{\ell})^2 } \, \frac{d^2 \Gamma(B_{(s)}  \to \pi (K) \ell \bar \nu_{\ell})}
{d q^2 \, d \cos \theta_{\ell} }
\nonumber \\
&=&  \left [ a_{\theta_{\ell}}(q^2) + c_{\theta_{\ell}}(q^2) \right ] :
\left [ a_{\theta_{\ell}}(q^2) + {1 \over 3} \, c_{\theta_{\ell}}(q^2) \right ]  \,.
\hspace{-1.0 cm}
\label{definition: flat term}
\end{eqnarray}
Evidently, both the normalized forward-backward asymmetries ${\cal A}_{\rm FB}^{B_{(s)}  \to \pi (K) \ell \bar \nu_{\ell}}$
and the $q^2$ differential flat terms ${\cal F}_{\rm H}^{B_{(s)}  \to \pi (K) \ell \bar \nu_{\ell}}$ \cite{Bobeth:2011nj}
will vanish in the massless lepton limit in the SM.
Another appropriate candidate for the potential  BSM probe can be introduced by investigating
the polarization asymmetry of the final-state lepton
\begin{eqnarray}
&& {\cal A}_{\lambda_{\ell}}^{B_{(s)}  \to \pi (K) \ell \bar \nu_{\ell}} (q^2)
= \left [ \frac{d \Gamma(B_{(s)}  \to \pi (K) \ell \bar \nu_{\ell})} {d q^2} \right ]^{-1} \,
\left [ \frac{d \Gamma^{\lambda_{\ell}= -1/2}}{d q^2}
- \frac{d \Gamma^{\lambda_{\ell}= + 1/2}}{d q^2} \right ]\left(B_{(s)}  \to \pi (K) \ell \bar \nu_{\ell} \right )
\nonumber \\
&& =  1 - {2 \over 3} \, \left \{ \left[ 3 \,  \left ( a_{\theta_{\ell}}(q^2) + c_{\theta_{\ell}}(q^2) \right )
+ {2 \, m_{\ell}^2 \over q^2 -  m_{\ell}^2} \, c_{\theta_{\ell}}(q^2)  \right ]
: \left [ a_{\theta_{\ell}}(q^2) + {1 \over 3} \, c_{\theta_{\ell}}(q^2) \right ] \right \}  \,,
\label{analytical result of polarization asymmetry}
\end{eqnarray}
which turns out to be sensitive to  helicity-violating NP interactions.
The  analytic structure of the above-mentioned expression of the lepton polarization fraction
(\ref{analytical result of polarization asymmetry})
can be actually understood from the $\ell$-helicity conservation of the semileptonic $b \to q \ell \bar \nu_{\ell}$ transition
in the massless lepton approximation in the SM.
In order to facilitate  the numerical comparisons with the future experimental measurements,
we collect our theory predictions for the three different classes of the
angular  observables ${\cal A}_{\rm FB}^{B_{(s)}  \to \pi (K) \ell \bar \nu_{\ell}}(q^2)$,
${\cal F}_{\rm H}^{B_{(s)}  \to \pi (K) \ell \bar \nu_{\ell}}(q^2)$ and
${\cal A}_{\lambda_{\ell}}^{B_{(s)}  \to \pi (K) \ell \bar \nu_{\ell}}(q^2)$
in Figure \ref{fig: three angular observables for B to pi l nu and Bs to K l nu}.
It is perhaps worthwhile to mention that our predictions for the normalized differential  forward-backward asymmetries
are  in excellent agreement with  the available lattice QCD simulation results from both the RBC/UKQCD \cite{Flynn:2015mha}
and  HPQCD (without  $B  \to \pi \ell \bar \nu_{\ell}$) \cite{Bouchard:2014ypa} Collaborations.
On the other hand, the resulting predictions for the lepton polarization fractions
of $B_{s}  \to K \ell \bar \nu_{\ell}$ coincide well with the previous HPQCD \cite{Bouchard:2014ypa}
and FNAL/MILC \cite{FermilabLattice:2019ikx} determinations
(see also \cite{Meissner:2013pba} for the numerical predictions with the vector and scalar $B_s \to K$ form factors
computed with the TMD factorization approach).
Moreover, we summarize our numerical predictions  for the aforementioned three distinct classes
of the integrated observables by employing the combined BCL expansion fit results of
the semileptonic heavy-to-light $B_{(s)} \to \pi (K)$ form factors
in Table \ref{table: angular observables for B to pi l nu and Bs to K l nu},
where we further confront our results with the previous determinations from the lattice QCD
and LCSR techniques for convenience.
Generally, our newly obtained results of the integrated angular observables
for the semileptonic  $B_{(s)}  \to \pi (K) \ell \bar \nu_{\ell}$ decays
are compatible with the available QCD determinations within the theory uncertainties,
but with the exceptions of the previously extracted intervals of the flat terms
${\cal F}_{\rm H}^{B  \to \pi \mu \bar \nu_{\mu}}$ and ${\cal F}_{\rm H}^{B  \to \pi \tau \bar \nu_{\tau}}$
from \cite{Leljak:2021vte}, which turn out to be approximately one fourth of our numerical predictions individually.
In order to better clarify such striking discrepancies, we can readily derive an exact but quite loose bound,
independent of the scalar form-factor ratio,
for the $q^2$ differential flat term by applying the explicit definition (\ref{definition: flat term})
as well as the kinematic constraint
$q^2 \in \left [m_{\ell}^2 \,,  (m_{B_{(s)} }  - m_{\pi (K)})^2 \right ]$.
For definiteness, we obtain
\begin{eqnarray}
{\cal F}_{\rm H}^{B_{(s)}  \to \pi (K) \ell \bar \nu_{\ell}} (q^2)
\geq   \frac{3}{1 + 2 \, y_{\ell, \, {\rm max}}} \,,
\qquad
y_{\ell, \, {\rm max}} \equiv  \left [  \frac{m_{B_{(s)} }  - m_{\pi (K)}} {m_{\ell}} \right ]^2 \,.
\label{an exact bound of FH}
\end{eqnarray}
Plugging the input values for the emerged hadron and lepton masses
summarized in Table \ref{table: theory inputs} into (\ref{an exact bound of FH})
immediately leads to the desired numerical bounds
\begin{eqnarray}
{\cal F}_{\rm H}^{B  \to \pi \mu \bar \nu_{\mu}}(q^2) \geq 6.337 \times 10^{-4},
& \qquad &
{\cal F}_{\rm H}^{B  \to \pi \tau \bar \nu_{\tau}}(q^2)  \geq  0.169   \,,
\nonumber \\
{\cal F}_{\rm H}^{B_s  \to K \mu \bar \nu_{\mu}}(q^2) \geq 7.049 \times 10^{-4},
& \qquad &
{\cal F}_{\rm H}^{B_s  \to K \tau \bar \nu_{\tau}} (q^2) \geq  0.187   \,,
\end{eqnarray}
which are well respected by our combined BCL $z$-fit results for the corresponding four observables
as collected in Table \ref{table: angular observables for B to pi l nu and Bs to K l nu}.

\begin{table}
\centering
\renewcommand{\arraystretch}{1.25}
\resizebox{0.85 \columnwidth}{!}{
\begin{tabular}{|c||c|c|c|}
\hline
\hline
Observables  & Lattice QCD & Lattice QCD $\oplus$ LCSR   & This work \\
\hline
\multirow{3}{*}{${\cal A}_{\rm FB}^{B  \to \pi \mu \bar \nu_{\mu}}$}
&
&    & $\left (3.99 \pm 0.35 \right )  \times 10^{-3} \big |_{N=3}$
\\
& $\left (4.4 \pm 1.3 \right ) \times 10^{-3}$ \cite{Flynn:2015mha}
& $\left (4.8 \pm 0.3 \right ) \times 10^{-3}$ \cite{Leljak:2021vte}  &
\\
&   &      & $ \left (3.72 \pm 0.51 \right )  \times 10^{-3}  \big |_{N=4}$
\\
\hline
\multirow{3}{*}{${\cal A}_{\rm FB}^{B  \to \pi \tau \bar \nu_{\tau}}$} &
&    & $0.248 \pm 0.005  \big |_{N=3}$
\\
& $0.252 \pm 0.012$ \cite{Flynn:2015mha}  & $0.259 \pm 0.004$ \cite{Leljak:2021vte}  &
\\
&   &      & $ 0.244 \pm 0.007  \big |_{N=4}$
\\
\hline
\multirow{3}{*}{${\cal A}_{\rm FB}^{B_s  \to K \mu \bar \nu_{\mu}}$}
&  $\left (3.9 \pm 1.1 \right )  \times 10^{-3} $ \cite{Flynn:2015mha}  &
& $\left (4.49 \pm 0.21 \right )  \times 10^{-3} \big |_{N=3}$
\\
& $\left (6.6 \pm 1.0\right ) \times 10^{-3}$ \cite{Bouchard:2014ypa}  & $-$  &
\\
& $\left (3.21 \pm 0.97 \right ) \times 10^{-3}$ \cite{FermilabLattice:2019ikx} &
& $ \left (5.14 \pm 0.33 \right ) \times 10^{-3}  \big |_{N=4}$
\\
\hline
\multirow{3}{*}{${\cal A}_{\rm FB}^{B_s  \to K \tau \bar \nu_{\tau}}$} & $0.2650 \pm 0.0079$ \cite{Flynn:2015mha}   &
& $0.267 \pm 0.002 \big |_{N=3}$
\\
&  $0.284 \pm 0.017$ \cite{Bouchard:2014ypa}  & $-$  &
\\
& $0.2536 \pm 0.0084$ \cite{FermilabLattice:2019ikx}  &    & $ 0.272 \pm 0.003  \big |_{N=4}$
\\
\hline
\hline
\multirow{3}{*}{${\cal F}_{\rm H}^{B  \to \pi \mu \bar \nu_{\mu}}$} &
&    & $\left (8.04 \pm 0.72 \right )  \times 10^{-3} \big |_{N=3}$
\\
& $-$ &  $\left (2.4  \pm 0.1 \right )  \times 10^{-3}$  \cite{Leljak:2021vte}&
\\
&   &      & $ \left (7.52 \pm 1.02 \right )  \times 10^{-3} \big |_{N=4}$
\\
\hline
\multirow{3}{*}{${\cal F}_{\rm H}^{B  \to \pi \tau \bar \nu_{\tau}}$} &
&    & $0.514 \pm 0.012 \big |_{N=3}$
\\
& $-$ & $0.134 \pm 0.003$  \cite{Leljak:2021vte}  &
\\
&   &      & $ 0.508 \pm 0.014   \big |_{N=4}$
\\
\hline
\multirow{3}{*}{${\cal F}_{\rm H}^{B_s  \to K \mu \bar \nu_{\mu}}$} &
&    & $\left (9.10 \pm 0.43 \right )  \times 10^{-3} \big |_{N=3}$
\\
& $-$  &  $-$  &
\\
&   &      & $ \left (10.36 \pm 0.67 \right )  \times 10^{-3}   \big |_{N=4}$
\\
\hline
\multirow{3}{*}{${\cal F}_{\rm H}^{B_s  \to K \tau \bar \nu_{\tau}}$} &
&    & $0.555 \pm 0.006   \big |_{N=3}$
\\
& $-$  &  $-$  &
\\
&   &      & $ 0.565 \pm 0.007  \big |_{N=4}$
\\
\hline
\hline
\multirow{3}{*}{${\cal A}_{\lambda_{\ell}}^{B  \to \pi \mu \bar \nu_{\mu}}$} &
&    & $ 0.988 \pm 0.001  \big |_{N=3}$
\\
& $-$ & $-$  &
\\
&   &      & $ 0.989 \pm 0.002  \big |_{N=4}$
\\
\hline
\multirow{3}{*}{${\cal A}_{\lambda_{\ell}}^{B  \to \pi \tau \bar \nu_{\tau}}$} &
&    & $ 0.266  \pm 0.029  \big |_{N=3}$
\\
& $-$ & $0.21 \pm 0.02$  \cite{Leljak:2021vte}  &
\\
&   &      & $ 0.272  \pm 0.032   \big |_{N=4}$
\\
\hline
\multirow{3}{*}{${\cal A}_{\lambda_{\ell}}^{B_s  \to K \mu \bar \nu_{\mu}}$} &
&    & $ 0.987 \pm 0.001  \big |_{N=3}$
\\
& $0.982^{+0.018}_{-0.079}$ \cite{Bouchard:2014ypa} & $-$  &
\\
&   &      & $ 0.985 \pm 0.001   \big |_{N=4}$
\\
\hline
\multirow{3}{*}{${\cal A}_{\lambda_{\ell}}^{B_s  \to K \tau \bar \nu_{\tau}}$} &
&    & $ 0.191 \pm 0.014  \big |_{N=3}$
\\
& $0.105 \pm 0.063$ \cite{Bouchard:2014ypa}  & $-$  &
\\
&   &      & $ 0.172 \pm 0.017  \big |_{N=4}$
\\
\hline
\hline
\end{tabular}
}
\vspace{0.3 cm}
\caption{Theory predictions for the three distinct classes of the integrated  observables
${\cal A}_{\rm FB}^{B_{(s)}  \to \pi (K) \ell \bar \nu_{\ell}}$,
${\cal F}_{\rm H}^{B_{(s)}  \to \pi (K) \ell \bar \nu_{\ell}}$ and
${\cal A}_{\lambda_{\ell}}^{B_{(s)}  \to \pi (K) \ell \bar \nu_{\ell}}$
obtained from the combined BCL $z$-series expansion fitting of the exclusive $B_{(s)} \to \pi (K)$ form factors against
the synthetic lattice data points and the newly obtained bottom-meson LCSR results.}
\label{table: angular observables for B to pi l nu and Bs to K l nu}
\end{table}


\subsection{Phenomenological analysis of the   $B \to K \nu_{\ell} \bar \nu_{\ell}$  observables }

\begin{figure}
\begin{center}
\includegraphics[width=0.45 \columnwidth]{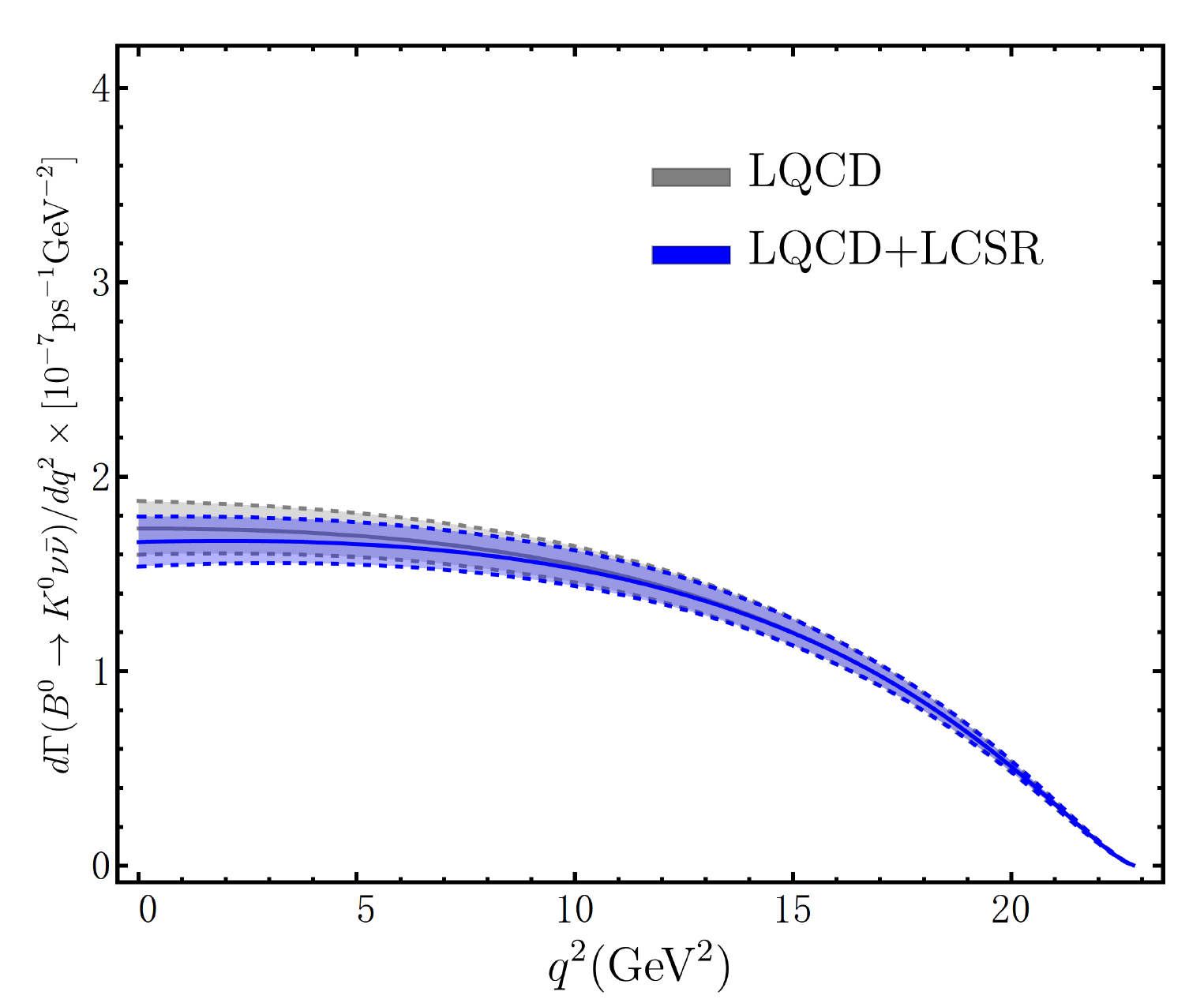}
\hspace{1.0 cm}
\includegraphics[width=0.45 \columnwidth]{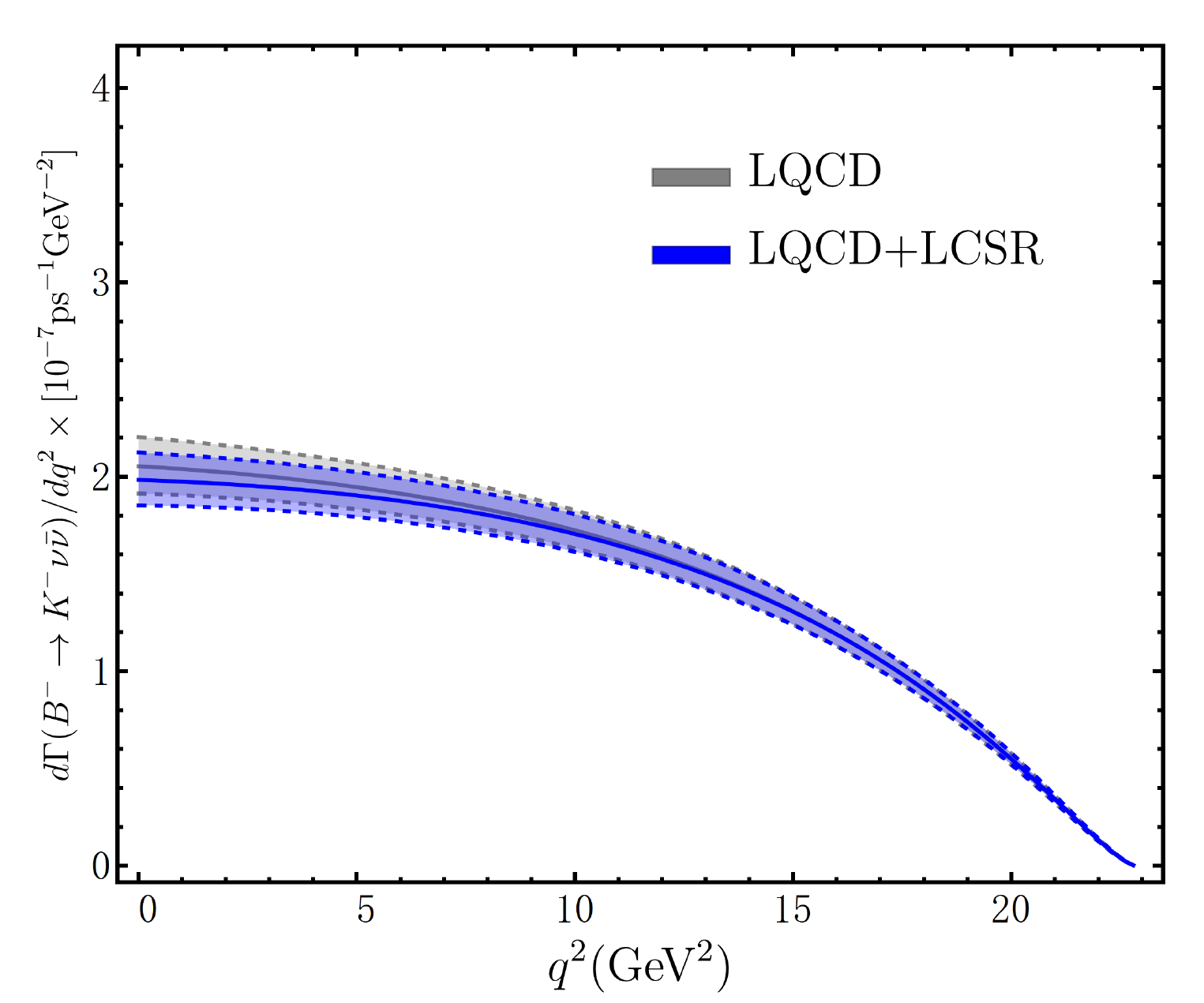}
\vspace*{0.1 cm}
\caption{Theory predictions for the differential decay distributions of
$B^{0} \to K^{0} \nu_{\ell} \bar \nu_{\ell}$ (left panel) and $B^{-} \to K^{-} \nu_{\ell} \bar \nu_{\ell}$
(right panel)  by applying the form factors determined from the two distinct  scenarios of the BCL $z$-series
expansion fitting strategies.}
\label{fig: differentialdecay rates for B to K nu nu}
\end{center}
\end{figure}

We are now in a position to explore  phenomenological implications of  the  newly determined $B \to K$ form factors
on the electroweak  penguin  $B \to K \nu_{\ell} \bar \nu_{\ell}$ decays,
which are expected to be observed with first $10 \, {\rm ab}^{-1}$ of the Belle II data \cite{Halder:2021sgd}
(see the earlier experimental searches  by the BaBar \cite{BaBar:2013npw},
Belle \cite{Belle:2017oht} and Belle II \cite{Belle-II:2021rof} Collaborations).
Importantly, the expected sensitivity of the total branching fraction for
$B \to K \nu_{\ell} \bar \nu_{\ell}$ (summing over neutrino flavours)
with $50 \, {\rm ab}^{-1}$  of integrated luminosity
has been estimated to be at the level of $10 \, \%$,
thus comparable to the current theoretical uncertainties of the  SM predictions \cite{Belle-II:2018jsg}.
It is straightforward to derive the differential decay width formula for the  theoretically
cleanest FCNC $B^{0} \to K^{0} \nu_{\ell} \bar \nu_{\ell}$  decay process of the neutral bottom meson
\cite{Altmannshofer:2009ma,Buras:2014fpa}
\begin{eqnarray}
\frac{d \Gamma(B^{0} \to K^{0} \nu_{\ell} \bar \nu_{\ell} )}
{d q^2 } &=&  {G_F^2 \, \alpha_{\rm em}^2 \over  256 \, \pi^5} \,
{\lambda^{3/2}(m_B^2, m_K^2, q^2) \over m_B^3 \, \sin^4 \theta_W } \,
|V_{tb} \, V_{ts}^{\ast}|^{2} \,
\left [ X_t \left ({m_t^2 \over m_W^2}, {m_H^2 \over m_W^2}, \sin \theta_W, \mu \right )   \right ]^2
\nonumber \\
&& \times \left |f_{BK}^{+}(q^2) \right |^2  \,,
\label{differential decay width of B0 to K0 nu nu}
\end{eqnarray}
where the CKM matrix elements $|V_{tb}|$ and $|V_{ts}^{\ast}|$ can be further evaluated from the four
Wolfenstein parameters collected in Table \ref{table: theory inputs} with the expanded matching relations
at the accuracy of ${\cal O}(\lambda^9)$ \cite{Charles:2004jd}.
The short-distance Wilson coefficient $X_t$ can be expanded perturbatively in terms of the SM gauge couplings
\begin{eqnarray}
X_t = X_t^{(0)} + {\alpha_s \over 4 \pi} \,  X_t^{{\rm QCD}(1)} +
 {\alpha_{\rm em} \over 4 \pi} \,  X_t^{{\rm EW} (1)} + ...  \,,
\end{eqnarray}
where the LO contribution $X_t^{(0)}$ \cite{Inami:1980fz},  the NLO QCD correction $X_t^{{\rm QCD}(1)}$
\cite{Buchalla:1992zm,Buchalla:1998ba,Misiak:1999yg}
and the two-loop electroweak correction $ X_t^{{\rm EW} (1)}$ \cite{Brod:2010hi} are already known analytically.
By contrast, there exists an additional long-distance contribution to the counterpart charged channel
$B^{-} \to K^{-} \nu_{\ell} \bar \nu_{\ell}$ due to the double charged-current interaction
$B^{-} \to \tau (\to K^{-} \nu_{\tau}) \, \bar \nu_{\tau}$ at tree level as originally discussed in \cite{Kamenik:2009kc}.
In the narrow $\tau$-lepton width approximation
($\Gamma_{\tau} \simeq 2.3 \times 10^{-3} \, {\rm eV}$ \cite{ParticleDataGroup:2022pth}),
we can readily derive the tree-level charged-current contribution to the exclusive rare
$B^{-} \to K^{-} \nu_{\tau} \bar \nu_{\tau}$ decay rate
\begin{eqnarray}
\frac{d \Gamma(B^{-} \to K^{-} \nu_{\ell} \bar \nu_{\ell} )}
{d q^2 } \bigg|_{\rm LD} &=&  {G_F^4  \, |V_{ub} V_{us}^{\ast}|^2 \over 64 \, \pi^2 \, m_{B^{-}}^3} \,
\left | f_{B^{-}} \, f_{K^{-}}  \right |^2  \,
\frac{m_{\tau}^3}{\Gamma_{\tau}} \,
\left [ ( m_{B^{-}}^2 - m_{\tau}^2 ) \, ( m_{\tau}^2  - m_{K^{-}}^2 ) -   m_{\tau}^2 \, q^2  \right ],
\nonumber \\
\label{tree contribution to B to K nu nu}
\end{eqnarray}
where  the invariant mass distribution of the two invisible particles
satisfies the  constraint \cite{ParticleDataGroup:2022pth}
\begin{eqnarray}
 0 \leq q^2 \leq  \frac{( m_{B^{-}}^2 - m_{\tau}^2 ) \, ( m_{\tau}^2  - m_{K^{-}}^2 )}{ m_{\tau}^2} \,.
\end{eqnarray}
At the first sight, this  new mechanism will be suppressed by an extra factor of $G_F^2$
in comparison with the customary penguin contribution presented in (\ref{differential decay width of B0 to K0 nu nu}).
However,  the very appearance of $1/\Gamma_{\tau}$ on the right-handed side of (\ref{tree contribution to B to K nu nu}),
due to the on-shell $\tau$-lepton enhancement, will be counted as ${\cal O}(G_F^{-2})$ parametrically,
thus compensating the observed suppression factor \cite{Du:2015tda}.
Moreover, the interference effect  between the tree and penguin amplitudes
turns out to be numerically negligible (estimated to be at the order of $10^{-11}$ \cite{Kamenik:2009kc})
on account of the extremely small  $\tau$-lepton width.
We display in Figure \ref{fig: differentialdecay rates for B to K nu nu}
the yielding results for the differential decay distributions of
$B^{0} \to K^{0} \nu_{\ell} \bar \nu_{\ell}$ and $B^{-} \to K^{-} \nu_{\ell} \bar \nu_{\ell}$
by employing  the form factors determined from  fitting the BCL $z$-series parameterizations
with two  distinct scenarios of the input data points:
I) only synthetic lattice data points,
II) synthetic lattice data points $\oplus$ LCSR results.

In order to confront our numerical predictions  with the anticipated measurements from the high-luminosity Belle II experiment,
we introduce  the following three $q^2$-binned observables for the semileptonic $B \to K \nu_{\ell} \bar \nu_{\ell}$ decays
\cite{Lu:2018cfc}
\begin{eqnarray}
&& \Delta {\cal BR}^{B^{0} \to K^{0} \nu_{\ell} \bar \nu_{\ell}}(q_1^2, \,  q_2^2)
= \tau_{B^{0}} \, \int_{q_1^2}^{q_2^2} \, d q^2 \,
\frac{d \Gamma(B^{0} \to K^{0} \nu_{\ell} \bar \nu_{\ell} )} {d q^2 } \,,
\nonumber \\
&& \Delta {\cal BR}^{B^{-} \to K^{-} \nu_{\ell} \bar \nu_{\ell}}(q_1^2, \,  q_2^2)
= \tau_{B^{-}} \, \int_{q_1^2}^{q_2^2} \, d q^2 \,
\frac{d \Gamma(B^{-} \to K^{-} \nu_{\ell} \bar \nu_{\ell} )} {d q^2 } \,,
\nonumber \\
&& {\cal R}_{K \pi}(q_1^2, \,  q_2^2) =  \left [ \int_{q_1^2}^{q_2^2} \, d q^2 \,
\frac{d \Gamma(B^{0} \to K^{0} \nu_{\ell} \bar \nu_{\ell} )} {d q^2 } \right ] :
\left [ \int_{q_1^2}^{q_2^2} \, d q^2 \,
\frac{d \Gamma(B^{0} \to \pi^{-} \bar \nu_{\mu} \nu_{\mu} )} {d q^2 } \right ]  \,,
\label{definitions: observables for B to K nu nu}
\end{eqnarray}
where the ratio  of partially integrated differential branching 
fractions  ${\cal R}_{K \pi}$  is expected to  suffer from the lower hadronic uncertainties
due to the  correlations between the exclusive $B \to \pi$ and $B \to K$ form factors.
We summarize our final  predictions for these three quantities with the choice of the $q^2$ intervals
following \cite{Belle-II:2018jsg} in Table \ref{table: three  binned distributions of B to K nu nu},
where we further display the previous theoretical determinations with the lattice simulation and LCSR methods
for convenience.

\begin{table}
\centering
\renewcommand{\arraystretch}{2.0}
\resizebox{\columnwidth}{!}{
\begin{tabular}{|c|c|c|c|}
  \hline
  \hline
  $[q_1^2, \, q_2^2]$  \,\, (${\rm in \,\, GeV}^2$)
  & $10^6 \times \Delta {\cal BR}^{B^{0} \to K^{0} \nu_{\ell} \bar \nu_{\ell}}(q_1^2, q_2^2)$
  & $10^6 \times \Delta {\cal BR}^{B^{-} \to K^{-} \nu_{\ell} \bar \nu_{\ell}} (q_1^2, q_2^2)$
   & $10^2 \times {\cal R}_{K \pi}(q_1^2, q_2^2)$  \\
  \hline
  $[0.0, 1.0]$ & $0.253^{+0.020}_{-0.019}$  & $0.324^{+0.023}_{-0.021}$  &  $3.949 \pm 0.319$
  \\
  $[1.0, 2.5]$ & $0.381^{+0.028}_{-0.027}$  & $0.482^{+0.033}_{-0.030}$  &  $3.932 \pm 0.303$
  \\
  $[2.5, 4.0]$ & $0.380^{+0.027}_{-0.025}$  & $0.476^{+0.031}_{-0.029}$  &  $3.904 \pm 0.289$
  \\
  $[4.0, 6.0]$ & $0.502^{+0.034}_{-0.032}$  & $0.622^{+0.039}_{-0.035}$  &  $3.859 \pm  0.278$
  \\
  $[6.0, 8.0]$ & $0.492^{+0.032}_{-0.030}$  & $0.602^{+0.037}_{-0.034}$  &  $3.788 \pm  0.270$
  \\
  $[8.0, 12.0]$ & $0.924^{+0.058}_{-0.053}$  & $1.112^{+0.066}_{-0.060}$  &  $3.626 \pm  0.262$
  \\
  $[12.0, 16.0]$ & $0.776^{+0.047}_{-0.043}$  & $0.916^{+0.053}_{-0.048}$  &  $3.245 \pm  0.238$
  \\
  $\left [16.0, \, (m_B-m_{K})^2 \right ]$ & $0.607^{+0.036}_{-0.032}$  & $0.705^{+0.040}_{-0.036}$  &  $1.918 \pm 0.129$
  \\
  \hline
  \hline
 \multirow{7}{*}{$\left [0.0, \, (m_B-m_{K})^2 \right ]$}  & $6.02^{+1.68}_{-1.76}$  \cite{Lu:2018cfc}
  &  $5.10 \pm 0.80$ \cite{Altmannshofer:2009ma}  &   $-$
 \\
               & $4.01 \pm 0.49$  \cite{Du:2015tda}  &  $3.98 \pm 0.47$ \cite{Buras:2014fpa}    &  $-$
 \\
               & $4.67 \pm 0.35$ \cite{Parrott:2022dnu} & $4.94 \pm 0.52$ \cite{Du:2015tda} &  $-$
 \\
               &  $4.1^{+1.3}_{-1.0}$ \cite{Wang:2012ab} & $5.67 \pm 0.38$ \cite{Parrott:2022dnu}   &   $-$
 \\
               & $4.4 \pm 1.5$ \cite{Bartsch:2009qp}  &  $4.53 \pm 0.64$ \cite{Buras:2021nns}  &   $-$
 \\
               & $-$  & $4.65 \pm 0.62$ \cite{Buras:2022wpw}   &  $-$
 \\
               & $4.315^{+0.271}_{-0.248}$  (this work)  & $5.239^{+0.311}_{-0.281}$  (this work)  &  $3.240 \pm 0.211$  (this work)
 \\
  \hline
  \hline
\end{tabular}
}
\vspace{0.3 cm}
\caption{Theory predictions for the three partially integrated differential observables
$\Delta {\cal BR}^{B^{0} \to K^{0} \nu_{\ell} \bar \nu_{\ell}}$,
$\Delta {\cal BR}^{B^{-} \to K^{-} \nu_{\ell} \bar \nu_{\ell}}$,
and ${\cal R}_{K \pi}$ (see (\ref{definitions: observables for B to K nu nu}) for their explicit definitions)
obtained from the combined BCL $z$-series expansion fitting of
the exclusive $B \to \pi, K$ form factors against
the synthetic lattice data points and the newly obtained bottom-meson LCSR results. }
\label{table: three  binned distributions of B to K nu nu}
\end{table}

\section{Conclusions}
\label{section: summary}

In the current paper we have carried out the improved computations of the semileptonic
$B_{d, s}  \to \pi, K$  decay form factors at large hadronic recoil,
which evidently belong to the most important hadronic quantities in heavy quark physics,
by employing the method of light-cone sum rules (LCSR) in soft-collinear effective theory (SCET)
with both the leading-twist and higher-twist bottom-meson distribution amplitudes.
In particular, we have computed for the first time the non-vanishing spectator-quark
mass corrections to these form factors at NLO in the strong coupling constant,
which appeared to preserve the approximated large-recoil symmetry relations for
the heavy-to-light transition form factors
and turned out to escape from an extra suppression of the powers of  $\Lambda_{\rm QCD}/m_b$
in heavy quark expansion.
Our explicit sum rules for these  spectator-quark mass corrections further implied that
evaluating such SU(3)-flavour symmetry breaking effects directly with the perturbative
factorization technique would result in the soft-collinear convolution integrals
with the notorious rapidity singularities.
Moreover, we have accomplished the complete NLL resummation for  the parametrically enhanced logarithms
of $m_b / \Lambda_{\rm QCD}$ entering in the factorized expressions of
the leading-power contributions to the considered  vacuum-to-bottom-meson correlation functions
displayed in (\ref{correlation_function}) by taking advantage of the standard renormalization-group
formalism in momentum space.
We then proceeded to explore the four distinct classes of the NLP contributions to
the exclusive $B_{d, s}  \to \pi, K$  form factors with the same LCSR technique at tree level:
I) the higher-order terms from  heavy quark expansion of the hard-collinear quark propagator,
II) the subleading power corrections from the effective matrix element of the ${\rm SCET_I}$ weak current
$(\bar \xi_{\rm hc} \, W_{\rm hc}) \, \gamma_{\mu} \,
\left [  i \,  {\slashed D}_{\top} /  \left (  2 \, m_b \right ) \right ] \,   h_v$,
III) the higher-twist corrections from the two-particle and three-particle
heavy quark effective theory (HQET) distribution amplitudes,
IV) the four-particle twist-five and twist-six contributions in the factorization approximation.
We have extensively used the nontrivial operator identities between the two-body and three-body light-cone HQET operators
due to the classical equations of motion in our constructions of the NLP sum rules.
Interestingly,  we observed that only the particular class-I NLP contribution
from the expanded hard-collinear propagator can generate the large-recoil symmetry violation effect
between the vector and scalar form factors,
while both the  class-I and class-II NLP corrections can yield the symmetry breaking effects
between the vector and tensor form factors at large hadronic recoil.

Having at our disposal the updated  SCET sum rules for the exclusive heavy-to-light
bottom-meson decay form factors, we turned to investigate the numerical implications
of the NLL resummation improved leading-power contributions and the newly obtained
four classes of the NLP corrections at tree level on the theory predictions for
the  semileptonic  $B_{d, s}  \to \pi, K$  form factors of our interest,
by employing the general three-parameter ansatz for the necessary HQET distribution amplitudes.
It has been explicitly shown that the most prominent subleading power contribution
arises from the two-particle twist-five off-light-cone  correction,
which can reduce the corresponding leading-power LCSR predictions  in the kinematic region
$0 \leq  q^2 \leq 8 \, {\rm GeV^2}$ by an amount of  $(25-30)\, \%$ numerically.
On the contrary, the yielding impacts from the four-particle higher-twist
bottom-meson distribution amplitudes have been demonstrated to be numerically  insignificant
in the factorization approximation owing to the smallness of the normalization constant
$|\langle \bar q q \rangle : (\lambda_B \, s_0)| \simeq 10 \, \%$
in the tree-level sum rules (\ref{LCSR for the 4P corrections}).
In addition, the higher-order QCD corrections to the short-distance
matching coefficients appeared  in  the leading-power SCET sum rules
can bring about consistently ${\cal O} (30 \, \%)$ reductions
of the counterpart leading-order LCSR predictions.
It remains important to remark that our numerical results
for the semileptonic bottom-meson decay form factors indicate the following
hierarchy relations
$f_{B K}^{+ \, (T)}(0) > f_{B_s K}^{+ \, (T)}(0) >  f_{B \pi}^{+ \, (T)}(0)$
in the maximal recoil limit.
Remarkably, we predicted very sizeable SU(3)-flavour symmetry violating  effects
between the exclusive $B \to \pi$  and $B \to K$ form factors
based upon the established LCSR with the HQET distribution amplitudes:
numerically ${\cal O} (30 \, \%)$ for the two form-factor ratios
${\cal R}_{\rm SU(3)}^{+, \, 0}$   and  ${\cal O} (40 \, \%)$
for the particular ratio ${\cal R}_{\rm SU(3)}^{T}$.

Subsequently,  we extrapolated the bottom-meson LCSR computations for
the exclusive $B_{d, s}  \to \pi, K$   form factors 
towards the large momentum transfer $q^2$ with the aid of
the Bourrely-Caprini-Lellouch (BCL) $z$-series parameterizations for these form factors.
It is interesting to note that including the newly obtained LCSR predictions at small momentum transfer
in our numerical fitting procedure turned out to be highly beneficial for
pinning down the theory uncertainties of all the three  $B \to \pi$ form factors
in the kinematic regime $0.10 \leq z(q^2) \leq 0.31$  significantly,
due to the yet non-negligible errors  of the lattice QCD results from the RBC/UKQCD Collaboration \cite{Flynn:2015mha}
(numerically at the level of  $(8.4-14.3) \, \%$  for the vector form factor
and  $(7.6-13.6) \, \%$ for the scalar form factor).
Furthermore, we have confronted our combined BCL fit results of the $B \to \pi$ form factors
with the theoretical expectations from the particular low-recoil symmetry relations
(\ref{soft-pion relation of fplus and fzero}) and  (\ref{soft-pion relation of fplus and fT}),
on account of the combination of the heavy quark spin symmetry and the so-called soft-pion approximation,
in Figures (\ref{fig: soft-pion relation of fplus and fzero}) and
(\ref{fig: soft-pion relation of fT and fplus}) manifestly.
Additionally, the available high-precision lattice QCD results of
the semileptonic $B \to K$ form factors from both the FNAL/MILC Collaboration  \cite{Bailey:2015dka}
and the HPQCD Collaboration \cite{Parrott:2022rgu} appeared to be in tension with
the strongly correlated predictions from the improved bottom-meson LCSR determinations,
when carrying out the numerical interpolations between the lattice simulation and LCSR results
with the standard BCL $z$-series expansions.
With regard to the flavour-changing charged-current $B_s \to K \ell \bar \nu_{\ell}$ form factors,
the yielding BCL fit results for the three coefficients $b_{0, 1, 2}^T$
in Table \ref{BCL fit for the Bs to K form factors} became more uncertain
in comparison with the counterpart results for the $z$-expansion parameters
of the tensor $B \to \pi$ form factor, due to the very absence  of
the lattice data points for $f_{B_s K}^{T}(q^2)$ in the lower recoil region.
Importantly, we indeed benefited from the combined BCL  $z$-expansion fit
to both the LCSR and lattice simulation results by  improving further
the theory  accuracy of the vector and scalar form factors $f_{B_s K}^{+, \, 0}(q^2)$,
numerically at the level of ${\cal O}(20 \, \%)$,
when compared with the counterpart BCL fitting procedure
with the ``{\it only lattice QCD}" data points.
It has been verified that  our BCL  $z$-fit predictions for  the scalar form-factor ratios
of both the exclusive  $B \to K$ and $B_s \to K$ form factors at the zero-recoil limit
differ from the expected Isgur-Wise relations due to the combined heavy quark and chiral symmetries enormously,
thus supporting the previous lattice simulation results  from the  FNAL/MILC Collaboration \cite{Bailey:2015dka}.

Performing the simultaneous BCL expansion fit to the SCET sum rules, lattice QCD and experimental data points
enabled us to extract the desired CKM  matrix element $|V_{ub}|$
from the ``golden"  exclusive process $B \to \pi \ell \bar \nu_{\ell}$ with the two distinct truncations
\begin{eqnarray}
\left |V_{ub} \right |_{B \to \pi \ell \bar \nu_{\ell}} = (3.76 \pm 0.13) \times \, 10^{-3} \,,
\qquad
({\rm BCL \,\, fit \,\, with}  \,\, N=3)
\nonumber \\
\left |V_{ub} \right |_{B \to \pi \ell \bar \nu_{\ell}} = (3.72 \pm 0.14) \times \, 10^{-3} \,.
\qquad
({\rm BCL \,\, fit \,\, with}  \,\, N=4)
\label{Vub value: N=3 and 4}
\end{eqnarray}
We are therefore led to conclude that  truncating the $z$-series expansions at the order $N=3$
in the numerical fit procedure can indeed be justified  for the practical purpose
due to the apparent stability against the truncation order.
Our numerical predictions for the  two  particular lepton-flavour-universality (LFU) ratios
for the exclusive $B_{(s)}  \to \pi (K) \ell \bar \nu_{\ell}$ decays
have been collected in Figure \ref{fig: differential Rpi and RK rarios}
and Table \ref{table: physical observables for B to pi l nu and Bs to Kl nu},
and we quote here the obtained results for such gold-plated quantities
\begin{eqnarray}
{\cal R}_{\pi} = 0.720 \pm 0.027  \,,
\qquad
{\cal R}_{K} = 0.700 \pm 0.016  \,,
\qquad
({\rm BCL \,\, fit \,\, with}  \,\, N=3)
\nonumber \\
{\cal R}_{\pi} = 0.746 \pm 0.039  \,,
\qquad
{\cal R}_{K} = 0.680 \pm 0.019  \,.
\qquad
({\rm BCL \,\, fit \,\, with}  \,\, N=4)
\label{Rpi and RK value: N=3 and 4}
\end{eqnarray}
The yielding results for the three different classes of the angular observables sensitive to
the potential Beyond the Standard Model (BSM) physics
${\cal A}_{\rm FB}^{B_{(s)}  \to \pi (K) \ell \bar \nu_{\ell}}$,
${\cal F}_{\rm H}^{B_{(s)}  \to \pi (K) \ell \bar \nu_{\ell}}$ and
${\cal A}_{\lambda_{\ell}}^{B_{(s)}  \to \pi (K) \ell \bar \nu_{\ell}}$
have been explicitly  displayed in Figure
\ref{fig: three angular observables for B to pi l nu and Bs to K l nu}
and Table \ref{table: angular observables for B to pi l nu and Bs to K l nu},
where the previous theory determinations from the lattice simulation and LCSR approaches
were further shown for a comparison.
We finally presented our numerical predictions for the three partially integrated differential observables
$\Delta {\cal BR}^{B^{0} \to K^{0} \nu_{\ell} \bar \nu_{\ell}}$,
$\Delta {\cal BR}^{B^{-} \to K^{-} \nu_{\ell} \bar \nu_{\ell}}$,
and ${\cal R}_{K \pi}$ (see (\ref{definitions: observables for B to K nu nu}) for their explicit definitions)
for the exclusive rare $B \to K \nu_{\ell} \bar \nu_{\ell}$ decays
in Table \ref{table: three  binned distributions of B to K nu nu}.

Future developments of the theory predictions for the heavy-to-light bottom-meson decays
beyond our work can be pursued forward in a variety of directions.
First, it would be interesting to perform the full two-loop QCD computations
of the semileptonic  $B_{d, s}  \to \pi, K$  decay form factors  at large hadronic recoil,
by employing  the SCET sum rules framework, in order to reduce further the current perturbative uncertainties
displayed in Figures \ref{fig: distinct pieces of B to pi and B to K form factors in LCSR}
and  \ref{fig: distinct pieces of Bs to K form factors in LCSR}.
Actually, the desired hard matching coefficients in the ${\rm SCET_I}$ representations
of the flavour-changing QCD currents $\bar q \, \Gamma_i \, b$ have been already
determined at the ${\cal O}(\alpha_s^2)$ accuracy.
The only missing ingredients for constructing the two-loop factorization formulae of
the very vacuum-to-bottom-meson correlation functions (\ref{correlation_function})
consist in the yet unknown short-distance Wilson coefficients
in the second-step ${\rm SCET_I} \to {\rm SCET_{II}}$ matching procedure.
Second, investigating the subleading-power contributions to the exclusive $B_{d, s}  \to \pi, K$   form factors
systematically with the effective field theory  techniques
and then evaluating the resulting (non)-local soft-collinear matrix elements
with the appropriate nonperturbative QCD methods will be evidently in high demand
from both the conceptual and phenomenological perspectives.
As a matter of fact, it would be of utmost importance to achieve the analytical regularization
of the unwanted end-point divergences entering  in the factorized expressions of
the  ${\rm SCET_{I}}$ matrix elements for the two-body $A$-type currents.
Third, advancing our knowledge of the poorly constrained bottom-meson distribution amplitudes
in HQET with model-independent techniques (for instance, along the line of \cite{Wang:2019msf})
will be indispensable for enhancing further our predictive power
of the exclusive bottom-meson decay matrix elements in the theory frameworks
of both QCD factorization and light-cone sum rules.
In particular the yielding noticeable uncertainties due to the two-particle $B_{d, s}$-meson distribution amplitudes
has already become the major stumbling block for accomplishing precision calculations of a wide range of
the interesting physical observables accessible at the LHCb and Belle II experiments.
Fourth, the established strategies of evaluating the subleading-power  $B_{d, s}  \to \pi, K$  matrix elements
can be further extended to compute the NLP corrections to the analogous heavy-to-light
$B \to \rho, \omega, K^{\ast}$ decay form factors  (thus going well beyond our previous work \cite{Gao:2019lta})
and to explore the delicate strong interaction mechanisms dictating
the Cabibbo favored semileptonic $B_{(s)} \to D_{(s)}^{\ast} \ell \bar \nu_{\ell}$ decay processes.

\section*{Acknowledgements}

We are grateful to Zi-Hao Mi for collaboration in the early stages of this project
and to Svenja Granderath and Yan-Bing Wei for illuminating discussions.
The research of Y.L.S. is supported by the  National Natural Science Foundation of China  with
Grant No. 12175218 and the Natural Science Foundation of Shandong with Grant No.  ZR2020MA093.
C.W is supported in part by the National Natural Science Foundation of China
with Grant No. 12105112 and  the Natural Science Foundation of
Jiangsu Education Committee with Grant No. 21KJB140027.
Y.M.W acknowledges support from
the  National Natural Science Foundation of China  with
Grant No. 11735010 and 12075125, and  the Natural Science Foundation of Tianjin
with Grant No. 19JCJQJC61100.


\end{document}